 \documentclass[manuscript]{emulateapj}

 \usepackage{apjfonts}

\slugcomment{to appear in the Astrophysical Journal, Supplement Series}

\shorttitle{Color-Magnitude Evolution of Classical Novae}
\shortauthors{Hachisu \& Kato}

\begin{document}

\title{The $UBV$ Color Evolution of Classical Novae. II. Color-Magnitude
Diagram}


\author{Izumi Hachisu}
\affil{Department of Earth Science and Astronomy, 
College of Arts and Sciences, The University of Tokyo,
3-8-1 Komaba, Meguro-ku, Tokyo 153-8902, Japan} 
\email{hachisu@ea.c.u-tokyo.ac.jp}

\and

\author{Mariko Kato}
\affil{Department of Astronomy, Keio University, 
Hiyoshi, Kouhoku-ku, Yokohama 223-8521, Japan} 
%%\email{mariko@educ.cc.keio.ac.jp}


%\altaffiltext{1}{Visiting Astronomer, Cerro Tololo Inter-American Observatory.
%CTIO is operated by AURA, Inc.\ under contract to the National Science
%\altaffiltext{2}{Society of Fellows, Harvard University.}
%\altaffiltext{3}{present address: Center for Astrophysics,
%    60 Garden Street, Cambridge, MA 02138}
%\altaffiltext{4}{Visiting Programmer, Space Telescope Science Institute}
%\altaffiltext{5}{Patron, Alonso's Bar and Grill}


\begin{abstract}
We have examined the outburst tracks of 40 novae in the color-magnitude
diagram (intrinsic $B-V$ color versus absolute $V$ magnitude).
After reaching the optical maximum, each nova generally evolves toward blue
from the upper-right to the lower-left and then turns back toward the right.
The 40 tracks are categorized into one of six templates:
very fast nova V1500~Cyg; fast novae V1668~Cyg, V1974~Cyg, 
and LV~Vul; moderately fast nova FH~Ser; and very slow nova PU~Vul.
These templates are located from the left (blue) to the right (red)
in this order, depending on the envelope mass and nova speed class. 
A bluer nova has a less massive envelope and faster nova
speed class.  In novae with multiple peaks, the track of the first
decay is more red than that of the second (or
third) decay, because a large part of the envelope mass had already
been ejected during the first peak.  
Thus, our newly obtained tracks in the color-magnitude diagram
provide useful information to understand the physics of classical novae.
We also found that the absolute magnitude at the beginning of
the nebular phase is almost similar among various novae.  
We are able to determine the absolute magnitude (or distance modulus)
by fitting the track of a target nova to the same classification of a
nova with a known distance.  This method for determining
nova distance has been applied to some recurrent novae and
their distances have been recalculated.
\end{abstract}




\section{Introduction}
A classical nova is a thermonuclear runaway event triggered by
unstable hydrogen-burning on a mass-accreting white dwarf (WD)
in a binary.  After an outburst begins, the hydrogen-rich envelope
expands to red-giant size and ejects mass.
The nova brightens up in the optical.
After maximum expansion of the pseudo-photosphere,
it begins to shrink owing to mass ejection.
The mass ejection process is well described
by the optically thick wind theory \citep[e.g.,][]{kat94h, hac06kb}. 
The optical brightness decays and
subsequently, the ultra-violet (UV) emission dominates the spectrum.
Finally, the super-soft X-ray emission increases. 
The nova outburst ends when the hydrogen shell burning is extinguished.

Several groups proposed various time-scaling laws 
to identify a common pattern among the nova light curves 
\citep[see, e.g.,][for a summary]{hac08kc}.
\citet{hac06kb} found a similarity in the optical and near-infrared (NIR)
light curves and calculated time-normalized light curves in terms of
free-free emission,
which are independent of the WD mass, chemical composition
of the ejecta, and wavelength.  
They called it ``the universal decline law.''
This decline law was examined 
in a number of classical novae 
\citep{hac07k,hac09ka,hac10k,hac15k,hac16k,hac06b,hac07kl,
hac08kc,kat09hc,kat12h}.
Based on the universal decline law, 
the maximum-magnitude versus rate-of-decline (MMRD) law was 
theoretically derived for classical novae \citep{hac10k, hac15k, hac16k}.
To summarize, nova optical and NIR light curves can be explained
theoretically in terms of free-free emission based on the optically
thick winds \citep{kat94h}.

The evolution of colors was also studied by many researchers 
\citep[see, e.g.,][]{due79, van87, mir88}.
If the optical fluxes in the $UBV$ bands are dominated by free-free
emission as derived by \citet{hac06kb},
its color is simply estimated to be $(B-V)_0 = +0.13$
and $(U-B)_0 = -0.82$ for optically thin free-free
($F_\nu \propto \nu^0$) emission, or to be $(B-V)_0 = -0.03$ 
and $(U-B)_0 = -0.97$ for optically thick free-free
($F_\nu \propto \nu^{2/3}$) emission \citep{wri75}.
Here, $F_\nu$ is the flux at the frequency $\nu$, $(B-V)_0$ 
and $(U-B)_0$ are the intrinsic colors of $B-V$ and $U-B$, respectively.  
However, the nova $B-V$ color does not always
stay long at these points but evolves further toward blue.
\citet[][hereafter, Paper I]{hac14k} examined the color-color evolutions
for a number of novae and identified a general course of classical
nova outbursts in the $B-V$ versus $U-B$ color-color diagram.
Matching the observed track of a target nova with this general course,
they obtained the color excess of the nova.  This is a new way to 
determine the color excess of a nova.
A part of the extinctions thus obtained
are summarized in Table \ref{extinction_various_novae}.

In the present paper, we examine if there is also a general track
of novae in the color-magnitude diagram.
To obtain the intrinsic colors and absolute magnitudes of novae,
we need to know the color excess (or extinction) for each nova.
For this purpose, we used the results obtained in Paper I.  
The present paper is organized as follows.
In Section \ref{color_magnitude_diagram}, we examine the color-magnitude
evolutions of ten well-observed novae, i.e., V1668~Cyg, LV~Vul, FH~Ser,
PW~Vul, V1500~Cyg, V1974~Cyg, PU~Vul, V723~Cas, HR~Del, and V5558~Sgr.
Using these data, we deduce six templates of outburst tracks
in the color-magnitude diagram. 
In Section \ref{application_to_novae},  we apply these templates
to 30 novae, and try to identify general trends
of novae.  Table \ref{extinction_various_novae}
lists the object name, outburst year,
color excess, distance modulus in the $V$ band, and distance of each nova,
including our target novae.
Discussion and conclusions follow in Sections
\ref{discussion} and \ref{conclusions}, respectively.

%Table 1
%\placetable{extinction_various_novae}

\begin{deluxetable}{llllll}
\tabletypesize{\scriptsize}
\tablecaption{Extinctions, distance moduli, and distances for selected novae
\label{extinction_various_novae}}
\tablewidth{0pt}
\tablehead{
\colhead{Object} & \colhead{Outburst} & \colhead{$E(B-V)$} 
& \colhead{$(m-M)_V$} & 
\colhead{$d$} & \colhead{Reference\tablenotemark{a}} \\
  & year &  &  &  (kpc) & 
} 
\startdata
OS~And & 1986 & 0.15 & 14.8 & 7.3 & 1 \\
CI~Aql & 2000 & 1.0 & 15.7 & 3.3 & 4 \\
V603~Aql & 1918 & 0.07 & 7.2 & 0.25 & 6 \\
V1370~Aql & 1982 & 0.35 & 16.5 & 12.0 & 4 \\
V1419~Aql & 1993 & 0.50 & 14.6 & 4.1 & 1 \\
V1493~Aql & 1999\#1 & 1.15 & 17.7 & 6.7 & 4 \\
V1494~Aql & 1999\#2 & 0.50 & 13.1 & 2.0 & 4 \\
V705~Cas & 1993 & 0.45 & 13.4 & 2.5 & 2 \\
V723~Cas & 1995 & 0.35 & 14.0 & 3.85 & 2 \\
V1065~Cen & 2007 & 0.45 & 15.3 & 6.0 & 4 \\
IV~Cep & 1971 & 0.65 & 14.7 & 3.4 & 1,4 \\
V693~CrA & 1981 & 0.05 & 14.4 & 7.1 & 3 \\
V1500~Cyg & 1975 & 0.45 & 12.3 & 1.5 & 4 \\
V1668~Cyg & 1978 & 0.30 & 14.6 & 5.4 & 3 \\
V1974~Cyg & 1992 & 0.30 & 12.2 & 1.8 & 3 \\
V2274~Cyg & 2001\#1 & 1.35 & 18.7 & 8.0 & 4 \\
V2275~Cyg & 2001\#2 & 1.05 & 16.3 & 4.1 & 4 \\
V2362~Cyg & 2006 & 0.60 & 15.9 & 6.4 & 4 \\
V2467~Cyg & 2007 & 1.40 & 16.2 & 2.4 & 4 \\
V2468~Cyg & 2008 & 0.75 & 15.6 & 4.5 & 4 \\
V2491~Cyg & 2008 & 0.23 & 16.5 & 14.0 & 4 \\
HR~Del & 1967 & 0.12 & 10.4 & 1.0 & 4 \\
DQ~Her & 1934 & 0.10 & 8.2 & 0.39 & 6 \\
V446~Her & 1960 & 0.40 & 11.7 & 1.23 & 1 \\
V533~Her & 1963 & 0.038 & 10.8 & 1.36 & 4 \\
GQ~Mus & 1983 & 0.45 & 15.7 & 7.3 & 2 \\
RS~Oph  & 1958 & 0.65 & 12.8 & 1.4 & 1 \\
V2615~Oph  & 2007 & 0.95 & 16.5 & 5.1 & 4 \\
GK~Per  & 1901 & 0.30 & 9.3 & 0.48 & 6 \\
RR~Pic  & 1925 & 0.04 & 8.7 & 0.52 & 6 \\
V351~Pup  & 1991 & 0.45 & 15.1 & 5.5 & 3 \\
T~Pyx  & 1966 & 0.25 & 14.2 & 4.8 & 1,7 \\
U~Sco & 2010 & 0.35 & 16.0 & 9.6 & 4 \\
V745~Sco & 2014 & 0.70 & 16.6 & 7.8 & 4 \\
V1280~Sco & 2007\#1 & 0.35 & 11.0 & 0.96 & 4 \\
V443~Sct & 1989 & 0.40 & 15.5 & 7.1 & 1 \\
V475~Sct & 2003 & 0.55 & 15.4 & 5.5 & 4 \\
V496~Sct & 2009 & 0.50 & 14.4 & 3.7 & 4 \\
FH~Ser & 1970 & 0.60 & 11.7 & 0.93 & 1 \\
V5114~Sgr & 2004 & 0.45 & 16.5 & 10.5 & 1,4 \\
V5558~Sgr & 2007 & 0.70 & 13.9 & 2.2 & 1 \\
V382~Vel & 1999 & 0.15 & 11.5 & 1.6 & 3 \\
LV~Vul & 1968\#1 & 0.60 & 11.9 & 1.0 & 1 \\
NQ~Vul & 1976 & 0.95 & 13.6 & 1.26 & 4 \\
PU~Vul & 1979 & 0.30 & 14.3 & 4.7 & 5 \\
PW~Vul & 1984\#1 & 0.55 & 13.0 & 1.8 & 2 \\
QU~Vul & 1984\#2 & 0.55 & 13.6 & 2.4 & 3 \\
QV~Vul & 1987 & 0.60 & 14.0 & 2.7 & 1 \\
V458~Vul & 2007\#1 & 0.50 & 15.3 & 5.6 & 4  
\enddata
\tablenotetext{a}{
1 - \citet{hac14k},
2 - \citet{hac15k},
3 - \citet{hac16k},
4 - present work,
5 - \citet{kat12mh}, 
6 - \citet{har13}, 
7 - \citet{sok13}. 
} 
\end{deluxetable}

\begin{figure}
%%\epsscale{0.75}
%%\epsscale{0.8}
%\epsscale{1.0}
\epsscale{1.15}
\plotone{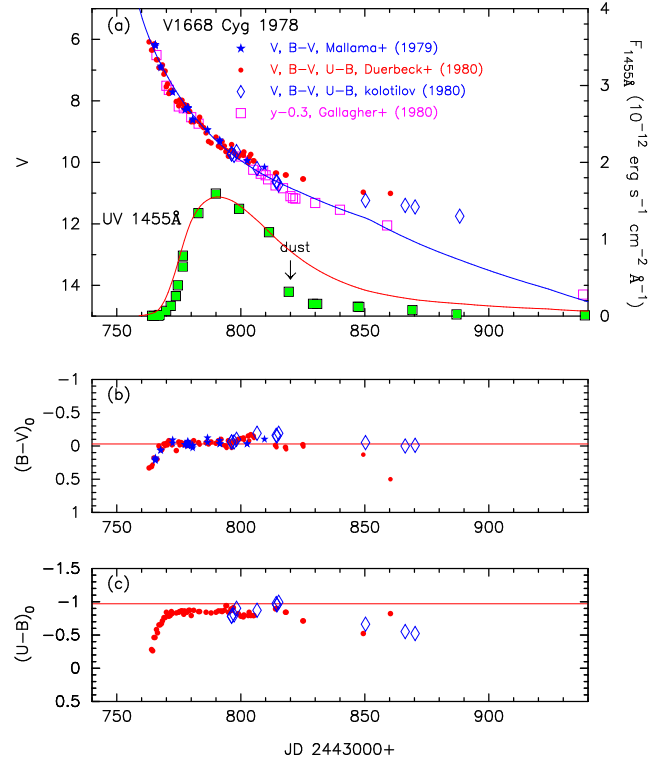}
%\plotone{v1668_cyg_v_bv_ub_color_curve.epsi}
%\plotfiddle{evolution1.ps}{5.0cm}{270}{0.4}{0.4}{-170}{220}
\caption{
(a) The $V$, $y$, and UV~1455 \AA\  light curves,
(b) $(B-V)_0$, and (c) $(U-B)_0$ color evolutions of V1668~Cyg.
The color data are de-reddened by Equations (\ref{dereddening_eq_bv}) 
and (\ref{dereddening_eq_ub}) with $E(B-V)=0.30$.
See the main text for the sources of the observational data.
In panel (a), the thin solid blue line denotes the model $V$ light curve
of a $0.98~M_\sun$ WD with the chemical composition of CO nova 3,
while the thin solid red line represents the model
UV~1455 \AA\  light curve of the same WD model, both of which are
taken from \citet{hac16k}.
In panel (b), the $B-V$ data of \citet{kol80}
are systematically shifted toward blue by 0.1 mag.  The horizontal solid red 
line denotes $(B-V)_0=-0.03$, which is the $B-V$ color of optically thick 
free-free emission.  In panel (c), the $U-B$ data of \citet{kol80} are
also systematically shifted toward blue by 0.1 mag.  The horizontal solid red
line denotes $(U-B)_0=-0.97$, which is the $U-B$ color of optically thick 
free-free emission.   
\label{v1668_cyg_v_bv_ub_color_curve}}
\end{figure}

%Fig.2
%\placefigure{color_color_v1668_cyg_distance_reddening_x45z02c15o20}

\begin{figure}
%\epsscale{0.40}
%%\epsscale{0.8}
\epsscale{1.0}
%\epsscale{1.15}
\plotone{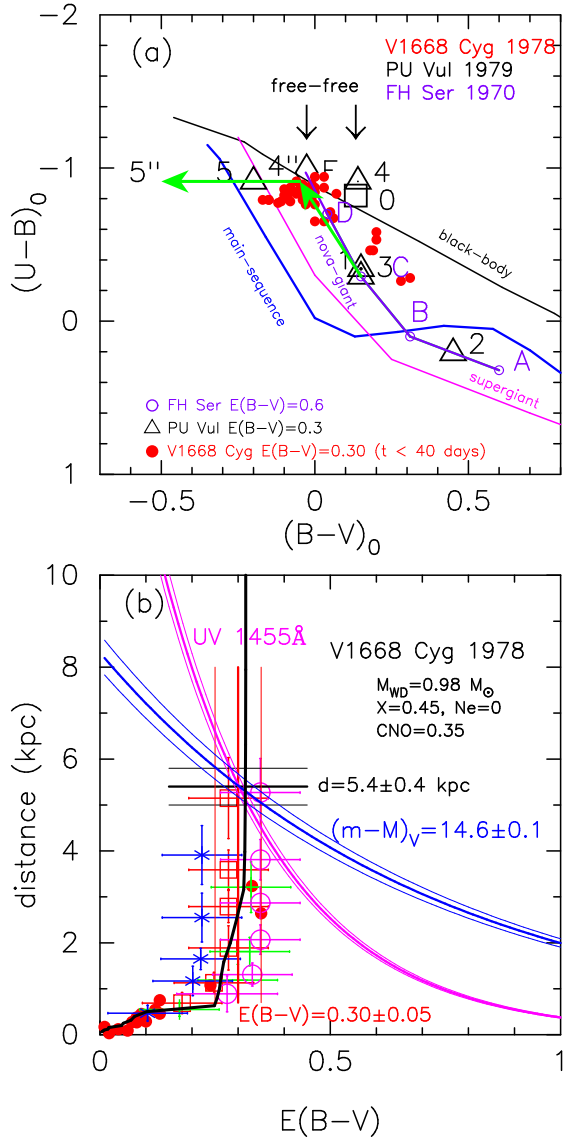}
%\plotone{color_color_v1668_cyg_distance_reddening_x45z02c15o20.epsi}
%\plotfiddle{evolution1.ps}{5.0cm}{270}{0.4}{0.4}{-170}{220}
\caption{
(a) The color-color diagram of V1668~Cyg, with the same data as
in Figure \ref{v1668_cyg_v_bv_ub_color_curve}.  The color data
are de-reddened with $E(B-V)=0.30$.
%%by Equations (\ref{dereddening_eq_bv}) and (\ref{dereddening_eq_ub}) 
The purple line shows a part of the nova-giant sequence and
the attached capitals, A, B, C, and D correspond to each stage of
the light curve of FH~Ser in Figure 
\ref{hr_diagram_fh_ser_pw_vul_v1500_cyg_v1974_cyg_outburst}(a).
Attached numbers, other symbols, and lines are the same as those
in Figures 4 and 8 of \citet{hac14k}.   
(b) Distance-reddening relations for V1668~Cyg,
$(l,b)=(90\fdg8373, -6\fdg7598)$.
A thick solid magenta line flanked by thin solid magenta lines
denotes the distance-reddening relation for UV~1455 \AA\  flux fitting
(about $\sim 10$\% 1$\sigma$ flux error).
A vertical solid red line flanked by thin solid red lines
show the reddening of $E(B-V)=0.30\pm0.05$.  A solid blue line flanked
by thin solid blue lines corresponds to a distance-reddening relation
calculated from the distance modulus of
$(m-M)_V=14.6\pm0.1$.  Filled red circles represent distance-reddening
relation data from \citet{slo79}.  The four sets of data with error bars
show Marshall et al.'s (2006) distance-reddening relations
in four directions close to V1668~Cyg:
$(l, b)=(90\fdg75,-6\fdg75)$ (open red squares),
$(91\fdg00, -6\fdg75)$ (filled green squares),
$(90\fdg75,  -7\fdg00)$ (blue asterisks),
and $(91\fdg00,  -7\fdg00)$ (open magenta circles).
The thick solid black line denotes the distance-reddening
relation given by \citet{gre15}.  
These trends/lines cross at $d\approx 5.4$~kpc and
$E(B-V)\approx0.30$.
\label{color_color_v1668_cyg_distance_reddening_x45z02c15o20}}
\end{figure}

%Fig.3
%\placefigure{hr_diagram_v1668_cyg_only_outburst}

%%\begin{figure*}
\begin{figure}
%\epsscale{0.45}
\epsscale{1.0}
%\epsscale{1.15}
\plotone{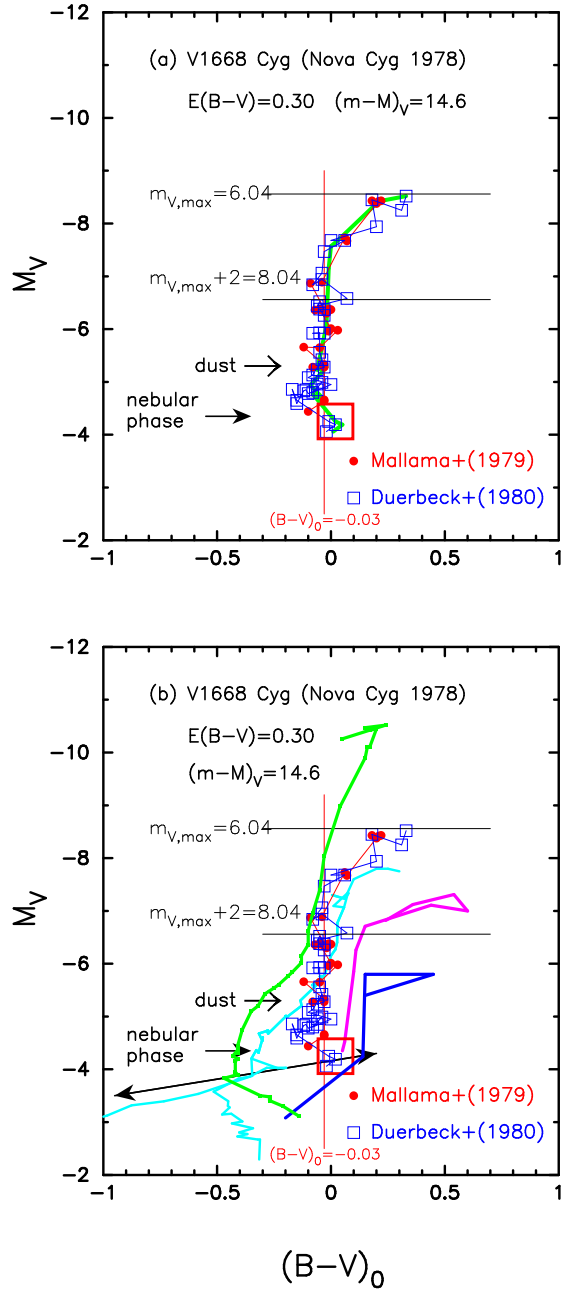}
%\plotone{hr_diagram_v1668_cyg_only_outburst.epsi}
%\plotfiddle{evolution1.ps}{5.0cm}{270}{0.4}{0.4}{-170}{220}
\caption{
The color-magnitude diagram of V1668~Cyg.  See the main text for the sources
of the observational data.  (a) The thick
solid green line denote a template light curve for V1668~Cyg.
The vertical solid red lines show the colors of $(B-V)_0=-0.03$ for 
optically thick free-free emission \citep[e.g.,][]{hac14k, hac15k}.
(b) Comparison of V1668~Cyg with V1500~Cyg (thick solid green line),
V1974~Cyg (thick solid cyan line), FH~Ser (thick solid magenta line), 
and PU~Vul (thick solid blue line).
The thick two-headed black arrow indicates 
Equation (\ref{absolute_magnitude_cusp}).
\label{hr_diagram_v1668_cyg_only_outburst}}
\end{figure}
%%\end{figure*}

\section{Color-magnitude Evolutions of Well-observed Novae}
\label{color_magnitude_diagram}
In this section, we study ten well-observed novae, V1668~Cyg,
LV~Vul, FH~Ser, PW~Vul, V1500~Cyg, V1974~Cyg, PU~Vul, V723~Cas, HR~Del,
and V5558~Sgr, in this order.  These ten novae were examined 
in detail in Paper I based on the color-color diagram.
Here, we examine each nova in the color-magnitude diagram.

\subsection{V1668~Cyg 1978}
\label{v1668_cyg_cmd}
Figure \ref{v1668_cyg_v_bv_ub_color_curve} shows (a) the $V$, $y$, and
UV~1455\ \AA\  light curves, (b) $(B-V)_0$, and (c) $(U-B)_0$ color 
evolutions of V1668~Cyg.  Here, $(B-V)_0$ and $(U-B)_0$
are the de-reddened colors of $B-V$ and $U-B$, i.e.,
\begin{equation}
(B-V)_0 = (B-V) - E(B-V),
\label{dereddening_eq_bv}
\end{equation}
\begin{equation}
(U-B)_0 = (U-B) - 0.64 E(B-V),
\label{dereddening_eq_ub}
\end{equation}
where the factor of $0.64$ is taken from \citet{rie85}.
The UV~1455 \AA\  band is designed to represent continuum
flux of UV light \citep[a narrow 20\ \AA\  width band at the center of
1455\AA,][]{cas02}.   The $UBV$ data of V1668~Cyg are taken from
\citet{due80} and \citet{kol80} whereas the $BV$ data are from 
\citet{mal79} and the $y$ data are from \citet{gal80}.  
The $V$ light curve of V1668~Cyg has $t_2=12.2$ and $t_3=24.3$ days 
\citep{mal79}.
\citet{hac16k} reanalyzed the light curves of V1668~Cyg on the basis
of model light curves, including the effects of both free-free emission
and photospheric emission.  
They redetermined the reddening as
$E(B-V)=0.30\pm0.05$ and the distance modulus in the $V$ band
as $\mu_V= (m-M)_V=14.6\pm0.1$.

Adopting their value of $E(B-V)=0.30$, we plot the color-color
diagram of V1668~Cyg in Figure 
\ref{color_color_v1668_cyg_distance_reddening_x45z02c15o20}(a).
Because the reddening of V1668~Cyg was updated to $E(B-V)=0.30$
in \citet{hac16k} from $E(B-V)=0.35$ in Paper I, 
we revised the color-color diagram (Figure 
\ref{color_color_v1668_cyg_distance_reddening_x45z02c15o20}(a)) and
conclude that the reddening value of $E(B-V)=0.30$ is still
consistent with the general tracks of novae (solid green lines).

Next, we examine the combination of the revised values of
$E(B-V)=0.30$ and $\mu_V= (m-M)_V=14.6$ 
in the distance-reddening relation for V1668~Cyg, whose
galactic coordinates are $(l,b)=(90\fdg8373, -6\fdg7598)$.
Figure \ref{color_color_v1668_cyg_distance_reddening_x45z02c15o20}(b)
shows various distance-reddening relations for V1668~Cyg.
\citet{mar06} published a three-dimensional extinction map
of our galaxy in the direction of $-100\fdg0 \le l \le 100\fdg0$
and $-10\fdg0 \le b \le +10\fdg0$ with grids of $\Delta l=0\fdg25$
and $\Delta b=0\fdg25$, where $(l,b)$ are the galactic coordinates.
Their results are shown
%in Figure \ref{color_color_v1668_cyg_distance_reddening_x45z02c15o20}(b)
by four directions close to V1668~Cyg.
We also plot the result given by \citet{slo79} (filled red circles).
Recently, \citet{gre15} published data for
the galactic extinction map, which covers a wider range of the galactic
coordinates (over three quarters of the sky) with much finer
grids of 3\farcm4 to 13\farcm7 and a maximum distance resolution of
25\%.  Their values of $E(B-V)$ could have an error of 0.05 -- 0.1 mag
compared with other two-dimensional dust extinction maps.
We added Green et al.'s distance-reddening line 
(the best fitted of their examples)
as the thick solid black line in 
Figure \ref{color_color_v1668_cyg_distance_reddening_x45z02c15o20}(b).

We also added our results of the model light curve fits of the $V$ 
(solid blue lines) and UV~1455 \AA\ (solid magenta lines) bands to
Figure \ref{color_color_v1668_cyg_distance_reddening_x45z02c15o20}(b).
These relations are calculated as follows:
Figure \ref{v1668_cyg_v_bv_ub_color_curve}(a) shows the theoretical light
curve taken from \citet{hac16k}, who calculated nova model light curves
for various chemical compositions and WD masses based on
free-free emission plus photospheric emission.  The solid blue line
shows the $V$ model light curve of a $0.98~M_\sun$ WD with
the chemical composition of ``CO nova 3'' \citep{hac16k}.
Here we adopt their value $(m-M)_V=14.6$ for V1668~Cyg. 
Then, the distance-reddening relation is calculated from 
\begin{equation}
(m-M)_V= 5 \log (d/{\rm 10~pc}) + 3.1 E(B-V),
\label{v_distance_modulus}
\end{equation}
together with $(m-M)_V=14.6\pm0.1$.  We plot Equation 
(\ref{v_distance_modulus}) by the blue thick solid
line flanked with thin solid blue lines in Figure
\ref{color_color_v1668_cyg_distance_reddening_x45z02c15o20}(b).
\citet{hac16k} also calculated the narrow band UV~1455 \AA\  flux
\citep{cas02} for the same WD model on the basis of blackbody emission,
which is shown in Figure \ref{v1668_cyg_v_bv_ub_color_curve}(a)
by the solid red line.  Fitting our model with the observed fluxes,
we also obtain a distance-reddening relation 
\begin{equation}
2.5\log \left( F_{\lambda}^{\rm obs}/F_{\lambda}^{\rm mod}\right)
= R_\lambda E(B-V) + 5\log \left( {{d} \over {10~{\rm kpc}}} \right),
\label{uv1455_distance_modulus}
\end{equation}
where $F_{\lambda}^{\rm mod}$ is the model flux at the distance of 
$d=10$~kpc, $F_{\lambda}^{\rm obs}$ is the observed flux,
the absorption is calculated from $A_\lambda=R_\lambda E(B-V)$, and
$R_\lambda=8.3$ for $\lambda=1455$ \AA\
\citep{sea79}.
For V1668~Cyg, $F_{1455}^{\rm obs}=4.0$ and $F_{1455}^{\rm mod}=11.75$
in units of $10^{-12}$~erg~cm$^{-2}$~s$^{-1}$~\AA$^{-1}$ at the upper bound
of Figure \ref{v1668_cyg_v_bv_ub_color_curve}(a).
This distance-reddening relation is plotted by the magenta lines
with a $\sim 10$\% flux 1$\sigma$ error in Figure
\ref{color_color_v1668_cyg_distance_reddening_x45z02c15o20}(b).
All the above trends consistently cross each other at 
$d\approx5.4$~kpc and $E(B-V)\approx0.30$ as shown in
Figure \ref{color_color_v1668_cyg_distance_reddening_x45z02c15o20}(b).

V1668~Cyg is located much below the galactic plane because its
galactic coordinates are $(l,b)=(90\fdg8373, -6\fdg7598)$.
It is far from the galactic plane ($-z > 0.6$~kpc) and much below
the galactic matter distribution \citep[e.g., $z\sim125$~pc,][]{mar06}
for a distance of $d\sim5.4$~kpc.
Therefore, the extinction for V1668~Cyg should be close to
the galactic dust extinction (two-dimensional map).
The NASA/IPAC Galactic dust absorption
map\footnote{http://irsa.ipac.caltech.edu/applications/DUST/},
which is based on the data of \citet{schl11},
 gives $E(B-V)=0.29\pm0.02$ for V1668~Cyg.
Thus, we confirmed that the adopted value of $E(B-V)=0.30$ is reasonable.
Our distance and reddening estimates appear in
%Table 1
Table \ref{extinction_various_novae}.

Using the new value of $E(B-V)=0.30$ together with $(m-M)_V=14.6$,
we plot the color-magnitude diagram of V1668~Cyg in
Figure \ref{hr_diagram_v1668_cyg_only_outburst}(a).  
The two horizontal solid lines indicate stages at the $V$ maximum, 
$m_{V,\rm max}$, and 2 mag below the $V$ maximum, $m_{V,\rm max}+2$.
After the optical maximum ($m_V\approx6$), V1668~Cyg goes down almost
along the line of $(B-V)_0=-0.03$, which is the intrinsic $B-V$ color
of optically thick free-free emission \citep{hac14k}.   
This is consistent with the theoretical light curve of Figure
\ref{v1668_cyg_v_bv_ub_color_curve}(a), in which the flux of
free-free emission dominates the photospheric emission.
An optically thin dust shell
formed at $m_V\sim10.5$ \citep{geh80} as indicated by an arrow.
After that, V1668~Cyg entered the nebular phase about 4 mag below
the maximum \citep[e.g.,][]{kla80}.
It moves rightward, i.e., toward red, around/after the start of
the nebular phase and then turns to the left, i.e., toward blue.
The position of this turning point is denoted by the large open red square
at $M_V=-4.19$ and $(B-V)_0=+0.02$.
We define a template of the color-magnitude track for V1668~Cyg
by a thick solid green line in Figure
\ref{hr_diagram_v1668_cyg_only_outburst}(a).
The orbital period of 3.32~hr was detected by \citet{kal90}.
%Table 2
Table \ref{color_magnitude_turning_point} lists the position ($(B-V)_0$,
$M_V$) of the turning point in the color-magnitude diagram, 
distance modulus in the $V$ band, orbital period (if it is known),
and type of the track in the color-magnitude diagram on the basis of our
classification introduced later in Section
\ref{summary_basic_properties_cmd}.

Figure \ref{hr_diagram_v1668_cyg_only_outburst}(b) compares
the V1668~Cyg track with other well-observed novae,
V1500~Cyg (thick solid green line), V1974~Cyg (thick solid cyan line),
FH~Ser (thick solid magenta line), and PU~Vul (thick solid blue line),
the data of which are taken from later sections corresponding to each nova.
These novae follow a similar path but their 
tracks are located from left to right depending on the nova speed class,
i.e., $t_2=2.4$, 12.2, 17, 42, and $\gtrsim 1500$~days for V1500~Cyg,
V1668~Cyg, V1974~Cyg, FH~Ser, and PU~Vul, respectively, 
except for the nebular phase.
We also added a two-headed arrow, which shows Equation 
(\ref{absolute_magnitude_cusp}).  We discuss these properties
later in Section \ref{properties_color_magnitude_diagram}.

%Fig.4
%\placefigure{lv_vul_v_bv_ub_color}

\begin{figure}
%\epsscale{0.75}
%%\epsscale{0.8}
%\epsscale{1.0}
\epsscale{1.15}
\plotone{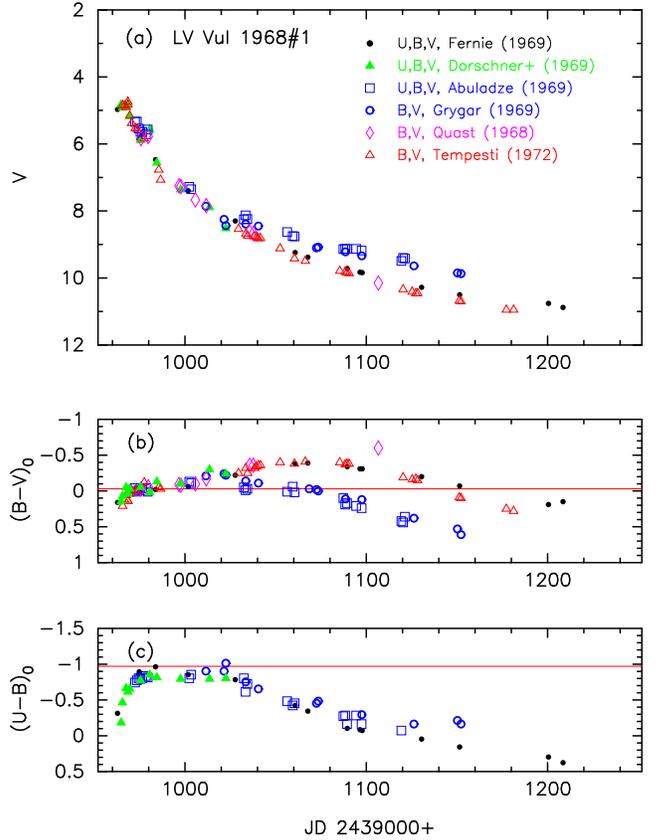}
%\plotone{lv_vul_v_bv_ub_color.epsi}
%\plotfiddle{evolution1.ps}{5.0cm}{270}{0.4}{0.4}{-170}{220}
\caption{
Same as Figure \ref{v1668_cyg_v_bv_ub_color_curve}, but for LV~Vul.
%%See the main text for the sources of observational data.
%The $UBV$ data are taken from \citet{fer69}, \citet{dor69},
%\citet{abu69}, and \citet{gry69}.  The $BV$ data are taken from
%\citet{qua68} and \citet{tem72}.  The $(B-V)_0$ and
%$(U-B)_0$ colors are dereddened with $E(B-V)=0.60$ \citep{hac14k}.
%%In panel (c), the $U-B$ data of \citet{dor69}
%%are 0.1 mag redder than the other data, so we systematically
%%shifted them toward blue by 0.1 mag.
\label{lv_vul_v_bv_ub_color}}
\end{figure}

%Fig.5 
%\placefigure{color_color_diagram_lv_vul_distance_reddening}

\begin{figure}
%\epsscale{0.40}
%%\epsscale{0.8}
\epsscale{1.0}
%\epsscale{1.15}
\plotone{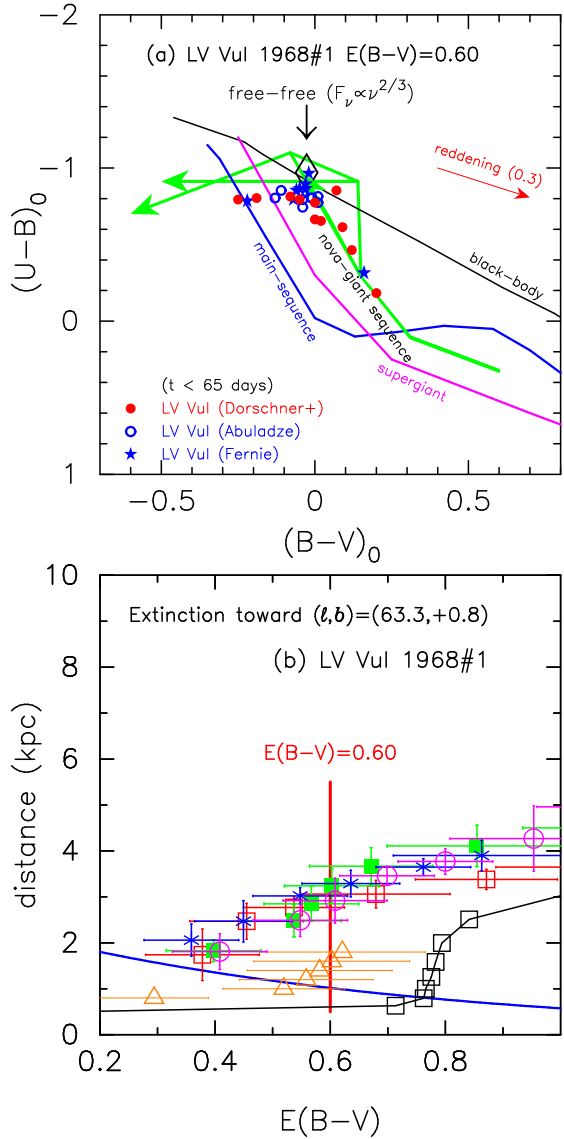}
%\plotone{color_color_diagram_lv_vul_distance_reddening.epsi}
%\plotfiddle{evolution1.ps}{5.0cm}{270}{0.4}{0.4}{-170}{220}
\caption{
(a) Color-color diagram of LV~Vul, with the same data as
those in Figure \ref{lv_vul_v_bv_ub_color}.  The color data
are de-reddened with $E(B-V)=0.60$.
(b) Distance-reddening relations for LV~Vul, 
$(l, b)=(63\fdg3024, +0\fdg8464)$.
The thick solid blue line denotes 
%%the distance-reddening relation of 
$(m-M)_V=11.9$.
%% and the vertical solid red line 
%%shows the reddening of $E(B-V)=0.60$.  
The four sets of data with error bars
show Marshall et al.'s (2006) distance-reddening relations
in four directions close to LV~Vul:
$(l, b)=(63\fdg25,0\fdg75)$ (open red squares),
$(63\fdg50,0\fdg75)$ (filled green squares),
$(63\fdg25, 1\fdg00)$ (blue asterisks),
and $(63\fdg50, 1\fdg00)$ (open magenta circles).
We also plot the results of \citet{hak97} 
(open orange triangles with error bars)
and \citet{gre15} (open black squares connected with a thin solid line).
\label{color_color_diagram_lv_vul_distance_reddening}}
\end{figure}

%Fig.6 
%\placefigure{hr_diagram_lv_vul_lv_vul_2fig_outburst}

%%\begin{figure*}
\begin{figure}
%\epsscale{0.5}
\epsscale{1.0}
%\epsscale{1.15}
\plotone{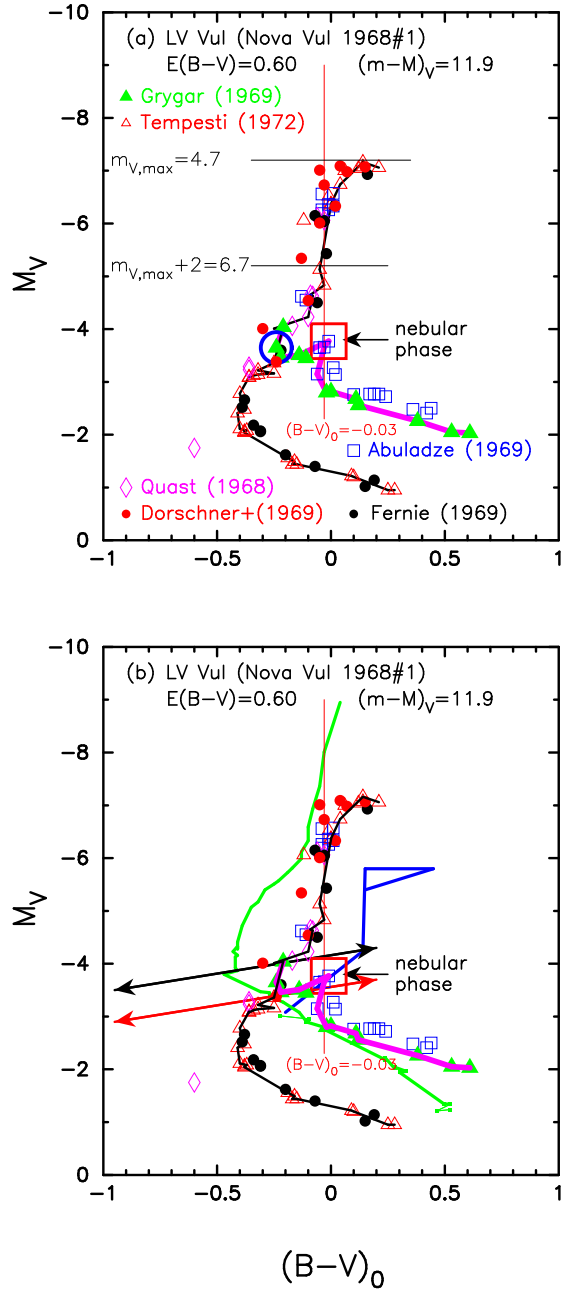}
%\plotone{hr_diagram_lv_vul_lv_vul_2fig_outburst.epsi}
%\plotfiddle{evolution1.ps}{5.0cm}{270}{0.4}{0.4}{-170}{220}
\caption{
Same as Figure 
\ref{hr_diagram_v1668_cyg_only_outburst}, but
for LV~Vul.  (a) 
%%LV~Vul 1968\#1 data and a template of LV~Vul.
The thick solid black and magenta lines denote a template track for LV~Vul.
The large open blue circle indicates the bifurcation point of these two
tracks.
%%See the main text for more detail.
(b) Comparison of LV~Vul with other well-observed novae, 
V1500~Cyg (thick solid green line) and PU~Vul (thick solid blue line).
The two-headed black arrow represents Equation
(\ref{absolute_magnitude_cusp}) from Section 
\ref{properties_color_magnitude_diagram} and the two-headed red arrow
shows a line 0.6 mag below the black one, i.e., Equation
(\ref{absolute_magnitude_cusp_low_red}) from Section 
\ref{properties_color_magnitude_diagram}.  
\label{hr_diagram_lv_vul_lv_vul_2fig_outburst}}
\end{figure}
%%\end{figure*}

%Fig.7
%\placefigure{hr_diagram_fh_ser_pw_vul_v1500_cyg_v1974_cyg_outburst}

\begin{figure*}
%\begin{figure}
%\epsscale{0.75}
\epsscale{0.8}
%%\epsscale{1.0}
\plotone{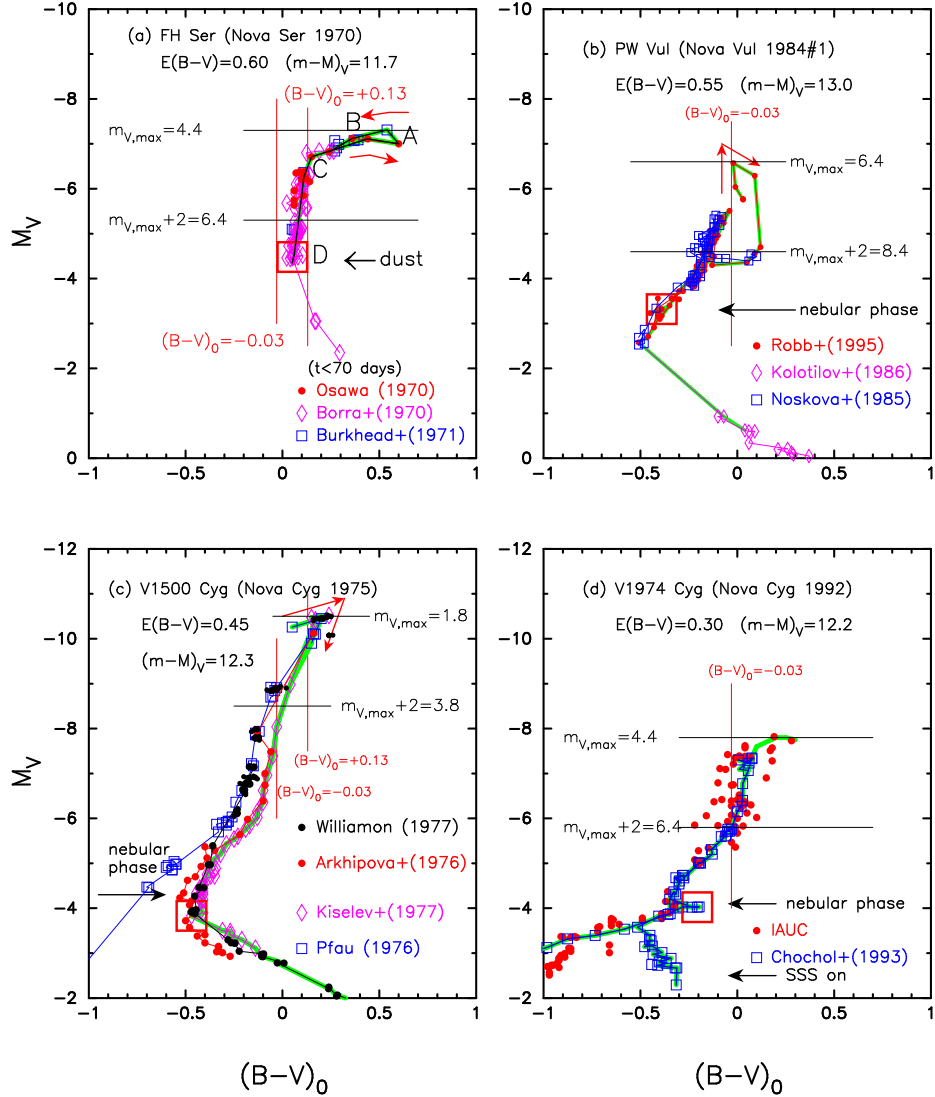}
%\plotone{hr_diagram_fh_ser_pw_vul_v1500_cyg_v1974_cyg_outburst.epsi}
%\plotfiddle{evolution1.ps}{5.0cm}{270}{0.4}{0.4}{-170}{220}
\caption{
Color-magnitude diagrams of (a) FH~Ser 1970, (b) PW~Vul 1984\#1, 
(c) V1500~Cyg 1975, and (d) V1974~Cyg 1992.  The thick solid green lines
show templates of the color-magnitude diagram for each nova.  The vertical
solid red lines show the colors of $(B-V)_0=+0.13$ for optically thin
free-free emission and of $(B-V)_0=-0.03$ for optically thick
free-free emission \citep[e.g.,][]{hac14k, hac15k}.
The red arrows show the direction of the evolution in the very early phase.
In panel (a), the solid black line denotes
a template light curve of FH~Ser and the attached capitals A, B, C,
and D correspond to the stages A, B, C, and D of
the FH~Ser light curves in Figure 2 of \citet{hac14k}.
In panel (d), we add the start of the supersoft X-ray source (SSS) 
phase by an arrow labeled ``SSS on.''
%%See the main text for the sources of the data.
\label{hr_diagram_fh_ser_pw_vul_v1500_cyg_v1974_cyg_outburst}}
%\end{figure}
\end{figure*}

%Fig.8
%\placefigure{hr_diagram_fh_ser_pw_vul_v1500_cyg_v1974_cyg_outburst_no2}

\begin{figure*}
%\begin{figure}
%\epsscale{0.75}
\epsscale{0.8}
%%\epsscale{1.0}
\plotone{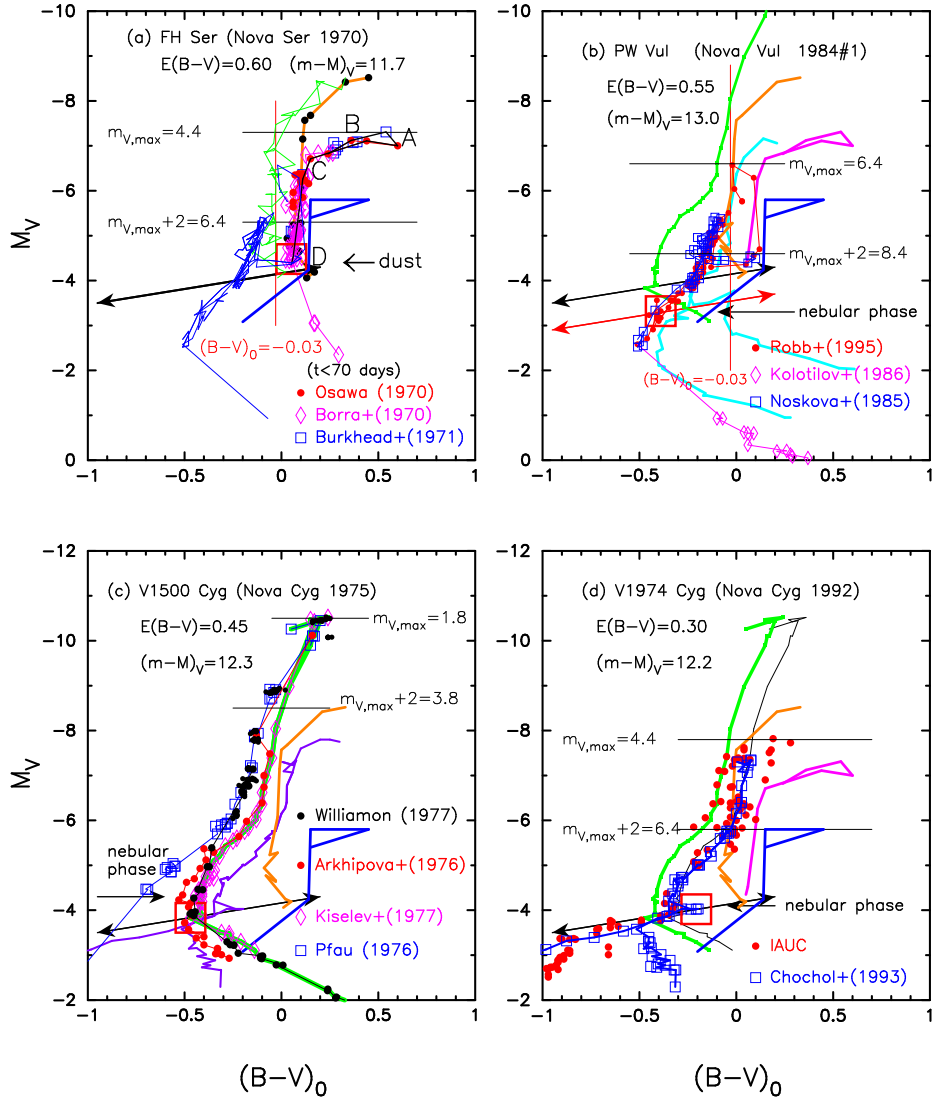}
%\plotone{hr_diagram_fh_ser_pw_vul_v1500_cyg_v1974_cyg_outburst_no2.epsi}
%\plotfiddle{evolution1.ps}{5.0cm}{270}{0.4}{0.4}{-170}{220}
\caption{
Same as Figure
\ref{hr_diagram_fh_ser_pw_vul_v1500_cyg_v1974_cyg_outburst}, but 
we also plot other nova templates in the same figure, i.e.,
V1500~Cyg (thick solid green lines), V1668~Cyg (thick solid orange lines),
V1974~Cyg (thick solid purple lines), LV~Vul (thick solid cyan lines),
FH~Ser (thick solid magenta lines), and PU~Vul (thick solid blue lines).
The two-headed black arrows indicate 
Equation (\ref{absolute_magnitude_cusp}) whereas the two-headed
red arrow, which is defined by Equation (\ref{absolute_magnitude_cusp_low_red})
from Section \ref{properties_color_magnitude_diagram}, denotes
a line 0.6 mag below the thick two-headed black one.
In panel (a), we add two detailed tracks for V1668~Cyg (thin
solid green lines) and PW~Vul (thin solid blue lines).  The thick 
solid orange line with black points indicates the track of V1668~Cyg
shifted by $\Delta(B-V)=0.12$ toward red.  In panel (d), we also add
a thin solid black line which is the V1500~Cyg track shifted 
by $\Delta(B-V)=0.12$ toward red.
\label{hr_diagram_fh_ser_pw_vul_v1500_cyg_v1974_cyg_outburst_no2}}
%\end{figure}
\end{figure*}

%Fig.9
%\placefigure{distance_reddening_fh_ser_pw_vul_v1500_cyg_v1974_cyg}

\begin{figure*}
%\begin{figure}
\epsscale{0.75}
%%\epsscale{0.8}
%%\epsscale{1.0}
%%\epsscale{1.15}
\plotone{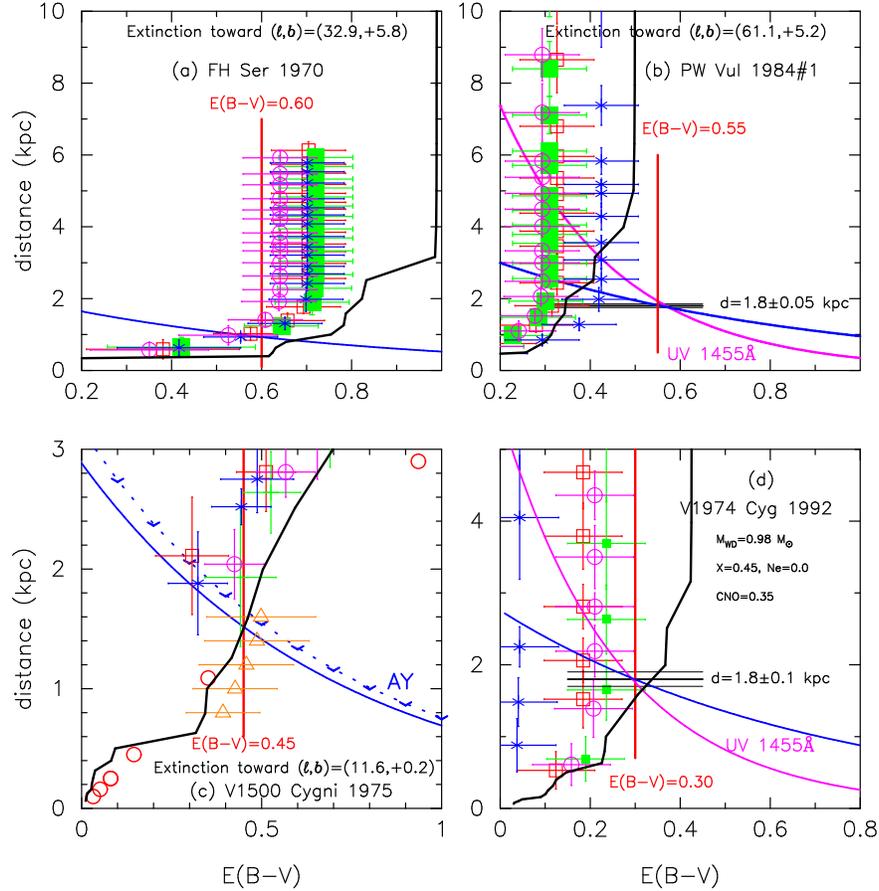}
%\plotone{distance_reddening_fh_ser_pw_vul_v1500_cyg_v1974_cyg.epsi}
%\plotfiddle{evolution1.ps}{5.0cm}{270}{0.4}{0.4}{-170}{220}
\caption{
Distance-reddening relation for (a) FH~Ser, (b) PW~Vul, 
(c) V1500~Cyg, and (d) V1974~Cyg.
The solid black lines represent the distance-reddening relations given
by \citet{gre15}.
The thick solid blue lines denote 
%%the distance-reddening relation calculated from the distance modulus
%% in the $V$ band, i.e., 
(a) $(m-M)_V=11.7$, (b) $(m-M)_V=13.0$, (c) $(m-M)_V=12.3$, 
and (d) $(m-M)_V=12.2$.
%The vertical solid red lines represent the color excesses of 
%(a) $E(B-V)=0.60$, (b) $E(B-V)=0.55$,
%(c) $E(B-V)=0.45$, and (d) $E(B-V)=0.30$.
%Four sets of data (open red squares, filled green squares, blue asterisks,
%magenta circles) with error bars show distance-reddening relations
%in four directions close to each nova, with data taken
%from \citet{mar06}.  
%The solid black lines denote the distance-reddening relation given
%by \citet{gre15}.
In panel (c), the dashed blue line with downward carets represents
the upper bound of the distance-modulus derived by \citet{and76} from
the galactic rotation.  The open red circles are distance-reddening 
relations of stars taken from \citet{you76}.
We also plot the results of \citet{hak97} (open orange triangles).
In panels (b) and (d), the thick solid magenta lines
denote the distance-reddening relations
calculated from the model fits of the UV~1455 \AA\  narrow band for
the $0.83~M_\sun$ WD with $X=0.55$, $Y=0.23$, $X_{\rm CNO}=0.20$, and $Z=0.02$
\citep{hac15k} 
and for the $0.98~M_\sun$ WD with $X=0.45$, $Y=0.18$, $X_{\rm CNO}=0.35$,
and $Z=0.02$ \citep{hac16k}, respectively.
%%%(b) at the upper bound, $F_{1455}^{\rm obs}=1.65$
%%%and $F_{1455}^{\rm mod}=30.0$ in units of
%%%$10^{-12}$~erg~cm$^{-2}$~s$^{-1}$~\AA$^{-1}$.
\label{distance_reddening_fh_ser_pw_vul_v1500_cyg_v1974_cyg}}
%\end{figure}
\end{figure*}

%Fig.10
%\placefigure{distance_reddening_pu_vul_v723_cas_hr_del_v5558_sgr}

\begin{figure*}
%\begin{figure}
\epsscale{0.75}
%%\epsscale{0.8}
%%\epsscale{1.0}
%%\epsscale{1.15}
\plotone{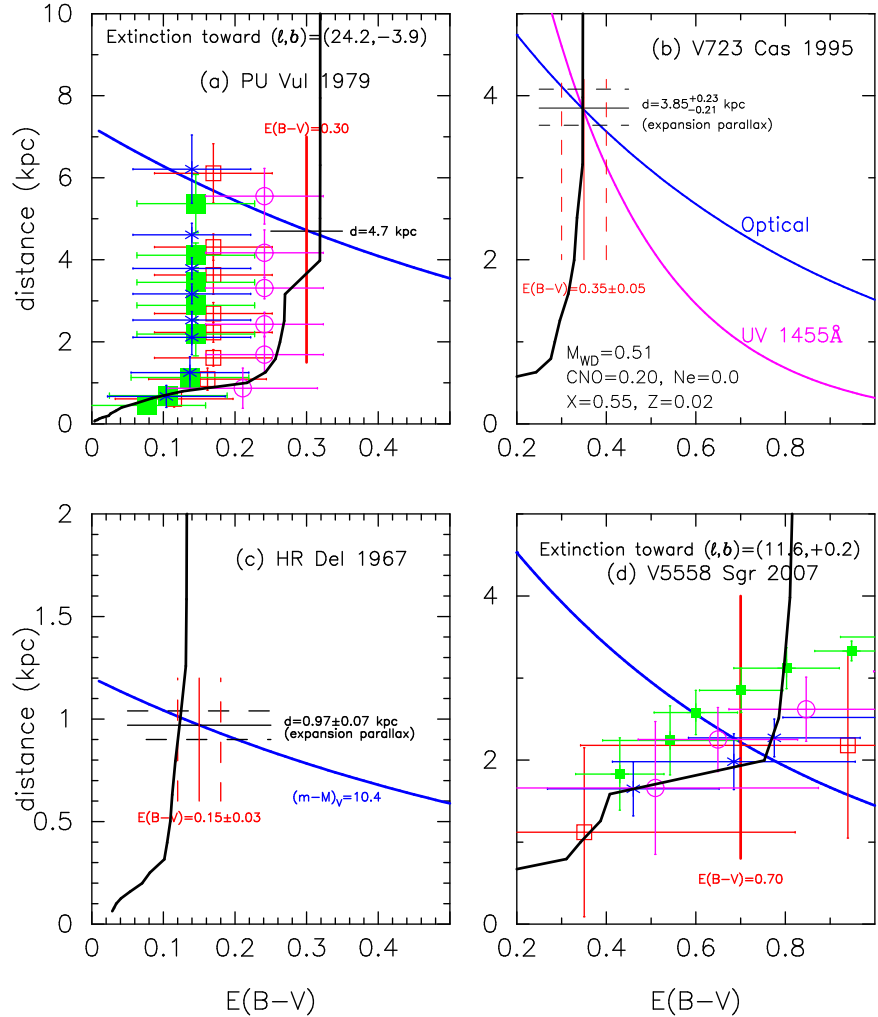}
%\plotone{distance_reddening_pu_vul_v723_cas_hr_del_v5558_sgr.epsi}
%\plotfiddle{evolution1.ps}{5.0cm}{270}{0.4}{0.4}{-170}{220}
\caption{
Same as Figure \ref{distance_reddening_fh_ser_pw_vul_v1500_cyg_v1974_cyg},
but for (a) PU~Vul, (b) V723~Cas, (c) HR~Del, and (d) V5558~Sgr.
The thick solid blue lines denote 
%%the distance-reddening relation calculated from the distance modulus in the 
%%$V$ band, i.e., 
(a) $(m-M)_V=14.3$, (b) $(m-M)_V=14.0$, (c) $(m-M)_V=10.4$,  
and (d) $(m-M)_V=13.9$.  
%%The vertical solid red lines represent
%%the color excesses of 
%(a) $E(B-V)=0.30$, (b) $E(B-V)=0.35\pm0.05$,
%(c) $E(B-V)=0.15\pm0.03$, and (d) $E(B-V)=0.70$.
%%The solid black lines denote the distance-reddening relation given
%%by \citet{gre15}.
In panel (b), the thick solid magenta line represents
the distance-reddening relation calculated from the UV~1455 \AA\  flux
fitting with the $0.51~M_\sun$ WD model \citep{hac15k}.
%In panels (a) and (d),
%four sets of data with error bars show distance-reddening relations
%in four directions close to each nova, with data taken
%from \citet{mar06}.  
%See the main text for more detail.
\label{distance_reddening_pu_vul_v723_cas_hr_del_v5558_sgr}}
%\end{figure}
\end{figure*}

%Fig.11 
%\placefigure{hr_diagram_pu_vul_v723_cas_hr_del_v5558_sgr_outburst}

\begin{figure*}
%\begin{figure}
%\epsscale{0.75}
\epsscale{0.8}
%%\epsscale{1.0}
\plotone{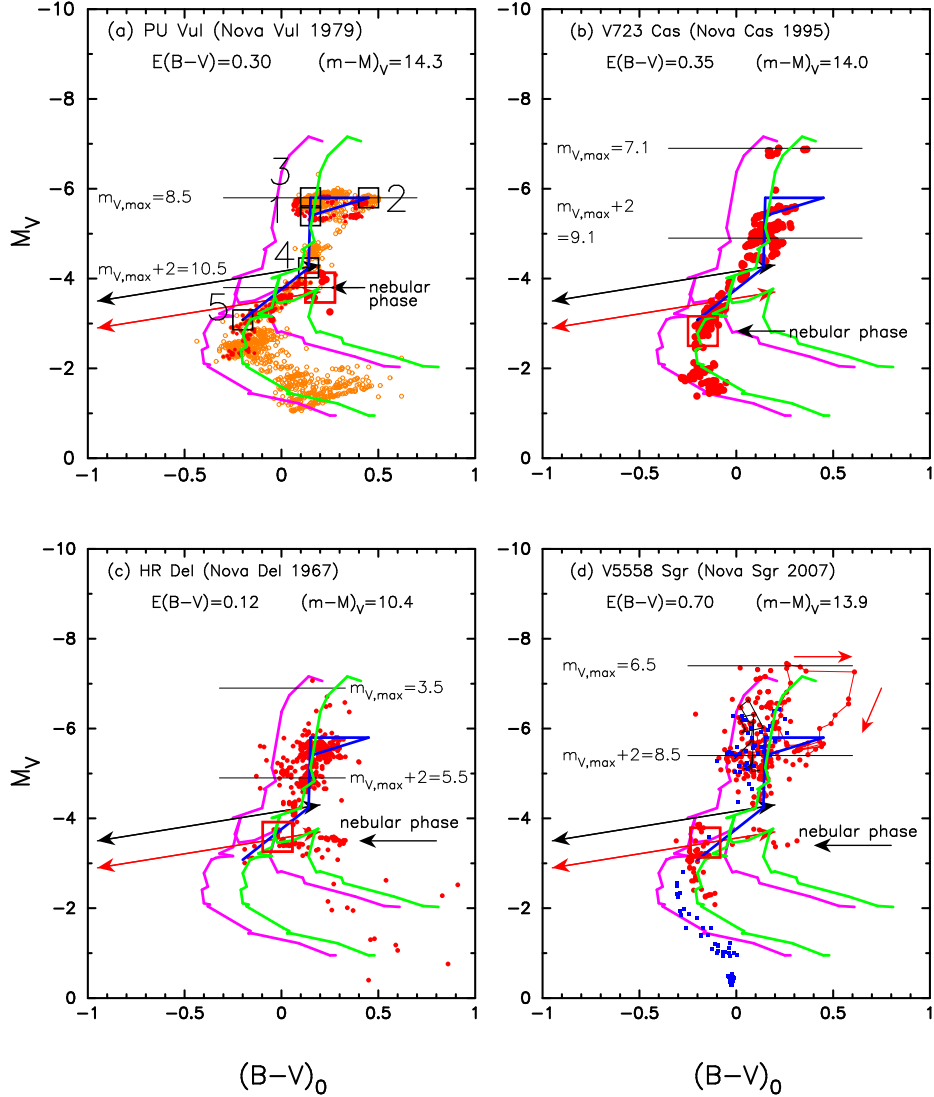}
%\plotone{hr_diagram_pu_vul_v723_cas_hr_del_v5558_sgr_outburst.epsi}
%\plotfiddle{evolution1.ps}{5.0cm}{270}{0.4}{0.4}{-170}{220}
\caption{
Same as Figure
\ref{hr_diagram_fh_ser_pw_vul_v1500_cyg_v1974_cyg_outburst_no2},
but for (a) PU~Vul 1979, (b) V723~Cas 1995, (c) HR~Del 1967,
and (d) V5558~Sgr 2007.
%The two-headed black arrows indicate 
%Equation (\ref{absolute_magnitude_cusp}), whereas the
%two-headed red arrows represent lines 0.6 mag below 
%the two-headed black arrows.
The attached numbers 1 -- 5 in panel (a) correspond to stages 1 -- 5 of
PU~Vul \citep[see Figure 15 of][]{hac14k}.
Thick magenta and solid blue lines denote
the template color-magnitude tracks of LV~Vul and
PU~Vul, respectively.   Thick solid green lines are those of
LV~Vul shifted toward red by $\Delta (B-V)=0.2$. 
%See the main text for the sources of observational data.
\label{hr_diagram_pu_vul_v723_cas_hr_del_v5558_sgr_outburst}}
%\end{figure}
\end{figure*}

%Fig.12 
%\placefigure{hr_diagram_6types_novae_one}

\begin{figure*}
%\begin{figure}
\epsscale{0.55}
%%\epsscale{1.0}
%\epsscale{1.15}
\plotone{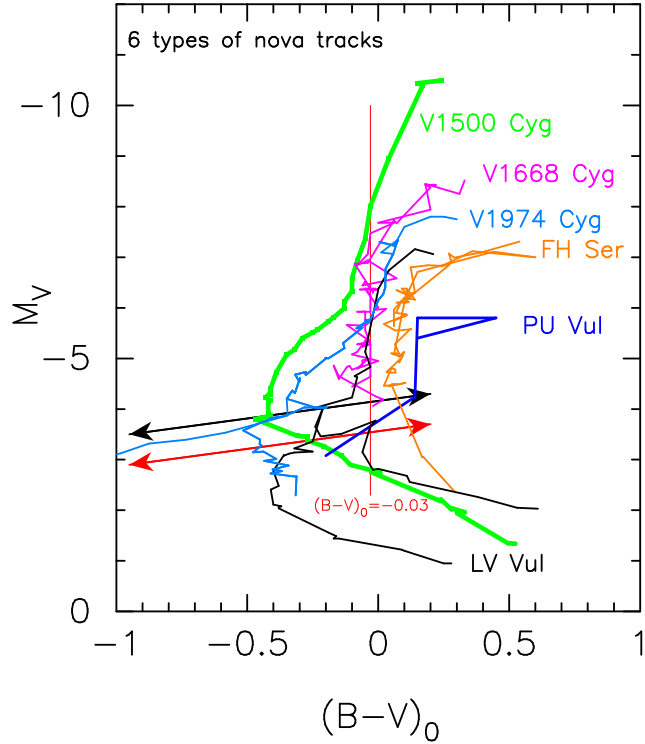}
%\plotone{hr_diagram_6types_novae_one.epsi}
%\plotfiddle{evolution1.ps}{5.0cm}{270}{0.4}{0.4}{-170}{220}
\caption{
Summary of the six typical nova tracks in the color-magnitude diagram:
from left to right:
V1500~Cyg (thick solid green lines), V1668~Cyg (solid magenta lines),
V1974~Cyg (solid sky-blue lines), LV~Vul (solid black lines), 
FH~Ser (solid orange lines), and PU~Vul (thick solid blue lines).
The two-headed black and red arrows represent Equations 
(\ref{absolute_magnitude_cusp}) and (\ref{absolute_magnitude_cusp_low_red}),
respectively.
A vertical solid red line indicates $(B-V)_0=-0.03$, the color
of optically thick free-free emission.
\label{hr_diagram_6types_novae_one}}
%\end{figure}
\end{figure*}

%Fig.13  
%\placefigure{rs_oph_v_bv_ub_color}

\begin{figure}
%\epsscale{0.75}
%%\epsscale{0.8}
%\epsscale{1.0}
\epsscale{1.15}
\plotone{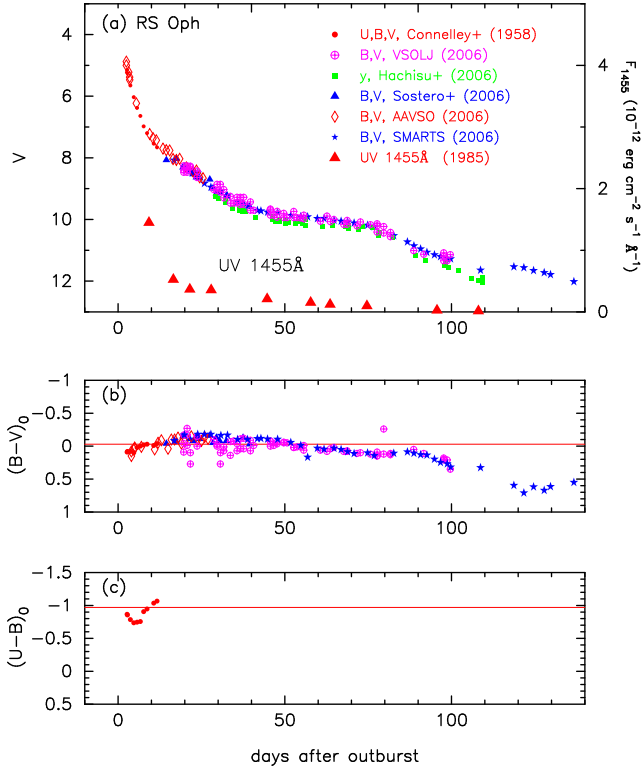}
%\plotone{rs_oph_v_bv_ub_color.epsi}
%\plotfiddle{evolution1.ps}{5.0cm}{270}{0.4}{0.4}{-170}{220}
\caption{
Same as Figure \ref{v1668_cyg_v_bv_ub_color_curve}, but for RS~Oph.
%%%(a) Light curve, (b) $(B-V)_0$ color, and (c) $(U-B)_0$ color curves of
%%%V533~Her. 
The $(B-V)_0$ and
$(U-B)_0$ are de-reddened with the extinction of $E(B-V)=0.65$.
In panel (a), we also plot the UV~1455 \AA\  data of the 1985 outburst
\citep{cas02} (filled red triangles). 
In panel (b), the $B-V$ data of the 2006 outburst are
shifted toward blue by 0.1 mag to match the data of the 1958 outburst
\citep{con58}.  The SMARTS $B-V$ data are shifted toward red by 0.05 mag. 
\label{rs_oph_v_bv_ub_color}}
\end{figure}

%Fig.14
%\placefigure{color_color_diagram_templ_rs_oph_v446_her_v533_her_no2}

\begin{figure*}
%\begin{figure}
\epsscale{0.65}
%\epsscale{0.75}
%%\epsscale{0.8}
%%\epsscale{1.0}
%%\epsscale{1.15}
\plotone{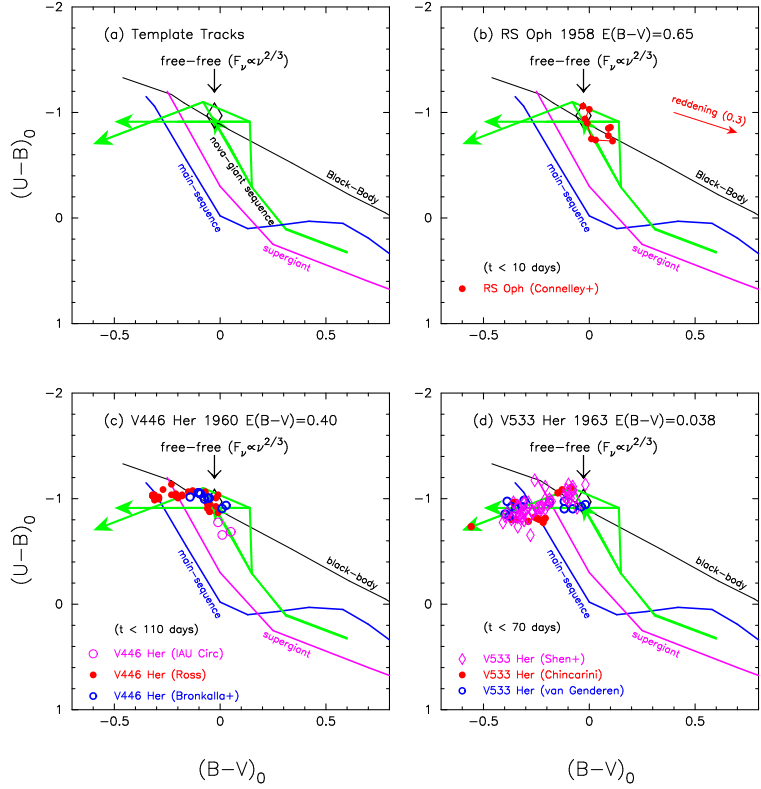}
%\plotone{color_color_diagram_templ_rs_oph_v446_her_v533_her_no2.epsi}
%\plotfiddle{evolution1.ps}{5.0cm}{270}{0.4}{0.4}{-170}{220}
\caption{
Color-color diagrams of (a) our templates for novae, 
(b) RS~Oph (1958), (c) V446~Her 1960, and (d) V533~Her 1963.
These color-color diagrams are similar to Figure 29 of \citet{hac14k},
but we reanalyzed the data. 
%%See the main text for the sources of observational data.
\label{color_color_diagram_templ_rs_oph_v446_her_v533_her_no2}}
%\end{figure}
\end{figure*}

%Fig.15 
%\placefigure{distance_reddening_rs_oph_v446_her_v533_her_iv_cep}

\begin{figure*}
%\begin{figure}
\epsscale{0.75}
%%\epsscale{0.8}
%%\epsscale{1.0}
%%\epsscale{1.15}
\plotone{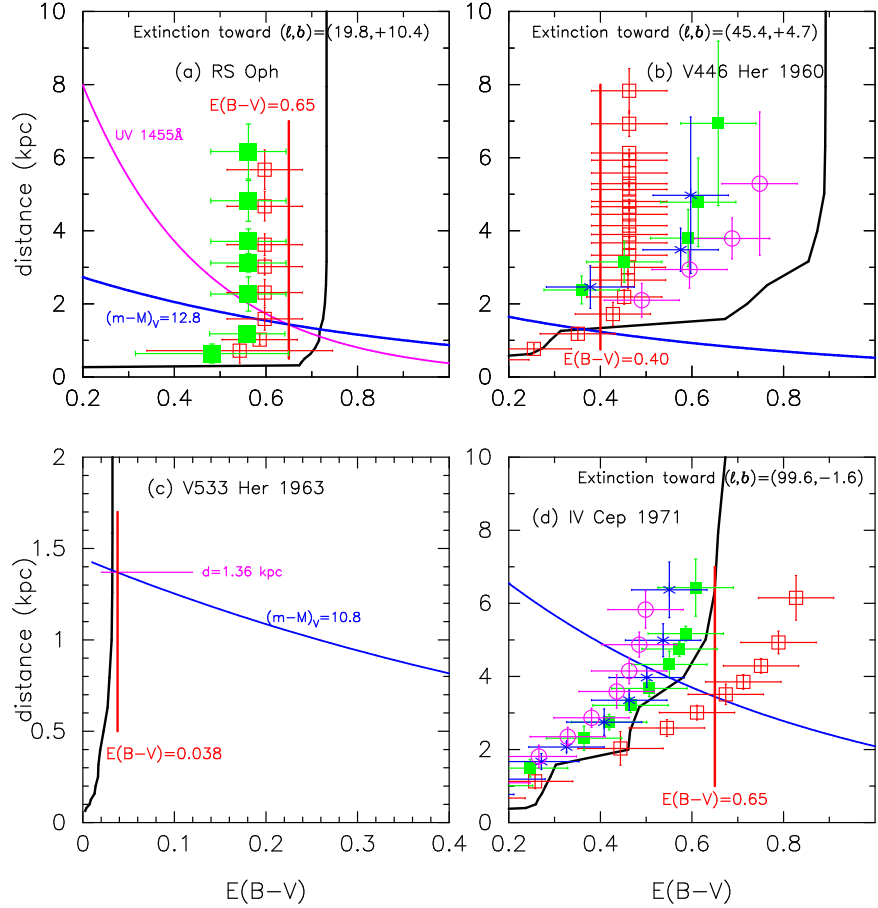}
%\plotone{distance_reddening_rs_oph_v446_her_v533_her_iv_cep.epsi}
%\plotfiddle{evolution1.ps}{5.0cm}{270}{0.4}{0.4}{-170}{220}
\caption{
Same as Figure \ref{distance_reddening_fh_ser_pw_vul_v1500_cyg_v1974_cyg},
but for (a) RS~Oph, (b) V446~Her, (c) V533~Her, and (d) IV~Cep.
The thick solid blue lines denote the distance modulus in the $V$ band, 
%%the distance-reddening relation calculated from 
i.e., (a) $(m-M)_V=12.8$, (b) $(m-M)_V=11.7$, (c) $(m-M)_V=10.8$, 
and (d) $(m-M)_V=14.7$.
%The vertical solid red lines represent the color excesses of 
%(a) $E(B-V)=0.65$, (b) $E(B-V)=0.40$,
%(c) $E(B-V)=0.038$, and (d) $E(B-V)=0.65$.
%%The solid black lines denote the distance-reddening relation given 
%%by \citet{gre15}.
In panel (a), the thick solid magenta line represents
%%the distance-reddening relation calculated from 
the UV~1455 \AA\  flux fitting with a $1.37~M_\sun$ WD model \citep{hac14k}.
%%In panels (a), (b), and (d), two or 
%%four sets of data with error bars show distance-reddening relations
%%in two or four directions close to each nova, the data of which are taken
%%from \citet{mar06}.  
%%See the main text for more detail.
\label{distance_reddening_rs_oph_v446_her_v533_her_iv_cep}}
%\end{figure}
\end{figure*}

%Fig.16 
%\placefigure{hr_diagram_rs_oph_v446_her_v533_her_t_pyx_outburst}

\begin{figure*}
%\begin{figure}
%\epsscale{0.75}
\epsscale{0.8}
%%\epsscale{1.0}
\plotone{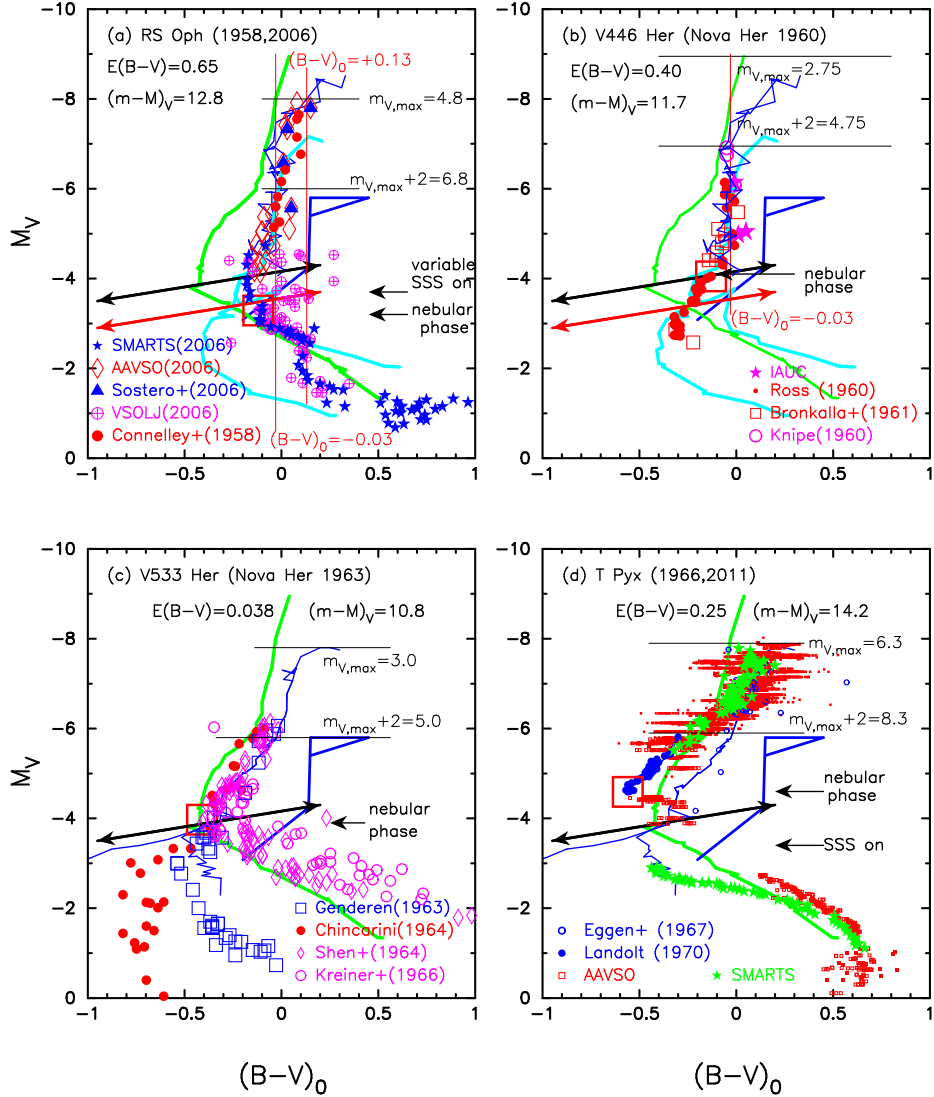}
%\plotone{hr_diagram_rs_oph_v446_her_v533_her_t_pyx_outburst.epsi}
%\plotfiddle{evolution1.ps}{5.0cm}{270}{0.4}{0.4}{-170}{220}
\caption{
Same as Figure 
\ref{hr_diagram_fh_ser_pw_vul_v1500_cyg_v1974_cyg_outburst_no2}, but
for (a) RS~Oph (1958, 2006), (b) V446~Her 1960, (c) V533~Her 1963,
and (d) T~Pyx (1966, 2011).
The thick green and blue lines in each figure denote
the tracks of V1500~Cyg and PU~Vul, respectively.
The thick cyan and thin blue lines in panels (a) and (b)
represent the tracks of LV~Vul and V1668~Cyg, respectively.
The thin blue lines in panels (c) and (d)
denote the track of V1974~Cyg.  
%%In panels (a) and (b), 
%%we add a two-headed red arrow to show a line 0.6 mag below 
%%the two-headed black arrow, which is related to the track of LV~Vul.
%%See Figure 
%%\ref{hr_diagram_lv_vul_lv_vul_2fig_outburst}(b) and Section
%%\ref{lv_vul_cmd} for LV~Vul.
\label{hr_diagram_rs_oph_v446_her_v533_her_t_pyx_outburst}}
%\end{figure}
\end{figure*}

%Fig.17  
%\placefigure{v446_her_v_bv_ub_color_curve}

\begin{figure}
%\epsscale{0.75}
%%\epsscale{0.8}
%\epsscale{1.0}
\epsscale{1.15}
\plotone{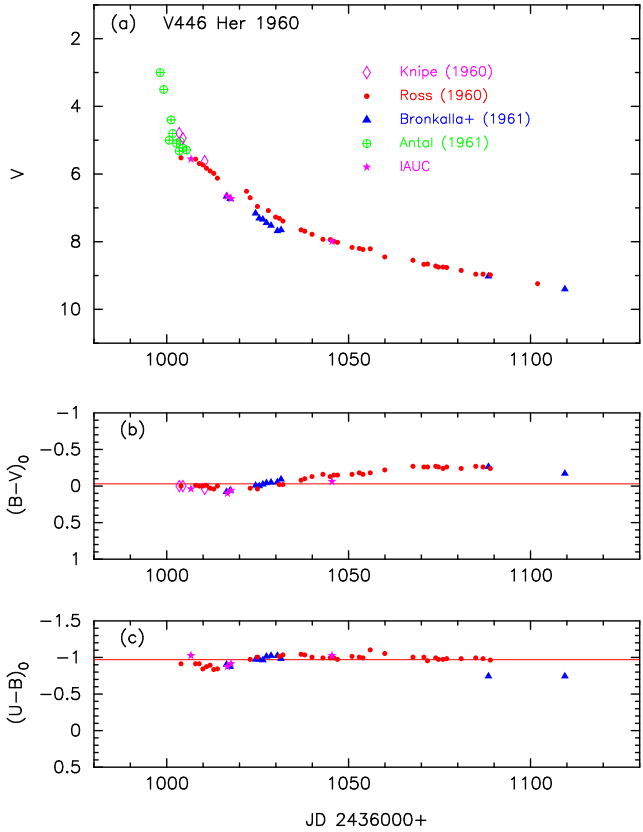}
%\plotone{v446_her_v_bv_ub_color_curve.epsi}
%\plotfiddle{evolution1.ps}{5.0cm}{270}{0.4}{0.4}{-170}{220}
\caption{
Same as Figure \ref{v1668_cyg_v_bv_ub_color_curve}, but for V446~Her.
%%%(a) Light curve, (b) $(B-V)_0$ color, and (c) $(U-B)_0$ color curves of
%%%V446~Her.  
The $(B-V)_0$ and $(U-B)_0$ are de-reddened with $E(B-V)=0.40$.
In panel (b), the $B-V$ data of IAU Circular No.\ 1730,
\citet{kni60}, and \citet{bro61} are systematically shifted toward blue
by 0.1 mag.
%%%  The horizontal red solid line denotes $(B-V)=-0.03$,
%%% which is the $B-V$ color of optically thick free-free emission.
In panel (c), the $U-B$ data of the IAU Circular and \citet{bro61}
are also systematically shifted toward blue by 0.1 mag.
%%% The horizontal red solid line denotes $(U-B)=-0.97$, which is 
%%% the $U-B$ color of optically thick free-free emission.   
\label{v446_her_v_bv_ub_color_curve}}
\end{figure}

%Fig.18 
%\placefigure{v533_her_v_bv_ub_color_curve}

\begin{figure}
%\epsscale{0.75}
%%\epsscale{0.8}
%\epsscale{1.0}
\epsscale{1.15}
\plotone{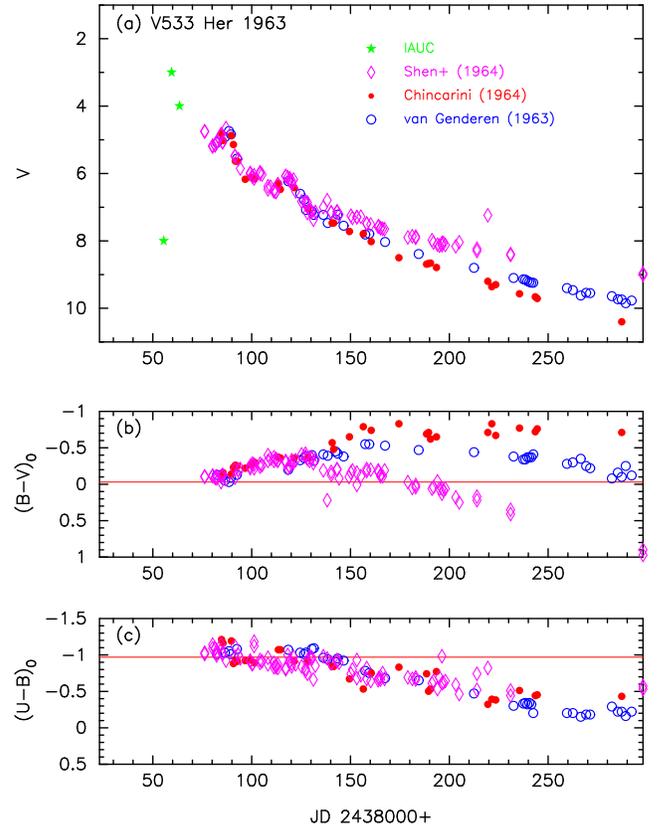}
%\plotone{v533_her_v_bv_ub_color_curve.epsi}
%\plotfiddle{evolution1.ps}{5.0cm}{270}{0.4}{0.4}{-170}{220}
\caption{
Same as Figure \ref{v446_her_v_bv_ub_color_curve}, but for V533~Her.
The $(B-V)_0$ and $(U-B)_0$ are de-reddened with $E(B-V)=0.038$.
%%%(a) Light curve, (b) $(B-V)_0$ color, and (c) $(U-B)_0$ color curves of
%%%V533~Her.  
In panel (c), the $U-B$ data of \citet{chi64} and \citet{gen63} are
systematically shifted toward blue by 0.3 mag.  
\label{v533_her_v_bv_ub_color_curve}}
\end{figure}

%Fig.19 
%\placefigure{t_pyx_v_bv_ub_color}

\begin{figure}
%\epsscale{0.75}
%%\epsscale{0.8}
%\epsscale{1.0}
\epsscale{1.15}
\plotone{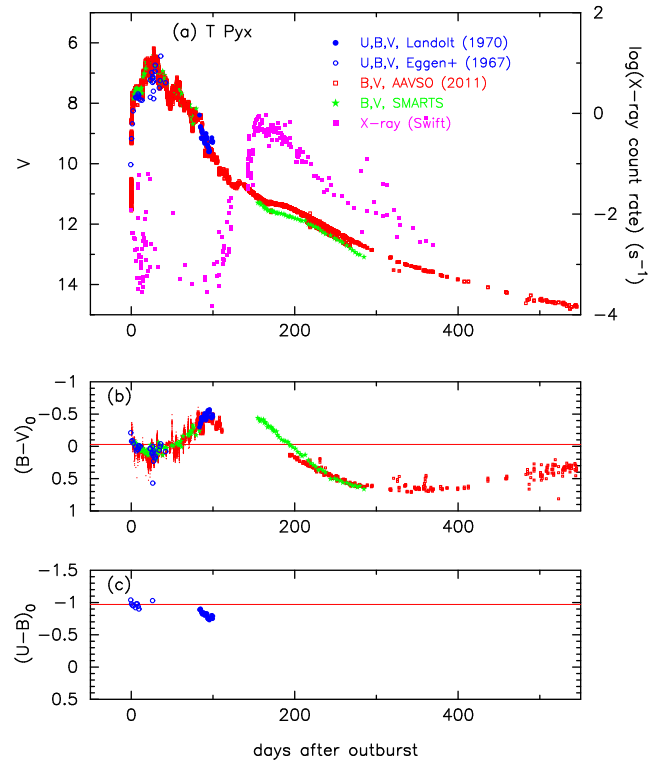}
%\plotone{t_pyx_v_bv_ub_color.epsi}
%\plotfiddle{evolution1.ps}{5.0cm}{270}{0.4}{0.4}{-170}{220}
\caption{
Same as Figure \ref{v446_her_v_bv_ub_color_curve}, but for T~Pyx.
%%The $UBV$ data of the 1966 outburst are taken from \citet{egg67} and
%%\citet{lan70}.  The $BV$ data of the 2011 outburst are taken from
%%the AAVSO and SMARTS archive.  
%%The X-ray data are taken from the {\it Swift} web page \citep{eva09}.
The $(B-V)_0$ and $(U-B)_0$ are de-reddened with $E(B-V)=0.25$.
\label{t_pyx_v_bv_ub_color}}
\end{figure}

%Fig.20  
%\placefigure{t_pyx_pw_vul_nq_vul_dq_her_v_bv_ub_color_logscale_no6}

\begin{figure}
%\epsscale{0.75}
%%\epsscale{0.8}
%\epsscale{1.0}
\epsscale{1.15}
\plotone{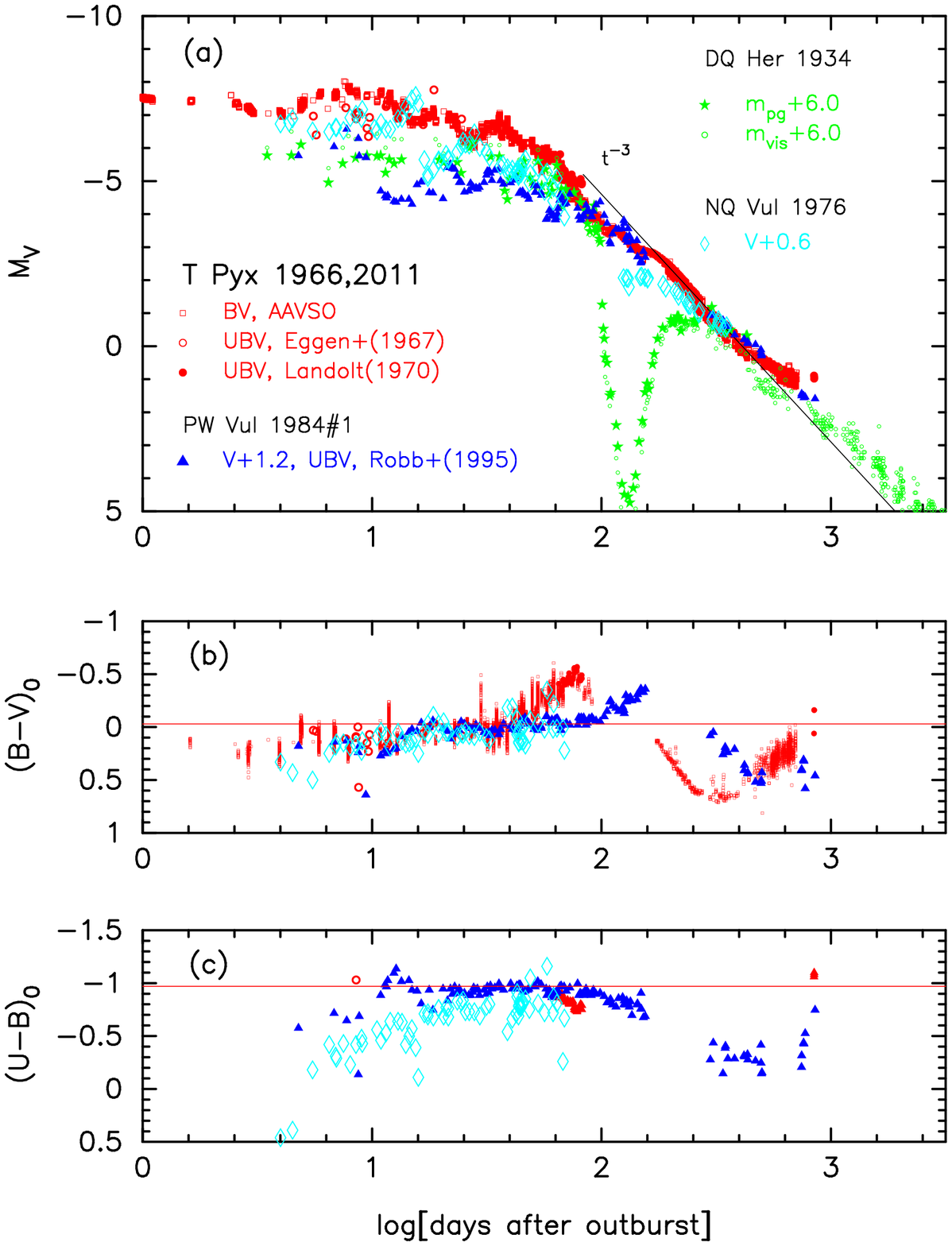}
%\plotone{t_pyx_pw_vul_nq_vul_dq_her_v_bv_ub_color_logscale_no6.epsi}
%\plotfiddle{evolution1.ps}{5.0cm}{270}{0.4}{0.4}{-170}{220}
\caption{
Comparison of the light curve of T~Pyx with those of
DQ~Her 1934, NQ~Vul 1976, and PW~Vul 1984\#1.
Note that the horizontal axis (time) is logarithmic.
The $UBV$ data of PW~Vul are taken
from \citet{rob95}. The $UBV$ data of NQ~Vul are taken from \citet{cha77},
\citet{lan77}, \citet{yam77}, and \citet{due79},
and the optical data of DQ~Her are from \citet{gap56}.  
%%See the main text for more detail.
\label{t_pyx_pw_vul_nq_vul_dq_her_v_bv_ub_color_logscale_no6}}
\end{figure}

\subsection{LV~Vul 1968\#1}
\label{lv_vul_cmd}
The reddening for LV~Vul was estimated to be $E(B-V)=0.60\pm0.05$
by matching the observed color-color track with the general course of
novae \citep{hac14k}.
Figure \ref{lv_vul_v_bv_ub_color} shows the $V$, $(B-V)_0$, 
and $(U-B)_0$ evolutions of LV~Vul.
The $(B-V)_0$ and $(U-B)_0$ colors are de-reddened with $E(B-V)=0.60$.
The $UBV$ data are taken from \citet{fer69}, \citet{dor69},
\citet{abu69}, and \citet{gry69}.  The $BV$ data are taken from
\citet{qua68} and \citet{tem72}.  
In Figure \ref{lv_vul_v_bv_ub_color}(b), the $B-V$ data of \citet{dor69}
are 0.05 mag redder than the other data, so we shifted them toward blue
by 0.05 mag.
LV~Vul reached its optical maximum at $m_{V,\rm max}=4.7$ on UT 1968 April 17.
The $V$ light curve of LV~Vul has $t_2=20.2$ and $t_3=37$ days \citep{tem72}.

We plot the color-color diagram of LV~Vul in Figure 
\ref{color_color_diagram_lv_vul_distance_reddening}(a),
using the data shown in Figure 
\ref{lv_vul_v_bv_ub_color}(b) and (c).
\citet{and86} recorded a spectrum at the pre-maximum phase 
and it showed a spectrum of an F-type supergiant star.  Therefore,
we expect that the early color evolution of LV~Vul follows
the nova-giant sequence in the color-color diagram.
We confirm that the adopted value of $E(B-V)=0.60$ is reasonable
because the observed track of LV~Vul is located on 
the general course of novae (solid green lines)
especially on the nova-giant sequence.

\citet{hac06kb} found that nova light curves follow a universal
decline law when free-free emission dominates the spectrum.
Using the universal decline law, \citet{hac10k} derived
that if two nova light curves overlap each other after one of
them  is squeezed/stretched by a factor
of $f_s$ ($t=t' \times f_s$) in the direction of time, one nova
brightness of ($m'_V, ~t'$) is related to the other nova brightness of
($m_V, ~t$) as $m_V = m'_V + 2.5 \log f_s$.
Using this result and the calibrated nova light curves, we are able to estimate
the absolute magnitude of a target nova.
They called this method the ``time-stretching method.''

\citet{hac14k} estimated the absolute magnitude of LV~Vul
as $(m-M)_V=11.9$, using the time-stretching method.
Then the distance is calculated to be $d=1.0$~kpc for $E(B-V)=0.60$,
which is consistent with $d=0.92\pm 0.08$~kpc obtained by \citet{sla95} 
from the expansion parallax method.
We plot the various distance-reddening relations for LV~Vul,
$(l, b)=(63\fdg3024, +0\fdg8464)$,
in Figure \ref{color_color_diagram_lv_vul_distance_reddening}(b).
Our set of $E(B-V)=0.60$ and $d=1.0$~kpc is located in a different area
than the relations of both Marshall et al. (2006) and Green et al. (2015).
Our result is roughly at midway between theirs.
We also plot the results of \citet{hak97}.  
Their distance-reddening relation is roughly
consistent with our set of $E(B-V)=0.60$ and $d=1.0$~kpc. 

Adopting $E(B-V)=0.60$ and $(m-M)_V=11.9$,
we plot the color-magnitude diagram of LV~Vul
in Figure \ref{hr_diagram_lv_vul_lv_vul_2fig_outburst}(a).
After the optical maximum, LV~Vul goes down almost
along the line of $(B-V)_0=-0.03$.
This is similar to the trend in V1668~Cyg. 
After that, the track of LV~Vul departs
into two tracks at $m_V=8.2$ (denoted by an open blue circle
at $M_V=8.2 - 11.9= -3.7$) in the color-magnitude diagram
of Figure \ref{hr_diagram_lv_vul_lv_vul_2fig_outburst}(a).
This reason will be clarified below
shortly after the explanation of the nebular phase.

The start of the nebular phase is identified by the first clear
appearance of the nebular emission lines [\ion{O}{3}] (or [\ion{Ne}{3}])
stronger than the permitted lines.   LV~Vul had already entered the nebular
phase on June 20 at $m_V=8.2$ \citep{hut70b}.  Thus, we specify the onset
of the nebular phase at $m_V=8.2$ when the track
departed into two tracks (solid black and magenta lines)
at the large open blue circle
in Figure \ref{hr_diagram_lv_vul_lv_vul_2fig_outburst}(a).
The presence of two tracks is due to the strong emission lines
of [\ion{O}{3}] contributing
to the blue edge of the $V$ filter.  Because the response of each $V$ filter
differs slightly at the shorter wavelength edge of the $V$ passband,
the resultant $V$ magnitude and color index $B-V$ is significantly
different among the different $V$ filters.
After the nebular phase started at $m_V=8.2~(M_V=-3.7)$,
this difference becomes more and more significant as shown in 
Figure \ref{lv_vul_v_bv_ub_color}(a) and 
\ref{lv_vul_v_bv_ub_color}(b).  Here,
one group (solid magenta line) are \citet{abu69}, \citet{dor69},
and \citet{gry69} and the other (solid black line)
are \citet{qua68}, \citet{fer69}, and \citet{tem72}.  These two trends
began to depart at $m_V=8.2~(M_V=-3.7)$
in the color-magnitude diagram of Figure 
\ref{hr_diagram_lv_vul_lv_vul_2fig_outburst}(a).
We also specify a turning (cusp) point at $(B-V)_0=-0.01$ and $M_V=-3.77$ 
(a large open red square) for the data of \citet{abu69} 
as shown in Figure \ref{hr_diagram_lv_vul_lv_vul_2fig_outburst}(a).

Figure \ref{hr_diagram_lv_vul_lv_vul_2fig_outburst}(b) compares
the track of LV~Vul with those of V1500~Cyg (thick solid green line)
and PU~Vul (thick solid blue line) in the color-magnitude diagram.
The track of LV~Vul ($t_2=20.2$~days) is between
V1500~Cyg ($t_2=2.4$~days) and PU~Vul ($t_2 > 1500$~days).
The tracks are located from left to right in the order of nova speed
class.  This is the same trend as that of Figure
\ref{hr_diagram_v1668_cyg_only_outburst}(b), as mentioned
in the previous section (Section \ref{v1668_cyg_cmd}).

The two-headed black arrow in Figure 
\ref{hr_diagram_lv_vul_lv_vul_2fig_outburst}(b)
is located 0.6 mag above the trend of the LV~Vul track just
after the nebular phase started.
Thus, we define another line by the two-headed red arrow 
in the same figure.
The two-headed black arrow is defined by Equation 
(\ref{absolute_magnitude_cusp}) and the two-headed red arrow 
is defined by Equation (\ref{absolute_magnitude_cusp_low_red}),
both of which will be discussed later in Section 
\ref{properties_color_magnitude_diagram}.  Their physical meaning 
will be clarified there.

\subsection{FH~Ser 1970}
\label{fh_ser_cmd}
FH~Ser shows a dust blackout.  The $V$ light curve of FH~Ser has
$t_2=42$ and $t_3=59$~days \citep[e.g.,][]{dow00}.
The light curves of FH~Ser were already analyzed in Paper I
based mainly on the color-color evolution.
\citet{hac14k} determined the reddening to be $E(B-V)=0.60$ and the
distance modulus to be $(m-M)_V=11.7$.  
Using the same data as those in Figure 2 of \citet{hac14k},
which showed the $V$, $(B-V)_0$, and $(U-B)_0$ evolutions
of FH~Ser, we plotted the color-magnitude diagram of FH~Ser in Figure
\ref{hr_diagram_fh_ser_pw_vul_v1500_cyg_v1974_cyg_outburst}(a).
We adopted the same reddening and distance modulus as \citet{hac14k}. 
The $UBV$ data are taken from \citet[][filled red circles]{osa70},
\citet[][open magenta diamonds]{bor70}, and \citet[][open blue squares]{bur71}.
The $B-V$ color of \citet{bor70} is systematically
$\sim0.2$~mag bluer than the others, 
while their $V$ and $U-B$ data are 
reasonably consistent with the others.  Therefore,
we shifted Borra \& Andersen's $B-V$ data 0.2 mag redder 
\citep[see also Figure 2 of][]{hac14k}. 
Connecting main observational points in the color-magnitude diagram, we 
define a template track (thick solid green line) for FH~Ser in 
Figure \ref{hr_diagram_fh_ser_pw_vul_v1500_cyg_v1974_cyg_outburst}(a). 

Figure \ref{hr_diagram_fh_ser_pw_vul_v1500_cyg_v1974_cyg_outburst}(a) 
also shows stages at the $V$ maximum, $m_{V,\rm max}$,
and 2 mag below the $V$ maximum, $m_{V,\rm max}+2$, by the
thin horizontal solid lines.  
FH~Ser first rises in the color-magnitude diagram and then turns
to the right.  It goes toward red up to $(B-V)_0 \approx +0.60$ at point A
in Figure \ref{hr_diagram_fh_ser_pw_vul_v1500_cyg_v1974_cyg_outburst}(a). 
Then, it turns back to the left, toward blue, and reaches maximum
at $m_V=4.4~(M_V=4.4-11.7=-7.3)$.  
Subsequently, it declines along the template track
from point B to D through C.
After point D, the nova suddenly darkened due to formation of an
optically thick dust shell.
We consider the start of the dust blackout (large open red square) to be
$(B-V)_0=+0.05$ and $M_V=-4.48$ as shown in Figure
\ref{hr_diagram_fh_ser_pw_vul_v1500_cyg_v1974_cyg_outburst}(a). 

Figure \ref{hr_diagram_fh_ser_pw_vul_v1500_cyg_v1974_cyg_outburst_no2}(a)
compares the color-magnitude diagram of FH~Ser with those of
V1668~Cyg (thin solid green lines), PW~Vul (thin solid blue lines), 
and PU~Vul (thick solid blue lines), which are analyzed in
Sections \ref{v1668_cyg_cmd}, \ref{pw_vul_cmd}, and \ref{pu_vul_cmd},
respectively.  
The location of FH~Ser ($t_2=42$~days) is between those of
V1668~Cyg ($t_2=12.2$~days) and PU~Vul ($t_2 \gtrsim 1500$~days).
V1668~Cyg declines vertically along $(B-V)_0=-0.03$
\citep[see, e.g.,][]{hac14k,hac16k},
which is the color of optically thick free-free emission calculated
from $F_\nu \propto \nu^{2/3}$, whereas PU~Vul goes down along 
$(B-V)_0=+0.13$  \citep[see, e.g.,][]{hac14k}, the color of optically
thin free-free emission calculated from $F_\nu \propto \nu^0$.
If we shift the track of V1668~Cyg toward red by $\Delta(B-V)=0.12$ mag
as shown by the thick solid orange line with black points, it
overlaps with that of FH~Ser between $M_V\sim-6.5$ and $M_V\sim-4$.

We examine the combination of $E(B-V)=0.60$ and $(m-M)_V=11.7$
in the distance-reddening relation for FH~Ser,
$(l, b)=(32\fdg9090,+5\fdg7860)$.
Figure \ref{distance_reddening_fh_ser_pw_vul_v1500_cyg_v1974_cyg}(a)
shows various distance-reddening relations for FH~Ser.
We show those given by \citet{mar06}, 
$(l, b)=(32\fdg75,5\fdg75)$ (open red squares),
$(33\fdg00,5\fdg75)$ (filled green squares),
$(32\fdg75,6\fdg00)$ (blue asterisks),
and $(33\fdg00,6\fdg00)$ (open magenta circles).
The closest one is that of the filled green squares.
The solid blue line represents the relation of $(m-M)_V=11.7$,
crossing the trend of the filled green squares of Marshall et al.\
at $d\approx0.93$~kpc and $E(B-V)\approx0.60$.  
This point is consistent with our adopted values.
This figure is essentially the same as Figure 3 of \citet{hac14k}, but
we added the distance-reddening relation (solid black line)
given by \citet{gre15}.  Green et al.'s relation is located at a slightly
lower position than that of Marshall et al.
Considering the ambiguity of Green et al.'s relation
(see Section \ref{v1668_cyg_cmd}), we may conclude that
the combination of $E(B-V)=0.60$ and $d=0.93$~kpc is still
reasonably consistent with these distance-reddening relations.

\subsection{PW~Vul 1984\#1}
\label{pw_vul_cmd}
The light curves of the moderately fast nova PW~Vul were studied
in detail by \citet{hac14k, hac15k}.
They determined the distance modulus to be $(m-M)_V = 13.0$
and the reddening to be $E(B-V)=0.55$. 
Figure \ref{hr_diagram_fh_ser_pw_vul_v1500_cyg_v1974_cyg_outburst}(b) 
shows the outburst track of PW~Vul in the color-magnitude diagram,
the data of which are taken from \citet[][open blue squares]{nos85}, 
\citet[][open magenta diamonds]{kol86}, 
and \citet[][filled red circles]{rob95}.
We define a template track by a thick
solid green line for PW~Vul almost along Robb \& Scarfe's
observation in Figure 
\ref{hr_diagram_fh_ser_pw_vul_v1500_cyg_v1974_cyg_outburst}(b).

The $V$ light curve of PW~Vul shows a wavy structure in the early
decline phase (see, e.g., Figure 6 of Paper I).
The brightness drops to $m_V=8.7$ immediately
after the $V$ maximum ($m_{V,\rm max}=6.4$).
Then it goes up to $m_V=7.5$ and repeats oscillations with 
smaller amplitudes of brightness.
The smoothed $V$ light curve of PW~Vul has $t_2=82$
and $t_3=126$~days \citep[e.g.,][]{dow00}. 
PW~Vul moves clockwise in the color-magnitude diagram
during this first brightness drop just after the $V$ maximum,
the movement direction of which is indicated by red arrows in
Figure \ref{hr_diagram_fh_ser_pw_vul_v1500_cyg_v1974_cyg_outburst}(b).
This clockwise movement is different from the usual nova decline, like
in Figure \ref{hr_diagram_fh_ser_pw_vul_v1500_cyg_v1974_cyg_outburst}(a),
and we will discuss it in more detail
in Sections \ref{v5558_sgr_cmd} and \ref{summary_basic_properties_cmd}.

\citet{ros87} reported that the nova entered the nebular phase
at mid January 1985 (at $m_V=9.8$) as indicated in Figure
\ref{hr_diagram_fh_ser_pw_vul_v1500_cyg_v1974_cyg_outburst}(b).
This onset corresponds to the large open red square on the track,
i.e., $M_V=-3.32$ and $(B-V)_0=-0.39$.
The track of PW~Vul shows a small bend globally and a tiny zigzag
motion locally near this point.
A possible orbital period of 5.13~hr was detected by \citet{hack87}.

Figure \ref{hr_diagram_fh_ser_pw_vul_v1500_cyg_v1974_cyg_outburst_no2}(b)
shows the position of PW~Vul among the tracks of other novae.
We also plot two-headed black and red arrows represented by 
Equations (\ref{absolute_magnitude_cusp}) and 
(\ref{absolute_magnitude_cusp_low_red}), respectively.
The onset of the nebular phase (large open red square)
is located on the two-headed red arrow.
The track of PW~Vul is very close to that of LV~Vul except for
the early clockwise circle.  We regard PW~Vul as the same type
of nova as LV~Vul in the color-magnitude diagram. 

The five template tracks are located from left to right depending on 
the nova speed class, that is, $t_2=2.4$, 12.2, 20.2, 42, 
and $\gtrsim 1500$~days for V1500~Cyg, V1668~Cyg, LV~Vul, FH~Ser,
and PU~Vul, respectively.  This trend is the same as 
in Section \ref{v1668_cyg_cmd}.  Note that the $t_2=82$~days of PW~Vul
is much longer than $t_2=20.2$~days of LV~Vul.  The reason is that
the early $V$ light curve of PW~Vul has a wavy structure with a large
amplitude of the $V$ magnitude and the $t_2$ time could not represent
the intrinsic nova speed class.  On the other hand, other novae
(V1500~Cyg, V1668~Cyg, LV~Vul, FH~Ser, and PU~Vul) show smooth declines
and their $t_2$ times could show their intrinsic nova speed class.   

We check the combination of $E(B-V)=0.55$ and $(m-M)_V=13.0$
in the distance-reddening relation for PW~Vul, 
$(l, b)=(61\fdg0983,+5\fdg1967)$.
Figure \ref{distance_reddening_fh_ser_pw_vul_v1500_cyg_v1974_cyg}(b) shows
various distance-reddening relations for PW~Vul.
Marshall et al.'s (2006) relations are plotted
in four directions close to PW~Vul:
$(l, b)=(61\fdg00,5\fdg00)$ (open red squares),
$(61\fdg25,5\fdg00)$ (filled green squares),
$(61\fdg00,5\fdg25)$ (blue asterisks), and
$(61\fdg25,5\fdg25)$ (open magenta circles).
The closest one is that of blue asterisks.
We also add Green et al.'s (2015) relation (thick solid black line).

\citet{hac15k} calculated model $V$ and UV~1455 \AA\  
light curves for various WD masses and chemical compositions
of the hydrogen-rich envelope and obtained a best fit model
for a $0.83~M_\sun$ WD with the chemical composition
of CO nova 4 \citep[see Figure 10 of][]{hac15k}.
Their UV~1455 \AA\  fit together with Equation
(\ref{uv1455_distance_modulus}) is plotted
by a solid magenta line.
The $V$ light curve fit is the same as our value of $(m-M)_V = 13.0$.
This relation is plotted by a solid blue line.

The three trends, UV~1455 \AA\  fit, $E(B-V)=0.55$, and $(m-M)_V=13.0$,
consistently cross at the point $E(B-V)\approx0.55$ and 
$d\approx1.8$~kpc, but this cross point is not consistent with the trends of
Marshall et al. and Green et al.  
If we adopt $d=1.8$~kpc, we obtain $E(B-V)\approx0.42$ from Marshall 
et al.'s relation of blue asterisks. 
This value is consistent with $E(B-V)=0.43\pm0.02$ calculated
from the NASA/IPAC dust map in the direction of PW~Vul.
Thus, our set of $E(B-V)=0.55$ and $(m-M)_V=13.0$ is not consistent with
the trends of the 3D dust map.

Therefore, we again discuss previous reddening estimates pinpointing
PW~Vul.  The reddening of PW~Vul was estimated as
$E(B-V)=A_V/3.1 = (1.78\pm0.05)/3.1 = 0.57\pm0.02$ from the
Pa$\beta$ and Pa$\gamma$ line strengths compared with H$\beta$ and
H$\gamma$ line strengths by \citet{wil96},
$E(B-V)=0.58 \pm 0.06$ from the \ion{He}{2} $\lambda1640/\lambda4686$
ratio, and $E(B-V)=0.55 \pm 0.1$ from the interstellar absorption feature
at 2200 \AA\ both by \citet{and91},
$E(B-V)=0.60 \pm 0.06$ according to \citet{sai91} from the \ion{He}{2}
$\lambda1640/\lambda4686$ ratio.
The simple arithmetic mean of these four values is $E(B-V)=0.57 \pm 0.1$.
The distance to PW~Vul was
estimated to be $d=1.8 \pm 0.05$~kpc by \citet{dow00} from the
expansion parallax method.  We plot this by horizontal black lines
in Figure \ref{distance_reddening_fh_ser_pw_vul_v1500_cyg_v1974_cyg}(b).
Then the distance modulus in the $V$ band is 
calculated to be $(m-M)_V = 5 \log (1800/10) + 3.1\times0.57= 13.04$.
Our combination of $d=1.8$~kpc and $(m-M)_V =13.0$ is consistent with
these estimates. 
The reddening trend of Marshall et al.'s blue asterisks suggests a large
deviation from the other three trends by $\Delta E(B-V)\approx0.1$, 
suggesting that the reddening distribution has a patchy structure
in this direction and a further deviation of $\Delta E(B-V)\sim0.1$
may be possible.  Thus, we use the set of $E(B-V)\approx0.55$ and 
$d\approx1.8$~kpc in this paper.

\subsection{V1500~Cyg 1975}
\label{v1500_cyg_cmd}
V1500~Cyg was identified as a neon nova by \citet{fer78a,fer78b}.
The $V$ light curve shows a very rapid decline with $t_2=2.4$ and 
$t_3=3.7$~days \citep[e.g.,][]{dow00}. 
The orbital period of 3.35~hr was detected by \citet{tem75}.
We already analyzed the nova light curves in Paper I, and determined
the reddening to be $E(B-V)=0.45\pm0.05$ and the distance modulus
to be $(m-M)_V=12.3\pm0.1$.  Adopting their values of  $E(B-V)=0.45$ and 
$(m-M)_V=12.3$, we plot the color-magnitude diagram of V1500~Cyg
in Figures \ref{hr_diagram_fh_ser_pw_vul_v1500_cyg_v1974_cyg_outburst}(c),
the data of which are taken from \citet{ark76}, \citet{pfa76},
\citet{kis77}, and \citet{wil77}.  

The spectrum energy distribution changed from blackbody emission
during the first 3 days to 
thermal bremsstrahlung emission on day $\sim 4-5$ \citep{gal76,enn77}.
Thus, we conclude that the nova enters a free-free emission phase
about 5 days after the outburst.
Optically thin free-free emission ($F_\nu \propto \nu^0$)
yields $(B-V)_0=+0.13$ whereas
optically thick free-free emission ($F_\nu \propto \nu^{2/3}$)
gives $(B-V)_0=-0.03$, both of which are indicated in Figure 
\ref{hr_diagram_fh_ser_pw_vul_v1500_cyg_v1974_cyg_outburst}(c).

In Figure \ref{hr_diagram_fh_ser_pw_vul_v1500_cyg_v1974_cyg_outburst}(c),
different observers obtained different tracks.
The data of \citet{pfa76} are $\sim0.2$ mag bluer than that
of the solid green line based mainly on the data of \citet{kis77}. 
This difference is partly attributed to slight difference in the response
of each color filter which is sensitive to strong emission lines on 
the edge and eventually makes the nova color significantly different
among the observers.
We define a template track of V1500~Cyg by the thick solid green line,
which is based mainly on the data taken from \citet{kis77}. 

The first feature of the nebular, forbidden, lines of 
[\ion{O}{3}] and [\ion{Ne}{3}], appeared on September 8, 1975, 
at $m_V=6.1$ and their intensities steadily increased and reached that 
of H$\beta$ on October 12.9, at $m_V=7.9$ \citep[e.g.,][]{wos75}.
We specify that the nebular phase started around October 13.0--14.0,
that is, $M_V=m_V-(m-M)_V=7.9-12.3=-4.4$.
The two tracks of \citet{wil77}
and \citet{pfa76} began to diverge at $M_V=m_V-(m-M)_V=6.1-12.3=-6.2$
as shown in Figure 
\ref{hr_diagram_fh_ser_pw_vul_v1500_cyg_v1974_cyg_outburst}(c).
This is due to the contribution of strong emission lines [\ion{O}{3}] 
close to the blue edge of the $V$ filter.  
After the nebular phase started at $M_V\sim-4.4$, this difference 
develops more and more.  The strong emission lines of [\ion{O}{3}] 
eventually made the nova color evolution turn to the right
for the three cases of \citet{wil77}, \citet{ark76}, and \citet{kis77}
near the cusp (or zigzag) point denoted by a large open red square.
This position is $M_V=-3.83$ ($m_V=8.5$) and $(B-V)_0=-0.47$
based on the data (filled red circles) of \citet{ark76}.

Figure \ref{hr_diagram_fh_ser_pw_vul_v1500_cyg_v1974_cyg_outburst_no2}(c)
shows the position of V1500~Cyg among other well-observed novae. 
The track of V1500~Cyg is located on the bluest side
in the color-magnitude diagram.  
The onset of the nebular phase (large open red square)
is located on this two-headed arrow.  

The distance to V1500~Cyg was discussed by many authors (see Paper I). 
We obtained the set of $E(B-V)=0.45$ and $(m-M)_V=12.3$ in Paper I.
Figure \ref{distance_reddening_fh_ser_pw_vul_v1500_cyg_v1974_cyg}(c)
shows various distance-reddening relations toward V1500~Cyg, 
$(l, b)=(89\fdg8233,-0\fdg0720)$.
We add Marshall et al.'s (2006) relations 
for four directions close to V1500~Cyg,
that is, $(l, b)=(89\fdg75, 0\fdg00)$ (open red squares),
$(90\fdg00,  0\fdg00)$ (filled green squares),
$(89\fdg75, -0\fdg25)$ (blue asterisks), and
$(90\fdg00, -0\fdg25)$ (open magenta circles).
We further add the relations of \citet{hak97} and \citet{gre15}. 
Green et al.'s relation gives $d=1.5$~kpc for $E(B-V)=0.45$.

We also plot the two relations $(m-M)_V=12.3$ (solid blue line)
and $E(B-V)=0.45$ (vertical solid red line), which are taken from
\citet{hac14k}.  These two relations and Green et al.'s relation cross 
each other at the same point of $d\approx1.5$~kpc and 
$E(B-V)\approx0.45$, consistent with the distance
estimate of $d=1.5\pm0.1$~kpc by the expansion parallax method.
This strongly supports Hachisu \& Kato's (2014) set of 
$(m-M)_V=12.3$ and $E(B-V)=0.45$.

\subsection{V1974~Cyg 1992}
\label{v1974_cyg_cmd}
V1974~Cyg was identified as a neon nova by \citet{hay92}.
The $V$ light curve shows a fast decline with 
$t_2=17$ and $t_3=37$~days \citep[e.g.,][]{dow00}. 
The orbital period of 1.95~hr was detected by \citet{dey94}.
Paper I and \citet{hac16k} already analyzed the nova light curves
on the basis of the universal decline law and determined
the reddening as $E(B-V)=0.30\pm0.05$ and the distance modulus as 
$(m-M)_V=12.2\pm0.1$.  Adopting their values of  $E(B-V)=0.30$ and 
$(m-M)_V=12.2$, we plot the color-magnitude diagram of V1974~Cyg
in Figure \ref{hr_diagram_fh_ser_pw_vul_v1500_cyg_v1974_cyg_outburst}(d).
The observational data are
taken from \citet{cho93} and IAU Circular Nos.\ 5455, 5457, 5459, 5460, 5463,
5467, 5475, 5479, 5482, 5487, 5490, 5520, 5526, 5537, 5552, 5571, and 5598.
We define a template track of V1974~Cyg by the thick solid green line,
based mainly on the data of \citet{cho93}.

The nova entered the nebular phase on April 20 
at $m_V=8.1$ \citep{raf95}, as indicated by an arrow in Figure 
\ref{hr_diagram_fh_ser_pw_vul_v1500_cyg_v1974_cyg_outburst}(d).
The position is denoted by a large open red square at
$M_V=-4.02$ and $(B-V)_0=-0.21$.
This point is taken from the observational points of \citet{cho93}
and corresponds to a cusp on the track.
We list the values at the cusp in Table \ref{color_magnitude_turning_point}.
We add the epoch when the supersoft X-ray source (SSS) phase started
at $m_V=9.7$ about 250 days after the outburst \citep{kra96}.

Figure \ref{hr_diagram_fh_ser_pw_vul_v1500_cyg_v1974_cyg_outburst_no2}(d)
compares the position of V1974~Cyg with other well-observed novae. 
The track of V1974~Cyg is located between those of V1500~Cyg and FH~Ser.
The data of V1974~Cyg taken from IAU Circulars (filled red circles) 
scatter slightly and their blue edge (blue side bound) coincides
with the template track of V1500~Cyg whereas their red side edge
almost coincides with the template track of FH~Ser.
We shift the track of V1500~Cyg toward red by $\Delta(B-V)=0.12$ mag
and plot it by a thin solid black line, which
overlaps with that of V1974~Cyg between $M_V=-8$ and $M_V=-4$.
The onset of the nebular phase (large open red square)
is located on the two-headed black arrow.

Figure \ref{distance_reddening_fh_ser_pw_vul_v1500_cyg_v1974_cyg}(d)
shows various distance-reddening relations for V1974~Cyg,
$(l, b)=(89\fdg1338, 7\fdg8193)$.
We plot our distance modulus of $(m-M)_V=12.2$ by a thick solid blue line
and the UV~1455 \AA\  light curve fit by a thick solid magenta line
\citep{hac16k}.  
We added Chochol et al.'s distance value as
$d=1.8\pm0.1$~kpc (horizontal straight solid black line flanked
by thin lines).  Marshall et al.'s (2006) relations are given for
$(l, b)=(89\fdg00,7\fdg75)$ (open red squares),
$(89\fdg25, 7\fdg75)$ (filled green squares),
$(89\fdg00,  8\fdg00)$ (blue asterisks),
and $(89\fdg25,  8\fdg00)$ (open magenta circles).
We also add Green et al.'s (2015) relation by a thick solid black line.
These trends almost cross at the same point of
$E(B-V)\approx0.30$ and $d\approx1.8$~kpc.
This agreement strongly supports our set of $(m-M)_V=12.2$ and
$E(B-V)=0.30$.

\subsection{PU~Vul 1979}
\label{pu_vul_cmd}
PU~Vul is a symbiotic nova with an orbital period of 13.46~yr 
\citep[e.g.,][]{kat12mh}.
Figure \ref{distance_reddening_pu_vul_v723_cas_hr_del_v5558_sgr}(a) shows
various distance-reddening relations for PU~Vul,
$(l, b)= (62\fdg5753, -8\fdg5317)$.
\citet{kat12mh} examined the distance-reddening relation for PU~Vul
with several different methods and determined the reddening as
$E(B-V)=0.30$, the apparent distance modulus in the $V$ band
as $(m-M)_V=14.3$, and the distance as $d=4.7$~kpc.
We plot these results in Figure
\ref{distance_reddening_pu_vul_v723_cas_hr_del_v5558_sgr}(a) by
the vertical solid red line, solid blue line, and horizontal black line,
respectively.  \citet{hac14k} reanalyzed the light curve and color-color 
evolution of PU~Vul and reached the same conclusion, i.e., $E(B-V)=0.30$ and
$(m-M)_V=14.3$.  
The NASA/IPAC galactic dust absorption map gives $E(B-V)=0.29 \pm 0.01$
in the direction toward PU~Vul. 
We also add Marshall et al.'s (2006) and Green et al.'s (2015) 
distance-reddening relations to Figure
\ref{distance_reddening_pu_vul_v723_cas_hr_del_v5558_sgr}(a).
Green et al.'s trend consistently crosses our solid lines at $d=4.7$~kpc
and $E(B-V)=0.30$, which supports the values of $E(B-V)=0.30$ and
$(m-M)_V=14.3$.

Figure \ref{hr_diagram_pu_vul_v723_cas_hr_del_v5558_sgr_outburst}(a) shows
the color-magnitude track of PU~Vul as well as the template track of 
LV~Vul (thick solid magenta line).  Here, we use $E(B-V)=0.30$
and $(m-M)_V=14.3$ for PU~Vul.  The solid green line represents
the track of LV~Vul, but is shifted toward red by $\Delta (B-V)=0.2$.
The data of the small open orange circles are taken from \citet{shu12}.
The other data of small filled red circles are taken from various
sources but are the same as in Figures 16 and 17 of Paper I.  
We define a template track of PU~Vul by a thick solid blue line.
The numbers 1 -- 5 attached to large open black squares on the solid blue line
correspond to the stages 1 -- 5 of PU~Vul as defined in Figure 15 of Paper I.
The figure also shows the stages at the $V$ maximum, $m_{V,\rm max}$, and 
2 mag below the $V$ maximum, $m_{V,\rm max}+2$,
by thin horizontal solid lines.  It is remarkable 
that the green shifted LV~Vul track almost coincides with the PU~Vul 
track except for the flat optical peak of
PU~Vul (see Figure 15 of Paper I for the $V$ light curve of PU~Vul).

\citet{vog92} reported that a distinct nebular spectrum
emerged between September 1989 ($m_V\approx10.4$, $M_V\approx -3.9$)
and November 1990 ($m_V\approx10.6$, $M_V\approx -3.7$).
We consider this to be the nova entering the nebular phase at 
$m_V\approx10.5$ ($M_V\approx -3.8$), which is denoted by an arrow
in Figure \ref{hr_diagram_pu_vul_v723_cas_hr_del_v5558_sgr_outburst}(a).
Here we specify this onset point by a large open red square
at $(B-V)_0=+0.20$ and $M_V=-3.80$.   
The onset of the nebular phase (large open red square)
is located on the two-headed red arrow.
This position accidentally coincides with the onset point of
the nebular phase on the green shifted LV~Vul track as shown in Figure
\ref{hr_diagram_lv_vul_lv_vul_2fig_outburst}(a).

\subsection{V723~Cas 1995}
\label{v723_cas_cmd}
V723~Cas is a very slow nova with an orbital period
of 16.64~hr \citep{gor00}.  
Figure \ref{distance_reddening_pu_vul_v723_cas_hr_del_v5558_sgr}(b) shows
various distance-reddening relations for V723~Cas,
$(l, b)= (124\fdg9606, -8\fdg8068)$.
The NASA/IPAC galactic dust absorption map gives $E(B-V)=0.34 \pm 0.01$
in the direction toward V723~Cas. 
We plot the distance-reddening relation given by \citet{gre15}
by a solid black line, the apparent distance modulus in the $V$ band
of $(m-M)_V=14.0$ \citep{hac15k} by a solid blue line,
the reddening of $E(B-V)=0.35\pm0.05$ from Paper I by a vertical
solid red line flanked with dashed red lines,
the distance of $d=3.85^{+0.23}_{-0.21}$~kpc
from the expansion parallax method \citep{lyk09} by a horizontal
solid black line flanked with dashed lines, and the UV~1455 \AA\  
model light curve fit \citep{hac15k} by a solid magenta line.
All these trends cross
each other at $E(B-V)\approx0.35$ and $d\approx3.85$~kpc.
Therefore, we adopt $E(B-V)=0.35$ and $(m-M)_V= 5 \log(3850/10) +
3.1\times0.35=14.0$ for V723~Cas after \citet{hac15k}. 

Figure \ref{hr_diagram_pu_vul_v723_cas_hr_del_v5558_sgr_outburst}(b) shows
the color-magnitude track of V723~Cas.
The data of V723~Cas are taken from \citet{cho97b}. 
The solid green line represents the LV~Vul track
shifted toward red by $\Delta (B-V)=0.2$.
The PU~Vul and green shifted LV~Vul tracks are remarkably similar
to that of V723~Cas except for the flaring pulses around optical maximum
of $M_V=-7$ to $-6$.  See Figures 19, 20, and 21 of Paper I
for the $V$ light curve and color curves.

The onset of the nebular phase was detected by \citet{iij06} 
between May 30 %JD 2450598.5 
and July 1, %JD 2450630.5
1997, at $m_V\approx11.3$,  % $m_V\approx11.4$,
as shown in Figure 
\ref{hr_diagram_pu_vul_v723_cas_hr_del_v5558_sgr_outburst}(b).
We specify the point $(B-V)_0=-0.17$ and $M_V=-2.83$
from the observational data in Figure 
\ref{hr_diagram_pu_vul_v723_cas_hr_del_v5558_sgr_outburst}(b), 
and denote it by a large open red square.
The start of the nebular phase is slightly below the line of
the two-headed red arrow.

\subsection{HR~Del 1967}
\label{hr_del_cmd}
HR~Del is a very slow nova with an orbital period of 
5.14~hr \citep{bru82, kur88}.  The light curve shape is very
similar to that of V723~Cas and V5558~Sgr
(see, e.g., Figures 19, 20, and 21 of Paper I).
Figure \ref{distance_reddening_pu_vul_v723_cas_hr_del_v5558_sgr}(c) shows
several distance-reddening relations,
the apparent distance modulus in the $V$ band of $(m-M)_V=10.4$
\citep{hac15k}, the reddening of $E(B-V)=0.15\pm0.03$ \citep{ver87},
and the distance of $d=0.97\pm0.07$~kpc \citep{har03}.
We also add Green et al.'s (2015) relation.
All the trends consistently cross at $E(B-V)\approx0.12$
and $d\approx1.0$~kpc.  The NASA/IPAC galactic dust absorption map
also gives $E(B-V)=0.112 \pm 0.006$ in the direction toward HR~Del, 
$(l, b)= (63\fdg4304, -13\fdg9721)$.
Therefore, we adopt $E(B-V)=0.12$ and $(m-M)_V=10.4$ for HR~Del.

Figure \ref{hr_diagram_pu_vul_v723_cas_hr_del_v5558_sgr_outburst}(c)
shows the color-magnitude track of HR~Del.
The data of HR~Del are taken from \citet{oco68}, \citet{man70}, 
\citet{bar70}, and \citet{ond68}.   
The track of HR~Del is very similar to that of PU~Vul 
except for the flaring pulses around the optical peak of HR~Del.
The track of HR~Del also follows the green LV~Vul track 
shifted toward red by $\Delta (B-V)=0.2$.

The start of the nebular phase was identified by \citet{hut70a} 
at $(B-V)_0=-0.02$ and $M_V=-3.59$, which is indicated by a large
open red square in Figure 
\ref{hr_diagram_pu_vul_v723_cas_hr_del_v5558_sgr_outburst}(c).
The track of HR~Del departs into two branches at this point,
depending on the different filter responses of various observers,
just as for LV~Vul. One of them turns to the right (toward red)
when the nebular phase started.
This departing point accidentally coincides with that of
the green shifted LV~Vul track. 
The start of nebular phase is almost on the line of
the two-headed red arrow.

\subsection{V5558~Sgr 2007}
\label{v5558_sgr_cmd}
V5558~Sgr is a very slow nova.  Its light curve shape is very
similar to that of V723~Cas and HR~Del
(see, e.g., Figures 19, 20, and 21 of Paper I). 
Figure \ref{distance_reddening_pu_vul_v723_cas_hr_del_v5558_sgr}(d) shows
various distance-reddening relations for V5558~Sgr,
$(l, b)= (11\fdg6107, -0\fdg2067)$.
This figure is the same as Figure 24 of Paper I, but we added Green
et al.'s (2015) relation.
All the trends consistently cross at $E(B-V)\approx0.70$ and
$d\approx2.2$~kpc.  Therefore, we adopt $E(B-V)=0.70$ and $(m-M)_V=13.9$,
the same values as in Paper I.

Figure \ref{hr_diagram_pu_vul_v723_cas_hr_del_v5558_sgr_outburst}(d) shows
the color-magnitude track of V5558~Sgr.
The data of V5558~Sgr are taken from the archives of the American
Association of Variable Star Observers (AAVSO, filled red circles),
the Variable Star Observers League of Japan (VSOLJ, filled red circles),
and SMARTS\footnote{http://www.astro.sunysb.edu/fwalter/SMARTS/NovaAtlas/}
\citep{wal12} (filled blue squares).

The track of V5558~Sgr is very similar to that of PU~Vul, 
except for the flaring pulses around the optical peak.
We connect the track of V5558~Sgr by a thin solid red line
during the first flaring pulse around the optical maximum.
Also, we connect the track during the second flaring pulse by a thin
solid black line.  The first flaring pulse shows a clockwise movement
in the color-magnitude diagram.  This clockwise behavior is very
similar to that of PW~Vul in Figure
\ref{hr_diagram_fh_ser_pw_vul_v1500_cyg_v1974_cyg_outburst}(b).
This clockwise movement, however, is very large in V5558~Sgr.
It rises vertically along $(B-V)_0\sim0.2$ up to the maximum brightness
($m_V\sim 6.5$, $M_V\sim-7.4$), then turns to the right (toward red) up to 
$(B-V)_0\sim0.6$, and then goes down to $M_V\sim-5.8$, coinciding
with the flat maximum brightness of PU~Vul.
This large clockwise circular movement
in the color-magnitude diagram is very different from the usual
tracks of novae such as FH~Ser and LV~Vul.
The track of the second flaring pulse follows that of
the first flaring pulse in the rising phase but does not go toward red.
It loops on the blue side of the first
flaring pulse.  The third and fourth flaring pulses follow
the second flaring pulse and these tracks almost overlap 
the track of the second flaring pulse.  They are
very similar to the loop of PW~Vul in Figure
\ref{hr_diagram_fh_ser_pw_vul_v1500_cyg_v1974_cyg_outburst}(b).

The start of the nebular phase was identified by \citet{pog12} at 
$(B-V)_0=-0.16$ and $M_V=-3.46$, which is denoted by a large open red square 
in Figure \ref{hr_diagram_pu_vul_v723_cas_hr_del_v5558_sgr_outburst}(d).
The track of V5558~Sgr departs into two branches at this point,
depending on the different filter responses of various observers
just as in LV~Vul and HR~Del. One of them turns to the right (toward red)
when the nebular phase started.
The start of the nebular phase is located on the line of
the two-headed red arrow.

\subsection{Categorization of color-magnitude tracks}
\label{summary_basic_properties_cmd}
It is clear that there is no single track common to all novae
in the color-magnitude diagram.  
This is in contrast to the general track in the color-color diagram.
In Paper I, we found that, in the color-color diagram, 
novae generally go down along the nova-giant sequence 
in the pre-maximum phase and then come back after the optical maximum
as shown in Figure
\ref{color_color_v1668_cyg_distance_reddening_x45z02c15o20}(a)
(see also Figures 4 and 8 of Paper I).  
Fast novae tend to have short excursions and slow novae 
tend to have long journeys to their peaks along the nova-giant sequence.

Figure \ref{hr_diagram_6types_novae_one} collects the templates of
the color-magnitude diagrams for the very fast nova V1500~Cyg,
fast novae V1668~Cyg, V1974~Cyg, and  LV~Vul, 
moderately fast nova FH~Ser, and symbiotic (very slow) nova PU~Vul. 
Because we excluded novae such as PW~Vul, which shows oscillatory behavior
around the optical peak, all these six novae show smooth declines
from their optical peaks.
The behaviors of these six novae in the color-magnitude diagram
are summarized as follows: 
(1)~After the optical peak, each nova generally evolves toward blue
from the upper-right to the lower-left along similar but different tracks.
(2)~Each track is located from the left (blue) to right (red) depending
on the nova speed class, i.e., $t_2=2.4$, 12.2, 17, 20.4, 42, 
$\gtrsim1500$~days for V1500~Cyg, V1668~Cyg, V1974~Cyg,
LV~Vul, FH~Ser, and PU~Vul, respectively.
Thus, we propose six templates of smooth decline nova tracks, i.e.,
V1500~Cyg, V1668~Cyg, V1974~Cyg, LV~Vul, FH~Ser, and PU~Vul.

In the previous subsections, we examined the behaviors of ten novae
in the color-magnitude diagram.
The LV~Vul track almost overlaps that of PW~Vul except for 
the early pulse (a loop).  The track of PU~Vul almost overlaps those
of V723~Cas, HR~Del, and V5558~Sgr except for the early pulses (loops).
We may call PW~Vul a LV~Vul type track because the track of
PW~Vul is very close to the template of LV~Vul.  We also call
V723~Cas, HR~Del, and V5558~Sgr PU~Vul type tracks.
Thus, we define six types of nova tracks in the color-magnitude
diagram, i.e., the V1500~Cyg, V1668~Cyg, V1974~Cyg, LV~Vul, FH~Ser,
and PU~Vul types. 

These six template novae are further grouped into three families 
by their similarities.
The track of V1500~Cyg overlaps that of V1974~Cyg if we shift it
by $\Delta (B-V)=0.12$ mag toward red, as shown in Figure
\ref{hr_diagram_fh_ser_pw_vul_v1500_cyg_v1974_cyg_outburst_no2}(d).  
V1668~Cyg shows a shallow dust blackout while FH~Ser does a deep
dust blackout.  The track of V1668~Cyg is $\Delta (B-V)=0.12$ mag bluer
than that of FH~Ser, as shown in Figure
\ref{hr_diagram_fh_ser_pw_vul_v1500_cyg_v1974_cyg_outburst_no2}(a).
The track of LV~Vul (PW~Vul) almost overlaps those of PU~Vul, 
V723~Cas, HR~Del, and V5558~Sgr, if we shift it by $\Delta (B-V)=0.2$
mag toward red.  Thus, we may categorize these ten novae into three
families, i.e., V1500~Cyg, V1668~Cyg, and LV~Vul families.
The V1500~Cyg family includes V1500~Cyg and V1974~Cyg.
The V1668~Cyg family includes V1668~Cyg and FH~Ser
and the LV~Vul family includes LV~Vul, PW~Vul, PU~Vul, V723~Cas, HR~Del,
and V5558~Sgr.

\subsection{Characteristic properties of color-magnitude tracks}
\label{properties_color_magnitude_diagram}
If optically thick free-free emission ($F_\nu\propto\nu^{2/3}$)
dominates the spectrum of a nova, its color should be $(B-V)_0=-0.03$
(Paper I),
the color of which is indicated by a vertical thin solid red line in 
Figure \ref{hr_diagram_6types_novae_one}.
If the spectrum is of optically thin free-free emission ($F_\nu\propto
\nu^0$), its color is $(B-V)_0=+0.13$ (not shown in 
Figure \ref{hr_diagram_6types_novae_one}).
V1668~Cyg and LV~Vul almost follows the line of $(B-V)_0=-0.03$ 
while V1500~Cyg and V1974~Cyg across this line and move further toward blue.
This is due to emission line effects (see Figure 10 and
discussion of Paper I).  In the later phase, these novae turn to the
right (toward red) due to the contribution of [\ion{O}{3}] lines
to the blue edge of the $V$ filter.  Therefore, this excursion toward red could
be prominent in the nebular phase.  
We have already marked the beginning of the nebular
phase in each figure.  For V1500~Cyg and V1974~Cyg,  the start of the
nebular phase agrees with the clear cusp in the color-magnitude
track as indicated by the large open red squares.  For V1668~Cyg, we also 
found a cusp (not clear, but a slight cusp) on the track as shown in
Figure \ref{hr_diagram_v1668_cyg_only_outburst}.  
For LV~Vul, we found that the track departs into two branches
at the onset of nebular phase, depending on the different response
of each $V$ filter, especially on the blue edge of the response function.
For the other novae, we already specify the onset of the nebular phase
on their tracks if it was detected and reported in the literature. 

In this way, we found that there is a special feature like a cusp,
bifurcation, or sharp turning point on the track near the onset
of nebular phase.  Such a cusp (sharp turning point) appears
at $M_V\approx-4$ for the three fast novae, 
V1668~Cyg, V1500~Cyg, and V1974~Cyg.  An inflection appears also 
at $M_V\approx-4$ for the symbiotic novae PU~Vul, and slow novae
V723~Cas and PW~Vul.  We found that the positions of
these cusps/inflections are located on the lines described by
\begin{equation}
M_V=-0.7(B-V)_0-4.17\pm0.1~(\mbox{two-headed black arrow}).
\label{absolute_magnitude_cusp}
\end{equation}
This line is plotted in Figure \ref{hr_diagram_6types_novae_one}
by the two-headed black arrow, which is obtained simply by connecting
the inflection of PU~Vul and the turning point of V1500~Cyg.
The $1\sigma$ error of 0.1 mag comes from our entire analysis
of the color-magnitude diagrams for 13 novae,
which will be discussed later from Section \ref{absolute_mag_novae}.  

For LV~Vul, HR~Del, and V5558~Sgr, on the other hand,
the start of the nebular phase is close to the line of the two-headed red arrow in Figure \ref{hr_diagram_6types_novae_one}, i.e., 
\begin{equation}
M_V=-0.7(B-V)_0-3.57\pm0.2~(\mbox{two-headed red arrow}).
\label{absolute_magnitude_cusp_low_red}
\end{equation}
This line is 0.6 mag below the line of the two-headed black arrow. 
The $1\sigma$ error of 0.2 mag comes from our entire analysis
of the color-magnitude diagrams for 11 novae,
which will be discussed later from Section \ref{absolute_mag_novae}.  
These two lines are empirically determined, so we do not have
a theoretical justification yet.

If the positions of cusps/inflections are always located on Equation
(\ref{absolute_magnitude_cusp}) or (\ref{absolute_magnitude_cusp_low_red}),
we are able to estimate the absolute magnitudes of novae by placing
the observed cusp/inflection on the line of 
Equation (\ref{absolute_magnitude_cusp}) or 
(\ref{absolute_magnitude_cusp_low_red}).
This could be a new method for obtaining the absolute magnitude of novae.

\clearpage

%Table 2
%\placetable{color_magnitude_turning_point}

\begin{deluxetable*}{llccclll}
%\begin{deluxetable}{llccclll}
\tabletypesize{\scriptsize}
\tablecaption{Start of nebular phase or dust formation
in the color-magnitude diagram
\label{color_magnitude_turning_point}}
\tablewidth{0pt}
\tablehead{
\colhead{Object} & \colhead{Outburst} & \colhead{$(B-V)_0$} 
& \colhead{$M_V$} & \colhead{$(m-M)_V$} & \colhead{$P_{\rm orb}$} 
& \colhead{Type\tablenotemark{a}} & \colhead{Comment\tablenotemark{b}} \\
\colhead{} & \colhead{year} & \colhead{} 
& \colhead{} & \colhead{} & \colhead{(hr)} 
& \colhead{} & \colhead{} 
} 
\startdata
OS~And & 1986 & ($+0.14$)\tablenotemark{c} & ($-4.59$) & 14.8 & --- & V1668~Cyg & dust \\  %%% 10.21 +0.29
CI~Aql & 2000 & $-0.52$ & $-3.84$ & 15.7 & 14.8 & V1500~Cyg & recurrent \\ %%% 11.85700    0.4830 
V1370~Aql & 1982 & $-0.35$ & $-3.10$ & 16.5 & --- & LV~Vul & dust, Ne nova \\ %%% 2445171.45347   7 20 22 53 13.40 0.00
V1419~Aql & 1993 & ($-0.13$) & ($-3.85$) & 14.6 & --- & V1668~Cyg & dust \\ %%% 10.75  +0.37
V1493~Aql & 1999\#1 & $-0.35$ & $-3.78$ & 17.7 & 3.74 & V1974~Cyg & 2nd max \\ %%% 14.66  +0.65
V1494~Aql & 1999\#2 & --- & --- & 13.1 & 3.23 & V1500~Cyg & transition oscillation \\ %%% 0.13 -0.38 9.74
V705~Cas & 1993 & ($-0.20$) & ($-4.09$) & 13.4 & 5.47 & V1668~Cyg & dust \\ %%% 9.31 -0.40 0.15
V723~Cas & 1996 & $-0.17$ & $-2.83$ & 14.0 & 16.6 & PU~Vul & multiple peaks \\ %%%
V1065~Cen & 2007 & $-0.35$ & $-3.42$ & 15.3 & --- & LV~Vul & dust, Ne nova \\ %%%  31.94   V=11.24       B=11.63   B-V=0.3900
IV~Cep & 1971 & $-0.37$ & $-3.23$ & 14.7 & --- & LV~Vul & \\ %%% 1971 9  24.18  11.47  0.28 -0.26  2441218.68
V1500~Cyg & 1975 & $-0.47$ & $-3.83$ & 12.3 & 3.35 & V1500~Cyg & super bright, Ne nova \\  %%% 8.47 -0.02 -0.52
V1668~Cyg & 1978 & $+0.02$ & $-4.19$ & 14.6 & 3.32 & V1668~Cyg & dust \\  %%% 818 10.41 0.32 -0.65
V1974~Cyg & 1992 & $-0.21$ & $-4.02$ & 12.2 & 1.95 & V1974~Cyg & Ne nova \\  %%% 8.176  0.093  -0.696
V2274~Cyg & 2001\#1 & --- & --- & 18.7 & --- & V1668~Cyg & dust \\
V2275~Cyg & 2001\#2 & --- & --- & 16.3 & 7.55 & V1500~Cyg & \\
V2362~Cyg & 2006 & $-0.51$ & $-3.83$ & 15.9 & 1.58 & V1500~Cyg & 2nd max \\ %%% 2453984.4741 12.067  0.090
V2467~Cyg & 2007 & $-0.53$ & $-3.88$ & 16.2 & 3.83 & V1974~Cyg & transition oscillation\\ %%% 12.32 0.87 0.61
V2468~Cyg & 2008 & $-0.38$ & $-3.80$ & 15.6 & 3.49 & V1500~Cyg & \\ %%% 11.80 0.366 2454590.9417
V2491~Cyg & 2008 & $-0.51$ & $-3.97$ & 16.5 & --- & V1500~Cyg & 2nd max \\ %%% 12.53 -0.28 
HR~Del & 1967 & $-0.02$ & $-3.59$ & 10.4 & 5.14 & PU~Vul & multiple peaks \\  %%% 2440064.9000    6.8100   0.1000  -0.7900
V446~Her & 1960 & $-0.13$ & $-4.05$ & 11.7 & 4.97 & V1668~Cyg & \\  %%% 1960 4 12.5 4 7.65 +0.27 -0.82
V533~Her & 1963 & $-0.40$ & $-3.97$ & 10.8 & 3.53 & V1974~Cyg & \\  %%%1963  4  05.82 -0.36 05.81 -0.81 05.82 6.83
%%%%V838~Her & 1991 & --- & --- & --- & 7.14 & --- & \\ 
GQ~Mus & 1983 & $-0.20$ & $-5.70$ & 15.7 & 1.43 & V1500~Cyg & \\ %%% 396.50 10.00 0.25  -0.70 1.34 1.31
RS~Oph  & 1958 & $-0.12$ & $-3.29$ & 12.8 & $1.1\times10^4$ & LV~Vul & recurrent \\  %%% 9.51000    0.63000   2453815.6857   Mhh
%%V2540~Oph  & 2002 & --- & --- & --- & 6.83 & --- & --- \\ 
V2615~Oph  & 2007 & ($-0.06$) & ($-4.58$) & 16.5 & 6.54 & FH~Ser & dust \\ %%% 2454247.8834 11.916 0.894 0.927 0.678 1.611
T~Pyx  & 1966 & $-0.56$ & $-4.59$  & 14.2 & 1.83 & V1500~Cyg & recurrent \\ %%% 2439564.6220 9.61 -0.31 -0.57
U~Sco & 2010 & $-0.64$ & $-3.72$ & 16.0 & 29.5 & V1500~Cyg & recurrent \\ %%% 7.63000    0.88252   -0.6400   -0.2900  2455231.83  11.35  11.99  12.28
V745~Sco & 2014 & --- & --- & 16.6 & --- & LV~Vul & recurrent \\
V1280~Sco & 2007\#1 & ($+0.01$) & ($-5.70$) & 11.0 & --- & FH~Ser & dust \\
V443~Sct & 1989 & --- & --- & 15.5 & --- & LV~Vul & \\
V475~Sct & 2003 & ($-0.15$) & ($-3.80$) & 15.4 & --- & V1668~Cyg & dust \\  %%% 11.59900    0.40100   -0.22700   52941.29
V496~Sct & 2009 & $-0.35$ & $-3.26$ & 14.4 & --- & LV~Vul & dust \\  %%% 2010  3 14.172 115.456 11.143 0.152
FH~Ser & 1970 & $(+0.05)$ & $(-4.48)$ & 11.7 & --- & FH~Ser & dust \\  %%%  0.45 7.22 (Borra)+0.20
V5114~Sgr & 2004 & $-0.50$ & $-3.92$ & 16.5 & --- & V1974~Cyg & transition oscillation \\ %%% 2453136.25 58.25 12.10 12.53 12.58 -0.05 -0.43 
V5558~Sgr & 2007 & $-0.16$ & $-3.46$ & 13.9 & --- & PU~Vul & multiple peaks \\ %%% 
V382~Vel & 1999 & $-0.29$ & $-4.04$ & 11.5 & 3.51 & V1974~Cyg & Ne nova \\ %%% 7.46  -0.94 -0.14
LV~Vul & 1968\#1 & $-0.01$ & $-3.77$ & 11.9 & --- & LV~Vul & \\ %%%    2440033.400      8.13 +0.59      -0.23   3  abuladze.data
NQ~Vul & 1976 & ($+0.01$) & ($-4.26$) & 13.6 & --- & FH~Ser & dust \\  %%% 12  25.41  9.34  0.05  1.06  0.08  0.28  0.28 2443138.88
PU~Vul & 1979 & $+0.20$ & $-3.80$ & 14.3 & $1.18\times10^5$ & PU~Vul & \\ %%% 2488027.550      44054.550          4.644     0.5000    -0.6300  48027.55 10.370 11.000 10.500 9.030 8.980
PW~Vul & 1984\#1 & $-0.39$ & $-3.32$ & 13.0 & 5.13 & LV~Vul & \\ %%% 6038.587      4    9.68    0.16   -0.58 
QU~Vul & 1984\#2 & $-0.31$ & $-4.01$ & 13.6 & 2.68 & V1974~Cyg & Ne nova \\ %%% 6121 9.59 9.73 +0.14
QV~Vul & 1987 & ($-0.03$) & ($-5.00$) & 14.0 & --- & FH~Ser & dust \\
V458~Vul & 2007\#1 & --- & --- & 15.3 & 1.63 & PU~Vul &  multiple peaks %%% 11.36  0.6170
%%%% neon nova
% Nova Mon 2012 is a Neon nova  2013ATel.4709....1M
% Strong Neon Emission in CSS081007:030559+054715: A Possible Oxygen/ Neon Nova
%   2008ATel.1835....1P
% 
\enddata
\tablenotetext{a}{Nova tracks are categorized into six types, 
depending on the shape and position of the track in the color-magnitude
diagram, i.e., V1500~Cyg, V1668~Cyg, V1974~Cyg, LV~Vul, FH~Ser, and PU~Vul.} 
\tablenotetext{b}{``dust'' indicates a dust formation nova,
``recurrent'' a recurrent nova, ``Ne nova'' a neon nova,
``2nd max'' a nova showing a single secondary maximum like V2362~Cyg,
``transition oscillation'' a nova showing a transition oscillation
like GK~Per, 
``multiple peaks'' means a nova with multiple-peaks like V5558~Sgr,
and ``super bright'' a super bright nova \citep{del91}.} 
\tablenotetext{c}{numbers in parenthesis show the position in the 
color-magnitude diagram where dust blackout starts.} 
%\end{deluxetable}
\end{deluxetable*}

%Fig.21
%\placefigure{iv_cep_v_bv_ub_color}

\begin{figure}
%\epsscale{0.75}
%%\epsscale{0.8}
%\epsscale{1.0}
\epsscale{1.15}
\plotone{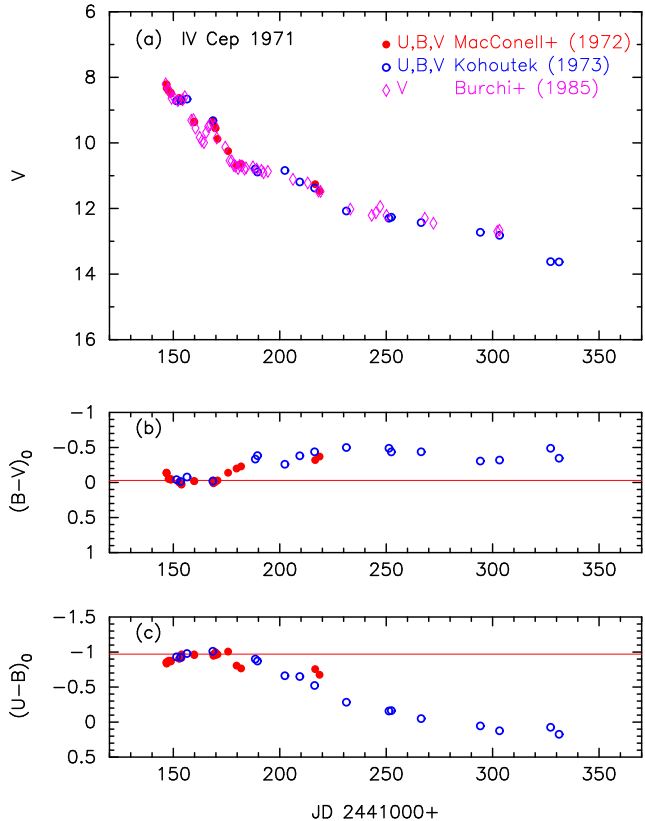}
%\plotone{iv_cep_v_bv_ub_color.epsi}
%\plotfiddle{evolution1.ps}{5.0cm}{270}{0.4}{0.4}{-170}{220}
\caption{
Same as Figure \ref{v446_her_v_bv_ub_color_curve}, but for IV~Cep.
%The $UBV$ data are taken from \citet{mac72} and \citet{koh73}.
%The $V$ data are taken from \citet{bur85}.  
The $(B-V)_0$ and $(U-B)_0$ are de-reddened with $E(B-V)=0.65$.
\label{iv_cep_v_bv_ub_color}}
\end{figure}

%Fig.22
%\placefigure{color_color_diagram_iv_cep_nq_vul_qu_vul_qv_vul_no2}

\begin{figure*}
%\begin{figure}
\epsscale{0.75}
%%\epsscale{0.8}
%%\epsscale{1.0}
%%\epsscale{1.15}
\plotone{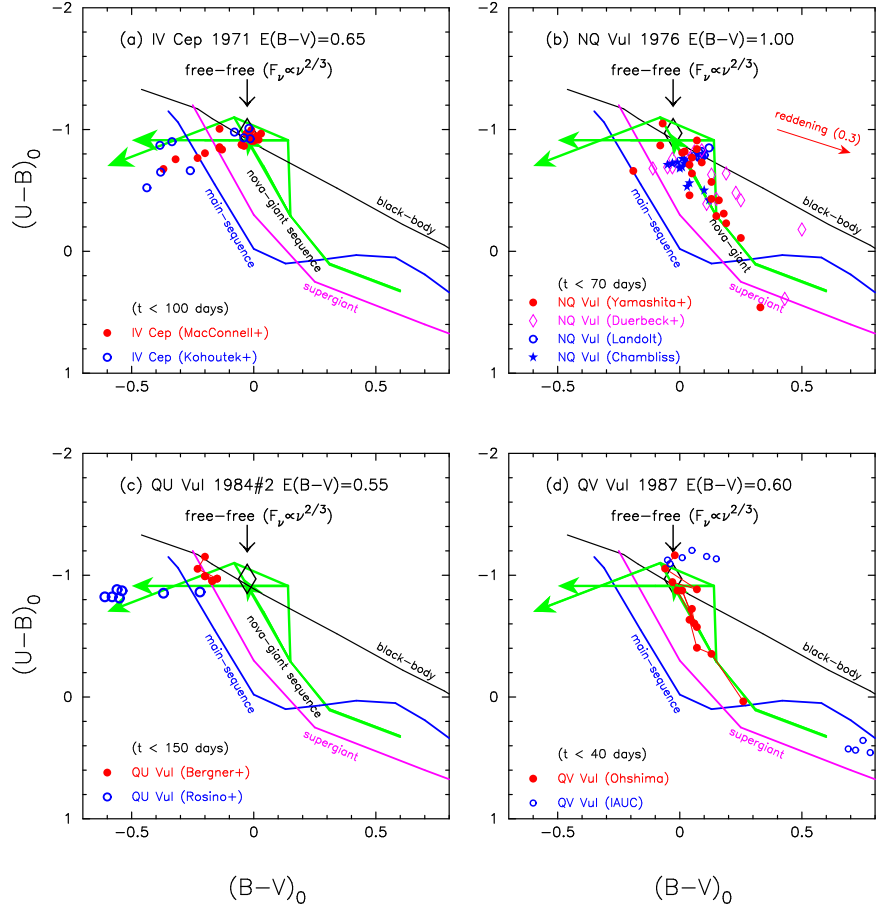}
%\plotone{color_color_diagram_iv_cep_nq_vul_qu_vul_qv_vul_no2.epsi}
%\plotfiddle{evolution1.ps}{5.0cm}{270}{0.4}{0.4}{-170}{220}
\caption{
Same as Figure \ref{color_color_diagram_templ_rs_oph_v446_her_v533_her_no2},
but for (a) IV~Cep 1971, (b) NQ~Vul 1976, (c) QU~Vul 1984\#2, 
and (d) QV~Vul 1987.
%See the main text for the sources of observational data.
\label{color_color_diagram_iv_cep_nq_vul_qu_vul_qv_vul_no2}}
%\end{figure}
\end{figure*}

%Fig.23 
%\placefigure{hr_diagram_iv_cep_nq_vul_v1370_aql_gq_mus_outburst}

\begin{figure*}
%\begin{figure}
%\epsscale{0.75}
\epsscale{0.8}
%%\epsscale{1.0}
\plotone{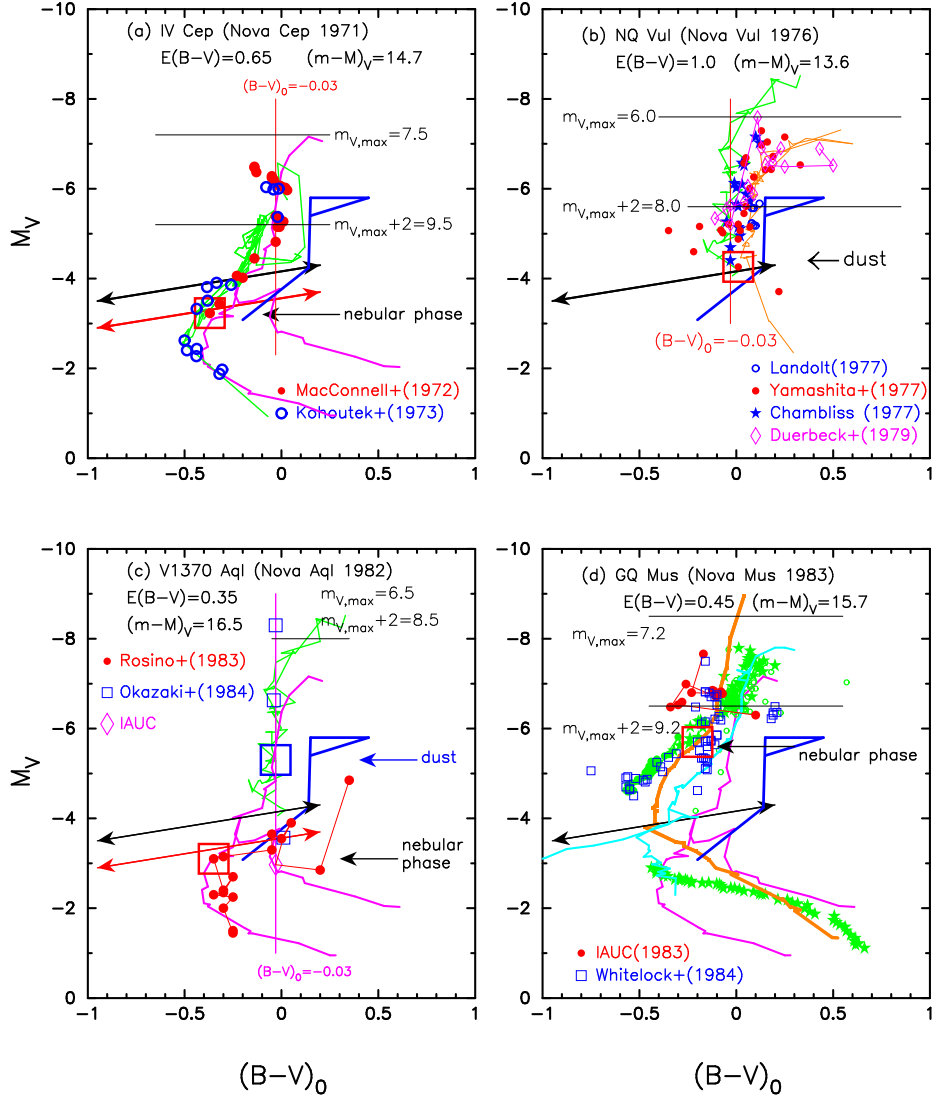}
%\plotone{hr_diagram_iv_cep_nq_vul_v1370_aql_gq_mus_outburst.epsi}
%\plotfiddle{evolution1.ps}{5.0cm}{270}{0.4}{0.4}{-170}{220}
\caption{
Same as Figure 
\ref{hr_diagram_rs_oph_v446_her_v533_her_t_pyx_outburst}, but
for (a) IV~Cep, (b) NQ~Vul, (c) V1370~Aql, and (d) GQ~Mus.
Solid magenta lines denote the template track of LV~Vul 
in panels (a), (c), and (d).
Thin solid green lines represent the track of PW~Vul in panel (a),
but that of V1668~Cyg in panels (b) and (c).  In panel (b), 
thin solid orange lines denote the track of FH~Ser.  In panel (d),
solid cyan line denotes the track of V1974~Cyg,
thick solid orange line represents the track of V1500~Cyg, and
green symbols correspond to T~Pyx.
\label{hr_diagram_iv_cep_nq_vul_v1370_aql_gq_mus_outburst}}
%\end{figure}
\end{figure*}

%Fig.24 
%\placefigure{nq_vul_v_bv_ub_color}

\begin{figure}
%\epsscale{0.75}
%%\epsscale{0.8}
%\epsscale{1.0}
\epsscale{1.15}
\plotone{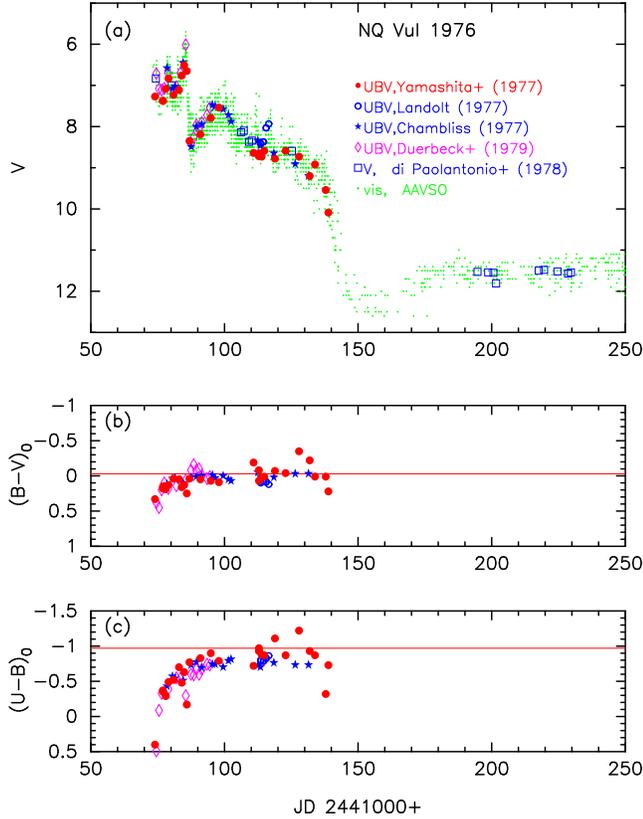}
%\plotone{nq_vul_v_bv_ub_color.epsi}
%\plotfiddle{evolution1.ps}{5.0cm}{270}{0.4}{0.4}{-170}{220}
\caption{
Same as Figure \ref{v446_her_v_bv_ub_color_curve}, but for NQ~Vul.
%%The $UBV$ data are taken from \citet{mac72} and \citet{koh73}.
%%The $V$ data are taken from \citet{bur85}.  
The $(B-V)_0$ and $(U-B)_0$ are de-reddened with $E(B-V)=1.0$.
\label{nq_vul_v_bv_ub_color}}
\end{figure}

%Fig.25 
%\placefigure{distance_reddening_nq_vul_v1370_aql_qu_vul_os_and}

\begin{figure*}
%\begin{figure}
\epsscale{0.75}
%%\epsscale{0.8}
%%\epsscale{1.0}
%%\epsscale{1.15}
\plotone{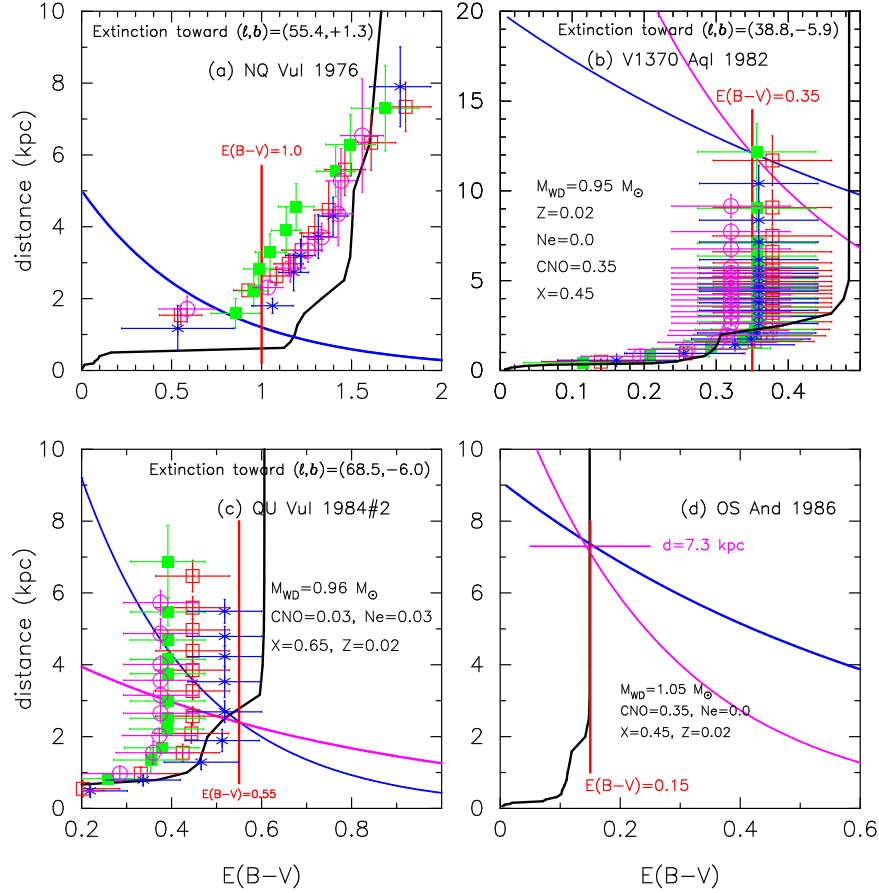}
%\plotone{distance_reddening_nq_vul_v1370_aql_qu_vul_os_and.epsi}
%\plotfiddle{evolution1.ps}{5.0cm}{270}{0.4}{0.4}{-170}{220}
\caption{
Same as Figure \ref{distance_reddening_fh_ser_pw_vul_v1500_cyg_v1974_cyg},
but for (a) NQ~Vul, (b) V1370~Aql, (c) QU~Vul, and (d) OS~And.
The thick solid blue lines denote 
%%the distance-reddening relation calculated from 
%%the distance modulus in the $V$ band, i.e., 
(a) $(m-M)_V=13.6$, (b) $(m-M)_V=16.5$, (c) $(m-M)_V=13.6$,  
and (d) $(m-M)_V=14.8$.
The thick solid magenta lines indicate the UV~1455 \AA\  flux fitting.
%%The vertical solid red lines represent the color excesses of 
%%(a) $E(B-V)=1.0$, (b) $E(B-V)=0.35$,
%%(c) $E(B-V)=0.55$, and (d) $E(B-V)=0.15$.
%The black solid lines denote the distance-reddening relation given
%by \citet{gre15}.
%In panel (a), the magenta thick solid line represents
%the distance-reddening relation calculated from the UV~1455 \AA\  flux
%fitting with the $0.51~M_\sun$ WD model \citep{hac15k}.
%In panels (a), (b), and (d), two or 
%four sets of data with error bars show distance-reddening relations
%in two or four directions close to each nova, the data of which are taken
%from \citet{mar06}.  
%%See the main text for more detail.
\label{distance_reddening_nq_vul_v1370_aql_qu_vul_os_and}}
%\end{figure}
\end{figure*}

%Fig.26 
%\placefigure{v1370_aql_v_bv_ub_color}

\begin{figure}
%\epsscale{0.75}
%%\epsscale{0.8}
%\epsscale{1.0}
\epsscale{1.15}
\plotone{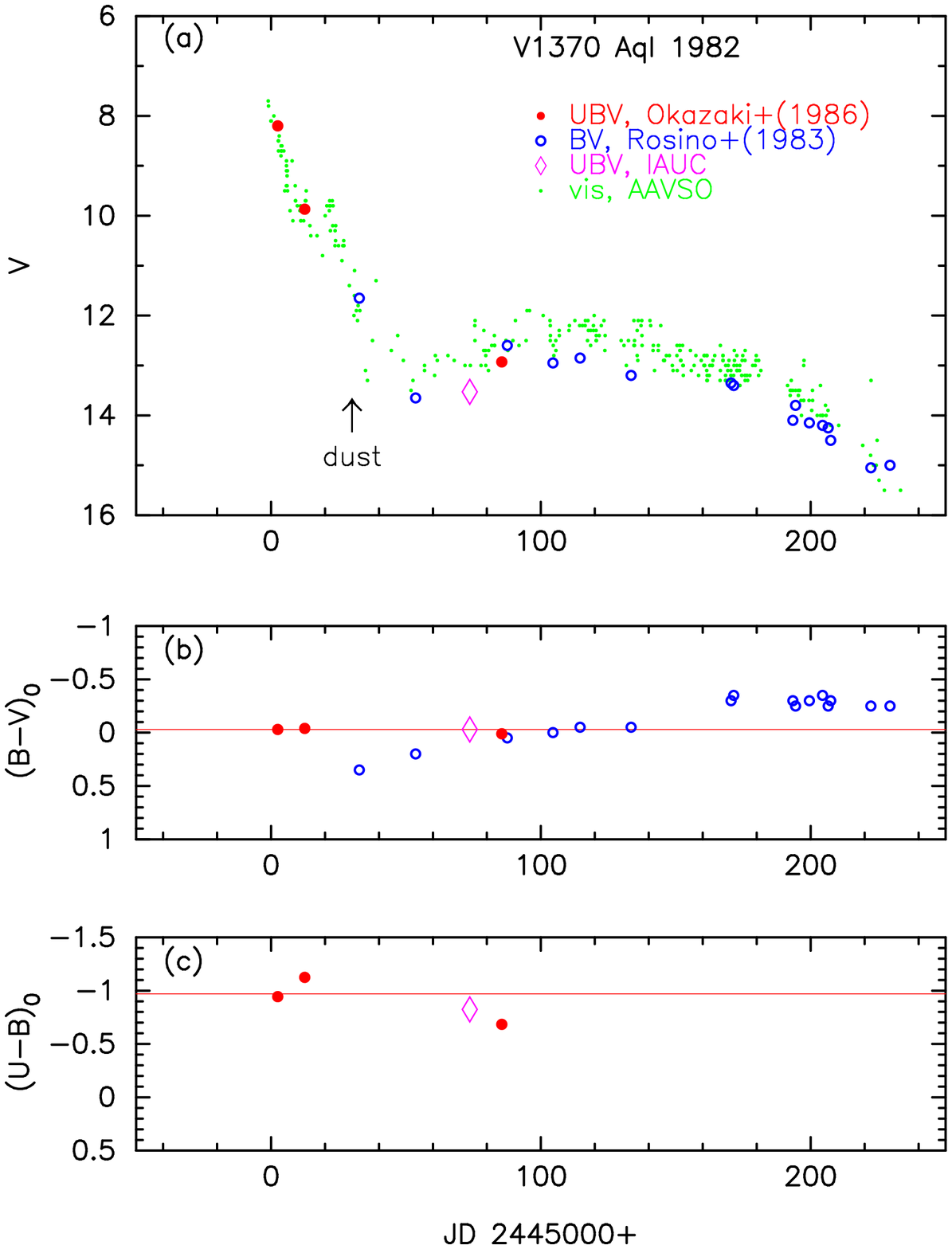}
%\plotone{v1370_aql_v_bv_ub_color.epsi}
%\plotfiddle{evolution1.ps}{5.0cm}{270}{0.4}{0.4}{-170}{220}
\caption{
Same as Figure \ref{v446_her_v_bv_ub_color_curve}, but for V1370~Aql.
%The $UBV$ data are taken from \citet{oka86} and IAU Circular No.3689.
%The $BV$ data are taken from \citet{ros83}.  
We de-reddened $(B-V)_0$ and $(U-B)_0$ colors with $E(B-V)=0.35$.
%%In panel (a), we also plot the visual magnitudes (green dots),
%%which are taken from the AAVSO archive.
\label{v1370_aql_v_bv_ub_color}}
\end{figure}

%Fig.27 
%\placefigure{v1370_aql_v1668_cyg_os_and_v_bv_ub_color_logscale_no2}

\begin{figure}
%\epsscale{0.75}
%%\epsscale{0.8}
%\epsscale{1.0}
\epsscale{1.15}
\plotone{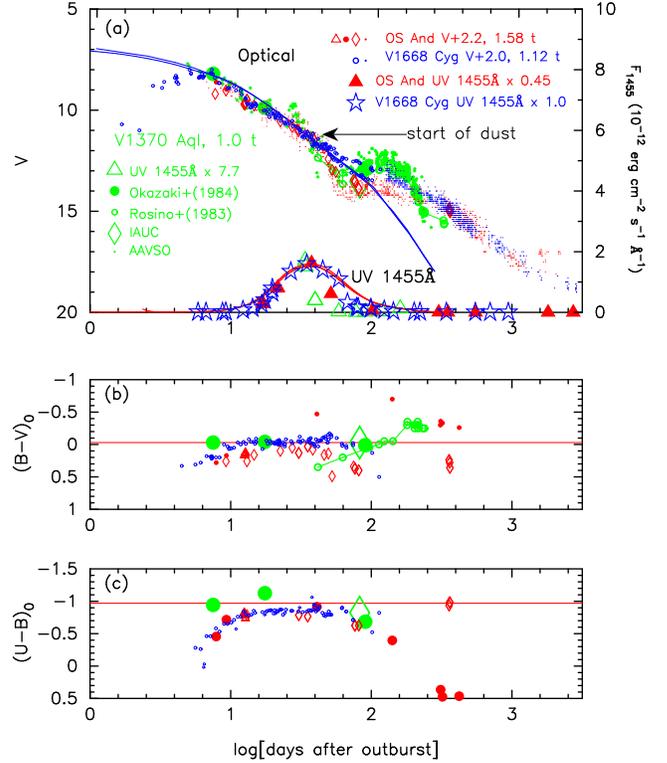}
%\plotone{v1370_aql_v1668_cyg_os_and_v_bv_ub_color_logscale_no2.epsi}
%\plotfiddle{evolution1.ps}{5.0cm}{270}{0.4}{0.4}{-170}{220}
\caption{
Comparison of 
%%%Same as Figure \ref{t_pyx_pw_vul_nq_vul_dq_her_v_bv_ub_color_logscale_no6},
V1370~Aql (green symbols) with V1668~Cyg (blue symbols)
and OS~And (red symbols).  In panel (a), we add a $0.95~M_\sun$
WD model with the chemical composition of CO nova 3 \citep{hac16k}
for the $V$ (thin solid blue line) and UV~1455 \AA\   
(thin solid red line) light curves.
%%See the main text for the sources of V1370~Aql data.
\label{v1370_aql_v1668_cyg_os_and_v_bv_ub_color_logscale_no2}}
\end{figure}

%Fig.28 
%\placefigure{gq_mus_v_bv_ub_color_curve}

\begin{figure}
%\epsscale{0.75}
%%\epsscale{0.8}
%\epsscale{1.0}
\epsscale{1.15}
\plotone{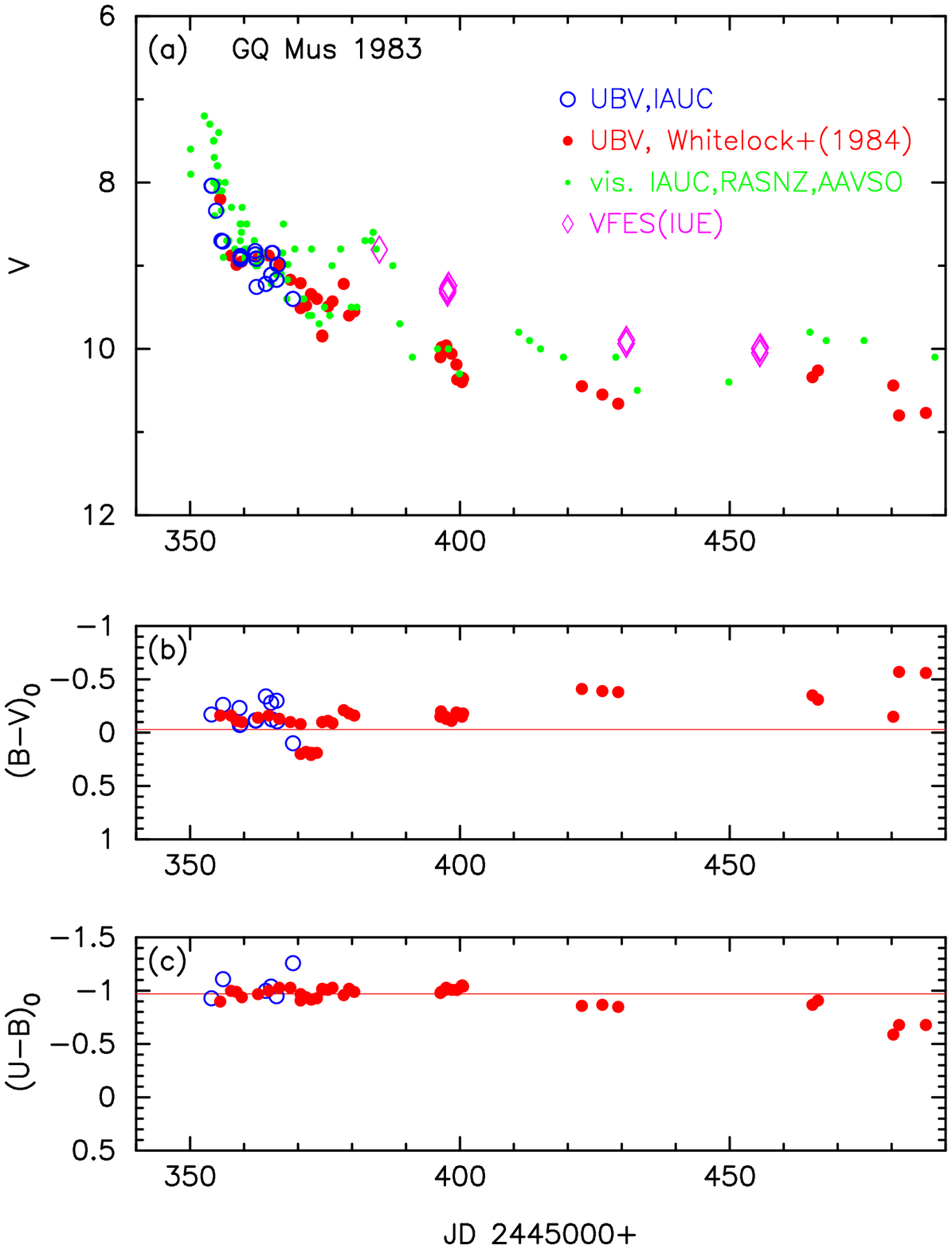}
%\plotone{gq_mus_v_bv_ub_color_curve.epsi}
%\plotfiddle{evolution1.ps}{5.0cm}{270}{0.4}{0.4}{-170}{220}
\caption{
Same as Figure \ref{v446_her_v_bv_ub_color_curve}, but for GQ~Mus.
We de-reddened $(B-V)_0$ and $(U-B)_0$ colors with $E(B-V)=0.45$.
%%See the main text for the sources of GQ~Mus data.
\label{gq_mus_v_bv_ub_color_curve}}
\end{figure}

%Fig.29
%\placefigure{qu_vul_v_bv_ub_color_curve}

\begin{figure}
%\epsscale{0.75}
%%\epsscale{0.8}
%\epsscale{1.0}
\epsscale{1.15}
\plotone{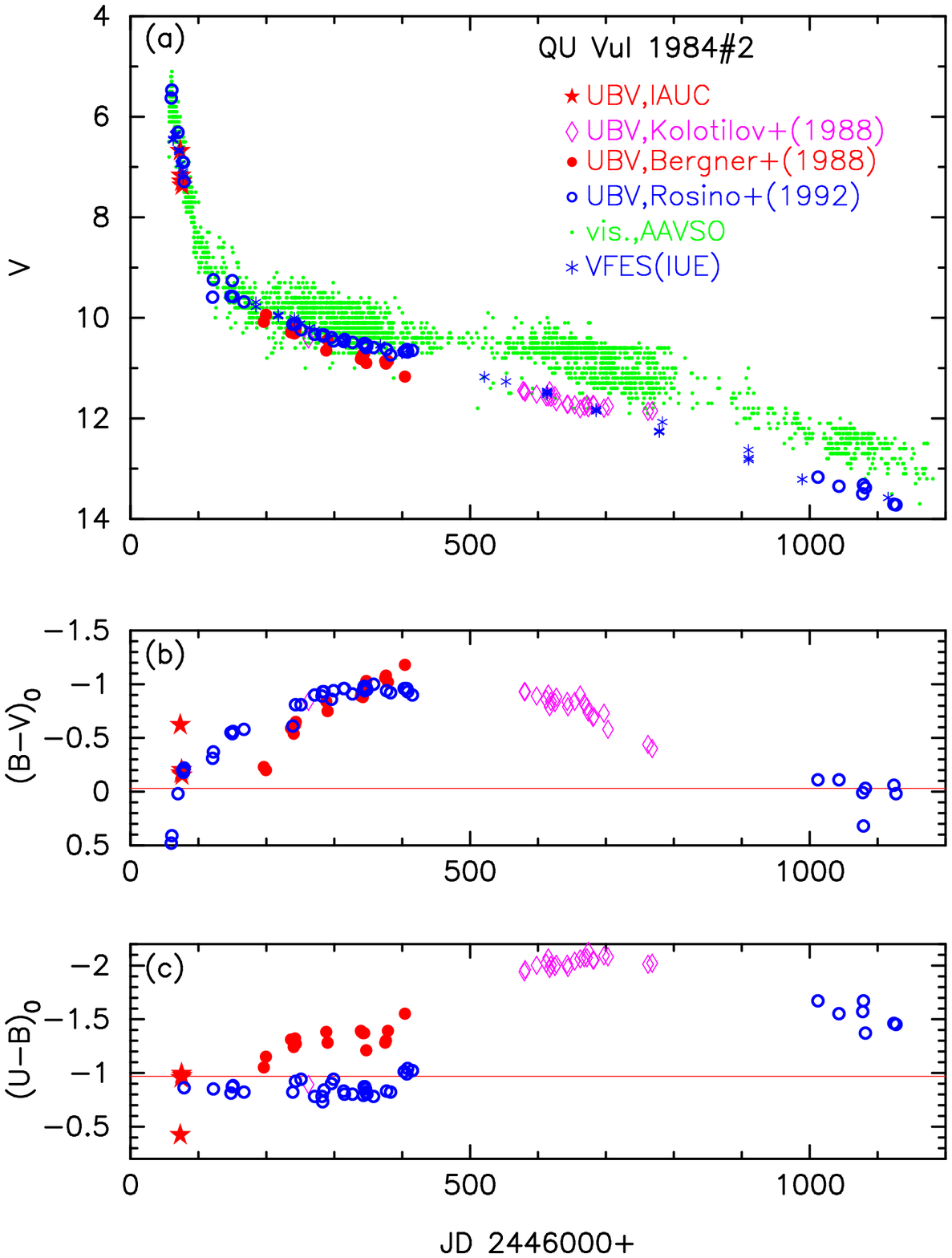}
%\plotone{qu_vul_v_bv_ub_color_curve.epsi}
%\plotfiddle{evolution1.ps}{5.0cm}{270}{0.4}{0.4}{-170}{220}
\caption{
Same as Figure \ref{v446_her_v_bv_ub_color_curve}, but for QU~Vul.
We de-reddened $(B-V)_0$ and $(U-B)_0$ colors with $E(B-V)=0.55$.
%%See the main text for the sources of QU~Vul data.
\label{qu_vul_v_bv_ub_color_curve}}
\end{figure}

%Fig.30
%\placefigure{hr_diagram_qu_vul_os_and_qv_vul_v443_sct_outburst}

\begin{figure*}
%\begin{figure}
%\epsscale{0.75}
\epsscale{0.8}
%%\epsscale{1.0}
\plotone{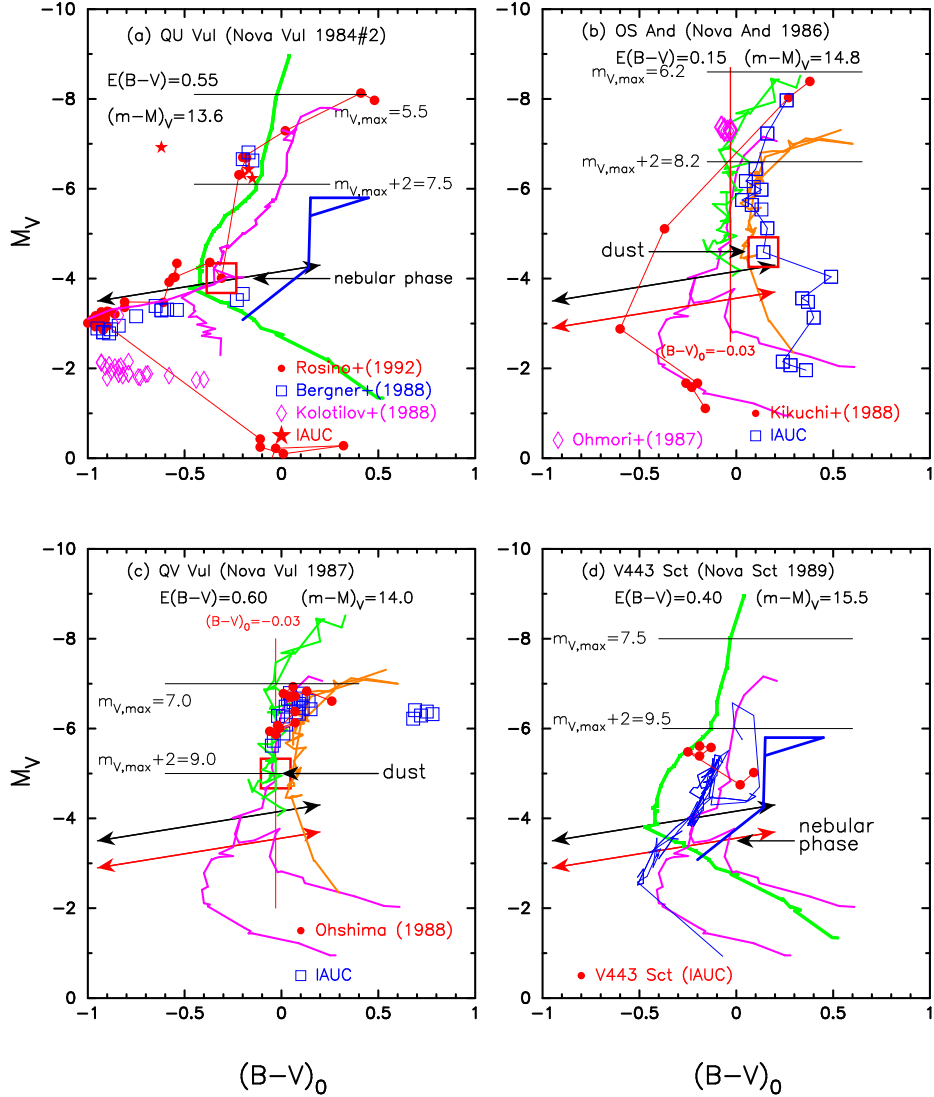}
%\plotone{hr_diagram_qu_vul_os_and_qv_vul_v443_sct_outburst.epsi}
%\plotfiddle{evolution1.ps}{5.0cm}{270}{0.4}{0.4}{-170}{220}
\caption{
Same as Figure 
\ref{hr_diagram_rs_oph_v446_her_v533_her_t_pyx_outburst}, but
for (a) QU~Vul, (b) OS~And, (c) QV~Vul, and (d) V443~Sct.
The thick green and blue lines denote the template tracks 
of V1500~Cyg and PU~Vul, respectively.  
Thin green and orange lines denote the tracks of V1668~Cyg and FH~Ser,
respectively, in panels (b) and (c).
Magenta lines denote the track of V1974~Cyg in panel (a) and
the tracks of LV~Vul in panels (b), (c), and (d).
Thin solid blue lines in panel (d)
denote the track of PW~Vul.
%%Vertical red lines in panels (b) and (c) represents $(B-V)_0=-0.03$
%%for optically thick free-free emission.  
%%See the main text for more detail. 
\label{hr_diagram_qu_vul_os_and_qv_vul_v443_sct_outburst}}
%\end{figure}
\end{figure*}

%Fig.31
%\placefigure{os_and_v_bv_ub_color_curve}

\begin{figure}
%\epsscale{0.75}
%%\epsscale{0.8}
%\epsscale{1.0}
\epsscale{1.15}
\plotone{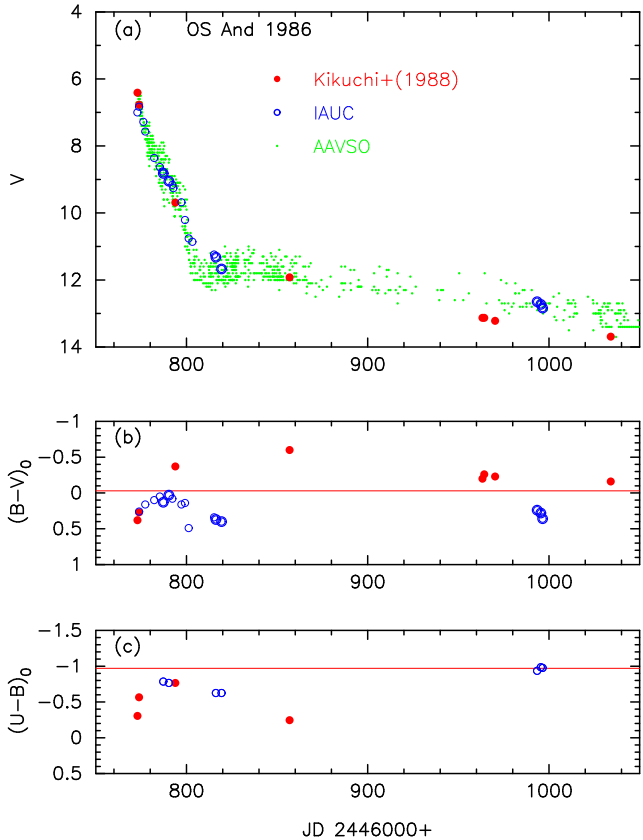}
%\plotone{os_and_v_bv_ub_color_curve.epsi}
%\plotfiddle{evolution1.ps}{5.0cm}{270}{0.4}{0.4}{-170}{220}
\caption{
Same as Figure \ref{v446_her_v_bv_ub_color_curve}, but for OS~And.
We de-reddened $(B-V)_0$ and $(U-B)_0$ colors with $E(B-V)=0.15$.
%%See the main text for the sources of OS~And data.
\label{os_and_v_bv_ub_color_curve}}
\end{figure}

%Fig.32 
%\placefigure{qv_vul_v_bv_ub_color_curve}

\begin{figure}
%\epsscale{0.75}
%%\epsscale{0.8}
%\epsscale{1.0}
\epsscale{1.15}
\plotone{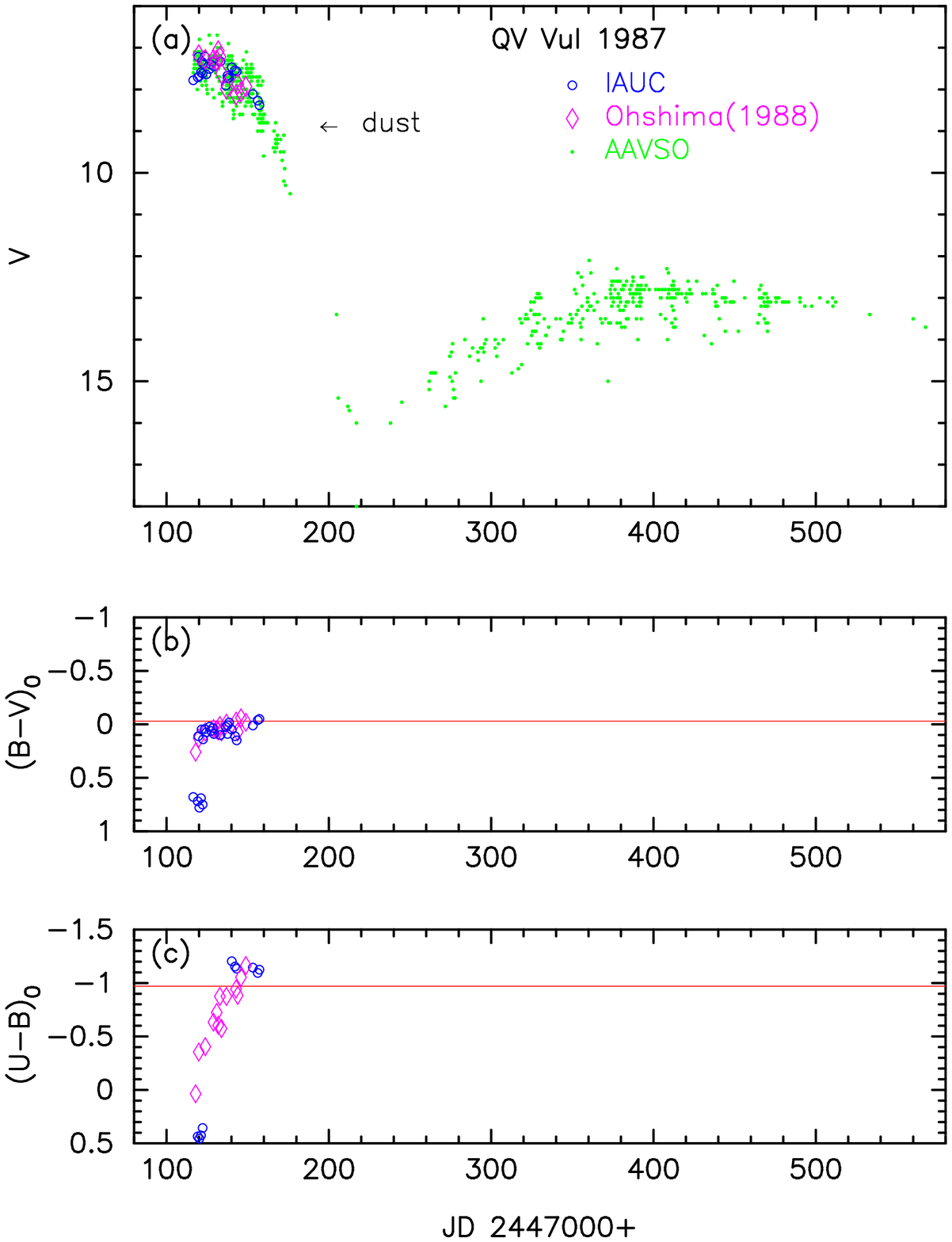}
%\plotone{qv_vul_v_bv_ub_color_curve.epsi}
%\plotfiddle{evolution1.ps}{5.0cm}{270}{0.4}{0.4}{-170}{220}
\caption{
Same as Figure \ref{v446_her_v_bv_ub_color_curve}, but for QV~Vul.
We de-reddened $(B-V)_0$ and $(U-B)_0$ colors with $E(B-V)=0.60$.
%%See the main text for the sources of QV~Vul data.
\label{qv_vul_v_bv_ub_color_curve}}
\end{figure}

%Fig.33
%\placefigure{distance_reddening_qv_vul_v443_sct_v1419_aql_v705_cas}

\begin{figure*}
%\begin{figure}
\epsscale{0.75}
%%\epsscale{0.8}
%%\epsscale{1.0}
%%\epsscale{1.15}
\plotone{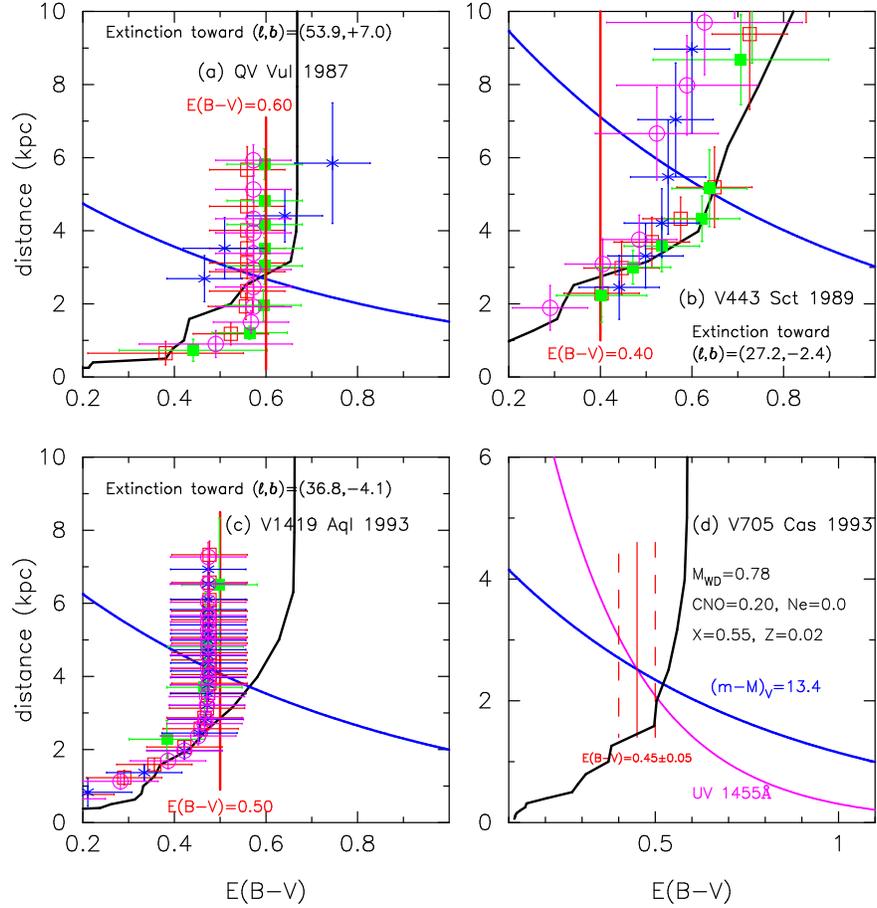}
%\plotone{distance_reddening_qv_vul_v443_sct_v1419_aql_v705_cas.epsi}
%\plotfiddle{evolution1.ps}{5.0cm}{270}{0.4}{0.4}{-170}{220}
\caption{
Same as Figure \ref{distance_reddening_fh_ser_pw_vul_v1500_cyg_v1974_cyg},
but for (a) QV~Vul, (b) V443~Sct, (c) V1419~Aql, and (d) V705~Cas.
The thick solid blue lines denote (a) $(m-M)_V=14.0$, 
(b) $(m-M)_V=15.5$, (c) $(m-M)_V=14.6$,  and (d) $(m-M)_V=13.4$.
%The vertical solid red lines represent the color excesses of 
%(a) $E(B-V)=0.60$, (b) $E(B-V)=0.40$,
%(c) $E(B-V)=0.50$, and (d) $E(B-V)=0.45$.
%The black solid lines denote the distance-reddening relation given
%by \citet{gre15}.
%In panel (a), the magenta thick solid line represents
%the distance-reddening relation calculated from the UV~1455 \AA\  flux
%fitting with the $0.51~M_\sun$ WD model \citep{hac15k}.
%In panels (a), (b), and (d), two or 
%four sets of data with error bars show distance-reddening relations
%in two or four directions close to each nova, the data of which are taken
%from \citet{mar06}.  
%%See the main text for more detail.
\label{distance_reddening_qv_vul_v443_sct_v1419_aql_v705_cas}}
%\end{figure}
\end{figure*}

%Fig.34 
%\placefigure{v443_sct_v_bv_ub_color_curve}

\begin{figure}
%\epsscale{0.75}
%%\epsscale{0.8}
%\epsscale{1.0}
\epsscale{1.15}
\plotone{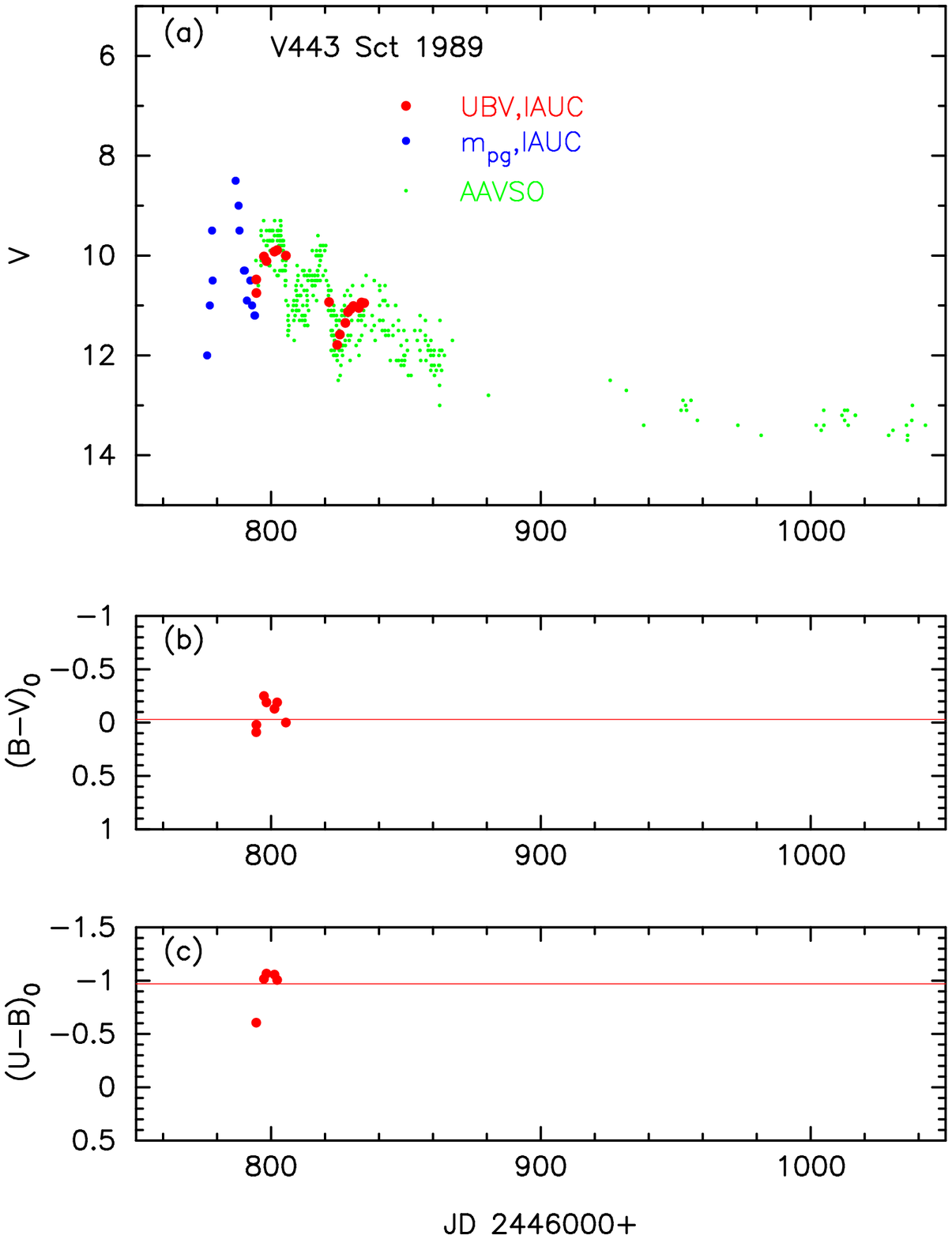}
%\plotone{v443_sct_v_bv_ub_color_curve.epsi}
%\plotfiddle{evolution1.ps}{5.0cm}{270}{0.4}{0.4}{-170}{220}
\caption{
Same as Figure \ref{v446_her_v_bv_ub_color_curve}, but for V443~Sct.
We de-reddened $(B-V)_0$ and $(U-B)_0$ colors with $E(B-V)=0.40$.
%%See the main text for the sources of V443~Sct data.
\label{v443_sct_v_bv_ub_color_curve}}
\end{figure}

\section{Color-magnitude Diagrams for Various Novae}
\label{application_to_novae}
In this section, we further examine various novae in the
color-magnitude diagram.
We have collected data from the literature for as many novae as possible 
that have a sufficient number of data points (usually more than a few tens).
We classify these 30 nova tracks into the previously discussed six types,
i.e., V1500~Cyg, V1668~Cyg, V1974~Cyg, FH~Ser, LV~Vul, and PU~Vul,
and discuss their physical properties in the color-magnitude diagram.  
These 30 novae are examined in the order of discovery.

\subsection{RS~Oph (1958,1985,2006)}
\label{rs_oph}
RS~Oph is a recurrent nova with six recorded outbursts 
in 1898, 1933, 1958, 1967, 1985, and 2006.
The orbital period of 456 days was obtained by \citet{fek00}.
Figure \ref{rs_oph_v_bv_ub_color} shows the $V$, $(B-V)_0$,
and $(U-B)_0$ evolutions of RS~Oph.  The observed data of RS~Oph
are taken from \citet{con58} (filled red circles) for the 1958 outburst,
and AAVSO (open red diamonds), VSOLJ (encircled magenta pluses),
SMARTS (blue stars),
\citet{sos06ga, sos06gb}, \citet{sos06gc} (filled blue triangles),
and \citet{hac08b} (filled green squares: data are tabulated 
only in arXiv:0807.1240)
for the 2006 outburst.  The $V$ light curve declined with $t_2=6.8$
and $t_3=14$~days \citep{sch10a}.
In panel (b) of Figure \ref{rs_oph_v_bv_ub_color},
the $B-V$ data of the 2006 outburst are systematically 0.1 mag redder
than those of the 1958 outburst \citep{con58}, so 
we shifted the $B-V$ data of the 2006 outburst by 0.1 mag up (toward blue)
to match them with the $B-V$ data of \citet{con58}.
The $B-V$ data of SMARTS are shifted by 0.05 mag down (toward red), however.

In Paper I, we determined the color excess as $E(B-V)=0.65\pm0.05$ and
the distance modulus as $(m-M)_V=12.8\pm0.2$ on the basis of the general
track in the color-color diagram and the time-stretching method,
respectively.  
Assuming that $E(B-V)=0.65$, we plot 
the color-color evolution of RS~Oph (1958) in 
Figure \ref{color_color_diagram_templ_rs_oph_v446_her_v533_her_no2}(b).
This color-color track of RS~Oph is the same as that in Figure 42 of
Paper I, but we reanalyze the data along the color evolution
in Figure \ref{rs_oph_v_bv_ub_color} and
confirmed that the color excess is $E(B-V)=0.65\pm0.05$
by matching the color-color track (filled red circles) 
with the general course of novae (green lines) in Figure
\ref{color_color_diagram_templ_rs_oph_v446_her_v533_her_no2}(a).

The distance to RS~Oph was already discussed in Paper I.
We plot various distance-reddening relations in Figure
\ref{distance_reddening_rs_oph_v446_her_v533_her_iv_cep}(a). 
This figure is the same as Figure 34(a) of Paper I, but we added
Green et al.'s (2015) relation (solid black line).
Three lines of $(m-M)_V=12.8$, UV~1455 \AA\  fit, and $E(B-V)=0.65$
cross consistently at $d\approx1.4$~kpc.
The cross point is midway between Marshall et al.'s (2006) 
and Green et al.'s (2015) relations.
Thus, we confirmed the values in Paper I, i.e.,
$(m-M)_V=12.8$, $E(B-V)=0.65$, and $d\approx1.4$~kpc.

Figure \ref{hr_diagram_rs_oph_v446_her_v533_her_t_pyx_outburst}(a) shows
the color-magnitude diagram of RS~Oph.
The track is very similar to that of V1668~Cyg
in the early phase, but turns to the right
and follows the right side of the LV~Vul track in the later phase.
The start of the nebular phase was identified from the 1985 outburst
observed by \citet{ros87} as shown
in Figure \ref{hr_diagram_rs_oph_v446_her_v533_her_t_pyx_outburst}(a).
We specify the point as $(B-V)_0=-0.12$ and $M_V=-3.29$ with a large
open red square in the figure.  
The starting point of the nebular phase is on the  
two-headed red arrow, so we identify RS~Oph as an LV~Vul type
rather than a V1668~Cyg type in the color magnitude diagram
as listed in Table \ref{color_magnitude_turning_point}.
We add the epoch when the variable SSS phase started
at $m_V=9.1$, about 30 days after the outburst \citep{hac08kl,osb11}.
The epoch, when the stable SSS phase started at
$m_V=9.6$, about 45 days after the outburst, is coincident with
the start of the nebular phase.
The track of RS~Oph follows the line of $(B-V)_0=-0.03$ in the
early phase, being consistent with optically thick
free-free emission ($F_\nu\propto \nu^{2/3}$).  After the stable
SSS phase started, it jumps to $(B-V)_0=+0.13$ and stays there 
for a while (between $M_V=-2.7$ and $M_V=-1.7$),
being consistent with the optically thin free-free
emission ($F_\nu\propto \nu^0$).  When the stable SSS phase started,
the ejecta had already become optically thin.
This agreement supports our adopted values of $E(B-V)=0.65$.

\subsection{V446~Her 1960}
\label{v446_her}
Figure \ref{v446_her_v_bv_ub_color_curve} shows the $V$,
$(B-V)_0$, and $(U-B)_0$ evolutions of V446~Her.
These light curve data are the same as those in Figure 42 of Paper I,
but we reanalyzed the data as mentioned below.
The data are taken from \citet{ant61}, \citet{bro61}, \citet{kni60}, 
\citet{ross60}, and IAU Circular No.1730. 
The orbital period of 4.97~hr was detected by \citet{thors00}.
The $B-V$ data of the IAU Circular, \citet{kni60}, and \citet{bro61}
are systematically 0.1 mag redder than those of \citet{ross60},
so we shift them by 0.1 mag toward blue.  As a result, the $B-V$ color curve
becomes smooth as shown in Figure \ref{v446_her_v_bv_ub_color_curve}(b).
We also shift the $U-B$ data of the IAU Circular and \citet{bro61}
by 0.1 mag toward blue, so the $U-B$ color curve also becomes
smooth as shown in Figure \ref{v446_her_v_bv_ub_color_curve}(c).

In Paper I, we determined the color excess as $E(B-V)=0.40\pm0.05$ 
on the basis of the general track in the color-color diagram and
the distance modulus as $(m-M)_V=11.7\pm0.2$ by the time-stretching method.
The reanalyzed data gives a consistent matching with the general tracks
of novae in the color-color diagram of V446~Her as shown in Figure 
\ref{color_color_diagram_templ_rs_oph_v446_her_v533_her_no2}(c) for the
same reddening value of $E(B-V)=0.40$ as in Paper I. 
We also plot various distance-reddening relations for V446~Her,
$(l, b)= (45\fdg4092, +4\fdg7075)$, in Figure
\ref{distance_reddening_rs_oph_v446_her_v533_her_iv_cep}(b).
This figure is the same as Figure 34(b) of Paper I, but we added
Green et al.'s (2015) relation (solid black line).
The lines cross consistently at the point of $E(B-V)=0.40$ and $d=1.23$~kpc.  
Thus, we confirmed the values of Paper I, i.e.,
$E(B-V)=0.40$, $(m-M)_V=11.7$, and $d=1.2$~kpc.

Using $E(B-V)=0.40$ and $(m-M)_V=11.7$, we plot 
the color-magnitude diagram of V446~Her in Figure
\ref{hr_diagram_rs_oph_v446_her_v533_her_t_pyx_outburst}(b).
We identified the start of the nebular phase as $\sim40$ days after
the outburst from Figure 11 of \citet{mei63}, at which the forbidden
lines of [\ion{O}{3}] surpassed the permitted lines. 
This epoch corresponds to $m_V=7.6$ ($M_V=-4.1$) as shown in Figure 
\ref{hr_diagram_rs_oph_v446_her_v533_her_t_pyx_outburst}(b).
We assign the start of the nebular phase to the observational point
$M_V=-4.05$ and $(B-V)_0=-0.13$, denoted
by a large open red square in Figure
\ref{hr_diagram_rs_oph_v446_her_v533_her_t_pyx_outburst}(b).
The track of V446~Her goes almost vertically down along the line of
$(B-V)_0=-0.03$ similarly to V1668~Cyg in the early phase,
and turns to the left (toward blue) near the onset of nebular phase
(large open red square), almost on the two-headed black arrow.  
Then the track turns to the right (toward red)
below the two-headed red arrow.  
Because the start of the nebular phase is located on the  
two-headed black arrow, we regard V446~Her as a V1668~Cyg type in the
color-magnitude diagram as listed in Table
\ref{color_magnitude_turning_point}.

\subsection{V533~Her 1963}
\label{v533_her}
Figure \ref{v533_her_v_bv_ub_color_curve} shows the $V$,
$(B-V)_0$, and $(U-B)_0$ evolutions of V533~Her, where the $UBV$ data are
taken from \citet{gen63}, \citet{chi64}, and \citet{she64}.   
The orbital period of 3.53~hr was obtained by \citet{thors00}.
The data of this figure is the same as those in Figure 41 of Paper I,
but we reanalyzed the data as mentioned below.
The $U-B$ data of \citet{chi64} and \citet{gen63} are systematically
0.3 mag redder than those of \citet{she64}, so we shifted them up
(toward blue) by 0.3 mag in Figure \ref{v533_her_v_bv_ub_color_curve}(c).

In Paper I, we determined the color excess as $E(B-V)=0.05\pm0.05$ 
on the basis of the general track in the color-color diagram, and
the distance modulus as $(m-M)_V=10.8\pm0.2$\footnote{It should be
noted that $(m-M)_V=11.1$ in Table 2 of Paper I is
a typographical error.}  by the time-stretching method
(see Paper I for other estimates of reddening and distance).
The NASA/IPAC galactic dust absorption map
gives $E(B-V)=0.038 \pm 0.002$ in the direction toward V533~Her,
$(l, b)= (69\fdg1887, +24\fdg2733)$.  
Here we adopt this smaller value of $E(B-V)=0.038$ and
we plot the color-color diagram of V533~Her in Figure 
\ref{color_color_diagram_templ_rs_oph_v446_her_v533_her_no2}(d),
resulting in a better matching with the general tracks of novae.

We plot three distance-reddening relations in Figure
\ref{distance_reddening_rs_oph_v446_her_v533_her_iv_cep}(c).
The lines of $(m-M)_V=10.8$ and $E(B-V)=0.038$ cross at $d=1.36$~kpc,
so we adopt $d=1.36$~kpc as the distance to V533~Her.
The distance was estimated also by \citet{coh85} to be $d\sim1.32$~kpc from
the expansion parallax method together with the nebular expansion velocity
of $v_{\rm exp}=1050$~km~s$^{-1}$.  \citet{gil00} obtained
$d=1.25\pm0.30$~kpc together with $v_{\rm exp}=850\pm150$~km~s$^{-1}$.
Our distance of $d=1.36$~kpc is consistent with the both values given
by \citet{coh85} and \citet{gil00}.
Green et al.'s (2015) distance-reddening relation
is also consistent with the value of $E(B-V)=0.038$.

Adopting $E(B-V)=0.038$ and $(m-M)_V=10.8$, we plot
the color-magnitude diagram in Figure
\ref{hr_diagram_rs_oph_v446_her_v533_her_t_pyx_outburst}(c).
We added the $BV$ data (open magenta circles) of \citet{kre66}.
The nebular phase started around UT 1963 April 19 at $m_V=7.2$
\citep{chi64r}.  After that, these three tracks branch off and diverge.
In the nebular phase, the [\ion{O}{3}] emission lines dominate the spectrum 
at the blue edge of the $V$ filter.  Because the response of
each $V$ filter is slightly different at the blue edge, 
the [\ion{O}{3}] emission lines make a large difference 
in the $V$ magnitude among the observers. 
This can be seen clearly in Figures 3, 4, and 5 of \citet{kre66},
in which the $V$ magnitude of each observer started to branch off
after UT 1963 April 10 (JD 2438129.5),
while the $B$ magnitude was essentially the same among the observers.
This is also shown clearly in
Figure \ref{v533_her_v_bv_ub_color_curve}(a) and 
\ref{v533_her_v_bv_ub_color_curve}(b).
In the color-magnitude diagram,
there is a sharp (cusp) turning point on the data of
\citet{she64} at $M_V=-3.97$ and $(B-V)_0=-0.40$.
We identify V533~Her as a V1974~Cyg type in the color-magnitude diagram
as listed in Table \ref{color_magnitude_turning_point}.

\subsection{T~Pyx (1966,2011)}
\label{t_pyx}
T~Pyx is a recurrent nova with six recorded outbursts in 1890,
1902, 1920, 1944, 1966, and 2011.  The orbital period of 1.83~hr was
obtained by \citet{uth10}.   Figure \ref{t_pyx_v_bv_ub_color} shows
the $V$, $(B-V)_0$, and $(U-B)_0$ evolutions
of the 1966 and 2011 outbursts, where the $UBV$ data are taken
from \citet{egg67} and \citet{lan70} and the $BV$ data are taken
from the SMARTS and AAVSO archives.  We also added the X-ray light curve
of T~Pyx taken from the {\it Swift} web
page\footnote{http://www.swift.ac.uk/} \citep{eva09}.
 
We adopt the distance of $4.8$~kpc after \citet{sok13} and
the extinction of $E(B-V)=0.25$ from Paper I 
(see Paper I for other estimates of the reddening and distance).  
Then, the distance modulus is $(m-M)_V=14.2$ for $E(B-V)=0.25$ and 
$d=4.8$~kpc.
Figure \ref{t_pyx_pw_vul_nq_vul_dq_her_v_bv_ub_color_logscale_no6}
compares four light curves of the novae, T~Pyx, DQ~Her, PW~Vul, and NQ~Vul.
The timescales of these four novae are almost the same except
the early fluctuations in the optical maximum phase. 
The light curves of these four novae almost overlap each other
in the later decline phase and decline as $t^{-3}$ 
(thin solid black line) except for the period of dust blackout of DQ~Her,
where $t$ is the time after the outburst.  
We obtained the following relation among them:
\begin{eqnarray}
(m-M)_{V, \rm T~Pyx} &=& 14.2 \cr
&=& (m-M+\Delta V)_{V,\rm DQ~Her} \cr
&=& 8.2 + (+6.0) =14.2 \cr
&=& (m-M+\Delta V)_{V,\rm PW~Vul} \cr
&=& 13.0 + (+1.2) =14.2 \cr
&=& (m-M+\Delta V)_{V,\rm NQ~Vul} \cr
&=& 13.6 + (+0.6) =14.2,
\label{distance_mudulus_t_pyx_dq_her_pw_vul_nq_vul}
\end{eqnarray}
where $\Delta V$ is the difference between the $V$ light curve of T~Pyx
and that of a target nova.  We shift the optical light curve of DQ~Her
down by $\Delta V=6.0$ mag, that of PW~Vul down by $\Delta V=1.2$ mag, 
and that of NQ~Vul down by $\Delta V=0.6$ mag against that of T~Pyx.
The distance of DQ~Her was obtained with the trigonometric
parallax method to be $d=386^{+33}_{-29}$~pc \citep{har13}.  
Adopting $A_V=3.1\times E(B-V)=0.31$ \citep{ver87}, we obtain
the distance modulus of DQ~Her as $(m-M)_V =8.24\pm0.18$.  
Thus, we use $(m-M)_{V,\rm DQ~Her} = 8.2$.
We have already obtained $(m-M)_{V,\rm PW~Vul} = 13.0$
in Section \ref{pw_vul_cmd}, and $(m-M)_{V,\rm NQ~Vul} = 13.6$
in Section \ref{nq_vul}.  The relations in
Equation (\ref{distance_mudulus_t_pyx_dq_her_pw_vul_nq_vul})
strongly suggest that the overlapping region of
$t^{-3}$ law has almost the same absolute brightness among novae having
similar timescales, although the peak absolute brightnesses are different.  

Using $(m-M)_V=14.2$ and $E(B-V)=0.25$, we plot
the color-magnitude diagram of T~Pyx in Figure 
\ref{hr_diagram_rs_oph_v446_her_v533_her_t_pyx_outburst}(d),
where the data are taken from \citet{lan70} for the 1966 outburst
and from the SMARTS and AAVSO archives for the 2011 outburst.
The nebular phase started around UT 1967 March 12, at $m_V\sim9.6$
\citep{cat69}, corresponding to the turning point of the track,
denoted by a large open red square at $M_V=-4.59$ and $(B-V)_0=-0.56$.
The track of T~Pyx is located near that of V1974~Cyg in the early phase,
and turns gradually to the left (toward blue) over the track of V1500~Cyg,
and then suddenly turns to the right (toward red) near the starting point
of the nebular phase. 
We add the epoch when the SSS phase started
at $m_V=10.8$ about 120 days after the outburst \citep{chom14}.
We identify T~Pyx as a V1500~Cyg type in the color-magnitude
diagram as tabulated in Table \ref{color_magnitude_turning_point}.

\subsection{IV~Cep 1971}
\label{iv_cep}
Figure \ref{iv_cep_v_bv_ub_color} shows the $V$, $(B-V)_0$, and $(U-B)_0$
evolutions of IV~Cep.  The $UBV$ data of IV~Cep are taken from 
\citet{mac72} and \citet{koh73}.  The $V$ data are taken from \citet{bur85}. 
IV~Cep is a fast nova with $t_2=25$ and $t_3=42$~days \citep[e.g.][]{koh73}.
In Paper I, we determined $E(B-V)=0.70\pm0.05$
from fitting in the color-color diagram, and 
$(m-M)_V=14.7\pm0.2$ by the time-stretching method
(see Paper I for other estimates of reddening and distance).
We reanalyzed the same data and redetermined 
the reddening as $E(B-V)= 0.65\pm0.05$, 
because, for $E(B-V)= 0.65$,
more color-color data of IV~Cep are concentrated on the general track
(especially on the open diamond) in the color-color diagram of Figure 
\ref{color_color_diagram_iv_cep_nq_vul_qu_vul_qv_vul_no2}(a).
Then, the distance is calculated to be $d=3.4$~kpc for $E(B-V)=0.65$.

We reanalyzed the distance-reddening relation with this new value of
$E(B-V)=0.65$ as shown in Figure 
\ref{distance_reddening_rs_oph_v446_her_v533_her_iv_cep}(d).
The data of this figure are the same as those in Figure 34(d) of
Paper I, but we added the distance-reddening relation (solid black line)
given by \citet{gre15}.
Because Marshall et al.'s $(l, b)=(99\fdg5,-1\fdg5)$, open red squares, 
is closest to the position of IV~Cep, $(l, b)= (99\fdg6137, -1\fdg6381)$, 
the three trends of distance-reddening relations, i.e.,
Marshall et al.'s data, $E(B-V)=0.65$, and $(m-M)_V = 14.7$,
cross each other at $E(B-V)\approx0.65$ and $d\approx3.4$~kpc.
Thus, we adopt $E(B-V)=0.65$, $(m-M)_V = 14.7$, and $d=3.4$~kpc.

Using $E(B-V)=0.65$ and $(m-M)_V = 14.7$,
we plot the color-magnitude diagram of IV~Cep in Figure 
\ref{hr_diagram_iv_cep_nq_vul_v1370_aql_gq_mus_outburst}(a).
The track almost follows that of PW~Vul and LV~Vul.
Thus, we regard IV~Cep as an LV~Vul type
in the color-magnitude diagram as listed in Table
\ref{color_magnitude_turning_point}.
Strong emission lines of [\ion{O}{3}] appeared between UT 1971 September 12
and 22 \citep{ros75}, which is an indication of the nebular phase.
We identify the start of the nebular phase
at $(B-V)_0=-0.37$ and $M_V=-3.23$, denoted by a large open red square in 
Figure \ref{hr_diagram_iv_cep_nq_vul_v1370_aql_gq_mus_outburst}(a).
This starting point is located on the line of two-headed red arrow.

\subsection{NQ~Vul 1976}
\label{nq_vul}
NQ~Vul belongs to the dust blackout type novae like FH~Ser.
Figure \ref{nq_vul_v_bv_ub_color} shows the $V$, $(B-V)_0$, and $(U-B)_0$
evolutions of NQ~Vul, where the $UBV$ data are taken from \citet{yam77}, 
\citet{lan77}, \citet{cha77}, and \citet{due79}, and
the $V$ and visual magnitudes are from \citet{dip78}
and the AAVSO archive, respectively.
The data of this figure are the same as those in Figure 43 of Paper I,
but we reanalyzed them as mentioned below.
The $B-V$ colors of \citet{due79} are systematically
bluer by 0.05 mag and of \citet{yam77} are redder by 0.05 mag
than the other data, so that we shift them down by 0.05 mag and
up by 0.05 mag, respectively.  The $U-B$ colors of \citet{due79} and
\citet{yam77} are systematically redder by 0.1 and bluer by 0.1 mag
than the other data, so we shift them up by 0.1 mag and down by 0.1 mag,
respectively.  
Using these color data, we fit the color-color evolution of NQ~Vul
with the general track of novae and obtain $E(B-V)=1.0\pm0.05$ as shown in
Figure \ref{color_color_diagram_iv_cep_nq_vul_qu_vul_qv_vul_no2}(b).
We reanalyzed the color data but obtained the same result as  
that in Paper I.

Figure \ref{distance_reddening_nq_vul_v1370_aql_qu_vul_os_and}(a) shows
various distance-reddening relations toward NQ~Vul,
$(l, b)= (55\fdg3552, +1\fdg2899)$.
The data are the same as those in Figure 35(a) of Paper I,
but we add Green et al.'s (2015) relation.
Our lines $(m-M)_V=13.6$ and $E(B-V)=1.0$ cross at the point 
$d=1.26$~kpc and the cross point is midway between the distance
reddening relations of \citet{mar06} and \citet{gre15}.
Thus, we confirmed the same values as in Paper I.

Using $(m-M)_V=13.6$ and $E(B-V)=1.0$, we plot
the color-magnitude diagram of NQ~Vul in Figure
\ref{hr_diagram_iv_cep_nq_vul_v1370_aql_gq_mus_outburst}(b).
The track of NQ~Vul is located closely to that of FH~Ser
(solid orange lines), although the data are scattered.
The large variation in the early phase data is partly owing to a few pulses
on the $V$ light curve in the pre-maximum phase.
We can see two small brightenings before the optical maximum 
in Figure \ref{nq_vul_v_bv_ub_color}(a).
The color-magnitude data obtained by \citet{due79}, 
which are connected by a thin solid magenta line,
show two clockwise movements that correspond to the two pulses
before the optical maximum in Figure
\ref{hr_diagram_iv_cep_nq_vul_v1370_aql_gq_mus_outburst}(b).
The first clockwise looping is close to the track of FH~Ser.
The second clockwise movement departs from
the track of FH~Ser and then approaches the track of V1668~Cyg.
Then, the track of NQ~Vul reaches its peak and 
goes down along between the tracks of
V1668~Cyg and FH~Ser after the optical maximum. 
The color of the track became bluer after a considerable part
of the envelope mass was ejected during these early pulses.
This strongly suggests that the bluer the nova color is the smaller
the envelope mass is. 
We regard NQ~Vul as a FH~Ser type in the color-magnitude diagram
as listed in Table \ref{color_magnitude_turning_point}.
The start of the dust blackout is denoted by a large open red square 
in Figure \ref{hr_diagram_iv_cep_nq_vul_v1370_aql_gq_mus_outburst}(b)
at $(B-V)_0=+0.01$ and $M_V=-4.26$.

\subsection{V1370~Aql 1982}
\label{v1370_aql}
V1370~Aql also shows a dust blackout, but its depth is much shallower
than those of FH~Ser and NQ~Vul.  Figure \ref{v1370_aql_v_bv_ub_color}
depicts the $V$ and visual, $(B-V)_0$, and $(U-B)_0$ evolutions
of V1370~Aql, where the $UBV$ data are very limited.
We found only the data of \citet{oka86} and IAU circular No.\ 3689
for the $UBV$ data and \citet{ros83} for the $BV$ data in addition to
the visual magnitudes from the AAVSO archive.
The $V$ light curve has $t_2=8$ and $t_3=13$~days \citep[e.g.,][]{wil84}.
V1370~Aql was identified as a neon nova by \citet{sni87b}.

In Paper I, we determined the color excess as $E(B-V)=0.35\pm0.05$ from 
the color-color diagram fit, and the distance modulus as 
$(m-M)_V=15.2\pm0.2$ by the time-stretching method
relative to the distance modulus of V1668~Cyg 
(see Paper I for other estimates of reddening and distance).
After that, \citet{hac16k} revised the distance modulus
of V1668~Cyg including the photospheric emission in addition to the 
free-free emission.  Therefore, we redetermine the distance modulus of
V1370~Aql based on the new estimate of the distance modulus of V1668~Cyg
(see also Section \ref{v1668_cyg_cmd}).

Figure \ref{v1370_aql_v1668_cyg_os_and_v_bv_ub_color_logscale_no2}
shows a comparison of V1370~Aql with V1668~Cyg and OS~And. 
We adopt the stretching factor as $f_s=1.12$ and 1.58 for V1668~Cyg
and OS~And against V1370~Aql, respectively.
These three nova light curves overlap each other. 
Then, the brightness difference is 
$\Delta V=2.0$ mag for V1668~Cyg and
$\Delta V=2.2$ mag for OS~And against that
of V1370~Aql.  Using the time-stretching method (see Section
\ref{lv_vul_cmd} for a short explanation of the time-stretching
method), we obtain the distance modulus of V1370~Aql as
\begin{eqnarray}
(m-M)_{V,\rm V1370~Aql} &=& (m-M+\Delta V)_{V,\rm V1668~Cyg} 
- 2.5\log 1.12 \cr 
&=& 14.6 + 2.0 - 0.12 = 16.48 \cr
&=& (m-M+\Delta V)_{V,\rm OS~And} - 2.5\log 1.58 \cr 
&=& 14.8 + 2.2 - 0.50  = 16.50,
\label{distance_modulus_v1370_aql_os_and}
\end{eqnarray}
where we use $(m-M)_{V,\rm V1668~Cyg} = 14.6$
from Sections \ref{v1668_cyg_cmd} and
$(m-M)_{V,\rm OS~And} = 14.8$ from Sections \ref{os_and}.
We adopt $(m-M)_{V,\rm V1370~Aql} = 16.5$. 
Then, the distance is calculated to be $d=12$~kpc for $E(B-V)=0.35$.

\citet{hac16k} calculated the absolute magnitudes of model light curves
of novae for various sets of chemical compositions.  Adopting their
chemical composition of CO nova 3, we obtained a best fit $V$ 
(thin solid blue line) and UV~1455 \AA\  (thin solid magenta line)
light curve model for a $0.95~M_\sun$ WD and plotted them in Figure
\ref{v1370_aql_v1668_cyg_os_and_v_bv_ub_color_logscale_no2}(a).
The fitting of the $V$ light curve gives a distance modulus of
$(m-M)_V=16.5$, being consistent with that obtained from the
time-stretching method mentioned above.  Therefore,
we plot the distance-reddening relations of the $V$ light curve fit
calculated from Equation (\ref{v_distance_modulus}) together with 
$(m-M)_V=16.5$ (solid blue line) and the UV~1455 \AA\  light curve
fitting calculated from Equation (\ref{uv1455_distance_modulus})
(solid magenta line) in Figure
\ref{distance_reddening_nq_vul_v1370_aql_qu_vul_os_and}(b).
We added other distance-reddening relations for V1370~Aql,
$(l, b)= (38\fdg8126,-5\fdg9465)$, in Figure 
\ref{distance_reddening_nq_vul_v1370_aql_qu_vul_os_and}(b),
that is, the relations given by \citet{mar06} and \citet{gre15}.
The three trends of the distance-reddening relations, i.e., 
Marshall et al.'s, and two distance moduli in the $V$ and
UV~1455 \AA\  band, cross each other at
$E(B-V)\approx 0.35$ and $d\approx 12$~kpc,
being consistent with our estimates for V1370~Aql.
Green et al.'s (2015) relation deviates largely from our value of
$E(B-V)\approx 0.35$. 

Using $E(B-V)=0.35$ and $(m-M)_V= 16.5$,
we plot the color-magnitude diagram for V1370~Aql
in Figure \ref{hr_diagram_iv_cep_nq_vul_v1370_aql_gq_mus_outburst}(c).
The peak $V$ magnitude reaches as bright as 
$M_V=m_V - (m-M)_V=6.5 - 16.5 = -10$.
We plot the starting position of the dust blackout in the color-magnitude
diagram by a large open blue square.
V1370~Aql experienced a relatively shallow dust blackout.  The $V$ magnitude
was about $m_V\approx11.2$ just before the dust blackout started as shown
in Figure \ref{v1370_aql_v_bv_ub_color}.
This corresponds to $M_V=11.2-16.5=-5.3$.
In the dust blackout type novae,
their $B-V$ colors are almost constant before the dust blackout as shown
in Figure 2(a) of Paper I for FH~Ser.
Therefore, we expect that $(B-V)_0=-0.03$ at this epoch.
This estimated point is indicated by a large open blue square
in Figure \ref{hr_diagram_iv_cep_nq_vul_v1370_aql_gq_mus_outburst}(c).
The color of $(B-V)_0=-0.03$ is just the same as that of
optically thick free-free emission.
\citet{ros83} concluded that [\ion{O}{3}] had already developed in
September and had been much stronger than H$\beta$.  Therefore, we
may conclude that the nova had already entered the nebular phase
in 1982 August at $m_V\sim14$.   We identify the start of the nebular
phase at $(B-V)_0=-0.35$ and $M_V=-3.10$,
and denote it by a large open red square in Figure 
\ref{hr_diagram_iv_cep_nq_vul_v1370_aql_gq_mus_outburst}(c). 
The track of V1370~Aql almost follows that of LV~Vul in the later phase
and the starting point of the nebular phase is close to, but a bit lower
than, the line of the two-headed red arrow. 
Therefore, we regard V1370~Aql as an LV~Vul type
in the color-magnitude diagram as listed in Table
\ref{color_magnitude_turning_point}.

\subsection{GQ~Mus 1983}
\label{gq_mus}
Figure \ref{gq_mus_v_bv_ub_color_curve} shows the $V$ and visual,
$(B-V)_0$, and $(U-B)_0$ evolutions of GQ~Mus.
The $UBV$ data of GQ~Mus are taken from \citet{whi84}
and IAU circular Nos.\ 3766, 3771, and 3853.  The $V$ data were
observed with the Fine Error Sensor (FES) monitor on board 
{\it IUE}, which are taken from the INES archive data
sever\footnote{http://sdc.cab.inta-csic.es/ines/index2.html}.
The visual photometric data are from the Royal
Astronomical Society of New Zealand (RASNZ) and by AAVSO
\citep[see][for more detail]{hac08kc}.
\citet{kra84} estimated the peak brightness to be $m_{V,\rm max}\approx7.0$ 
(or $m_{V,\rm max}<7.3$).  \citet{hac08kc} adopted
$m_{V,\rm max}\approx7.2$ after \citet{war95}.
GQ~Mus declined with $t_2=15$ and $t_3=40$~days \citep[e.g.,][]{war95}.
The orbital period of 1.43~hr was detected by \citet{dia89}.

Paper I and \citet{hac15k} analyzed the light curve of GQ~Mus and determined
the reddening as $E(B-V)=0.45\pm0.05$ from the color-color diagram fit, 
and the distance modulus in the $V$ band as $(m-M)_V=15.7\pm0.2$
by the time-stretching method (see Paper I and \citet{hac15k} for details).
Here, we adopt $E(B-V)=0.45$ and $(m-M)_V=15.7$ for GQ~Mus after Paper I
and \citet{hac15k}.

Using $E(B-V)=0.45$ and $(m-M)_V=15.7$, we plot the color-magnitude
diagram of GQ~Mus in Figure 
\ref{hr_diagram_iv_cep_nq_vul_v1370_aql_gq_mus_outburst}(d).
We superpose the data of T~Pyx (green symbols) on the figure.
These two tracks almost overlap each other in the middle part of the tracks.
We regard GQ~Mus as a V1500~Cyg type in the color-magnitude diagram
as listed in Table \ref{color_magnitude_turning_point},
because T~Pyx was identified as a V1500~Cyg type.
The nova entered the nebular phase no later than UT 1983 March 4,
at $m_V\approx10.2$ \citep{dre84}, which is denoted by an arrow in Figure 
\ref{hr_diagram_iv_cep_nq_vul_v1370_aql_gq_mus_outburst}(d), i.e.,
near the point of $M_V=-5.70$ and $(B-V)_0=-0.25$. 
This starting point of the nebular phase is much ($\sim 1.7$~mag) above 
the line of two-headed black arrow.

\subsection{QU~Vul 1984\#2}
\label{qu_vul}
Figure \ref{qu_vul_v_bv_ub_color_curve} shows the $V$ and
visual, $(B-V)_0$, and $(U-B)_0$ evolutions of QU~Vul.
The $UBV$ data of QU~Vul are taken from IAU Circular No.\ 
4033, \citet{kol88}, \citet{ber88}, and \citet{ros92}.
The visual data are taken from the AAVSO archive. 
The $B-V$ colors of \citet{ros92} are systematically
bluer by 0.1 mag, so we shift them down by 0.1 mag 
in Figure \ref{qu_vul_v_bv_ub_color_curve}(b).  
\citet{geh85} identified QU~Vul as a neon nova, and
\citet{shaf95} detected the orbital period of 2.68~hr.

Paper I and \citet{hac16k} analyzed the light curve of QU~Vul
and determined the reddening as $E(B-V)=0.55\pm0.05$ from 
the color-color diagram fit
and the distance modulus in the $V$ band as $(m-M)_V=13.6\pm0.2$
by the time-stretching method (see Paper I and \citet{hac16k}
for other estimates of reddening and distance).
Assuming $E(B-V)=0.55$, we plot the color-color diagram of QU~Vul in Figure 
\ref{color_color_diagram_iv_cep_nq_vul_qu_vul_qv_vul_no2}(c).
This figure is the same as Figure 31(c) of Paper I,
but we reanalyzed the color data
as mentioned above in Figure \ref{qu_vul_v_bv_ub_color_curve}(b). 
The color-color evolution is consistent with the general tracks of novae.
Therefore, we adopt $E(B-V)=0.55$ for QU~Vul.

We plot various distance-reddening relations for QU~Vul,
$(l,b)=(68\fdg5108,-6\fdg0263)$, in Figure
\ref{distance_reddening_nq_vul_v1370_aql_qu_vul_os_and}(c).
The thick solid blue line denotes $(m-M)_V=13.6$.
The solid magenta line is the relation calculated from
the model UV~1455 \AA\  light curve fitting of a $0.96~M_\sun$ WD model
with the chemical composition of Ne nova 3 \citep{hac16k}.
We also plot the four distance-reddening relations of Marshall et al. (2006). 
The solid black line is Green et al.'s (2015) relation.
These distance-reddening relations cross consistently at/near
the point of $d=2.4$~kpc and $E(B-V)=0.55$.
Thus, we adopt a set of $E(B-V)=0.55$, $d=2.4$~kpc, and $(m-M)_V=13.6$
for QU~Vul after \citet{hac16k}.  

Adopting $E(B-V)=0.55$ and $(m-M)_V=13.6$, we plot the color-magnitude
diagram of QU~Vul 
in Figure \ref{hr_diagram_qu_vul_os_and_qv_vul_v443_sct_outburst}(a).
The track of QU~Vul roughly overlaps with that of V1974~Cyg, so
we regard QU~Vul as a V1974~Cyg type in the color-magnitude diagram
as listed in Table \ref{color_magnitude_turning_point}.
The nova entered the nebular phase in April 1985 at $m_V\approx9.7$ 
\citep{ros87,ros92} as denoted by an arrow in the figure.
We obtain the starting position of the nebular phase 
at $M_V=-4.01$ and $(B-V)_0=-0.31$
as denoted by a large open red square.  
This point is located on the line of the two-headed black arrow. 
Then the track once made an excursion toward blue up to $(B-V)_0\sim-1.0$ 
followed by the final excursion toward red.

\subsection{OS~And 1986}
\label{os_and}
Figure \ref{os_and_v_bv_ub_color_curve} shows the visual and $V$,
$(B-V)_0$, and $(U-B)_0$ evolutions of OS~And on a linear timescale.
We have already plotted the same visual and $V$, $(B-V)_0$, 
and $(U-B)_0$ light curves in Figure
\ref{v1370_aql_v1668_cyg_os_and_v_bv_ub_color_logscale_no2},
but on a logarithmic timescale.  The $UBV$ data of OS~And
are taken from \citet{kik88}, \citet{ohm87}, and 
IAU Circular Nos.\ 4306, 4342, and 4452.  

In Paper I, we determined the reddening as $E(B-V)=0.15\pm0.05$ from 
the color-color diagram fit and the distance modulus in the $V$ band
as $(m-M)_V=14.7\pm0.2$ by the time-stretching method (see Paper I 
for other estimates of reddening and distance).
We reanalyzed the time-stretching method for V1668~Cyg, V1370~Aql,
and OS~And in Equation (\ref{distance_modulus_v1370_aql_os_and})
because the distance modulus of V1668~Cyg was revised in \citet{hac16k}.
The new distance modulus of OS~And is $(m-M)_V=14.8\pm0.2$.
Then the distance is calculated to be $d=7.3$~kpc for $E(B-V)=0.15$.

We fit our model light curves with the OS~And observation, i.e.,
$V$ and UV~1455 \AA\  light curves.  We adopt a $1.05~M_\sun$ WD model
of the CO nova 3 chemical composition \citep{hac16k}.  This model fits
well both the $V$ and UV~1455 \AA\  light curves as shown in Figure
\ref{v1370_aql_v1668_cyg_os_and_v_bv_ub_color_logscale_no2}(a).
Here, we plot the three model light curves for V1370~Aql ($0.95~M_\sun$), 
V1668~Cyg ($0.98~M_\sun$), and OS~And ($1.05~M_\sun$) with appropriate
time-stretching factors depicted in the figure.
The solid blue/red lines are almost the same for V1370~Aql,
V1668~Cyg, and OS~And,
because these model light curves have a universal shape.
The $V$ light curve fit gives a relation of $(m-M)_V=14.8$ for OS~And.
We plot the distance-reddening relations in Figure
\ref{distance_reddening_nq_vul_v1370_aql_qu_vul_os_and}(d), i.e.,
the lines for $(m-M)_V=14.8$ (solid blue line), given by \citet{gre15}
(solid black line), and UV~1455 \AA\  model light curve fit 
(a $1.05~M_\sun$ WD model with the chemical composition of CO nova 3).
These lines cross at/near the point of $d=7.3$~kpc
and $E(B-V)=0.15$.  Therefore, we adopt
$(m-M)_V=14.8$ and $E(B-V)=0.15$ in this paper.

Using $E(B-V)=0.15$ and $(m-M)_V=14.8$, we plot the color-magnitude
diagram of OS~And in Figure
\ref{hr_diagram_qu_vul_os_and_qv_vul_v443_sct_outburst}(b).
The track denoted by open blue squares (data from IAU Circulars)
is close to that of FH~Ser (solid orange lines),
while that denoted by filled red circles (data from Kikuchi et al.)
is close to that of V1668~Cyg.
We regard OS~And as a V1668~Cyg type in the color-magnitude diagram
and list it in Table \ref{color_magnitude_turning_point}.
The dust-blackout starts at $M_V=-4.59$ 
and $(B-V)_0=+0.14$, denoted by a large open red square.

\subsection{QV~Vul 1987}
\label{qv_vul}
Figure \ref{qv_vul_v_bv_ub_color_curve} shows the visual and $V$,
$(B-V)_0$, and $(U-B)_0$ evolutions of QV~Vul.
The $UBV$ data are taken from \citet{ohs88} and IAU Circular Nos.\ 
4493, 4511, and 4524, and the visual data are from the AAVSO archive.  
We shift the $B-V$ and $U-B$ data of IAU Circulars down
by 0.1 and 0.2 mag, respectively, to match these data to
those of \citet{ohs88}. 
QV~Vul is a dust-blackout type moderately fast nova
with $t_2=50$ and $t_3=53$~days \citep{dow00}. 

In Paper I, we determined the reddening as $E(B-V)=0.60\pm0.05$ from 
the color-color diagram fit
and the distance modulus in the $V$ band as $(m-M)_V=14.0\pm0.2$
from the time-stretching method (see Paper I 
for other estimates of reddening and distance).
Using $E(B-V)=0.60$, we plot the color-color track of QV~Vul
in Figure \ref{color_color_diagram_iv_cep_nq_vul_qu_vul_qv_vul_no2}(d).
The figure is the same as Figure 32(a) of Paper I, but we added
the data of IAU Circulars (open blue circles).
The track of QV~Vul is consistent with the general track of novae
(solid green line), so we adopt $E(B-V)=0.60$.

We plot various distance-reddening relations for QV~Vul,
$(l, b)= (53\fdg8585,+6\fdg9741)$, in Figure
\ref{distance_reddening_qv_vul_v443_sct_v1419_aql_v705_cas}(a), i.e., 
Marshall et al.'s (2006) four relations, Green et al.'s (2015)
relation, $E(B-V)=0.60$, and $(m-M)_V=14.0$.
The various distance-reddening relations cross each other at 
$E(B-V)=0.60$ and $d=2.7$~kpc.
Therefore, we adopt $E(B-V)=0.60$ and $(m-M)_V=14.0$ for QV~Vul.

Using $(m-M)_V=14.0$ and $E(B-V)=0.60$,
we plot the color-magnitude diagram of QV~Vul in Figure
\ref{hr_diagram_qu_vul_os_and_qv_vul_v443_sct_outburst}(c).
The track is similar to, but slightly bluer than, that of FH~Ser.
We regard QV~Vul as a FH~Ser type in the color-magnitude diagram.
The dust blackout started at $m_V\approx9.0$ as indicated by an
arrow and large open red square at $(B-V)_0=-0.03$ and $M_V=-5.0$.

\subsection{V443~Sct 1989}
\label{v443_sct}
Figure \ref{v443_sct_v_bv_ub_color_curve} shows the visual and $V$,
$(B-V)_0$, and $(U-B)_0$ evolutions of V443~Sct.
The $UBV$ data are taken from IAU Circular Nos.\ 
4862, 4865, 4868, 4873, and 4902.
The photographic magnitudes are taken from IAU Circulars Nos.\ 
4862 and 4868. 
V443~Sct possibly reached 7.5~mag at maximum \citep{ros91}.  
The visual magnitudes are from the AAVSO archive.
The global shape of the light curve is similar to that of PW~Vul 
(see Figure 52 of Paper I).

In Paper I, we determined the reddening as $E(B-V)=0.40\pm0.05$ from 
the color-color diagram fit and the distance modulus in the $V$ band
as $(m-M)_V=15.5\pm0.2$ from the time-stretching method (see Paper I 
for other estimates of reddening and distance).
Then, the distance is calculated to be $d=7.1$~kpc for $E(B-V)=0.40$.
We plot various distance-reddening relations in Figure
\ref{distance_reddening_qv_vul_v443_sct_v1419_aql_v705_cas}(b).
The two lines of $E(B-V)=0.40$ and $(m-M)_V=15.5$ cross at $d=7.1$~kpc.
On the other hand, Green et al.'s (2015) (solid black line) and
Marshall et al.'s (2006) four relations cross the line of $(m-M)_V=15.5$
at $E(B-V)\approx0.64$ and $d\approx5.2$~kpc.
\citet{ros91} estimated the reddening as $E(B-V)=0.3$ from the \ion{He}{1}
line ratios.
\citet{anu92} obtained the reddening of $E(B-V)=0.4$ from the Balmer/Paschen
line ratios, being consistent with our value.
\citet{and94} estimated the reddening as $E(B-V)=0.30$ from the H and He
recombination line ratios. 
Here, we adopt $E(B-V)=0.40$ and $(m-M)_V=15.5$
because the $1\sigma$ error of reddening is rather large for 
this direction in the Marshall et al.'s relations.

Using $E(B-V)=0.40$ and $(m-M)_V=15.5$, we plot the color-magnitude
diagram of V443~Sct in Figure
\ref{hr_diagram_qu_vul_os_and_qv_vul_v443_sct_outburst}(d).
The track of V443~Sct is similar to that of PW~Vul,
which is an LV~Vul type, although the number of data
points are small.  Thus, we regard V443~Sct as an LV~Vul type as listed
in Table \ref{color_magnitude_turning_point}.
\citet{ros91} reported that the nova had already
entered the nebular phase in March 1990 after a period of seasonal 
invisibility.  They also wrote that the nova was approaching the nebular
phase in November 1989.  Therefore, we specify that the nova entered
the nebular phase at $m_V\approx12.0$, which is denoted by an arrow
in Figure \ref{hr_diagram_qu_vul_os_and_qv_vul_v443_sct_outburst}(d).

%Fig.35 
%\placefigure{v1419_aql_v_bv_ub_color_curve}

\begin{figure}
%\epsscale{0.75}
%%\epsscale{0.8}
%\epsscale{1.0}
\epsscale{1.15}
\plotone{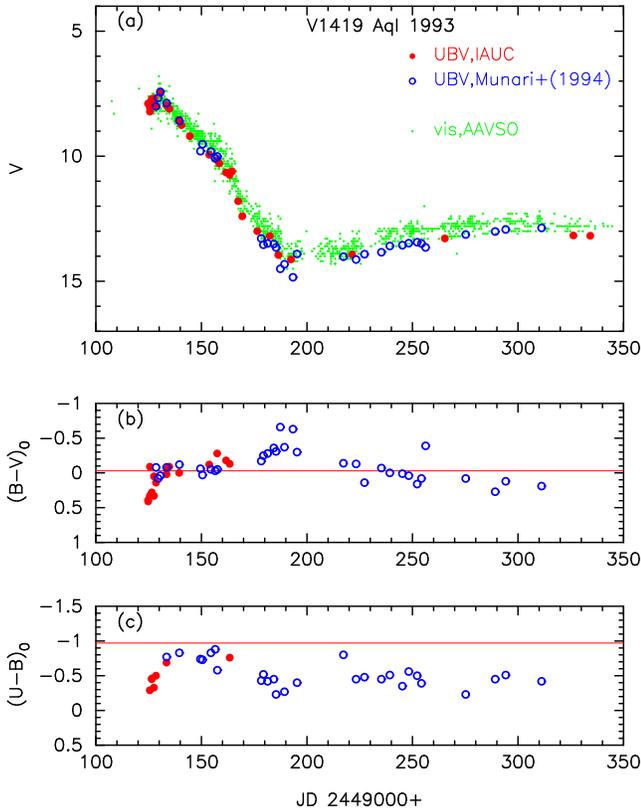}
%\plotone{v1419_aql_v_bv_ub_color_curve.epsi}
%\plotfiddle{evolution1.ps}{5.0cm}{270}{0.4}{0.4}{-170}{220}
\caption{
Same as Figure \ref{v446_her_v_bv_ub_color_curve}, but for V1419~Aql.
We de-reddened $(B-V)_0$ and $(U-B)_0$ colors with $E(B-V)=0.50$.
%%See the main text for the sources of V1419~Aql data.
\label{v1419_aql_v_bv_ub_color_curve}}
\end{figure}

%Fig.36
%\placefigure{color_color_diagram_v1419_aql_v705_cas_u_sco_v745_sco_no2}

\begin{figure*}
%\begin{figure}
%%\epsscale{0.35}
\epsscale{0.75}
%%\epsscale{0.8}
%%\epsscale{1.0}
%%\epsscale{1.15}
\plotone{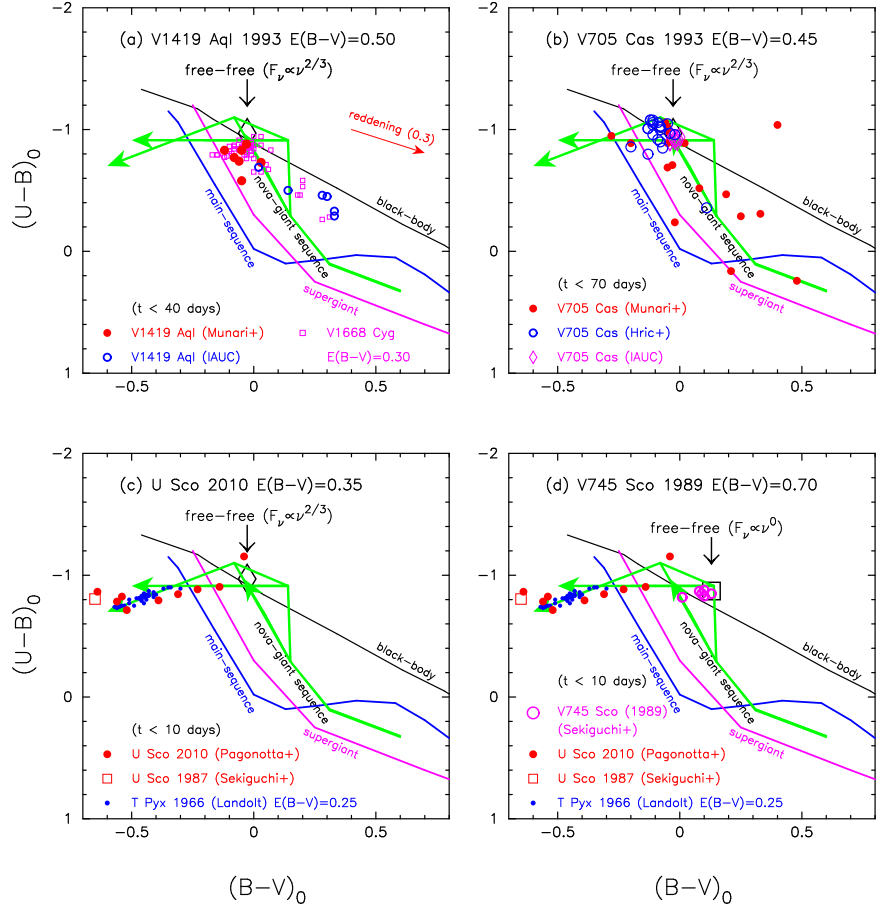}
%\plotone{color_color_diagram_v1419_aql_v705_cas_u_sco_v745_sco_no2.epsi}
%\plotfiddle{evolution1.ps}{5.0cm}{270}{0.4}{0.4}{-170}{220}
\caption{
Same as Figure \ref{color_color_diagram_templ_rs_oph_v446_her_v533_her_no2},
but for (a) V1419~Aql 1993,  (b) V705~Cas 1993, (c) U~Sco (2010),
and (d) V745~Sco (1989). 
%% See the main text for the sources of observational data.
\label{color_color_diagram_v1419_aql_v705_cas_u_sco_v745_sco_no2}}
%\end{figure}
\end{figure*}

%Fig.37 
%\placefigure{hr_diagram_v1419_aql_v705_cas_v382_vel_v1493_aql_outburst}

\begin{figure*}
%\begin{figure}
%\epsscale{0.75}
\epsscale{0.8}
%%\epsscale{1.0}
\plotone{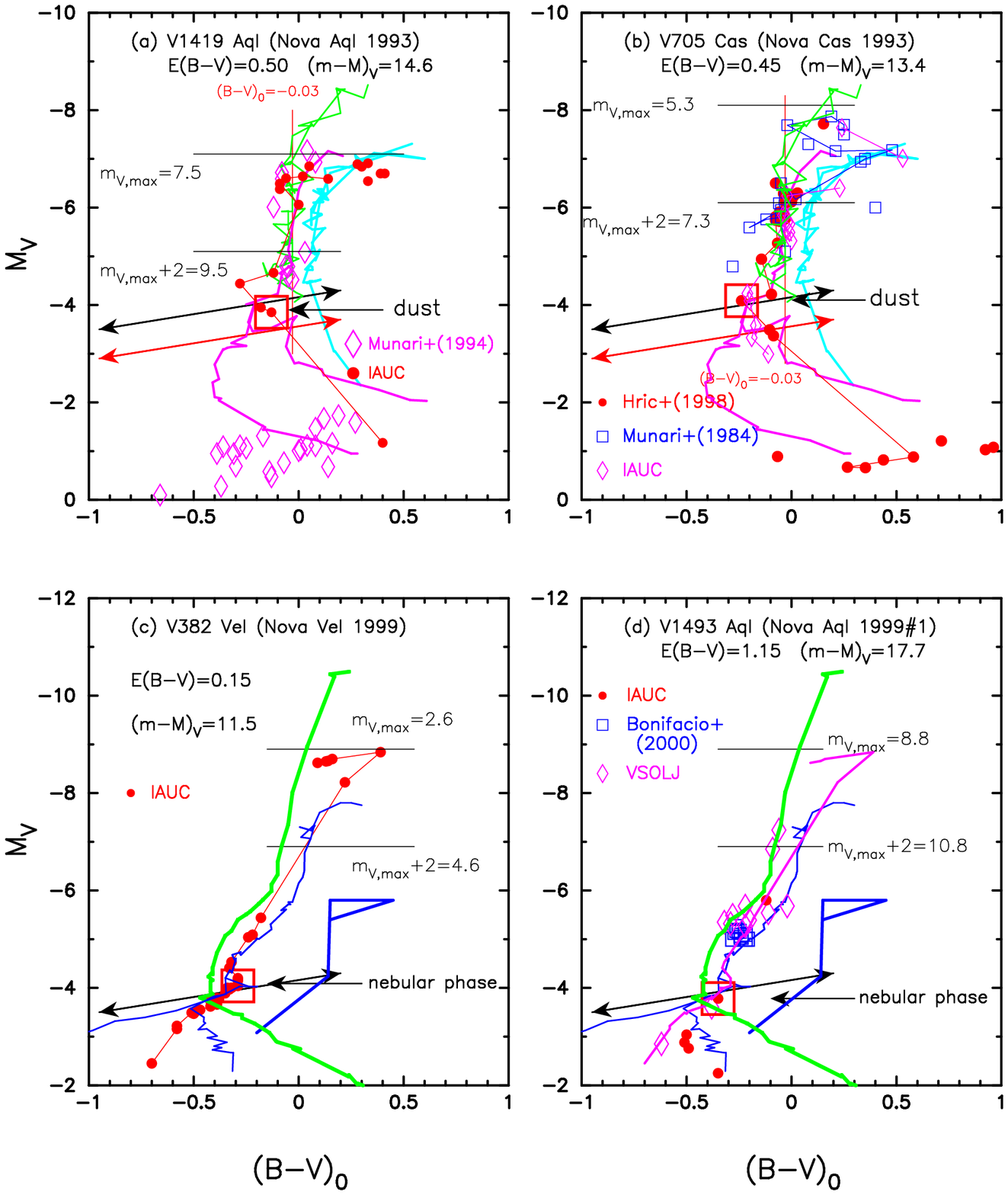}
%\plotone{hr_diagram_v1419_aql_v705_cas_v382_vel_v1493_aql_outburst.epsi}
%\plotfiddle{evolution1.ps}{5.0cm}{270}{0.4}{0.4}{-170}{220}
\caption{
Same as Figure 
\ref{hr_diagram_rs_oph_v446_her_v533_her_t_pyx_outburst}, but
for (a) V1419~Aql, (b) V705~Cas, (c) V382~Vel, and (d) V1493~Aql.
Thin solid green lines in panels (a) and (b) denote the track of V1668~Cyg.
Thin solid magenta lines in panels (a) and (b) denote the tracks
of LV~Vul, but the track of V382~Vel in panel (d).
Thin solid cyan lines in panels (a) and (b) denote the track of FH~Ser.
Thin solid blue lines represent the track of V1974~Cyg in panels (c) and (d).
\label{hr_diagram_v1419_aql_v705_cas_v382_vel_v1493_aql_outburst}}
%\end{figure}
\end{figure*}

\subsection{V1419~Aql 1993}
\label{v1419_aql}
Figure \ref{v1419_aql_v_bv_ub_color_curve} shows the visual and $V$,
$(B-V)_0$, and $(U-B)_0$ evolutions of V1419~Aql.
The $UBV$ data are taken from \citet{mun94a} and IAU Circular Nos.\ 
5794, 5802, 5807, and 5829.  
The visual magnitudes are from the AAVSO archive.
V1419~Aql is a dust-blackout type fast nova
with $t_2=17$ and $t_3=31$~days \citep{dow00}. 

In Paper I, we determined the reddening as $E(B-V)=0.50\pm0.05$ from 
the color-color diagram fit and the distance modulus in the $V$ band
as $(m-M)_V=14.6\pm0.2$ from the time-stretching method (see Paper I 
for other estimates of reddening and distance).
Then, the distance
is calculated to be $d=4.1$~kpc for $(m-M)_V=14.6$ and $E(B-V)=0.50$.
We plot various distance-reddening relations in Figure
\ref{distance_reddening_qv_vul_v443_sct_v1419_aql_v705_cas}(c),
which is the same as Figure 36(d) of Paper I, but we added Green et al.'s
relation (solid black line).
These trends, i.e., $(m-M)_V=14.6$, $E(B-V)=0.50$, Marshall et al.'s
(2006), and Green et al.'s (2015), roughly cross each other
at the point of $E(B-V)\approx0.50$, $(m-M)_V\approx14.6$,
and $d\approx4.1$~kpc, the same values as in Paper I.

Adopting $E(B-V)=0.50$, we plot the color-color diagram of V1419~Aql
in Figure \ref{color_color_diagram_v1419_aql_v705_cas_u_sco_v745_sco_no2}(a).
The figure is the same as Figure 32(c) of Paper I,
but we added more data points from \citet{mun94a} and the track of 
V1668~Cyg (small open magenta squares).
The color-color evolution of V1419~Aql almost overlaps that of V1668~Cyg.
This again confirms that our value of $E(B-V)=0.50\pm0.05$ is reasonable.

Using $E(B-V)=0.50$ and $(m-M)_V=14.6$, we plot the color-magnitude
diagram of V1419~Aql in Figure  
\ref{hr_diagram_v1419_aql_v705_cas_v382_vel_v1493_aql_outburst}(a).
The color-magnitude track of V1419~Aql is also very similar to
that of V1668~Cyg except for the maximum brightness.  The maximum
brightness is close to those of LV~Vul and FH~Ser.
We regard V1419~Aql as a V1668~Cyg type as listed in Table
\ref{color_magnitude_turning_point}.
In the color-magnitude diagram, the dust blackout started at
$(B-V)_0=-0.13$ and $M_V=-3.85$ (a large open red square).

%Fig.38 
%\placefigure{v705_cas_v_bv_ub_color_curve}

\begin{figure}
%\epsscale{0.75}
%%\epsscale{0.8}
%\epsscale{1.0}
\epsscale{1.15}
\plotone{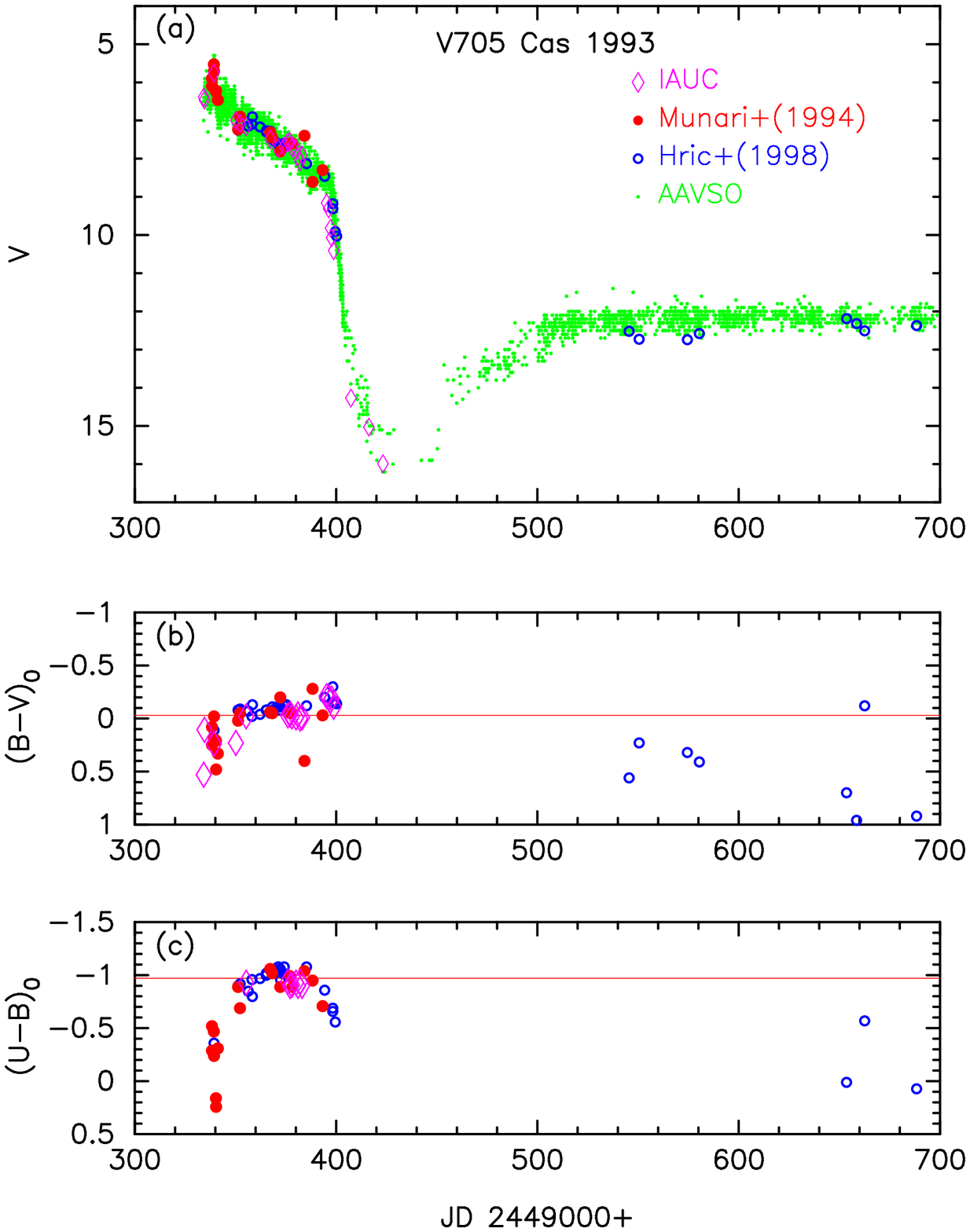}
%\plotone{v705_cas_v_bv_ub_color_curve.epsi}
%\plotfiddle{evolution1.ps}{5.0cm}{270}{0.4}{0.4}{-170}{220}
\caption{
Same as Figure \ref{v446_her_v_bv_ub_color_curve}, but for V705~Cas.
We de-reddened $(B-V)_0$ and $(U-B)_0$ colors with $E(B-V)=0.45$.
%%See the main text for the sources of V705~Cas data.
\label{v705_cas_v_bv_ub_color_curve}}
\end{figure}

\subsection{V705~Cas 1993}
\label{v705_cas}
Figure \ref{v705_cas_v_bv_ub_color_curve} shows the visual and $V$,
$(B-V)_0$, and $(U-B)_0$ evolutions of V705~Cas.
The $UBV$ data are taken from \citet{mun94b}, \citet{hri98},
and IAU Circular Nos.\ 
5905, 5912, 5914, 5920, 5928, 5929, 5945, and 5957.
The visual data are from the AAVSO archive.
V705~Cas is a dust-blackout type nova with $t_2=33$ and $t_3=61$~days
\citep{hri98}.  The orbital period of 5.47~hr was obtained by \citet{ret95}.

Paper I and \citet{hac15k} determined the reddening as $E(B-V)=0.45\pm0.05$ 
and the distance modulus in the $V$ band
as $(m-M)_V=13.4\pm0.1$, both from the model light curve fitting
(see Paper I and \citet{hac15k}
for other estimates of reddening and distance).
We plot three distance-reddening relations in Figure
\ref{distance_reddening_qv_vul_v443_sct_v1419_aql_v705_cas}(d).
\citet{hac15k} fitted their model light curves of a $0.78~M_\sun$ WD
of CO nova 4 chemical composition
with the V705~Cas observation.  The $V$ light curve fit gives
$(m-M)_V=13.4$, and the UV~1455 \AA\  light curve fit also 
yields that of the solid magenta line in Figure
\ref{distance_reddening_qv_vul_v443_sct_v1419_aql_v705_cas}(d).
The solid black line is the distance-reddening relation given
by \citet{gre15}.
These trends roughly cross each other at the point of
$d\approx2.5$~kpc, $E(B-V)\approx0.45$, and $(m-M)_V\approx 13.4$.
Therefore, we adopt $E(B-V)=0.45$ and $(m-M)_V= 13.4$, the same as those
in Paper I and \citet{hac15k}.

Using $E(B-V)=0.45$, we plot the color-color diagram of V705~Cas in 
Figure \ref{color_color_diagram_v1419_aql_v705_cas_u_sco_v745_sco_no2}(b).
Because the dust-blackout started about 60 days after discovery,
we plot only the data for $t < 70$ days.  This figure is the same
as Figure 32(d) of Paper I, but we reanalyzed the color data.
The color-color evolution is consistent with the general track of
novae (solid green lines).
This again confirms our value of $E(B-V)=0.45\pm0.05$.

Using $E(B-V)=0.45$ and $(m-M)_V= 13.4$, we plot
the color-magnitude diagram of V705~Cas in Figure
\ref{hr_diagram_v1419_aql_v705_cas_v382_vel_v1493_aql_outburst}(b).
V705~Cas moves along a flat circle anti-clockwise in the very early phase
near maximum.  After that, it follows the track of FH~Ser 
and then goes along the track of V1668~Cyg.  
We regard V705~Cas as a V1668~Cyg type.
The nova entered a dust blackout phase at the position of 
$(B-V)_0=-0.20$ and $M_V=-4.09$.  % 9.31 -0.40 0.15

%Fig.39 
%\placefigure{v382_vel_v_bv_ub_color_curve}

\begin{figure}
%\epsscale{0.75}
%%\epsscale{0.8}
%\epsscale{1.0}
\epsscale{1.15}
\plotone{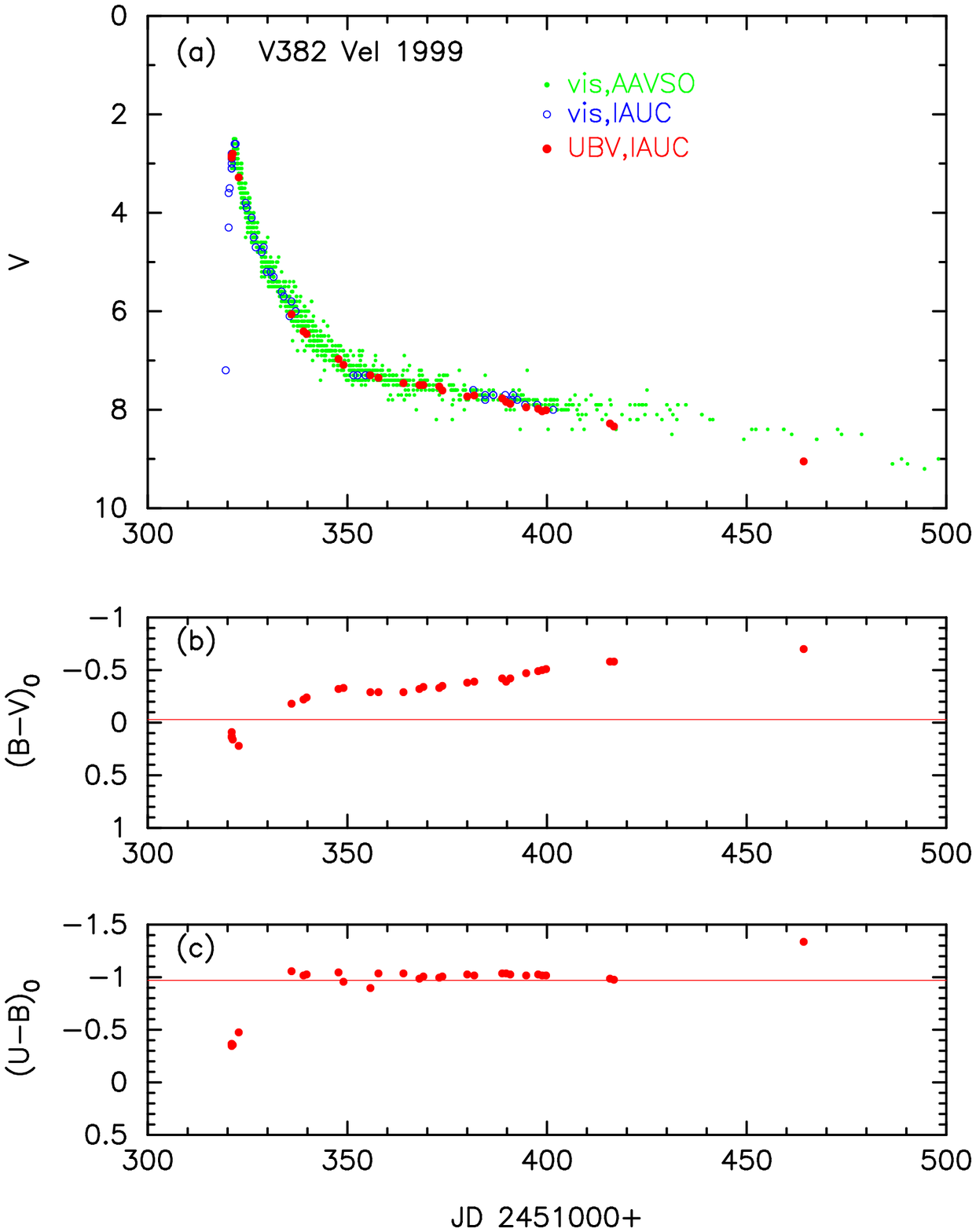}
%\plotone{v382_vel_v_bv_ub_color_curve.epsi}
%\plotfiddle{evolution1.ps}{5.0cm}{270}{0.4}{0.4}{-170}{220}
\caption{
Same as Figure \ref{v446_her_v_bv_ub_color_curve}, but for V382~Vel.
We de-reddened $(B-V)_0$ and $(U-B)_0$ colors with $E(B-V)=0.15$.
%%See the main text for the sources of V382~Vel data.
%%The photographic magnitudes are also take from IAU Circular Nos. xx.
%%The visual data are taken from the archive of AAVSO.
%%The $B-V$ and $U-B$ data of IAU Circulars are shifted down by 0.1 and
%%0.2 mag, respectively, in order to match the data of \citet{ohs88}.
%%See the main text for more detail.
\label{v382_vel_v_bv_ub_color_curve}}
\end{figure}

\subsection{V382~Vel 1999}
\label{v382_vel}
Figure \ref{v382_vel_v_bv_ub_color_curve} shows the visual and $V$,
$(B-V)_0$, and $(U-B)_0$ evolutions of V382~Vel.
The $UBV$ data are taken from IAU Circular Nos.\
7176, 7179, 7196, 7209, 7216, 7226, 7232, and 7277.
The visual magnitudes are taken from IAU Circulars and AAVSO archive.  
V382~Vel was identified as a very fast neon nova \citep{woo99}.
The orbital period of 3.5~hr was detected by \citet{bal06}.

In Paper I and \citet{hac16k}, 
we determined the reddening as $E(B-V)=0.15\pm0.05$ 
and the distance modulus in the $V$ band
as $(m-M)_V=11.5\pm0.1$, both from the model light curve fitting
(see Paper I and \citet{hac16k}
for other estimates of reddening and distance).
Using $E(B-V)=0.15$ and $(m-M)_V=11.5$,
we plot the color-magnitude diagram of V382~Vel in Figure
\ref{hr_diagram_v1419_aql_v705_cas_v382_vel_v1493_aql_outburst}(c).
The color-magnitude diagram of V382~Vel almost coincides with the track of
V1974~Cyg.  Therefore, we regard V382~Vel as a V1974~Cyg type in the
color-magnitude diagram.
The distance is calculated to be $d=1.6$~kpc
for $E(B-V)=0.15$ and $(m-M)_V=11.5$.

\citet{del02} reported that the nebular phase started at least 
by the end of 1999 June, i.e., $\sim40$ days after the optical maximum.
We plot this phase ($M_V=m_V-(m-M)_V\approx7.4-11.5=-4.1$) 
by an arrow in Figure
\ref{hr_diagram_v1419_aql_v705_cas_v382_vel_v1493_aql_outburst}(c).
Then we can specify the position of a cusp point to 
$(B-V)_0=-0.29$ and $M_V=-4.04$ as denoted by a large open red square.

%Fig.40 
%\placefigure{v1493_aql_v_bv_ub_color_curve}

\begin{figure}
%\epsscale{0.75}
%%\epsscale{0.8}
%\epsscale{1.0}
\epsscale{1.15}
\plotone{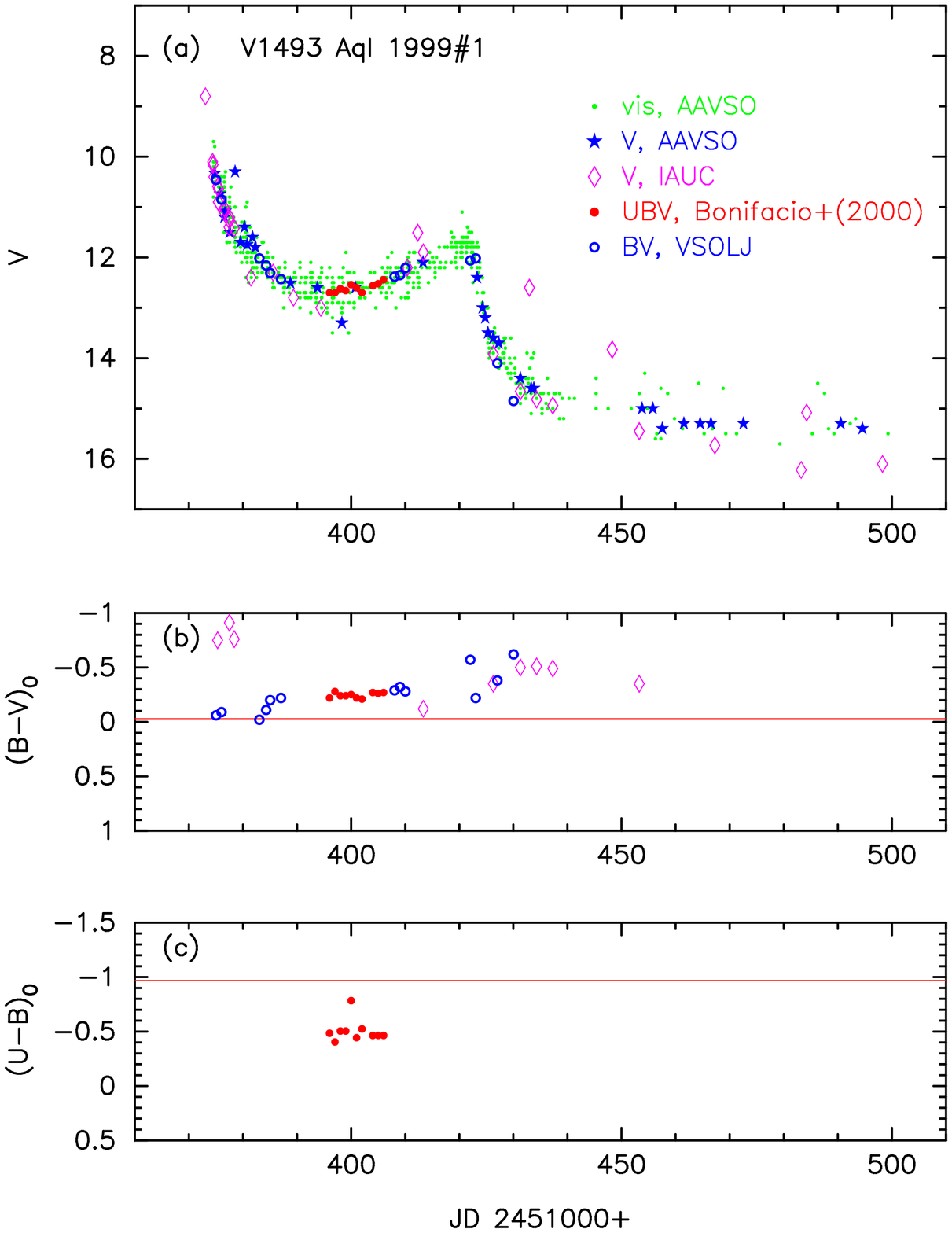}
%\plotone{v1493_aql_v_bv_ub_color_curve.epsi}
%\plotfiddle{evolution1.ps}{5.0cm}{270}{0.4}{0.4}{-170}{220}
\caption{
Same as Figure \ref{v446_her_v_bv_ub_color_curve}, but for V1493~Aql.
We de-reddened $(B-V)_0$ and $(U-B)_0$ colors with $E(B-V)=1.15$.
%%See the main text for the sources of V1493~Aql data.
%%The photographic magnitudes are also take from IAU Circular Nos. xx.
%%The visual data are taken from the archive of AAVSO.
%%The $B-V$ and $U-B$ data of IAU Circulars are shifted down by 0.1 and
%%0.2 mag, respectively, in order to match the data of \citet{ohs88}.
%%See the main text for more detail.
\label{v1493_aql_v_bv_ub_color_curve}}
\end{figure}

%Fig.41 
%\placefigure{v1493_aql_v2275_cyg_v382_vel_v1500_cyg_lv_vul_v_bv_ub_color_logscale}

\begin{figure}
%\epsscale{0.75}
%%\epsscale{0.8}
%\epsscale{1.0}
\epsscale{1.15}
\plotone{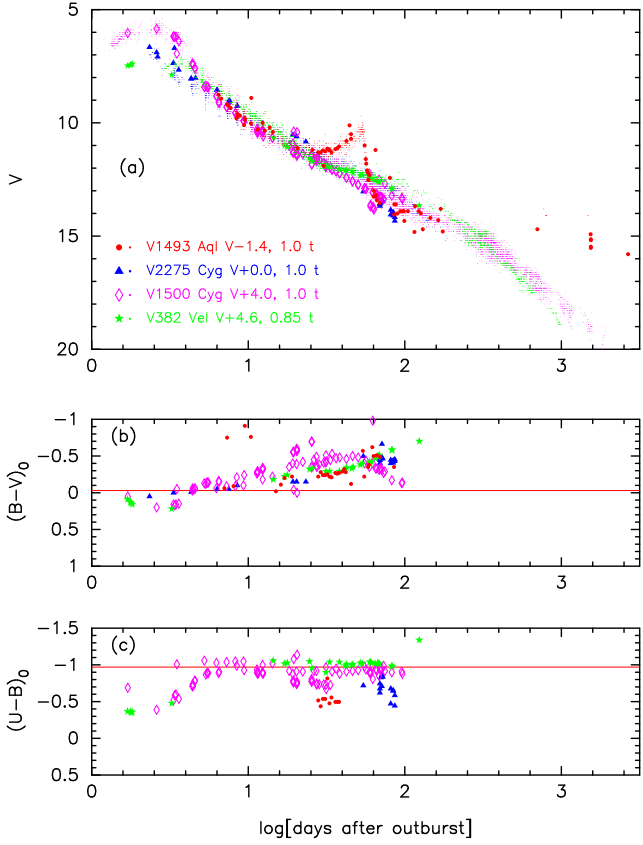}
%\plotone{v1493_aql_v2275_cyg_v382_vel_v1500_cyg_lv_vul_v_bv_ub_color_logscale.epsi}
%\plotfiddle{evolution1.ps}{5.0cm}{270}{0.4}{0.4}{-170}{220}
\caption{
Same as Figure \ref{t_pyx_pw_vul_nq_vul_dq_her_v_bv_ub_color_logscale_no6},
but for V1493~Aql and V2275~Cyg.
The light curves for V382~Vel and V1500~Cyg are added for comparison.
The data for V1493~Aql are the same as those in Figure
\ref{v1493_aql_v_bv_ub_color_curve}.  
%%See the main text for the sources of V2275~Cyg, V382~Vel, and V1500~Cyg data.
\label{v1493_aql_v2275_cyg_v382_vel_v1500_cyg_lv_vul_v_bv_ub_color_logscale}}
\end{figure}

%Fig.42
%\placefigure{distance_reddening_v1493_aql_v1494_aql_v2274_cyg_v2275_cyg}

\begin{figure*}
%\begin{figure}
\epsscale{0.75}
%%\epsscale{0.8}
%%\epsscale{1.0}
%%\epsscale{1.15}
\plotone{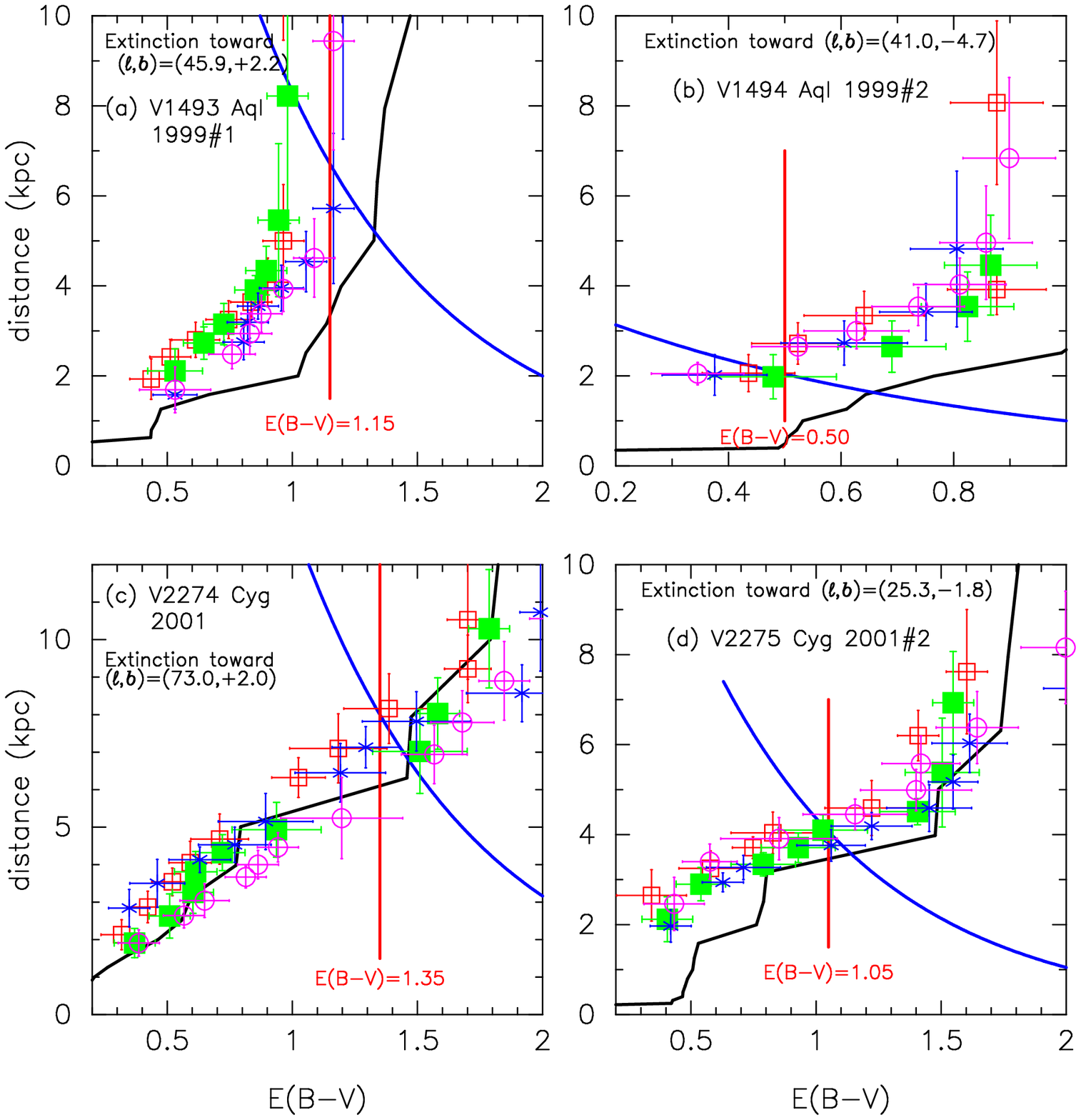}
%\plotone{distance_reddening_v1493_aql_v1494_aql_v2274_cyg_v2275_cyg.epsi}
%\plotfiddle{evolution1.ps}{5.0cm}{270}{0.4}{0.4}{-170}{220}
\caption{
Same as Figure \ref{distance_reddening_fh_ser_pw_vul_v1500_cyg_v1974_cyg},
but for (a) V1493~Aql, (b) V1494~Aql, (c) V2274~Cyg, and (d) V2275~Cyg.
The thick solid blue lines denote (a) $(m-M)_V=17.7$, 
(b) $(m-M)_V=13.1$, (c) $(m-M)_V=18.7$,  and (d) $(m-M)_V=16.3$.
%%The vertical solid red lines represent the color excesses of 
%%(a) $E(B-V)=1.15$, (b) $E(B-V)=0.50$,
%%(c) $E(B-V)=1.35$, and (d) $E(B-V)=1.05$.
%The black solid lines denote the distance-reddening relation given
%by \citet{gre15}.
%In panel (a), the magenta thick solid line represents
%the distance-reddening relation calculated from the UV~1455 \AA\  flux
%fitting with the $0.51~M_\sun$ WD model \citep{hac15k}.
%In panels (a), (b), and (d), two or 
%four sets of data with error bars show distance-reddening relations
%in two or four directions close to each nova, the data of which are taken
%from \citet{mar06}.  
%%See the main text for more detail.
\label{distance_reddening_v1493_aql_v1494_aql_v2274_cyg_v2275_cyg}}
%\end{figure}
\end{figure*}

\subsection{V1493~Aql 1999\#1}
This nova is not studied in Paper I.
Figure \ref{v1493_aql_v_bv_ub_color_curve} shows the visual and $V$, 
$(B-V)_0$, and $(U-B)_0$ evolutions of V1493~Aql.
The $UBV$ data are taken from \citet{bon00}.  
The $BV$ data are from the VSOLJ archive
and IAU Circular Nos.\ 7228, 7254, 7273, and 7313.
The peak could be missed, so we may assume $m_{V,\rm max}=8.8$
for this nova.  It quickly decayed with $t_2=3$ days \citep{bon00}
or with $t_2=7$ and $t_3=24$ days \citep{ven04}.  Therefore,
V1493~Aql belongs to the class of very fast novae.  
The light curve of V1493~Aql has a prominent secondary maximum
about 50 days after the optical maximum \citep[e.g.,][]{bon00, hac09ka}.
\citet{dob06} obtained an orbital period of 3.74 hr.  
The observational period of the $U$ band is too short to derive 
$E(B-V)$ from the general course of nova tracks
(Figure \ref{v1493_aql_v_bv_ub_color_curve}(c)).
Therefore, we could not estimate the extinction from the color-color diagram
of V1493~Aql.
We show the $V$ light curve, $(B-V)_0$, and $(U-B)_0$ color
evolutions in a logarithmic timescale in Figure 
\ref{v1493_aql_v2275_cyg_v382_vel_v1500_cyg_lv_vul_v_bv_ub_color_logscale}
together with the three fast novae, V2275~Cyg, V382~Vel, and V1500~Cyg,
which have similar decline rates.  

The reddening for V1493~Aql was estimated as $E(B-V)=0.33\pm0.1$ 
\citep{bon00} from the intrinsic $B-V$ color at $t_2$,
to be $E(B-V)=1.5$ \citep{ark02} from 
the Balmer decrement (mostly first Balmer lines), and to be 
$E(B-V)=0.57\pm0.14$ \citep{ven04} from the dust reddening curve 
of \citet{dra89}.  These three values are very different from each other.
Therefore, we made a new estimate 
assuming that the intrinsic $B-V$ color of V1493~Aql is
similar to those of V1500~Cyg, V382~Vel, and V2275~Cyg as shown in Figure
\ref{v1493_aql_v2275_cyg_v382_vel_v1500_cyg_lv_vul_v_bv_ub_color_logscale}(b).
Then we obtain the color excess of $E(B-V)=1.15\pm0.05$.

Using the time-stretching method as plotted in Figure
\ref{v1493_aql_v2275_cyg_v382_vel_v1500_cyg_lv_vul_v_bv_ub_color_logscale}(a),
we obtain the apparent distance modulus in $V$ as follows:
\begin{eqnarray}
(m-M)_{V,\rm V1493~Aql} &=& 17.7 \cr
&=& (m-M+\Delta V)_{V,\rm V1500~Cyg} - 2.5 \log 1.0/1.0\cr 
&=& 12.3 + (+4.0 + 1.4) + 0.0 = 17.7 \cr
&=& (m-M+\Delta V)_{V,\rm V382~Vel} - 2.5 \log 0.85/1.0\cr 
&\approx& 11.5 + (+4.6 + 1.4) + 0.18 = 17.68 \cr
&=& (m-M+\Delta V)_{V,\rm V2275~Cyg} - 2.5 \log 1.0/1.0\cr 
&=& 16.3 + (+0.0 + 1.4) + 0.0 = 17.7,
\end{eqnarray}
where we use $(m-M)_{V,\rm V1500~Cyg}=12.3$ from Section 
\ref{v1500_cyg_cmd},
$(m-M)_{V,\rm V382~Vel}=11.5$ from Section \ref{v382_vel},
and $(m-M)_{V,\rm V2275~Cyg}=16.3$ from Section \ref{v2275_cyg}.
Therefore, we adopt $(m-M)_V=17.7$ for V1493~Aql.

Figure \ref{distance_reddening_v1493_aql_v1494_aql_v2274_cyg_v2275_cyg}(a)
shows various distance-reddening relations for V1493~Aql,
$(l, b)=(45\fdg9080,+2\fdg1551)$.
Here we plot four nearby directions calculated by \citet{mar06},
i.e., $(l, b)= (45\fdg75,2\fdg25)$ denoted by open red squares,
$(46\fdg00,2\fdg25)$ by filled green squares, 
$(45\fdg75,2\fdg00)$ by blue asterisks, and
$(46\fdg00,2\fdg00)$ by open magenta circles.
We also plot the distance-reddening relation (solid black line)
given by \citet{gre15}.
The two lines of $E(B-V)=1.15$ and $(m-M)_V=17.7$ cross at the point 
$d=6.7$~kpc and $E(B-V)=1.15$.  
This position is consistent with the distance-reddening relation of
\citet{mar06} but deviates slightly from that of \citet{gre15}.

Using $E(B-V)=1.15$ and $(m-M)_V=17.7$, we plot the color-magnitude
diagram of V1493~Aql in Figure
\ref{hr_diagram_v1419_aql_v705_cas_v382_vel_v1493_aql_outburst}(d).
The track of V1493~Aql is located near the tracks of V382~Vel and V1974~Cyg.
We regard V1493~Aql as a V1974~Cyg type in the color-magnitude diagram.
The nova was clearly in the nebular phase on UT 1999 September 16 
at $m_V\sim15$, i.e., about 65 days after the discovery \citep{ark02},
while the nebular lines began to grow on UT 1999 August 4 ($m_V\sim13$).
We could specify a possible start ($m_V\sim14$) of the nebular phase
at the point $(B-V)_0=-0.35$ and $M_V=-3.78$,
which is denoted by a large open square in Figure
\ref{hr_diagram_v1419_aql_v705_cas_v382_vel_v1493_aql_outburst}(d).

%Fig.43
%\placefigure{v1494_aql_v_bv_ub_color_curve}

\begin{figure}
%\epsscale{0.75}
%%\epsscale{0.8}
\epsscale{1.0}
%\epsscale{1.15}
\plotone{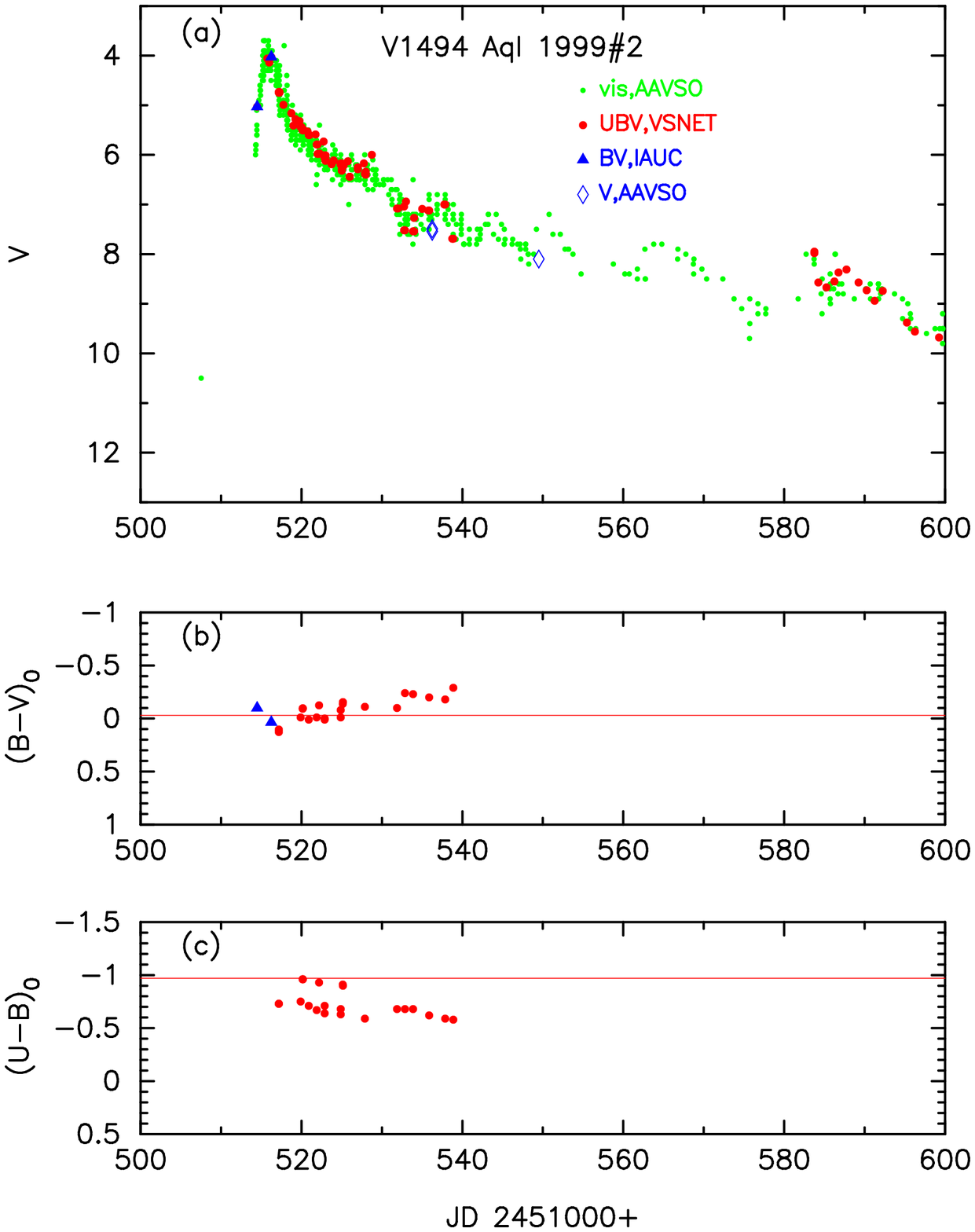}
%\plotone{v1494_aql_v_bv_ub_color_curve.epsi}
%\plotfiddle{evolution1.ps}{5.0cm}{270}{0.4}{0.4}{-170}{220}
\caption{
Same as Figure \ref{v446_her_v_bv_ub_color_curve}, but for V1494~Aql.
We de-reddened the $(B-V)_0$ and $(U-B)_0$ colors with $E(B-V)=0.50$.
The sources of V1494~Aql data are taken from the VSNET and 
AAVSO archives and IAU Circular Nos.\ 7324 and 7327.
%%The photographic magnitudes are also take from IAU Circular Nos. xx.
%%The visual data are taken from the archive of AAVSO.
%%The $B-V$ and $U-B$ data of IAU Circulars are shifted down by 0.1 and
%%0.2 mag, respectively, in order to match the data of \citet{ohs88}.
%%See the main text for more detail.
\label{v1494_aql_v_bv_ub_color_curve}}
\end{figure}

%Fig.44
%\placefigure{v5114_sgr_v1500_cyg_v1494_aql_iv_cep_lv_vul_v_color_logscale}

\begin{figure}
%\epsscale{0.75}
%%\epsscale{0.8}
\epsscale{1.0}
%\epsscale{1.15}
\plotone{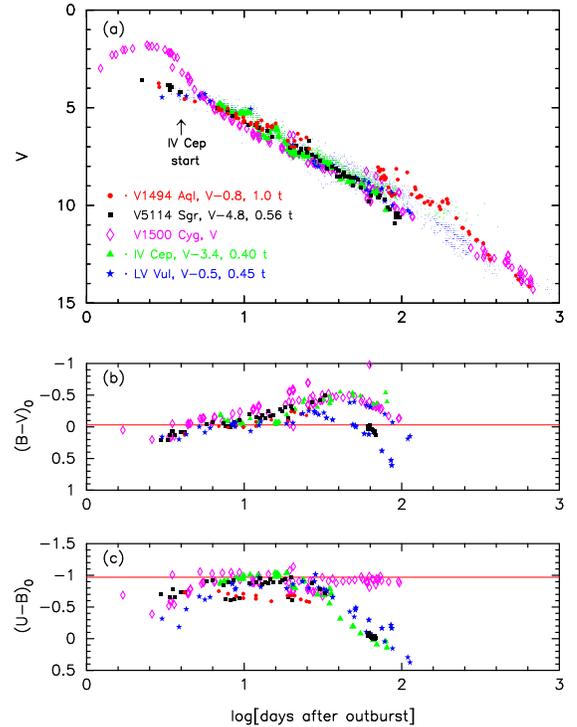}
%\plotone{v5114_sgr_v1500_cyg_v1494_aql_iv_cep_lv_vul_v_color_logscale.epsi}
%\plotfiddle{evolution1.ps}{5.0cm}{270}{0.4}{0.4}{-170}{220}
\caption{
Same as Figure \ref{t_pyx_pw_vul_nq_vul_dq_her_v_bv_ub_color_logscale_no6},
but for V1494~Aql and V5114~Sgr.
The light curves of IV~Cep, LV~Vul, and V1500~Cyg are added for comparison.
The data of V1494~Aql are the same as those in Figure
\ref{v1494_aql_v_bv_ub_color_curve}.  The data of V5114~Cyg
are the same as those in Figure \ref{v5114_sgr_v_bv_ub_color_curve}. 
%In order to make them overlap in the early decline phase,
%we stretched the light curves of LV~Vul, IV~Cep, V1500~Cyg,
%V1494~Aql, and V5114~Sgr by 0.45, 0.40, 1.0, 1.0, 0.56
%and shifted their magnitudes up by $0.5$,  $3.4$, $0.0$,
%$0.8$, and $4.8$~mag, respectively, as indicated in the figure.
\label{v5114_sgr_v1500_cyg_v1494_aql_iv_cep_lv_vul_v_color_logscale}}
\end{figure}

%Fig.45 
%\placefigure{hr_diagram_v1494_aql_v2274_cyg_v2275_cyg_v475_sct}

\begin{figure*}
%\begin{figure}
%\epsscale{0.75}
\epsscale{0.8}
%%\epsscale{1.0}
\plotone{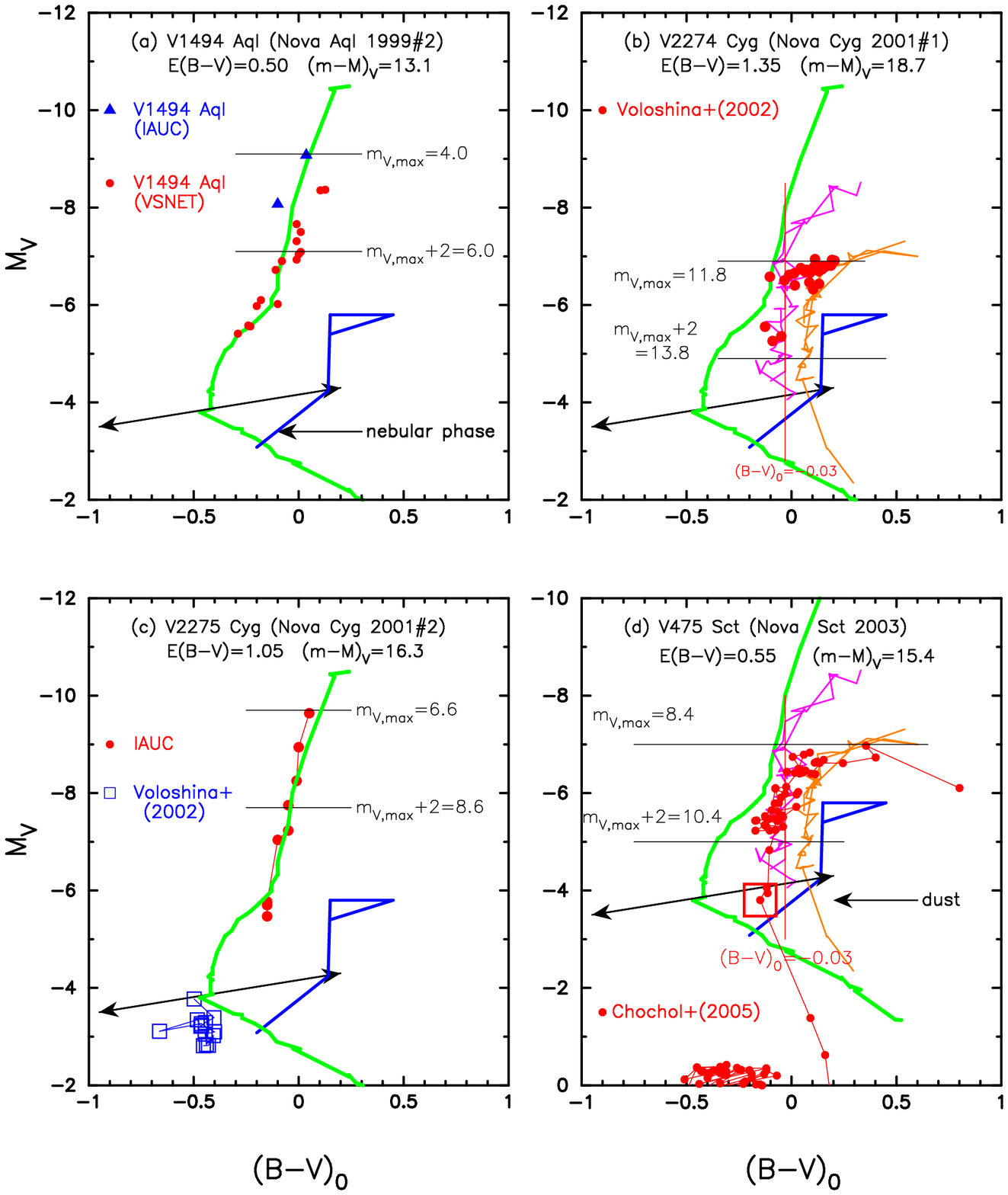}
%\plotone{hr_diagram_v1494_aql_v2274_cyg_v2275_cyg_v475_sct.epsi}
%\plotfiddle{evolution1.ps}{5.0cm}{270}{0.4}{0.4}{-170}{220}
\caption{
Same as Figure 
\ref{hr_diagram_rs_oph_v446_her_v533_her_t_pyx_outburst}, but
for (a) V1494~Aql, (b) V2274~Cyg, (c) V2275~Cyg, and (d) V475~Sct. 
In panels (b) and (d), solid orange lines denote the track of FH~Ser
and solid magenta lines represent the track of V1668~Cyg.
\label{hr_diagram_v1494_aql_v2274_cyg_v2275_cyg_v475_sct}}
%\end{figure}
\end{figure*}

\subsection{V1494~Aql 1999\#2}
\label{v1494_aql}
This nova is not studied in Paper I.
Figure \ref{v1494_aql_v_bv_ub_color_curve} shows the visual and $V$,
$(B-V)_0$, and $(U-B)_0$ evolutions of V1494~Aql.
The $UBV$ data are taken from the VSNET archive.  
The $BV$ data are from IAU Circulars Nos.\ 7324 and 7327.  
The visual data are from the AAVSO archive.
V1494~Aql reached its maximum of $m_{V, \rm max}=4.0$
on UT 1999 December 3.4.  V1494~Aql is a very fast nova with 
$t_2=6.6\pm0.5$ and $t_3=16\pm0.5$ days \citep{kis00}.  
The orbital period of 3.23~hr was detected by \citet{ret00}
and \citet{bar03}.

The reddening for V1494~Aql was estimated by \citet{iij03} to be
$E(B-V)=0.6\pm0.1$ from the interstellar \ion{Na}{1} D1 and D2 lines.
We could not estimate the extinction from 
the color-color diagram of V1494~Aql because there is a
large scatter in the $U-B$ data.
Instead, we use Figure
\ref{v5114_sgr_v1500_cyg_v1494_aql_iv_cep_lv_vul_v_color_logscale}(b)
and obtain $E(B-V)=0.50\pm0.05$ assuming that the intrinsic $(B-V)_0$
color evolution of V1494~Aql is similar to those of the other
color evolutions.  This value is consistent with that
obtained by \citet{iij03}. 

Using the time-stretching method (see Figure
\ref{v5114_sgr_v1500_cyg_v1494_aql_iv_cep_lv_vul_v_color_logscale}(a)), 
we obtain
\begin{eqnarray}
(m-M)_{V,\rm V1494~Aql} &=& 13.1\cr
&=& (m-M+\Delta V)_{V,\rm V1500~Cyg} - 2.5 \log 1.0/1.0\cr 
&=& 12.3 + (-0.0 + 0.8) + 0.0 \approx 13.10 \cr
&=& (m-M+\Delta V)_{V,\rm IV~Cep} - 2.5 \log 0.40/1.0\cr 
&\approx& 14.7 + (-3.4 + 0.8) + 0.99 \approx 13.09 \cr
&=& (m-M+\Delta V)_{V,\rm LV~Vul} - 2.5 \log 0.45/1.0\cr 
&\approx& 11.9 + (-0.5 + 0.8) + 0.87 \approx 13.07 \cr
&=& (m-M+\Delta V)_{V,\rm V5114~Sgr} - 2.5 \log 0.56/1.0\cr 
&\approx& 16.5 + (-4.8 + 0.8) + 0.63 \approx 13.13,
\end{eqnarray}
where we use $(m-M)_{V,\rm V1500~Cyg}=12.3$ 
from Section \ref{v1500_cyg_cmd},
$(m-M)_{V,\rm IV~Cep}=14.7$ from Section \ref{iv_cep},
$(m-M)_{V,\rm LV~Vul}=11.9$ from Section \ref{lv_vul_cmd},
and $(m-M)_{V,\rm V5114~Sgr}=16.5$ from Section \ref{v5114_sgr}.
Because these values are consistent with each other, 
we adopt $(m-M)_V=13.1$ for V1494~Aql.

Figure 
\ref{distance_reddening_v1493_aql_v1494_aql_v2274_cyg_v2275_cyg}(b)
shows various distance-reddening relations for V1494~Aql,
$(l, b)=(40\fdg9735,-4\fdg7422)$.
Here we plot the distance-reddening relations calculated by \citet{mar06},
i.e., for nearby directions, 
$(l, b)= (40\fdg75,-4\fdg50)$ denoted by open red squares,
$(41\fdg00,-4\fdg50)$ by filled green squares, 
$(40\fdg75,-4\fdg75)$ by blue asterisks, and
$(41\fdg00,-4\fdg75)$ by open magenta circles.
The closest one is that denoted by open magenta circles.
We also plot the distance-reddening relation (solid black line)
given by \citet{gre15}. 
The two lines of $E(B-V)=0.50$ and $(m-M)_V=13.1$ cross at the
distance of $d=2.0$~kpc and $E(B-V)=0.50$. 
This cross point is consistent with Marshall et al.'s relation,
but slightly different from that of Green et al.'s.

Using $E(B-V)=0.50$ and $(m-M)_V=13.1$, we
plot the color-magnitude diagram of V1494~Aql in Figure 
\ref{hr_diagram_v1494_aql_v2274_cyg_v2275_cyg_v475_sct}(a).
The track is very similar to that of V1500~Cyg in the middle part
of the whole track.  We regard V1494~Aql as a V1500~Cyg type
in the color-magnitude diagram. 
The nebular phase started at least before the middle of 2000 April,
i.e., around $m_V\sim10$ \citep{iij03}.  We plot this phase 
by an arrow in Figure
\ref{hr_diagram_v1494_aql_v2274_cyg_v2275_cyg_v475_sct}(a).
However, there are no $B-V$ data around the starting point 
of the nebular phase.

%Fig.46
%\placefigure{v2274_cyg_v_bv_ub_color_curve}

\begin{figure}
%\epsscale{0.75}
%%\epsscale{0.8}
%\epsscale{1.0}
\epsscale{1.15}
\plotone{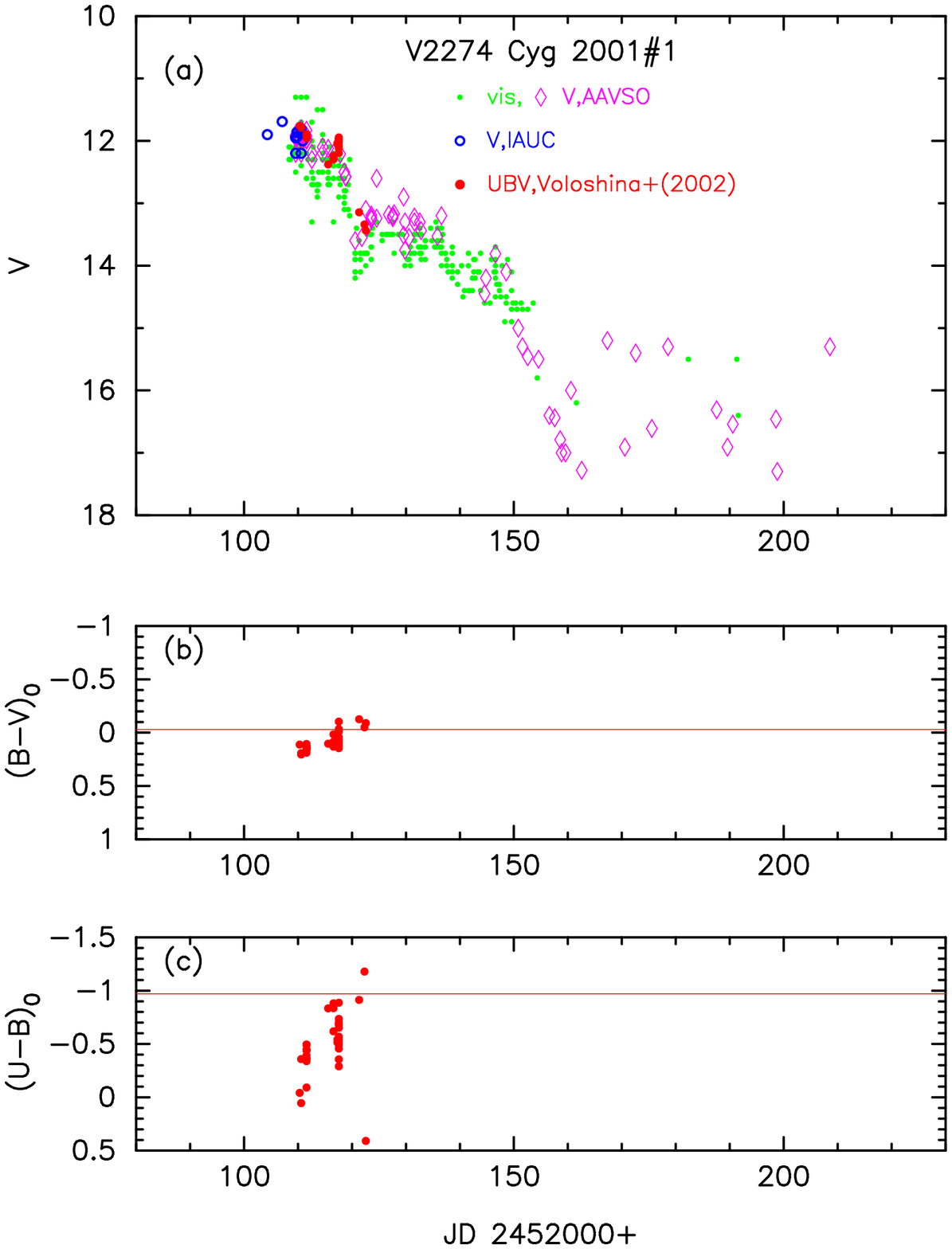}
%\plotone{v2274_cyg_v_bv_ub_color_curve.epsi}
%\plotfiddle{evolution1.ps}{5.0cm}{270}{0.4}{0.4}{-170}{220}
\caption{
Same as Figure \ref{v446_her_v_bv_ub_color_curve}, but for V2274~Cyg.
We de-reddened $(B-V)_0$ and $(U-B)_0$ colors with $E(B-V)=1.35$.
%%See the main text for the sources of V2274~Cyg data.
\label{v2274_cyg_v_bv_ub_color_curve}}
\end{figure}

%Fig.47
%\placefigure{v2274_cyg_v496_sct_fh_ser_nq_vul_v_bv_ub_color_logscale}

\begin{figure}
%\epsscale{0.75}
%%\epsscale{0.8}
%\epsscale{1.0}
\epsscale{1.15}
\plotone{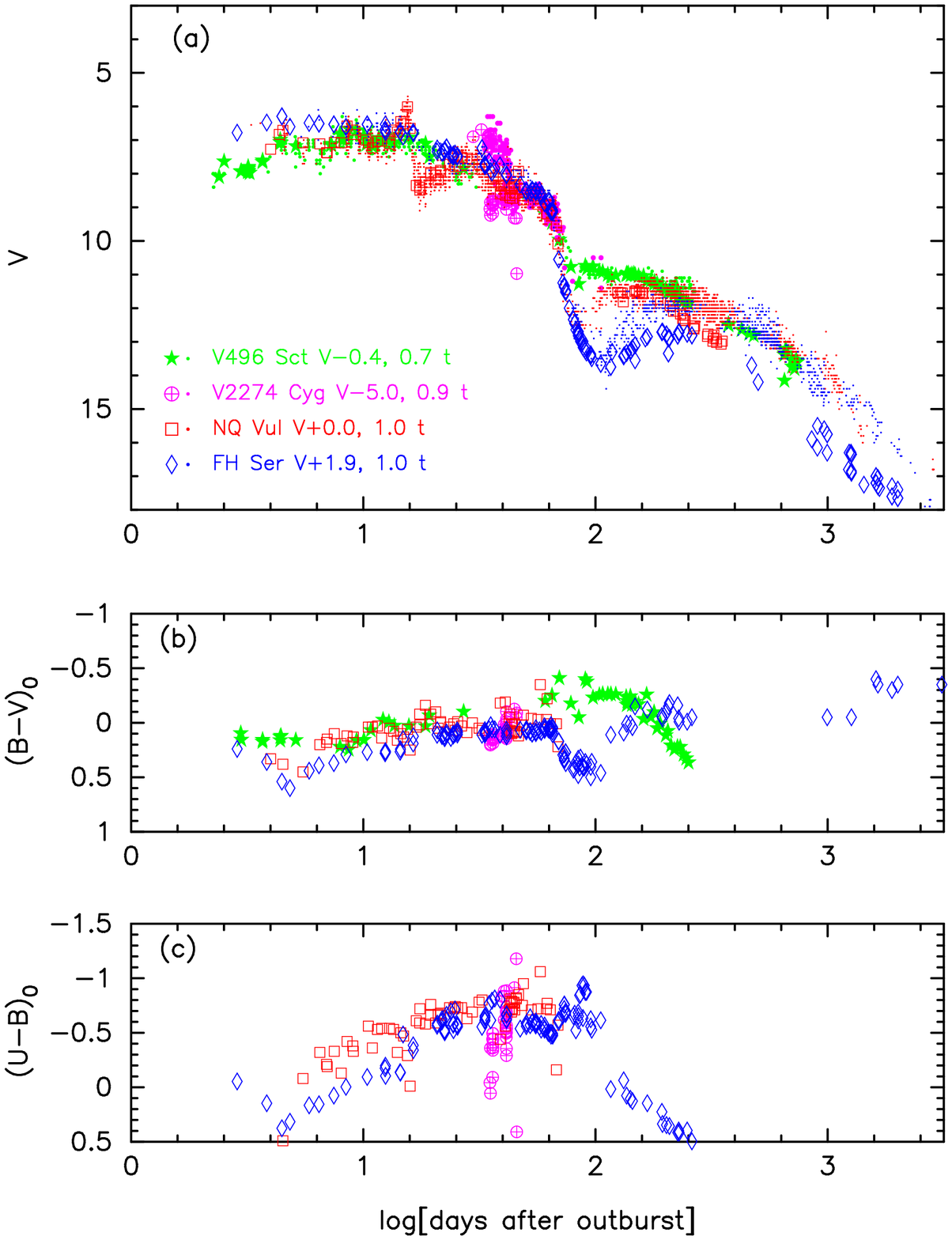}
%\plotone{v2274_cyg_v496_sct_fh_ser_nq_vul_v_bv_ub_color_logscale.epsi}
%\plotfiddle{evolution1.ps}{5.0cm}{270}{0.4}{0.4}{-170}{220}
\caption{
Same as Figure \ref{t_pyx_pw_vul_nq_vul_dq_her_v_bv_ub_color_logscale_no6},
but for V2274~Cyg and V496~Sct.
The light curves of FH~Ser and NQ~Vul are added for comparison.
The data of V2274~Cyg are the same as those in Figure
\ref{v2274_cyg_v_bv_ub_color_curve}.  The data of V496~Sct
are the same as those in Figure 
\ref{v496_sct_v_bv_ub_color_curve}. 
%In order to make them overlap in the early decline phase,
%we shift the light curves of V2274~Cyg, V496~Sct, and FH~Ser up by 5.1,
%$-0.7$, and $-1.9$ mag, respectively, against that of NQ~Vul.
\label{v2274_cyg_v496_sct_fh_ser_nq_vul_v_bv_ub_color_logscale}}
\end{figure}

\subsection{V2274~Cyg 2001\#1}
\label{v2274_cyg}
Figure \ref{v2274_cyg_v_bv_ub_color_curve} shows the visual and $V$,
$(B-V)_0$, and $(U-B)_0$ evolutions of V2274~Cyg.
The $UBV$ data are taken from \citet{vol02a}, $V$ data are
from the AAVSO archive and IAU Circular Nos.\ 7666 and 7668.

In Paper I, we determined the reddening as $E(B-V)=1.35\pm0.10$ from 
the color-color diagram fit and the distance modulus in the $V$ band
as $(m-M)_V=18.7\pm0.2$ from the time-stretching method (see Paper I 
for other estimates of reddening and distance).

The light curve of V2274~Cyg is similar to those of FH~Ser and NQ~Vul
and these three nova light curves have similar decline trends just before 
the dust blackout as shown in Figure
\ref{v2274_cyg_v496_sct_fh_ser_nq_vul_v_bv_ub_color_logscale}(a).
From the time-stretching method,
we obtain the apparent distance modulus of V2274~Cyg:
\begin{eqnarray}
(m-M)_{V,\rm V2274~Cyg} &=& (m-M+\Delta V)_{V,\rm FH~Ser} -2.5\log 0.9 \cr
&=& 11.7 + (+1.9 + 5.0) + 0.11 = 18.71 \cr
&=& (m-M+\Delta V)_{V,\rm NQ~Vul} - 2.5\log 0.9 \cr 
&=& 13.6 + (+0.0 + 5.0) + 0.11 = 18.71,
\end{eqnarray}
where we use $(m-M)_{V,\rm FH~Ser}=11.7$ from Section \ref{fh_ser_cmd} and
$(m-M)_{V,\rm NQ~Vul}=13.6$ from Section \ref{nq_vul}.
We adopt this distance modulus of $(m-M)_{V,\rm V2274~Cyg}=18.7$.
The distance is estimated as $d=8.0$~kpc for $E(B-V)=1.35$.

Figure \ref{distance_reddening_v1493_aql_v1494_aql_v2274_cyg_v2275_cyg}(c)
shows various distance-reddening relations for V2274~Cyg, 
$(l, b)=(73\fdg0415,+1\fdg9910)$.
This figure is the same as Figure 37(b) of Paper I, but
we added the distance-reddening relations given by \citet{gre15}.
The two lines of $E(B-V)=1.35$ and $(m-M)_V=18.7$ 
cross at the distance of $d=8.0$~kpc, which is consistent with
Marshall et al.'s and Green et al.'s relations.

Using $E(B-V)=1.35$ and $(m-M)_V=18.7$, we plot the color-magnitude
diagram of V2274~Cyg in Figure 
\ref{hr_diagram_v1494_aql_v2274_cyg_v2275_cyg_v475_sct}(b).
The track of V2274~Cyg is close to those of FH~Ser and V1668~Cyg.
We regard V2274~Cyg as a V1668~Cyg type in the color-magnitude diagram,
because V2274~Cyg goes down along the track of V1668~Cyg.

%Fig.48 
%\placefigure{v2275_cyg_v_bv_ub_color_curve}

\begin{figure}
%\epsscale{0.75}
%%\epsscale{0.8}
%\epsscale{1.0}
\epsscale{1.15}
\plotone{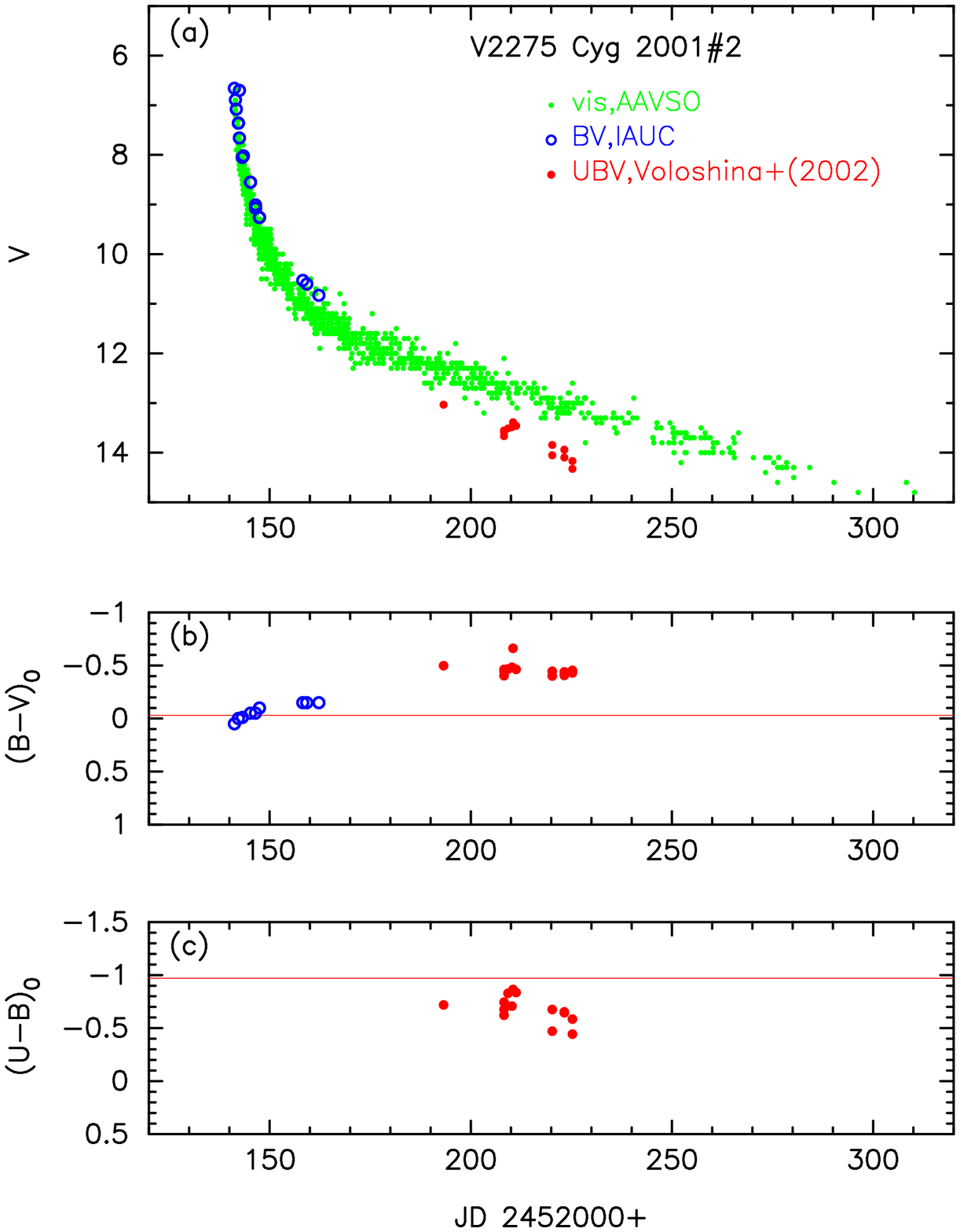}
%\plotone{v2275_cyg_v_bv_ub_color_curve.epsi}
%\plotfiddle{evolution1.ps}{5.0cm}{270}{0.4}{0.4}{-170}{220}
\caption{
Same as Figure \ref{v446_her_v_bv_ub_color_curve}, but for V2275~Cyg.
We de-reddened $(B-V)_0$ and $(U-B)_0$ colors with $E(B-V)=1.05$.
%%See the main text for the sources of V2275~Cyg data.
\label{v2275_cyg_v_bv_ub_color_curve}}
\end{figure}

\subsection{V2275~Cyg 2001\#2}
\label{v2275_cyg}
This nova is not studied in Paper I.
Figure \ref{v2275_cyg_v_bv_ub_color_curve} shows the visual and $V$,
$(B-V)_0$, and $(U-B)_0$ evolutions of V2275~Cyg.
The $V$ maximum is $m_{V,\rm max}=6.6$ \citep[e.g.,][]{sos01}.
Then it gradually declined with $t_2=2.9$ and $t_3=7$~days
\citep{kis02}.  The $V$ light curve of V2275~Cyg and its decay rate are
very similar to those of V1500~Cyg.  We could not estimate the extinction
from the color-color diagram of V2275~Cyg because no $UBV$ data
in the early phase are available.  $UBV$ data were
secured only in the nebular phase \citep{vol02b}.
\citet{bal05} suggested an orbital period of 7.55~hr from photometric
orbital variations.

The $V$ light curve shape and global timescale of V2275~Cyg is very
similar to that of V1500~Cyg.  Assuming that these two novae have
the same brightness in the free-free emission phase, i.e., applying
the time-stretching method to Figure
\ref{v1493_aql_v2275_cyg_v382_vel_v1500_cyg_lv_vul_v_bv_ub_color_logscale}(a),
we obtain
\begin{eqnarray}
(m-M)_{V,\rm V2275~Cyg} &=& (m-M+\Delta V)_{V,\rm V1500~Cyg} \cr 
&=& 12.3 + 4.0 = 16.3, 
\end{eqnarray}
where we use $(m-M)_{V,\rm V1500~Cyg}=12.3$ from Section 
\ref{v1500_cyg_cmd}.  We adopt this value of 
$(m-M)_{V,\rm V2275~Cyg}=16.3$ in the present paper.

\citet{kis02} estimated the absolute magnitude at maximum to be
$M_{V, \rm max}= -9.7$ from the MMRD relations and other empirical
relations together with $t_2=2.9\pm0.5$ days
and $t_3=7\pm1$ days.  Then we calculate the apparent distance modulus
of $(m-M)_V= 6.6 - (-9.7)=16.3$, being coincident with our value of
$(m-M)_V=16.3$.  

\citet{kis02} also discussed that V2275~Cyg closely resembles, in some
respects, the well-studied very fast nova V1500~Cyg.  Assuming that
the intrinsic $B-V$ color is almost the same for V2275~Cyg and V1500~Cyg,
we determine the reddening to be $E(B-V)=1.05\pm0.05$ by overlapping them
as shown in Figure 
\ref{v1493_aql_v2275_cyg_v382_vel_v1500_cyg_lv_vul_v_bv_ub_color_logscale}(b).
This estimate is also consistent with the Kiss et al.\ value of
$E(B-V)=1.0\pm0.1$, obtained with a few interstellar absorption laws.
We adopt $E(B-V)=1.05\pm0.05$ in this paper. 

Figure \ref{distance_reddening_v1493_aql_v1494_aql_v2274_cyg_v2275_cyg}(d)
shows various distance-reddening relations for V2275~Cyg, 
$(l, b)=(89\fdg3170, +1\fdg3905)$.
Here we plot distance-reddening relations calculated by \citet{mar06},
i.e., four nearby directions, $(l, b)= (89\fdg25,+1\fdg25)$
denoted by open red squares,
$(89\fdg50,+1\fdg25)$ by filled green squares, 
$(89\fdg25,+1\fdg50)$ by blue asterisks, and
$(89\fdg50,+1\fdg50)$ by open magenta circles.
The closest one is that denoted by blue asterisks.
We also plot Green et al.'s (2015) distance-reddening relation in the
same figure.
These four trends, Marshall et al.'s (blue asterisks), Green et al.'s
(solid black line), $(m-M)_V=16.3$
(solid blue line), and $E(B-V)=1.05$ (vertical solid black line), 
cross at the point $E(B-V)\approx1.05$ and $d\approx4.1$~kpc.
Thus, we confirm that our estimates of $E(B-V)=1.05$ and
$(m-M)_V=16.3$ are reasonable.

Using $E(B-V)=1.05$ and $(m-M)_V=16.3$,
we plot the color-magnitude diagram of V2275~Cyg in Figure
\ref{hr_diagram_v1494_aql_v2274_cyg_v2275_cyg_v475_sct}(c).
The two nova tracks of V2275~Cyg and V1500~Cyg
overlap each other in the color-magnitude diagram.
Therefore, we regard V2275~Cyg as a V1500~Cyg type in the
color-magnitude diagram.

%Fig.49 
%\placefigure{v475_sct_v_bv_ub_color_curve}

\begin{figure}
%\epsscale{0.75}
%%\epsscale{0.8}
%\epsscale{1.0}
\epsscale{1.15}
\plotone{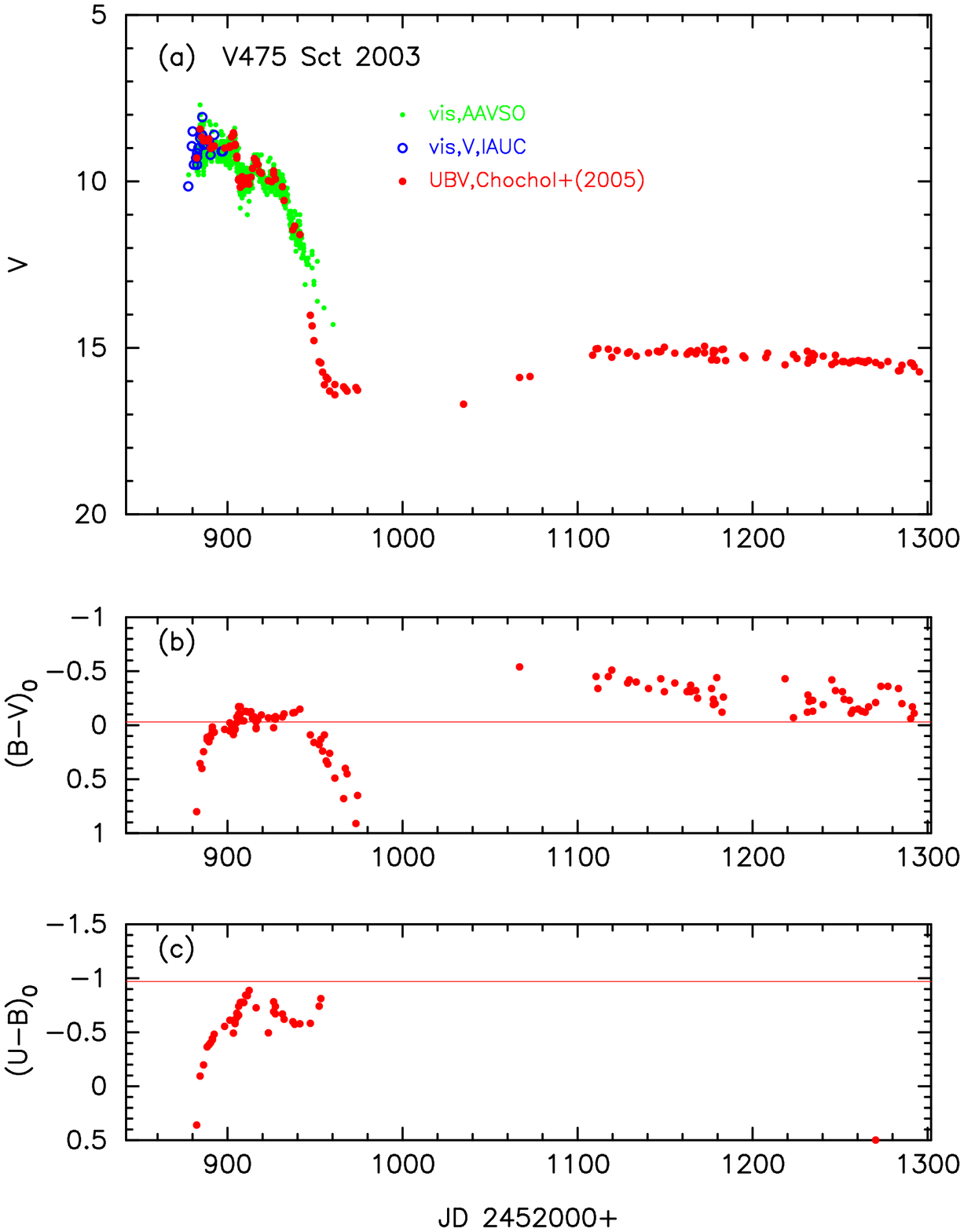}
%\plotone{v475_sct_v_bv_ub_color_curve.epsi}
%\plotfiddle{evolution1.ps}{5.0cm}{270}{0.4}{0.4}{-170}{220}
\caption{
Same as Figure \ref{v446_her_v_bv_ub_color_curve}, but for V475~Sct.
We de-reddened $(B-V)_0$ and $(U-B)_0$ colors with $E(B-V)=0.55$.
%%See the main text for the sources of V475~Sct data.
\label{v475_sct_v_bv_ub_color_curve}}
\end{figure}

%Fig.50 
%\placefigure{v475_sct_pw_vul_nq_vul_dq_her_v_bv_ub_color_logscale_no6}

\begin{figure}
%\epsscale{0.75}
%%\epsscale{0.8}
%\epsscale{1.0}
\epsscale{1.15}
\plotone{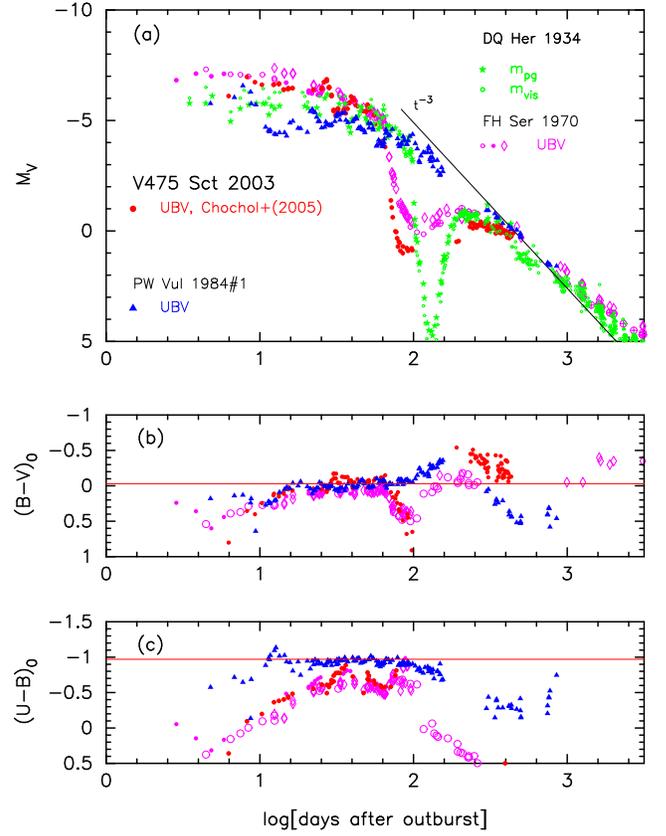}
%\plotone{v475_sct_pw_vul_nq_vul_dq_her_v_bv_ub_color_logscale_no6.epsi}
%\plotfiddle{evolution1.ps}{5.0cm}{270}{0.4}{0.4}{-170}{220}
\caption{
Same as Figure \ref{t_pyx_pw_vul_nq_vul_dq_her_v_bv_ub_color_logscale_no6},
but for V475~Sct.  The sources of V475~Sct data are the same as those in
Figure \ref{v475_sct_v_bv_ub_color_curve}.
Here we adopt $(m-M)_V=15.4$, 13.0, 8.2, and 11.7 
for V475~Sct, PW~Vul, DQ~Her, and FH~Ser, respectively.
%%See the main text for more detail.
\label{v475_sct_pw_vul_nq_vul_dq_her_v_bv_ub_color_logscale_no6}}
\end{figure}

%Fig.51 
%\placefigure{distance_reddening_v475_sct_v5114_sgr_v2362_cyg_v1065_cen}

\begin{figure*}
%\begin{figure}
\epsscale{0.75}
%%\epsscale{0.8}
%%\epsscale{1.0}
%%\epsscale{1.15}
\plotone{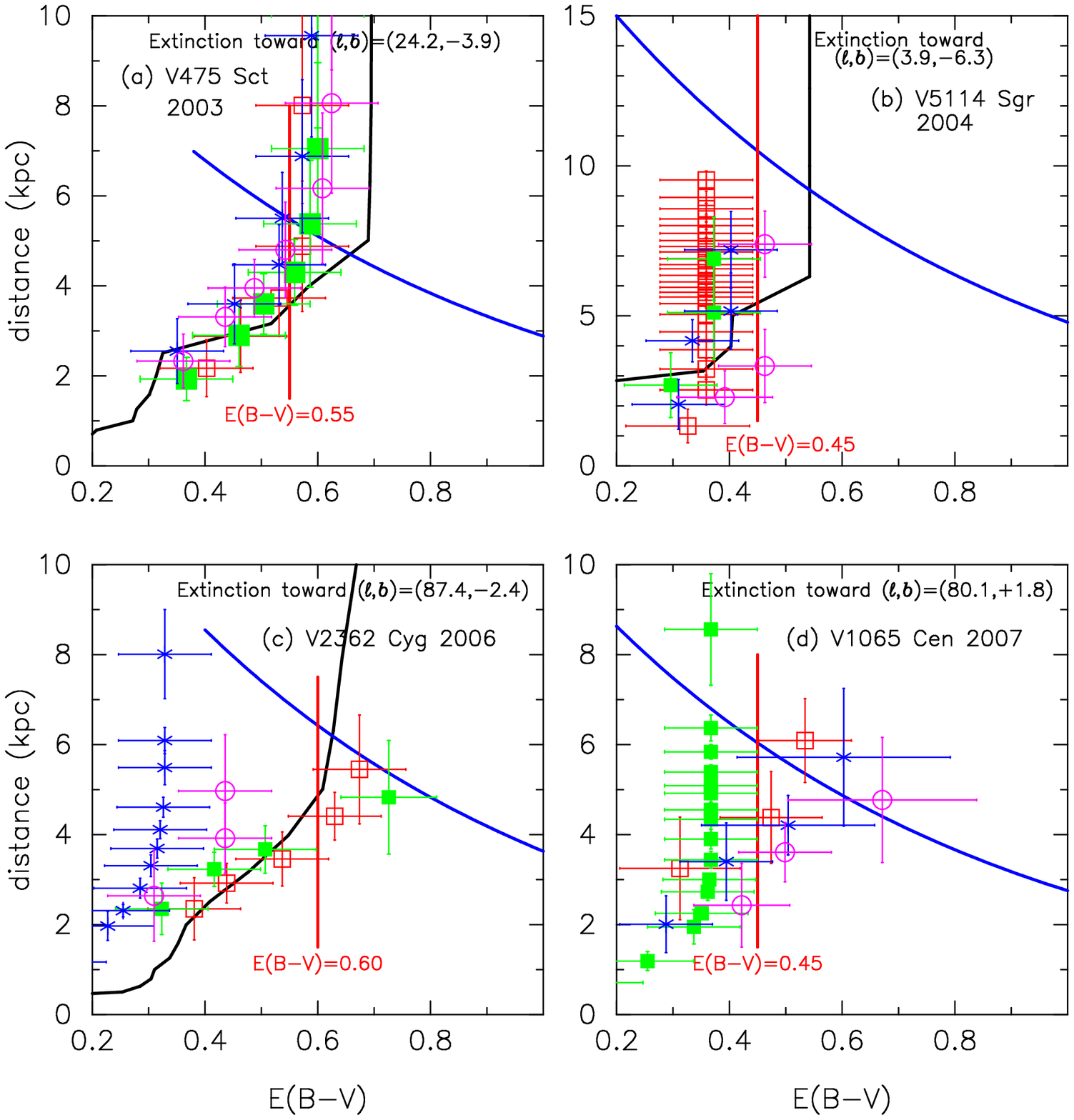}
%\plotone{distance_reddening_v475_sct_v5114_sgr_v2362_cyg_v1065_cen.epsi}
%\plotfiddle{evolution1.ps}{5.0cm}{270}{0.4}{0.4}{-170}{220}
\caption{
Same as Figure \ref{distance_reddening_fh_ser_pw_vul_v1500_cyg_v1974_cyg},
but for (a) V475~Sct, (b) V5114~Sgr, (c) V2362~Cyg, and (d) V1065~Cen.
The thick solid blue lines denote (a) $(m-M)_V=15.4$, 
(b) $(m-M)_V=16.5$, (c) $(m-M)_V=15.9$,  and (d) $(m-M)_V=15.3$.
%The vertical solid red lines represent the color excesses of 
%(a) $E(B-V)=0.55$, (b) $E(B-V)=0.45$,
%(c) $E(B-V)=0.60$, and (d) $E(B-V)=0.45$.
%The black solid lines denote the distance-reddening relation given
%by \citet{gre15}.
%In panel (a), the magenta thick solid line represents
%the distance-reddening relation calculated from the UV~1455 \AA\  flux
%fitting with the $0.51~M_\sun$ WD model \citep{hac15k}.
%In panels (a), (b), and (d), two or 
%four sets of data with error bars show distance-reddening relations
%in two or four directions close to each nova, the data of which are taken
%from \citet{mar06}.  
%See the main text for more detail.
\label{distance_reddening_v475_sct_v5114_sgr_v2362_cyg_v1065_cen}}
%\end{figure}
\end{figure*}

\subsection{V475~Sct 2003}
\label{v475_sct}
Figure \ref{v475_sct_v_bv_ub_color_curve} shows the visual and $V$,
$(B-V)_0$, and $(U-B)_0$ evolutions of V475~Sct.
The $UBV$ data are taken from \citet{cho05}, $V$ and visual data are 
from IAU Circular Nos.\ 8190 and 8200, and the AAVSO archive.

In Paper I, we determined the reddening $E(B-V)=0.55\pm0.10$ from 
the color-color diagram fit and the distance modulus in the $V$ band
as $(m-M)_V=15.6\pm0.2$ by assuming that the absolute magnitude
of the light curve is the same for FH~Ser and V475~Sct
(see Paper I for other estimates of reddening and distance).
To confirm this similarity, 
we plot the light curves of V475~Sct, FH~Ser, PW~Vul, and DQ~Her in Figure
\ref{v475_sct_pw_vul_nq_vul_dq_her_v_bv_ub_color_logscale_no6}, but
we adopt $(m-M)_V=15.4$ rather than the $(m-M)_V=15.6$ of Paper I for V475~Sct.
If we use $(m-M)_V=15.4$, the absolute brightness 
of V475~Sct and FH~Ser almost overlap.

Figure \ref{distance_reddening_v475_sct_v5114_sgr_v2362_cyg_v1065_cen}(a)
shows various distance-reddening relations for V475~Sct,
$(l, b) = (24\fdg2015, -3\fdg9466)$, i.e.,
the relations given by \citet{mar06} and \citet{gre15}.
Our new estimates of $(m-M)_V=15.4$ and $E(B-V)=0.55$ cross at $d=5.5$~kpc,
which is roughly consistent with trend of Marshall et al., but 
deviates from that of Green et al.
Thus, we adopt $E(B-V)=0.55$ and $(m-M)_V=15.4$ in this paper.

The color-magnitude diagram of V475~Sct is plotted in Figure
\ref{hr_diagram_v1494_aql_v2274_cyg_v2275_cyg_v475_sct}(d).
The track is similar to that of FH~Ser in the early phase, but
to that of V1668~Cyg in the later phase.
We regard V475~Sct as a V1668~Cyg type in the color-magnitude diagram.
We specify a starting point of dust blackout, that is, 
$(B-V)_0=-0.15$ and $M_V=-3.80$ denoted by a large open red square in Figure 
\ref{hr_diagram_v1494_aql_v2274_cyg_v2275_cyg_v475_sct}(d).

%Fig.52
%\placefigure{v5114_sgr_v_bv_ub_color_curve}

\begin{figure}
%\epsscale{0.75}
%%\epsscale{0.8}
%\epsscale{1.0}
\epsscale{1.05}
\plotone{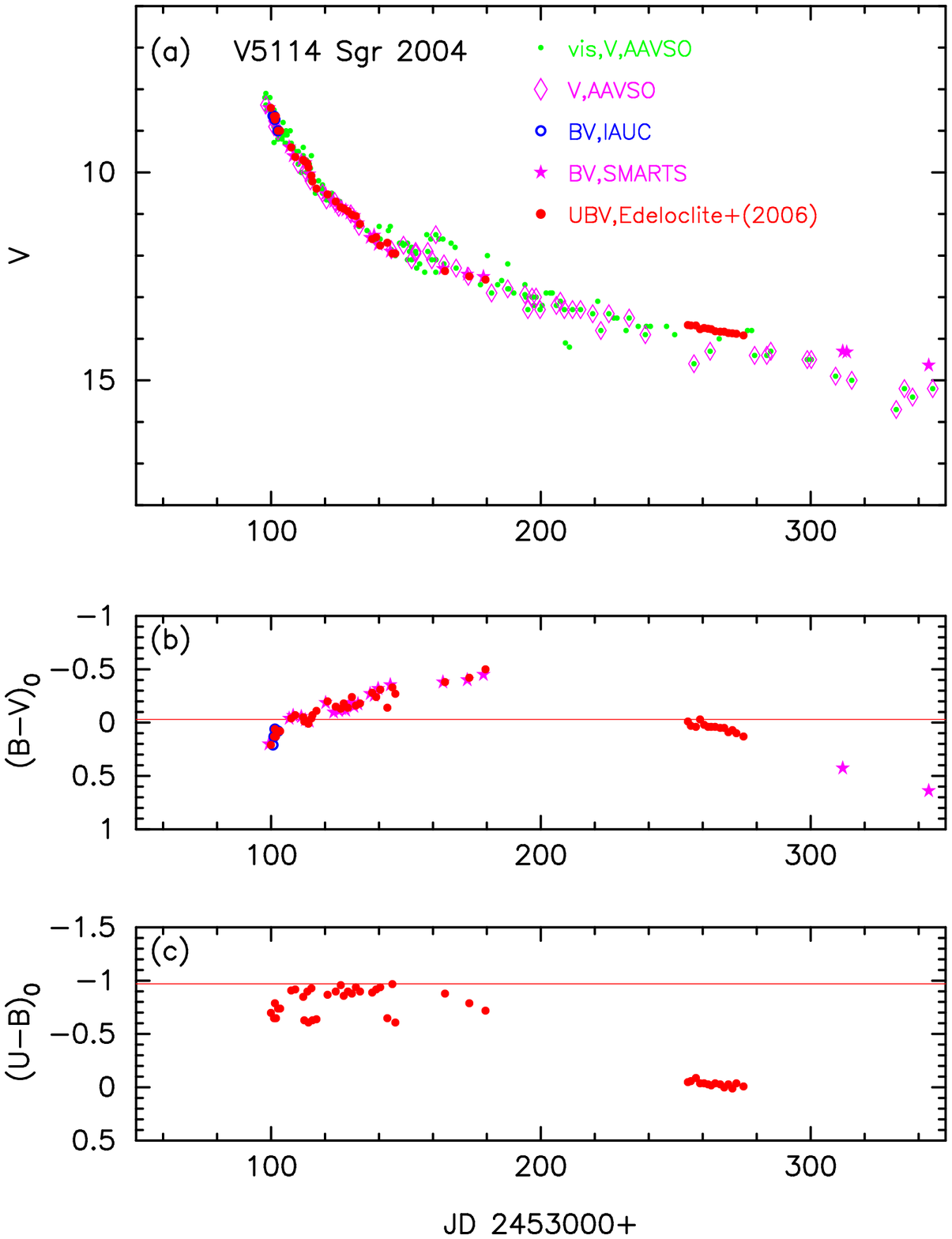}
%\plotone{v5114_sgr_v_bv_ub_color_curve.epsi}
%\plotfiddle{evolution1.ps}{5.0cm}{270}{0.4}{0.4}{-170}{220}
\caption{
Same as Figure \ref{v446_her_v_bv_ub_color_curve}, but for V5114~Sgr.
We de-reddened $(B-V)_0$ and $(U-B)_0$ colors with $E(B-V)=0.45$.
%%See the main text for the sources of V5114~Sgr data.
\label{v5114_sgr_v_bv_ub_color_curve}}
\end{figure}

%Fig.53
%\placefigure{hr_diagram_v5114_sgr_v2362_cyg_v1065_cen_v1280_sco_outburst}

\begin{figure*}
%\begin{figure}
%\epsscale{0.75}
\epsscale{0.8}
%%\epsscale{1.0}
\plotone{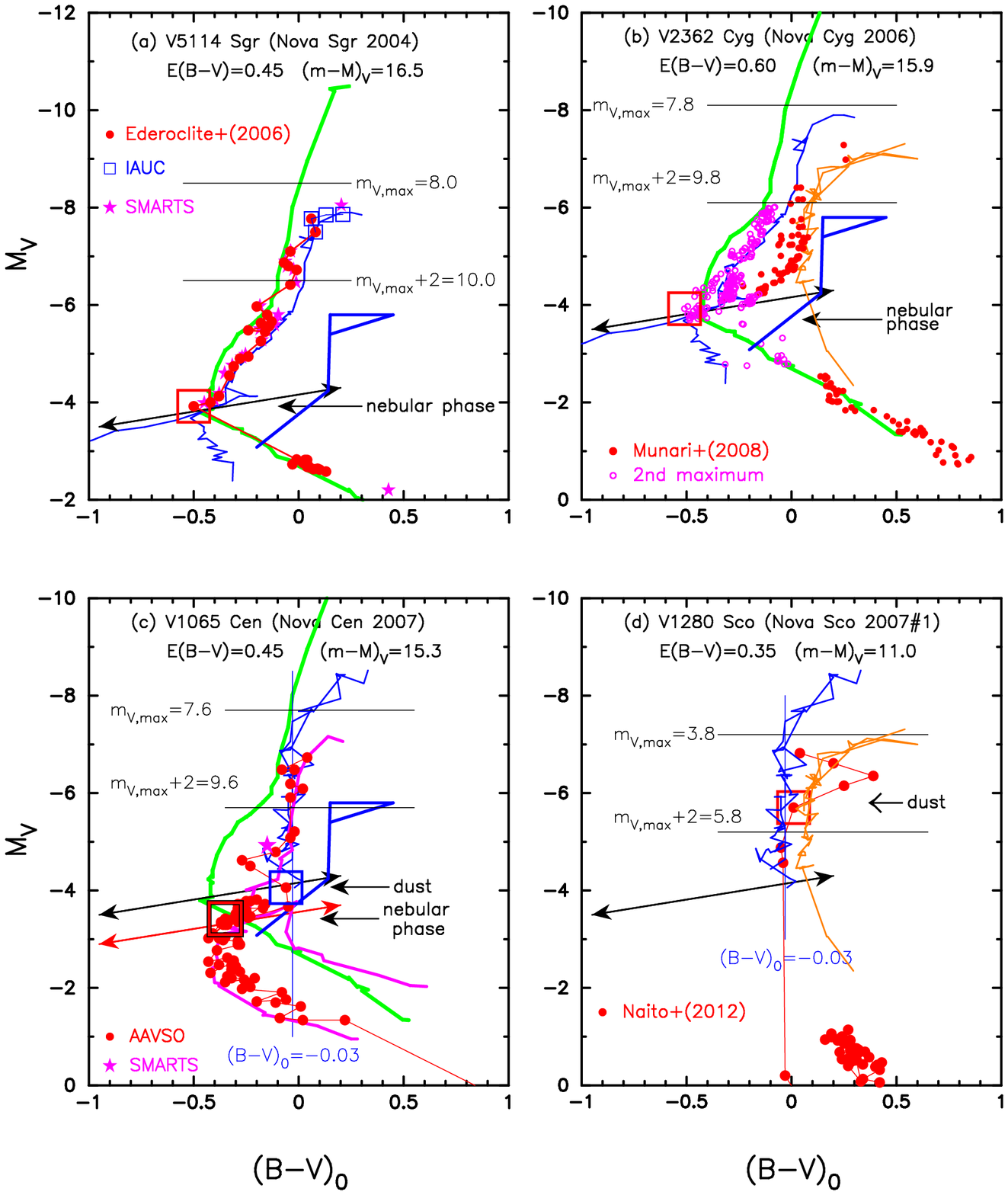}
%\plotone{hr_diagram_v5114_sgr_v2362_cyg_v1065_cen_v1280_sco_outburst.epsi}
%\plotfiddle{evolution1.ps}{5.0cm}{270}{0.4}{0.4}{-170}{220}
\caption{
Same as Figure 
\ref{hr_diagram_rs_oph_v446_her_v533_her_t_pyx_outburst}, but
for (a) V5114~Sgr, (b) V2362~Cyg, (c) V1065~Cen, and (d) V1280~Sco.
Thick solid green lines represent the track of V1500~Cyg and 
thick solid blue lines the track of PU~Vul.
Thin solid blue lines denote the track of V1974~Cyg in panels (a) and (b),
but that of V1668~Cyg in panels (c) and (d).
Solid orange lines in panels (b) and (d) denote the track of FH~Ser.
Solid magenta lines in panel (c) represent the track of LV~Vul.
\label{hr_diagram_v5114_sgr_v2362_cyg_v1065_cen_v1280_sco_outburst}}
%\end{figure}
\end{figure*}

\subsection{V5114~Sgr 2004}
\label{v5114_sgr}
Figure \ref{v5114_sgr_v_bv_ub_color_curve} shows the visual and $V$,
$(B-V)_0$, and $(U-B)_0$ evolutions of V5114~Sgr.
The $UBV$ data are taken from \citet{ede06}, $BV$ data
from the SMARTS archive and IAU Circular Nos.\ 8306 and 8310,
and visual and $V$ data from the AAVSO archive.

In Paper I, we determined the reddening as $E(B-V)=0.45\pm0.05$ from 
the color-color diagram fit and the distance modulus in the $V$ band
as $(m-M)_V=16.5\pm0.2$ from the time-stretching method
(see Paper I for other estimates of reddening and distance).
Then, the distance is calculated to be $d=10.5$~kpc.
Figure \ref{distance_reddening_v475_sct_v5114_sgr_v2362_cyg_v1065_cen}(b)
shows various distance-reddening relations for V5114~Sgr,
$(l, b)= (3\fdg9429,-6\fdg3121)$.
This figure is the same as Figure 37(d) of Paper I, but we added
the distance-reddening relation given by \citet{gre15}.  
The three trends, Marshall et al.'s trend,
$(m-M)_V=16.5$, and $E(B-V)=0.45$, consistently cross at 
$d\sim10.5$~kpc and $E(B-V)\sim0.45$, although
Green et al.'s relation deviates from this cross point.  
Thus, we adopt the same values of $(m-M)_V=16.5$ and $E(B-V)=0.45$
as those in Paper I.

Using $E(B-V)=0.45$ and $(m-M)_V=16.5$,
we plot the color-magnitude diagram of V5114~Sgr in Figure
\ref{hr_diagram_v5114_sgr_v2362_cyg_v1065_cen_v1280_sco_outburst}(a).
The track is close to those of V1500~Cyg (thick solid green line)
and V1974~Cyg (thin solid blue line).
This matching also supports our values of $E(B-V)=0.45$ and $(m-M)_V=16.5$.
We regard V5114~Sgr as a V1974~Cyg type in the color-magnitude diagram.
We also specify a turning point of $(B-V)_0=-0.50$ and $M_V=-3.92$,
which is denoted by a large open red square in Figure 
\ref{hr_diagram_v5114_sgr_v2362_cyg_v1065_cen_v1280_sco_outburst}(a).
The nebular phase started around 57 days after optical maximum,
when the [\ion{O}{3}] lines became stronger than the permitted lines
\citep[see Figure 3 of][]{ede06}.

% Fig.54
%\placefigure{v2362_cyg_v_bv_ub_color_curve}

\begin{figure}
%\epsscale{0.75}
%%\epsscale{0.8}
%\epsscale{1.0}
\epsscale{1.15}
\plotone{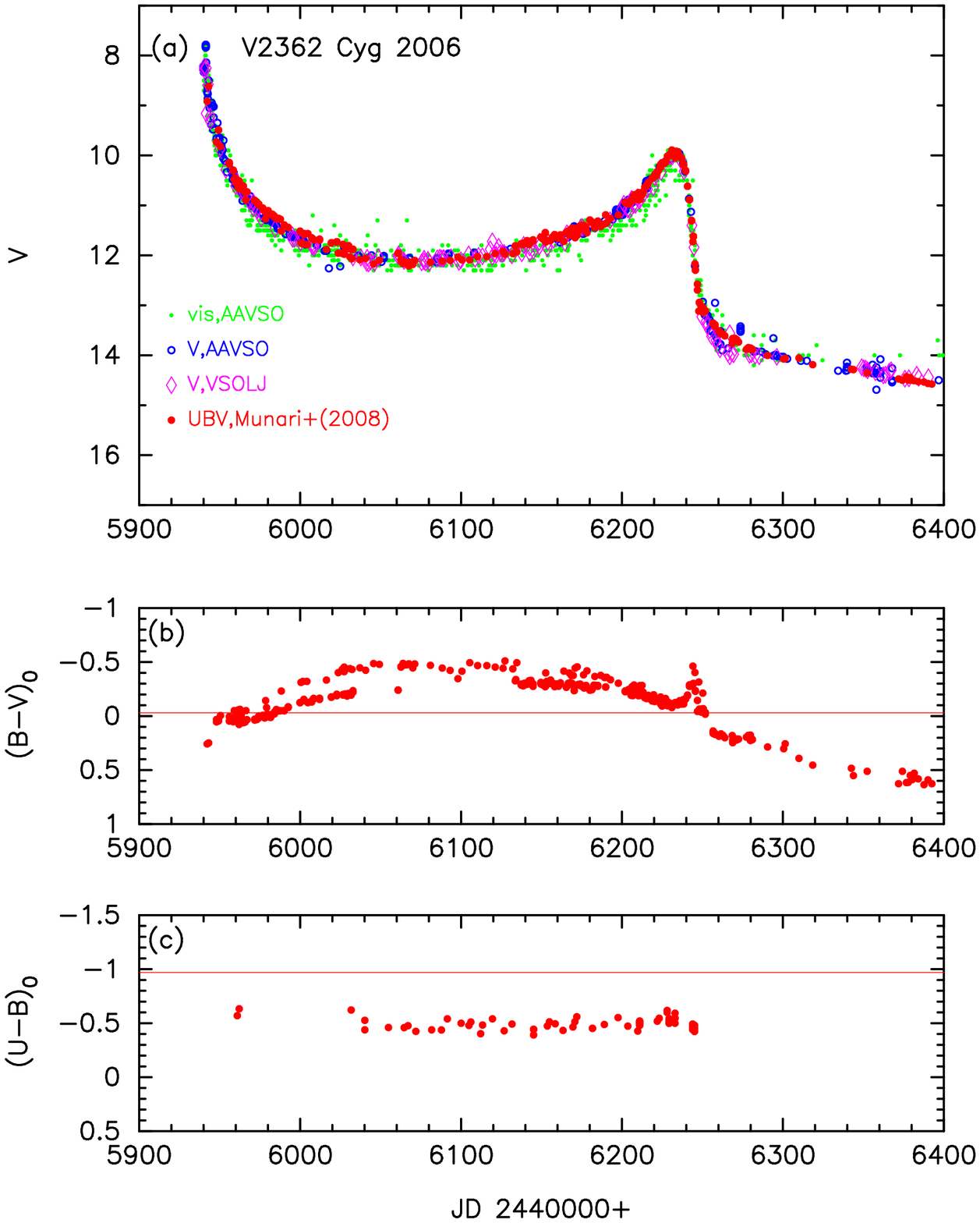}
%\plotone{v2362_cyg_v_bv_ub_color_curve.epsi}
%\plotfiddle{evolution1.ps}{5.0cm}{270}{0.4}{0.4}{-170}{220}
\caption{
Same as Figure \ref{v446_her_v_bv_ub_color_curve}, but for V2362~Cyg.
We de-reddened $(B-V)_0$ and $(U-B)_0$ colors with $E(B-V)=0.60$.
%%See the main text for the sources of V2362~Cyg data.
\label{v2362_cyg_v_bv_ub_color_curve}}
\end{figure}

% Fig.55 
%\placefigure{v2362_cyg_v2468_cyg_v1500_cyg_v1668_cyg_v_bv_ub_color_logscale}

\begin{figure}
%\epsscale{0.75}
%%\epsscale{0.8}
%\epsscale{1.0}
\epsscale{1.15}
\plotone{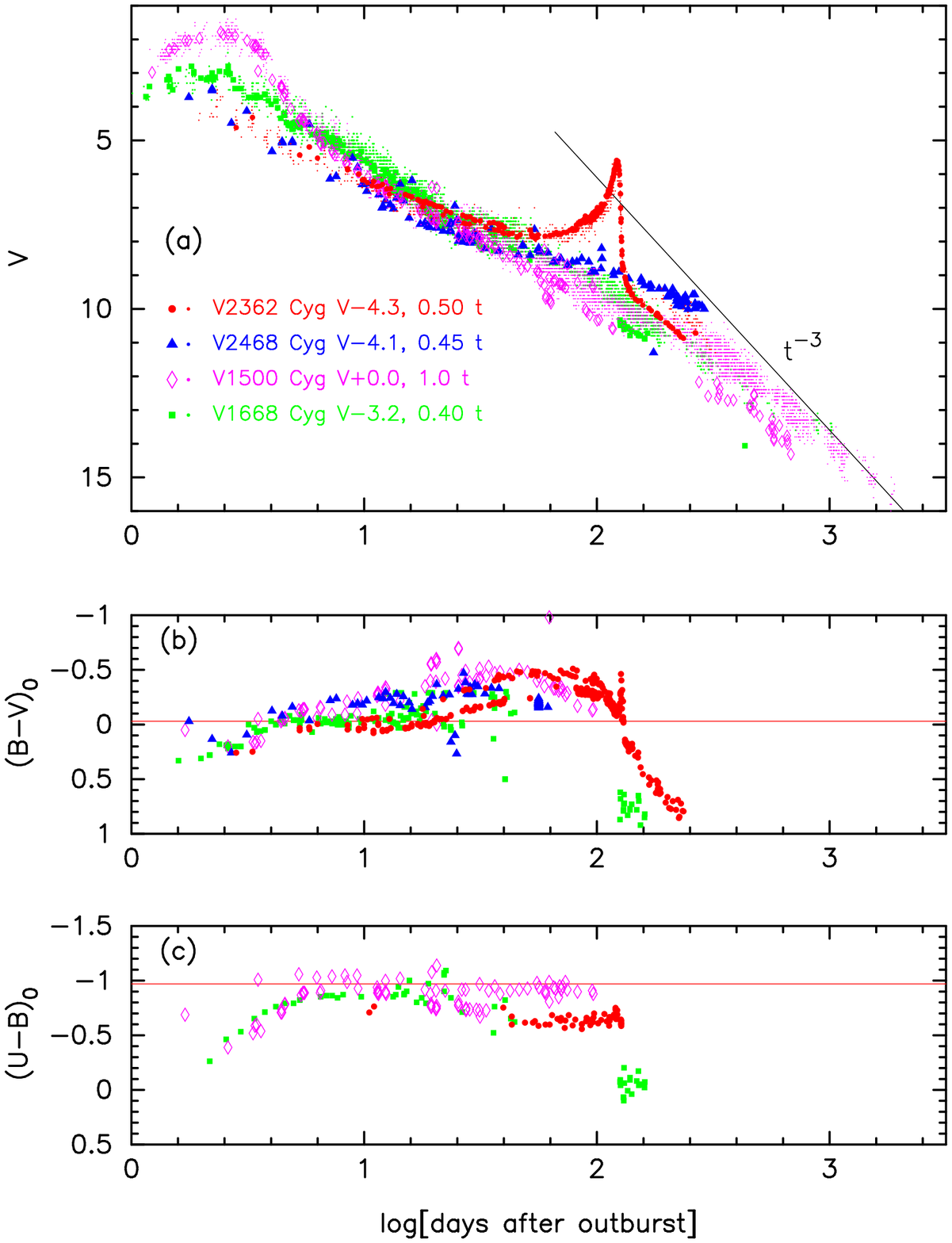}
%\plotone{v2362_cyg_v2468_cyg_v1500_cyg_v1668_cyg_v_bv_ub_color_logscale.epsi}
%\plotfiddle{evolution1.ps}{5.0cm}{270}{0.4}{0.4}{-170}{220}
\caption{
Same as Figure \ref{t_pyx_pw_vul_nq_vul_dq_her_v_bv_ub_color_logscale_no6},
but for V2362~Cyg (filled red circles) and V2468~Cyg (filled blue triangles).
We also add the light curves of V1500~Cyg (open magenta diamonds)
and V1668~Cyg (filled green squares).
%The $UBV$ data of V2362~Cyg and the $BV$ data of V2468~Cyg are taken 
%from \citet{mun08b} and the AAVSO archive, respectively.  
%The $V$ light curves of V2362~Cyg and V2468~Cyg are shifted 
%up by 4.3 and 4.1 mag, and the times of V2362~Cyg and V2468~Cyg are stretched
%by a factor of 0.50 and 0.45, respectively, against that of V1500~Cyg.
%%Here we assume that $(m-M)_V=15.6$, 13.0, 8.2, and 11.7 
%%for V475~Sct, PW~Vul, DQ~Her, and FH~Ser, respectively.
%See the main text for more detail.
\label{v2362_cyg_v2468_cyg_v1500_cyg_v1668_cyg_v_bv_ub_color_logscale}}
\end{figure}

\subsection{V2362~Cyg 2006}
\label{v2362_cyg}
This nova is not studied in Paper I.
Figure \ref{v2362_cyg_v_bv_ub_color_curve} shows the visual and $V$,
$(B-V)_0$, and $(U-B)_0$ evolutions of V2362~Cyg.
The $UBV$ data are taken from \citet{mun08b}, $V$ and visual from
the AAVSO and VSOLJ archives.
V2362~Cyg reached $m_{V,\rm max}=7.8$ at optical maximum 
and then declined with $t_2=9.0\pm0.5$
and $t_3=21.0\pm0.5$~days \citep{kim08}.  For the first 60 days, 
the decline down to 12th magnitude is smooth and resembles a power-law decline. 
Then it rose up again to 10th magnitude
(secondary maximum) about 240 days after the discovery,
and suddenly dropped to 13th magnitude in $\sim20$ days,
followed by the formation of an optically thin dust shell \citep{ara10},
and again slowly declined resembling a power law
before the secondary maximum \citep[e.g.,][]{kim08}.
An orbital period of 1.58~hr was suggested by \citet{bal09}.

The reddening of V2362~Cyg was obtained to be 
$E(B-V)= (B-V)_{t2} - (B-V)_{0, t2} 
= 0.58\pm 0.03 - (-0.02\pm 0.12) = 0.6\pm 0.1$ \citep{kim08},
$E(B-V)=0.58\pm0.04$ \citep{lyn08a} from Lyman $\beta$ 
fluoresced \ion{O}{1} lines, $E(B-V)=0.56$ 
from the equivalent width of interstellar lines 
\ion{Na}{1}~D1 and D2, and $E(B-V)= (B-V)_{t2} - (B-V)_{0, t2} 
= 0.54 - (-0.02) = 0.56$ \citep{mun08b}.
The NASA/IPAC galactic dust absorption map gives $E(B-V)=0.65 \pm 0.03$
in the direction toward V2362~Cyg, $(l, b)= (87\fdg3724,-2\fdg3574)$.
Because there are not enough $U$ data in the first decline phase as
shown in Figure \ref{v2362_cyg_v_bv_ub_color_curve}(c), 
we cannot accurately determine the color excess from the general track
fitting in the color-color diagram.  
Instead, we obtained $E(B-V)= 0.60\pm0.05$ by averaging the above
four estimates.  Assuming that $E(B-V)= 0.60$, we plot the $(B-V)_0$
and $(U-B)_0$ color evolutions of V2362~Cyg in Figure 
\ref{v2362_cyg_v_bv_ub_color_curve}(b) and 
\ref{v2362_cyg_v_bv_ub_color_curve}(c), respectively, and in Figure
\ref{v2362_cyg_v2468_cyg_v1500_cyg_v1668_cyg_v_bv_ub_color_logscale}(b) and
\ref{v2362_cyg_v2468_cyg_v1500_cyg_v1668_cyg_v_bv_ub_color_logscale}(c)
together with V1500~Cyg, V1668~Cyg, and V2468~Cyg.
The color evolution is similar to each other.
Therefore, we adopt $E(B-V)= 0.60\pm0.05$ in this paper.

Using the time-stretching method (Figure
\ref{v2362_cyg_v2468_cyg_v1500_cyg_v1668_cyg_v_bv_ub_color_logscale}(a)),
we obtain the apparent distance modulus of V2362~Cyg
\begin{eqnarray}
(m-M)_{V,\rm V2362~Cyg} &=& 15.9 \cr
&=& (m-M+ \Delta V)_{V,\rm V2468~Cyg}  - 2.5 \log 0.45/0.50 \cr 
&\approx& 15.6 + (-4.1 + 4.3) + 0.11 = 15.91 \cr
&=& (m-M+ \Delta V)_{V,\rm V1500~Cyg} - 2.5 \log 1.0/0.50 \cr 
&\approx& 12.3 + (-0.0 + 4.3) - 0.75 = 15.85 \cr
&=& (m-M+ \Delta V)_{V,\rm V1668~Cyg} - 2.5 \log 0.40/0.50 \cr 
&\approx& 14.6 + (-3.2 + 4.3) + 0.24 = 15.94,
\label{v2362_cyg_v2468_cyg_v1500_cyg_v1668_cyg}
\end{eqnarray}
where we use 
$(m-M)_{V,\rm V2468~Cyg}=15.6$ from Section \ref{v2468_cyg},
$(m-M)_{V,\rm V1500~Cyg}=12.3$ from Section \ref{v1500_cyg_cmd}, and 
$(m-M)_{V,\rm V1668~Cyg}=14.6$ from Section \ref{v1668_cyg_cmd}.
We adopt $(m-M)_{V,\rm V2362~Cyg}=15.9\pm0.2$ in this paper. 
Then the distance is estimated as $d=6.4$~kpc from $(m-M)_V=15.9$ 
and $E(B-V)=0.60$.
The distance to V2362~Cyg was also estimated by \citet{kim08}
as $d=7.5^{+3.0}_{-2.5}$~kpc from a simple average of
the MMRD relation, $M_{V,15}=-5.44$ at 15 days after maximum, and 
an assumed luminosity at maximum \citep{bon00}, and
by \citet{mun08b} as $d=7.2\pm0.2$~kpc from various methods
including the MMRD and $M_{V,15}$ relations.  \citet{mun08b} obtained
the apparent distance modulus in the $V$ band as $(m-M)_V=16.0$, which
is consistent with our value of $(m-M)_V=15.9\pm0.2$.

Figure \ref{distance_reddening_v475_sct_v5114_sgr_v2362_cyg_v1065_cen}(c)
shows various distance-reddening relations for V2362~Cyg.
Here we plot the distance-reddening relations given by \citet{mar06},
i.e., four nearby directions, $(l, b)= (87\fdg25,-2\fdg25)$
denoted by open red squares,
$(87\fdg50,-2\fdg25)$ by filled green squares, 
$(87\fdg25,-2\fdg50)$ by blue asterisks, and
$(87\fdg50,-2\fdg50)$ by open magenta circles.
These four relations show a large scatter.
We also added the distance-reddening relation given by \citet{gre15}.
These trends cross roughly at the point $d\approx6.4$~kpc
and $E(B-V)\approx0.60$, being consistent with our estimates
of $E(B-V)= 0.60$ and $(m-M)_V=15.9$.

Using $E(B-V)= 0.60$ and $(m-M)_V=15.9$, we plot the color-magnitude
diagram of V2362~Cyg in Figure
\ref{hr_diagram_v5114_sgr_v2362_cyg_v1065_cen_v1280_sco_outburst}(b).
The location of the color-magnitude track differs
between the first and second peaks, 
which is interesting and very suggestive.  
We depict the phase of the secondary maximum by open magenta circles
in order to distinguish it from the first maximum (filled red circles).
The track is close to that of FH~Ser in the first decline phase
and then moves to that of V1974~Cyg (or even to that of V1500~Cyg) 
during the secondary maximum phase (open magenta circles).
Thus, we regard V2362~Cyg as a V1500~Cyg type in the color-magnitude diagram.
This transition from a FH~Ser type to a V1500~Cyg type between
the first and second maxima shows that a massive mass ejection
had occurred between the two peaks.  FH~Ser is located
in the red side to V1500~Cyg because FH~Ser had a more
massive envelope.  Thus, V2362~Cyg undergoes a transition
from a redder track to a bluer after the massive mass ejection.
We also point out that the movement in the color-magnitude diagram
from the first to the second maxima 
is clockwise like in the track of PW~Vul (see Figure
\ref{hr_diagram_fh_ser_pw_vul_v1500_cyg_v1974_cyg_outburst}(b)).
We specify a turning point $(B-V)_0=-0.51$ and $M_V=-3.83$
by a large open red square as shown in Figure 
\ref{hr_diagram_v5114_sgr_v2362_cyg_v1065_cen_v1280_sco_outburst}(b).
This turning point corresponds to the beginning of the nebular phase.
The nebular phase started around 250 days after the optical maximum
\citep{mun08b}.

%Fig.56
%\placefigure{v1065_cen_v_bv_ub_color_curve}

\begin{figure}
%\epsscale{0.75}
%%\epsscale{0.8}
%\epsscale{1.0}
\epsscale{1.15}
\plotone{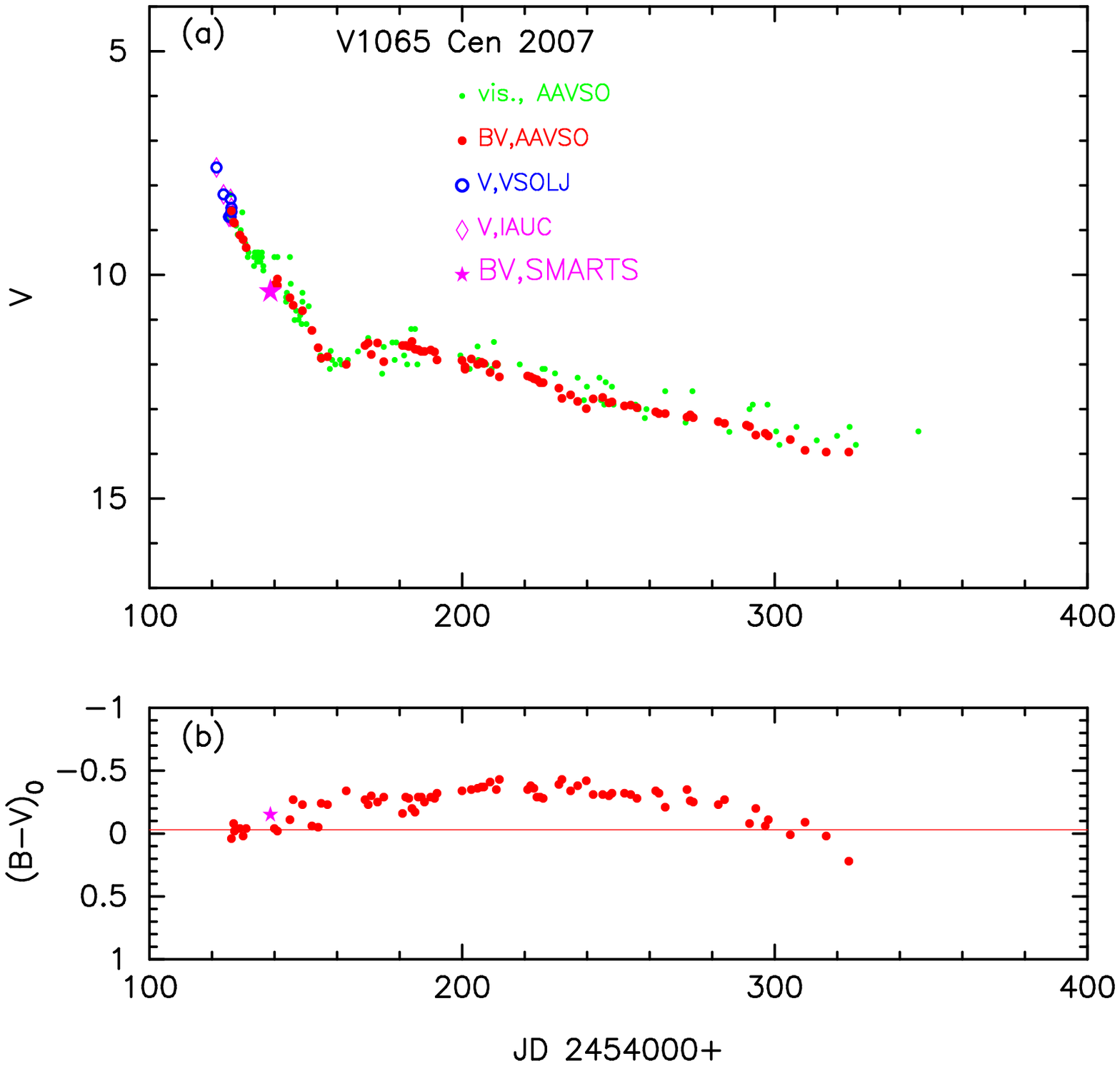}
%\plotone{v1065_cen_v_bv_ub_color_curve.epsi}
%\plotfiddle{evolution1.ps}{5.0cm}{270}{0.4}{0.4}{-170}{220}
\caption{
Same as Figure \ref{v446_her_v_bv_ub_color_curve}, but for V1065~Cen.
We omit the $(U-B)_0$ color evolution because no $U$ observations are 
found in the literature.  We de-reddened $(B-V)_0$ with $E(B-V)=0.45$.
%%See the main text for the sources of V1065~Cen data.
\label{v1065_cen_v_bv_ub_color_curve}}
\end{figure}

%Fig.57
%\placefigure{v496_sct_v1065_cen_v1419_aql_v1668_cyg_lv_vul_v_bv_ub_color_logscale}

\begin{figure}
%\epsscale{0.75}
%%\epsscale{0.8}
%\epsscale{1.0}
\epsscale{1.15}
\plotone{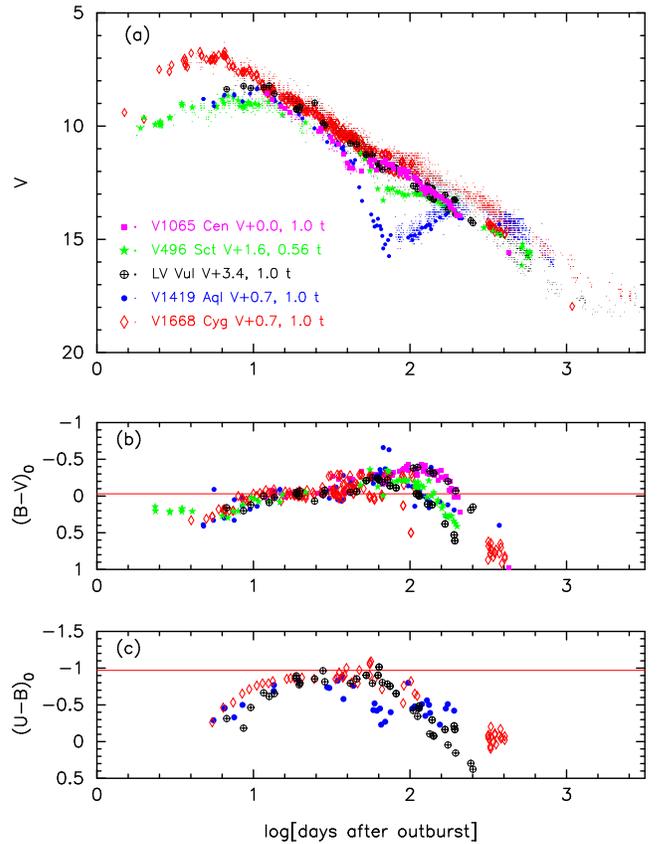}
%\plotone{v496_sct_v1065_cen_v1419_aql_v1668_cyg_lv_vul_v_bv_ub_color_logscale.epsi}
%\plotfiddle{evolution1.ps}{5.0cm}{270}{0.4}{0.4}{-170}{220}
\caption{
Same as Figure \ref{t_pyx_pw_vul_nq_vul_dq_her_v_bv_ub_color_logscale_no6},
but for V1065~Cen (filled magenta squares) and V496~Sct (filled green 
stars).  
We also add the light curves of LV~Vul (open black circles with plus sign),
V1419~Aql (filled blue circles), and V1668~Cyg (open red diamonds).
%%The $BV$ data of V1065~Cen and V496~Sct are taken from the AAVSO archive.
%The $V$ light curves of LV~Vul, V496~Sct, V1419~Aql, and V1668~Cyg are
%shifted down by 3.6, 1.6, 0.6, and 1.6 mag, 
%and their times stretched by a factor of 1.0, 0.50, 0.79, and 1.26,
%respectively, against that of V1065~Cen.
%See the main text for more detail.
\label{v496_sct_v1065_cen_v1419_aql_v1668_cyg_lv_vul_v_bv_ub_color_logscale}}
\end{figure}

\subsection{V1065~Cen 2007}
\label{v1065_cen}
This nova is not studied in Paper I.
Figure \ref{v1065_cen_v_bv_ub_color_curve} shows the visual and $V$,
and $(B-V)_0$ evolutions of V1065~Cen.
The $BV$ data are taken from the AAVSO archive, and $V$ data
are from the VSOLJ archive and IAU Circular Nos.\ 8800 and 8801.  V1065~Cen 
reached $m_{V, \rm max}=7.6\pm0.2$ at maximum on UT 2007 January 21.
Then it smoothly declined with $t_2=11$ and $t_3=26$~days \citep{hel10}.  
A shallow dust black out started about 30 days after the optical maximum.
V1065~Cen was identified as a neon nova by \citet{hel10}.

The reddening for V1065~Cen was obtained as $E(B-V)=0.50\pm0.10$ 
\citep{hel10} from an average of three estimates, i.e.,
$E(B-V)= (B-V)_{\rm max} - (B-V)_{0, \rm max} = 
0.52\pm0.04 - (0.23 \pm 0.06) = 0.29\pm 0.07$,
$E(B-V)= (B-V)_{t2} - (B-V)_{0, t2} = 0.41\pm0.05 - (-0.02\pm 0.04) 
= 0.43\pm 0.06$, and $E(B-V)=0.79\pm0.01$ from the Balmer decrement 
(H$\alpha$/H$\beta$).  
\citet{hel10} also estimated the apparent distance modulus in the $V$ band
as $(m-M)_V= 7.6\pm0.2 - (-8.6\pm0.5)= 16.2\pm0.6$ from the MMRD
relation together with $t_2=11$ days.  This gives a distance of
$d=8.7^{+2.8}_{-2.1}$~kpc.
Assuming that the intrinsic $(B-V)_0$ color evolution of V1065~Cen 
is identical with that for similar types of novae, i.e., LV~Vul, V1668~Cyg, 
V1419~Aql, and V496~Sct, we obtain $E(B-V)=0.45\pm0.05$ from Figure
\ref{v496_sct_v1065_cen_v1419_aql_v1668_cyg_lv_vul_v_bv_ub_color_logscale}(b),
which is consistent with the estimate of Helton et al.

Using the time-stretching method (Figure 
\ref{v496_sct_v1065_cen_v1419_aql_v1668_cyg_lv_vul_v_bv_ub_color_logscale}(a)),
we estimate the apparent distance modulus in the $V$ band, i.e.,
\begin{eqnarray}
(m-M)_{V,\rm V1065~Cen} &=& 15.3 \cr
&=& (m-M+ \Delta V)_{V,\rm V1668~Cyg}  + 2.5 \log 1.0/1.0\cr 
&\approx& 14.6 + (+0.7 - 0.0) + 0.0 = 15.3 \cr
&=& (m-M+ \Delta V)_{V,\rm LV~Vul} + 2.5 \log 1.0/1.0\cr 
&=& 11.9 + (+3.4 - 0.0) + 0.0 = 15.3 \cr
&=& (m-M+ \Delta V)_{V,\rm V1419~Aql} + 2.5 \log 1.0/1.0\cr 
&\approx& 14.6 + (+0.7 - 0.0) + 0.0 = 15.3 \cr
&=& (m-M+ \Delta V)_{V,\rm V496~Sct} + 2.5 \log 0.56/1.0\cr 
&\approx& 14.4 + (+1.6 - 0.0) - 0.625 = 15.375,
\label{v1065_cen_v1668_cyg_lv_vul_v1419_aql_v496_sct}
\end{eqnarray}
where we use $(m-M)_{V,\rm V1668~Cyg}=14.6$ 
from Section \ref{v1668_cyg_cmd},
$(m-M)_{V,\rm LV~Vul}=11.9$ from Section \ref{lv_vul_cmd},
$(m-M)_{V,\rm V1419~Aql}=14.6$ from Section \ref{v1419_aql}, and
$(m-M)_{V,\rm V496~Sct}=14.4$ from Section \ref{v496_sct}.
Then, we obtain the distance of $d=6.0$~kpc for $E(B-V)=0.45$
and $(m-M)_V=15.3$.

Figure \ref{distance_reddening_v475_sct_v5114_sgr_v2362_cyg_v1065_cen}(d)
shows various distance-reddening relations for V1065~Cen,
$(l, b)= (293\fdg9841,+3\fdg6130)$.
Here we plot the distance-reddening relation given by \citet{mar06},
i.e., four nearby directions, $(l, b)= (293\fdg75,3\fdg75)$
denoted by open red squares,
$(294\fdg00,3\fdg75)$ by filled green squares, 
$(293\fdg75,3\fdg50)$ by blue asterisks, and
$(294\fdg00,3\fdg50)$ by open magenta circles.
The closest ones are those denoted by filled green squares and
open magenta circles.
Although these two relations differ for $d\gtrsim3$~kpc, our values of $d=6.0$~kpc 
and $E(B-V)=0.45$ are midway between them. Thus, we think our cross
point is consistent with the relation of Marshall et al.
The relation of Green et al.\ (2015) is not available for these galactic
coordinates.

Using $E(B-V)=0.45$ and $(m-M)_V=15.3$,
we plot the color-magnitude diagram of V1065~Cen in Figure
\ref{hr_diagram_v5114_sgr_v2362_cyg_v1065_cen_v1280_sco_outburst}(c).
The track of V1065~Cen almost follows that of LV~Vul (solid magenta lines).
Thus, we regard V1065~Cen as an LV~Vul type in the color-magnitude diagram.
This coincidence with LV~Vul strongly supports our values of $(m-M)_V=15.3$
and $E(B-V)=0.45$.
We specify the starting point of dust blackout \citep{hel10} 
at $(B-V)_0=-0.06$ and $M_V=-4.06$, denoted by a large open blue
square in Figure 
\ref{hr_diagram_v5114_sgr_v2362_cyg_v1065_cen_v1280_sco_outburst}(c).
\citet{hel10} pointed out that the nova entered the early nebular 
phase at $m_V\approx12$, about 70 days after maximum.  We denote
this phase by a large open red square, at $(B-V)_0=-0.35$
and $M_V=-3.42$.  This point is close to the two-headed
red arrow.

%Fig.58 
%\placefigure{v1280_sco_qv_vul_v_bv_ub_color_curve}

\begin{figure}
%%\epsscale{0.60}
%\epsscale{0.75}
%\epsscale{1.0}
\epsscale{1.15}
\plotone{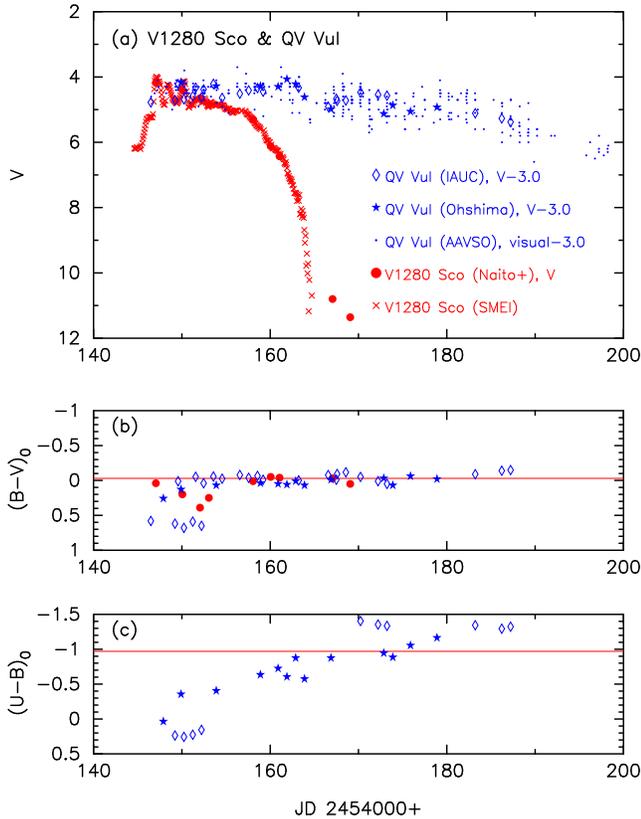}
%\plotone{v1280_sco_qv_vul_v_bv_ub_color_curve.epsi}
%\plotfiddle{evolution1.ps}{5.0cm}{270}{0.4}{0.4}{-170}{220}
\caption{
(a) visual and $V$ bands, (b) $(B-V)_0$, and (c) $(U-B)_0$ light curves
for V1280~Sco (red symbols) and QV~Vul (blue symbols) on
a linear timescale.  The light curve of V1280~Sco observed with
{\it SMEI} is also plotted in panel (a).  We shift the $V$ magnitudes 
of QV~Vul up by $3.0$ mag and rightward by about 20 years 
against those of V1280~Sco
and overlap their $V$ magnitudes near the peak.
We de-reddened $(B-V)_0$ and $(U-B)_0$ colors with $E(B-V)=0.35$
and $E(B-V)=0.60$ for V1280~Sco and QV~Vul, respectively.
\label{v1280_sco_qv_vul_v_bv_ub_color_curve}}
\end{figure}

%Fig.59
%\placefigure{distance_reddening_v1280_sco_v2467_cyg_v2615_oph_v458_vul}

\begin{figure*}
%\begin{figure}
\epsscale{0.75}
%%\epsscale{0.8}
%%\epsscale{1.0}
%%\epsscale{1.15}
\plotone{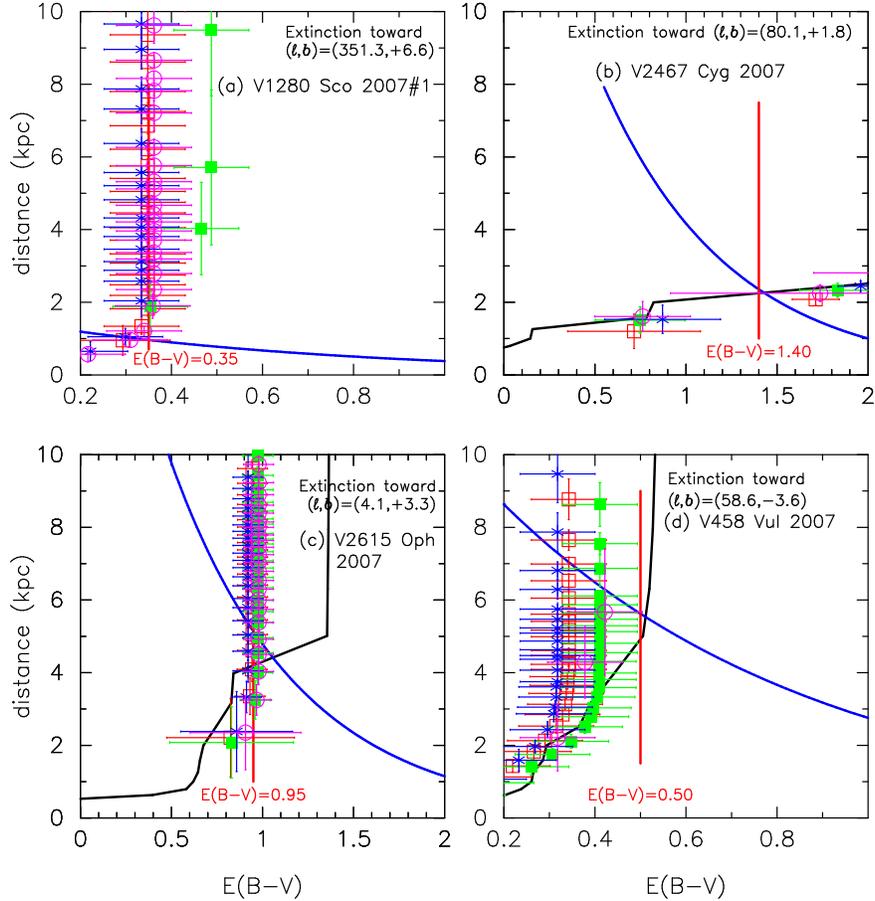}
%\plotone{distance_reddening_v1280_sco_v2467_cyg_v2615_oph_v458_vul.epsi}
%\plotfiddle{evolution1.ps}{5.0cm}{270}{0.4}{0.4}{-170}{220}
\caption{
Same as Figure \ref{distance_reddening_fh_ser_pw_vul_v1500_cyg_v1974_cyg},
but for (a) V1280~Sco, (b) V2467~Cyg, (c) V2615~Oph, and (d) V458~Vul.
The thick solid blue lines denote (a) $(m-M)_V=11.0$, 
(b) $(m-M)_V=16.2$, (c) $(m-M)_V=16.5$,  and (d) $(m-M)_V=15.3$.
%The vertical solid red lines represent the color excesses of 
%(a) $E(B-V)=0.35$, (b) $E(B-V)=1.40$,
%(c) $E(B-V)=0.95$, and (d) $E(B-V)=0.50$.
%The black solid lines denote the distance-reddening relation given
%by \citet{gre15}.
%In panel (a), the magenta thick solid line represents
%the distance-reddening relation calculated from the UV~1455 \AA\  flux
%fitting with the $0.51~M_\sun$ WD model \citep{hac15k}.
%In panels (a), (b), and (d), two or 
%four sets of data with error bars show distance-reddening relations
%in two or four directions close to each nova, the data of which are taken
%from \citet{mar06}.  
%See the main text for more detail.
\label{distance_reddening_v1280_sco_v2467_cyg_v2615_oph_v458_vul}}
%\end{figure}
\end{figure*}

\subsection{V1280~Sco 2007\#1}
\label{v1280_sco}
This nova is not studied in Paper I.
Figure \ref{v1280_sco_qv_vul_v_bv_ub_color_curve} shows the $V$ and
{\it SMEI} light curves and $(B-V)_0$ color evolution of V1280~Sco.
The $UBV$ data for QV~Vul (Figure \ref{qv_vul_v_bv_ub_color_curve})
are added for comparison.
The {\it SMEI} data for V1280~Sco are from \citet{hou10}, and
the $V$ data from \citet{nai12}.  
V1280~Sco reached $m_{V,\rm max}=3.8$
at optical maximum on UT 2007 February 16.19 \citep{mun07a}.
It experienced two major episodes of rebrightening peaking at 
UT February 16.15 and 19.18 \citep{hou10}.  On UT February 26.4
its decline rate changed rapidly \citep[e.g.,][]{hou10}, 
indicating formation of an optically thick dust shell \citep{das08}. 

The reddening for V1280~Sco was estimated by \citet{das08} as
$E(B-V)=A_V/R_V=1.2/3.1\approx0.4$ from Marshall et al.'s (2006) 3D dust map.
The distance was obtained by \citet{nai12} as $d=1.1\pm0.5$~kpc
from the expansion parallax of a dust shell \citep{che08}
together with the expansion velocity of $350\pm160$~km~s$^{-1}$.

In Figure \ref{v1280_sco_qv_vul_v_bv_ub_color_curve}(a),
we shift the $V$ light curve of QV~Vul up by 3.0 mag
to fit its peak with that of V1280~Sco. 
In the early phase of these outbursts, the brightness shows 
similar fluctuations (both the amplitude and period).
Therefore, we expect that the peak brightness is the same for
V1280~Sco and QV~Vul.  Then we obtain the apparent distance modulus of
$(m-M)_{V,\rm V1280~Sco}= (m-M)_{V,\rm QV~Vul}-3.0=14.0-3.0=11.0$, where
$(m-M)_{V,\rm QV~Vul}=14.0$ was already obtained in Section \ref{qv_vul}.
We also determined that $E(B-V)=0.35$ by assuming that the intrinsic $(B-V)_0$ 
color is the same for both V1280~Sco and QV~Vul (see Figure 
\ref{v1280_sco_qv_vul_v_bv_ub_color_curve}(b)). 
Then the distance is calculated as $d=0.96$~kpc for $E(B-V)=0.35$ and 
$(m-M)_V=11.0$, being consistent with
Naito et al.'s (2012) estimate of $d=1.1\pm0.5$~kpc.

Figure \ref{distance_reddening_v1280_sco_v2467_cyg_v2615_oph_v458_vul}(a)
shows various distance-reddening relations for V1280~Sco,
$(l, b)= (351\fdg3311, +6\fdg5534)$.  We plot Marshall et al.'s
(2006) four relations, i.e., 
$(l, b)= (351\fdg25,6\fdg50)$ denoted by open red squares,
$(351\fdg50,6\fdg50)$ by filled green squares, 
$(351\fdg25,6\fdg75)$ by blue asterisks, and
$(351\fdg50,6\fdg75)$ by open magenta circles.
The closer ones are those denoted by open red squares and
filled green squares.  The thick solid blue line indicates $(m-M)_V=11.0$.
These trends, i.e., Marshall et al.'s relations, $(m-M)_V=11.0$,
and $E(B-V)=0.35$ (vertical solid red line), cross 
at the point $d\approx0.96$~kpc and $E(B-V)\approx0.35$.
Thus we adopt $(m-M)_V=11.0$ and $E(B-V)=0.35$.

Using $E(B-V)=0.35$ and $(m-M)_V=11.0$, we plot the color-magnitude
diagram of V1280~Sco in Figure 
\ref{hr_diagram_v5114_sgr_v2362_cyg_v1065_cen_v1280_sco_outburst}(d).
The track in the early phase looks similar to that of V705~Cas in Figure
\ref{hr_diagram_v1419_aql_v705_cas_v382_vel_v1493_aql_outburst}(b),
although the peak brightness is $\sim1$ mag fainter than that.  
We regard V1280~Sco as a V1668~Cyg type because
the track goes down along that of V1668~Cyg.
A large excursion toward red on the track in the early phase corresponds to
the pulsation brightenings of the light curve.  
The dust formation occurred earlier when the nova was still in the
brighter phase of the outburst (as indicated by a large
open red square at $(B-V)_0=+0.01$ and $M_V=-5.70$)
compared with other cases such as FH~Ser and QV~Vul.

%Fig.60
%\placefigure{v2467_cyg_v_bv_ub_color_curve}

\begin{figure}
%%\epsscale{0.60}
%%\epsscale{0.75}
%\epsscale{1.0}
\epsscale{1.15}
\plotone{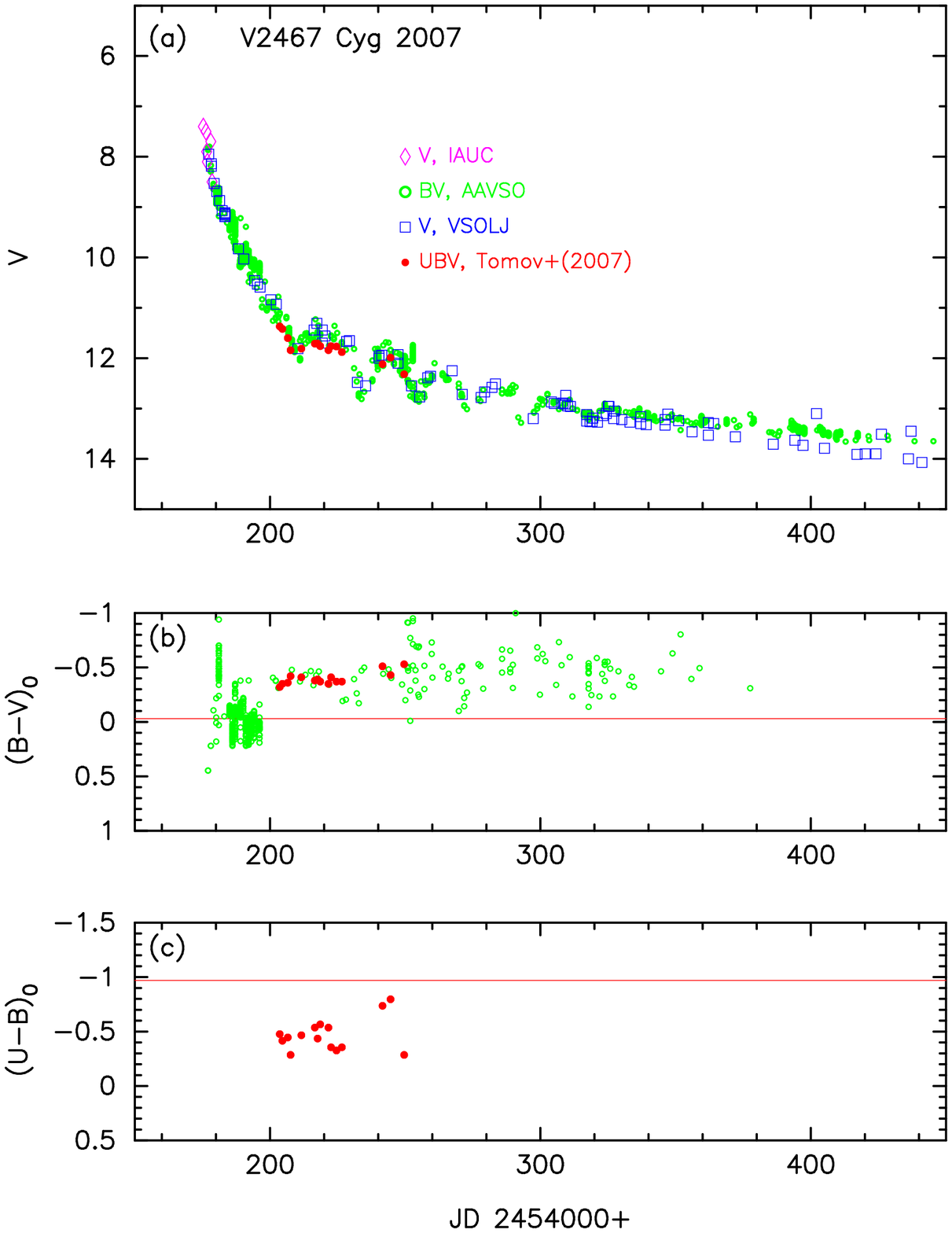}
%\plotone{v2467_cyg_v_bv_ub_color_curve.epsi}
%\plotfiddle{evolution1.ps}{5.0cm}{270}{0.4}{0.4}{-170}{220}
\caption{
Same as Figure \ref{v446_her_v_bv_ub_color_curve}, but for V2467~Cyg.
We de-reddened $(B-V)_0$ and $(U-B)_0$ colors with $E(B-V)=1.40$.
%%See the main text for more detail.
\label{v2467_cyg_v_bv_ub_color_curve}}
\end{figure}

%Fig.61  
%\placefigure{v2467_cyg_v2468_cyg_v1668_cyg_iv_cep_v_color_logscale}

\begin{figure}
%\epsscale{0.75}
%%\epsscale{0.8}
%\epsscale{1.0}
\epsscale{1.15}
\plotone{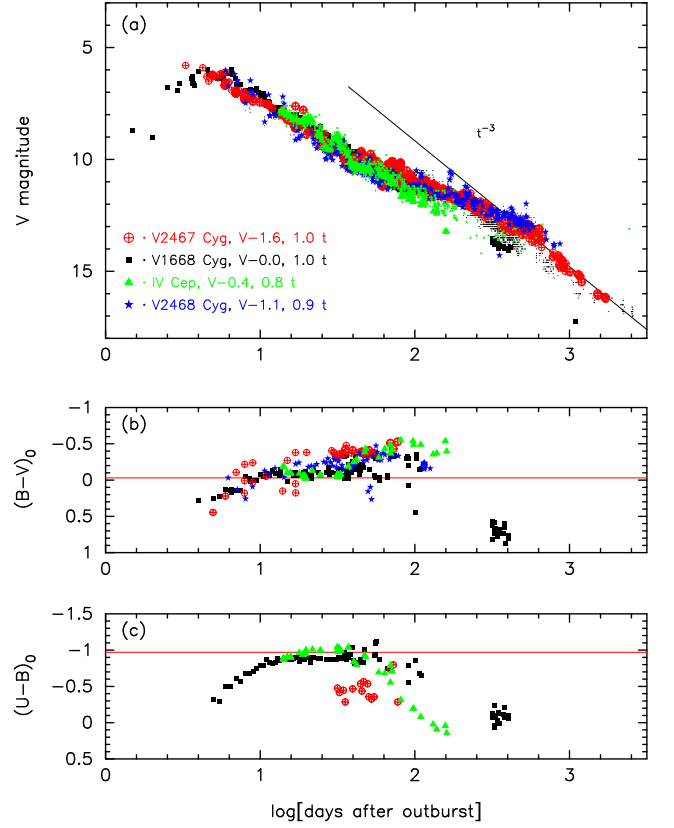}
%\plotone{v2467_cyg_v2468_cyg_v1668_cyg_iv_cep_v_color_logscale.epsi}
%\plotfiddle{evolution1.ps}{5.0cm}{270}{0.4}{0.4}{-170}{220}
\caption{
Same as Figure \ref{t_pyx_pw_vul_nq_vul_dq_her_v_bv_ub_color_logscale_no6},
but for V2467~Cyg (open red circles with plus sign) 
and V2468~Cyg (filled blue stars).  
We also add the light curves of IV~Cep (filled green triangles)
and V1668~Cyg (filled black squares).
%and V2468~Cyg are shifted up by 1.6, $-0.4$, and 1.1 mag, 
%and their times are stretched by a factor of 1.0, 0.8, and 0.9,
%respectively, against that of V1668~Cyg (filled black squares).
%%See the main text for more detail.
\label{v2467_cyg_v2468_cyg_v1668_cyg_iv_cep_v_color_logscale}}
\end{figure}

%Fig.62
%\placefigure{hr_diagram_v2467_cyg_v2615_oph_v458_vul_v2468_cyg_outburst}

\begin{figure*}
%\begin{figure}
\epsscale{0.75}
%\epsscale{0.8}
%%\epsscale{1.0}
\plotone{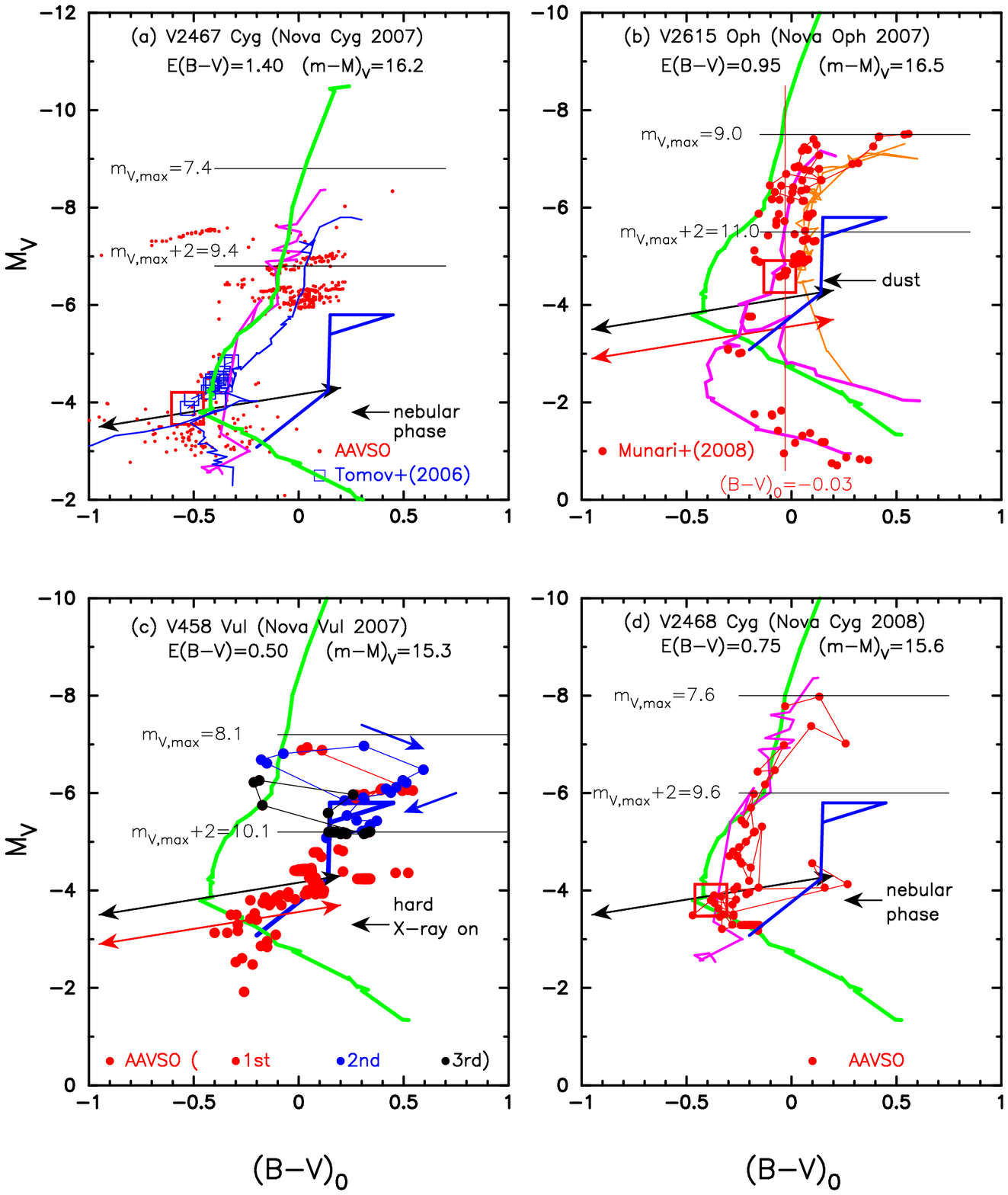}
%\plotone{hr_diagram_v2467_cyg_v2615_oph_v458_vul_v2468_cyg_outburst.epsi}
%\plotfiddle{evolution1.ps}{5.0cm}{270}{0.4}{0.4}{-170}{220}
\caption{
Same as Figure 
\ref{hr_diagram_rs_oph_v446_her_v533_her_t_pyx_outburst}, but
for (a) V2467~Cyg, (b) V2615~Oph, (c) V458~Vul, and (d) V2468~Cyg.
In panel (a), 
%%thick solid green lines denote the track of V1500~Cyg,
thin solid blue lines indicate the track of V1974~Cyg, 
solid magenta lines that of V1494~Aql.
%% and thick solid blue lines that of PU~Vul.
In panel (b), thick solid magenta lines that of LV~Vul,
thin solid orange lines that of FH~Ser.  
In panel (c), blue arrows denote the direction of clockwise movement
of the track (blue symbols connected by thin solid blue lines).  
In panel (d), thick solid magenta lines indicate the track of V1494~Aql.
\label{hr_diagram_v2467_cyg_v2615_oph_v458_vul_v2468_cyg_outburst}}
%\end{figure}
\end{figure*}

\subsection{V2467~Cyg 2007}
\label{v2467_cyg}
This nova is not studied in Paper I.
Figure \ref{v2467_cyg_v_bv_ub_color_curve} shows the $V$, $(B-V)_0$,
and $(U-B)_0$ evolutions of V2467~Cyg.
The $UBV$ data of V2467~Cyg are taken from \citet{tom07},
$BV$ data from the AAVSO archive, and $V$ data from
IAU Circular No.\ 8821, and the AAVSO and VSOLJ archives.
V2467~Cyg smoothly declines with $t_2=7.6\pm3.0$ and 
$t_3=14.6\pm3.5$~days \citep[e.g.,][]{pog09}.
Then the $V$ brightness shows some quasi-periodic oscillations
with a period of 20--30 days and an amplitude of 0.7 mag
during the transition phase as shown in Figure
\ref{v2467_cyg_v_bv_ub_color_curve}(a).
These kinds of transition oscillations are similar to those of
GK~Per, V603~Aql, and V1494~Aql.
The orbital period was estimated as $P_{\rm orb}=3.83$~hr
\citep{shu10}.  \citet{swi10} suggested that V2467~Cyg is an
intermediate polar with a spin period of 34.5 min.

The reddening and distance for V2467~Cyg were estimated
as $E(B-V)=1.0$--1.5 and $d\sim2$~kpc \citep{ste07} from
the brightness and color of the progenitor, $E(B-V)=0.31$ 
\citep{mun07b} from the \ion{Na}{1}~D2 equivalent width, 
$E(B-V)=1.5$ \citep{maz07} from \ion{O}{1} lines, and
$E(B-V)=1.7$ \citep{rus07b} also from \ion{O}{1} lines.
In addition, the estimate $E(B-V)=1.16\pm0.12$ and $d=2.6$--3.6~kpc \citep{pog09} came from 
$E(B-V)= (B-V)_{t2}- (B-V)_{0, t2} = 1.14 - (-0.02\pm 0.12) = 1.16\pm 0.12$
and the distance modulus of $(m-M)_V=15.9$--16.5, 
derived from various empirical relations of nova light curves,
and $(m-M)_V=16.4\pm0.2$ and $d=2.2\pm0.2$~kpc \citep{hac10k}  
together with $E(B-V)=1.5$ \citep{maz07} from their free-free 
model light curve fitting, and finally $E(B-V)=1.38\pm0.12$ 
and $d=2.5\pm0.3$ \citep{shu10}
from the absolute magnitude at maximum, $B-V$ color relations,
and interstellar extinction.  

We are not able to estimate the color excess from fitting
in the color-color diagram because the $UBV$ data of V2467~Cyg
obtained by \citet{tom07} are for the transition oscillation
phase, and hence inappropriate for the general track of novae,
i.e., too late to derive the reddening.
Using the time-stretching method for the light curves
in Figure
\ref{v2467_cyg_v2468_cyg_v1668_cyg_iv_cep_v_color_logscale}(a),
we obtain the distance modulus in the $V$ band, i.e.,
\begin{eqnarray}
(m-M)_{V,\rm V2467~Cyg} &=& 16.2\cr
&=& (m-M+ \Delta V)_{V,\rm V1668~Cyg}  - 2.5 \log 1.0/1.0\cr 
&\approx& 14.6 + (-0.0 + 1.6) - 0.0 = 16.2 \cr
&=& (m-M+ \Delta V)_{V,\rm IV~Cep} - 2.5 \log 0.80/1.0\cr 
&\approx& 14.7 + (-0.4 + 1.6) + 0.24 = 16.14 \cr
&=& (m-M+ \Delta V)_{V,\rm V2468~Cyg} - 2.5 \log 0.90/1.0\cr 
&\approx& 15.6 + (-1.1 + 1.6) + 0.11 = 16.21,
\label{v2467_cyg_v1668_cyg_iv_cep_v2468_cyg}
\end{eqnarray}
where we use 
$(m-M)_{V,\rm V1668~Cyg}=14.6$ from Section \ref{v1668_cyg_cmd},
$(m-M)_{V,\rm IV~Cep}=14.7$ from Section \ref{iv_cep},
and $(m-M)_{V,\rm V2468~Cyg}=15.6$ from Section \ref{v2468_cyg}.
Here, we squeeze the times of V2467~Cyg, IV~Cep, and V2468~Cyg
by a factor of 1.0, 0.80, and 0.90, and shift the $V$ light curves 
up by 1.6, 0.4, and 1.1 mag, respectively, against V1668~Cyg.

We also obtained $E(B-V)=1.40\pm0.05$ by assuming that
the intrinsic $(B-V)_0$ colors are similar among V2467~Cyg, V1668~Cyg, 
IV~Cep, and V2468~Cyg as shown in Figure
\ref{v2467_cyg_v2468_cyg_v1668_cyg_iv_cep_v_color_logscale}(b).
Here, we de-reddened the $B-V$ and $U-B$ colors of V2467~Cyg
with $E(B-V)=1.40$.  

Figure \ref{distance_reddening_v1280_sco_v2467_cyg_v2615_oph_v458_vul}(b)
shows various distance-reddening relations for V2467~Cyg,
$(l, b)= (80\fdg0690,+1\fdg8417)$.
Here we plot the distance-reddening relations given by \citet{mar06},
i.e., $(l, b)= (80\fdg00,2\fdg00)$ denoted by open red squares,
$(80\fdg25,2\fdg00)$ by filled green squares, 
$(80\fdg00,1\fdg75)$ by blue asterisks, and
$(80\fdg25,1\fdg75)$ by open magenta circles.
The closer ones are those denoted by open red squares and
blue asterisks.  
We also added the distance-reddening relation given by \citet{gre15}.
These four trends, i.e., Marshall et al.'s relations, 
Green et al.'s relation (solid black line), 
$(m-M)_V=16.2$ (thick solid blue line),
and $E(B-V)=1.40$ (vertical solid red line), cross consistently
at the point of $d\sim2.4$~kpc and $E(B-V)\sim1.40$.
Thus, we adopt $E(B-V)=1.40$ and $(m-M)_V=16.2$.

Using $E(B-V)=1.40$ and $(m-M)_V=16.2$, 
we plot the color-magnitude diagram of V2467~Cyg in Figure
\ref{hr_diagram_v2467_cyg_v2615_oph_v458_vul_v2468_cyg_outburst}(a)
as well as that of V1494~Aql (solid magenta line).
The track of V2467~Cyg is very similar to those of V1500~Cyg
and V1974~Cyg.  Therefore, we regard V2467~Cyg as a V1974~Cyg type
in the color-magnitude diagram.
The nova entered the nebular phase at least on 2007 May 18, i.e.,
at $m_V=12.5$ \citep{pog09}.  
We specify a possible start of the nebular phase at the point
of $(B-V)_0=-0.53$ and $M_V=-3.88$
as indicated by a large open red square in Figure 
\ref{hr_diagram_v2467_cyg_v2615_oph_v458_vul_v2468_cyg_outburst}(a).

\subsection{V2615~Oph 2007}
\label{v2615_oph}
\begin{figure}
%%\epsscale{0.60}
%\epsscale{0.75}
%\epsscale{1.0}
\epsscale{1.15}
\plotone{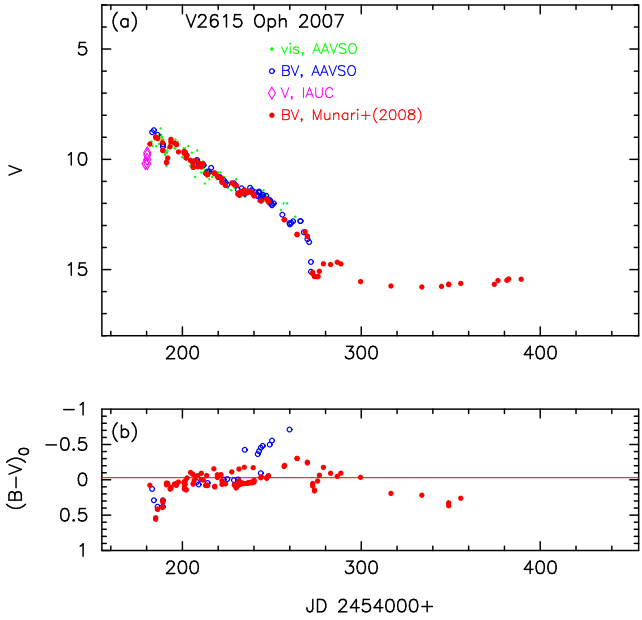}
%\plotone{v2615_oph_v_bv_ub_color_curve.epsi}
%\plotfiddle{evolution1.ps}{5.0cm}{270}{0.4}{0.4}{-170}{220}
\caption{
Same as Figure \ref{v1065_cen_v_bv_ub_color_curve}, but
%%(a) $V$ band and visual light curves and (b) $(B-V)_0$ color evolution 
for V2615~Oph.  
%The data are taken from \citet{mun08a}, the AAVSO archive, 
%and IAU Circular No.\ 8824.
We de-reddened $(B-V)_0$ color with $E(B-V)=0.95$.
\label{v2615_oph_v_bv_ub_color_curve}}
\end{figure}

This nova is not studied in Paper I.
Figure \ref{v2615_oph_v_bv_ub_color_curve} shows the visual and $V$,
and $(B-V)_0$ evolutions of V2615~Oph on a linear timescale.
The $BV$ data are taken from \citet{mun08a} and the AAVSO archive, 
$V$ data are from IAU Circular No.\ 8824, visual data are 
from the AAVSO archive.
V2615~Oph reached $m_{V,\rm max}=9.0$ at optical maximum
on UT 2007 March 25.48 \citep{mun08a}.   
Then it displayed an oscillatory behavior like PW~Vul and
gradually declined with $t_2=26.5$ and $t_3=48.5$~days \citep{mun08a}.
A dust shell formed about 60 days after optical maximum to
make a shallow dust blackout.
The orbital period of 6.54~hr was detected by \citet{mro15}.

The reddening and distance for V2615~Oph were estimated
as $E(B-V)=1.0$--1.3 \citep{rud07a} from \ion{O}{1} lines,
as $E(B-V)=0.90$ and $d=3.7\pm0.2$~kpc \citep{mun08a} from an average of
$E(B-V)= (B-V)_{\rm max} - (B-V)_{0, \rm max} = 1.12 - (0.23 \pm 0.06)
= 0.89\pm 0.06$ and $E(B-V)= (B-V)_{t2} - (B-V)_{0, t2} 
= 0.89 - (-0.02\pm 0.04) = 0.91\pm 0.04$ and from $(m-M)_V=15.7$ calculated
by various empirical formulae including the MMRD and $M_{V,\rm 15}$ relations.

\begin{figure}
%\epsscale{0.75}
%%\epsscale{0.8}
%\epsscale{1.0}
\epsscale{1.15}
\plotone{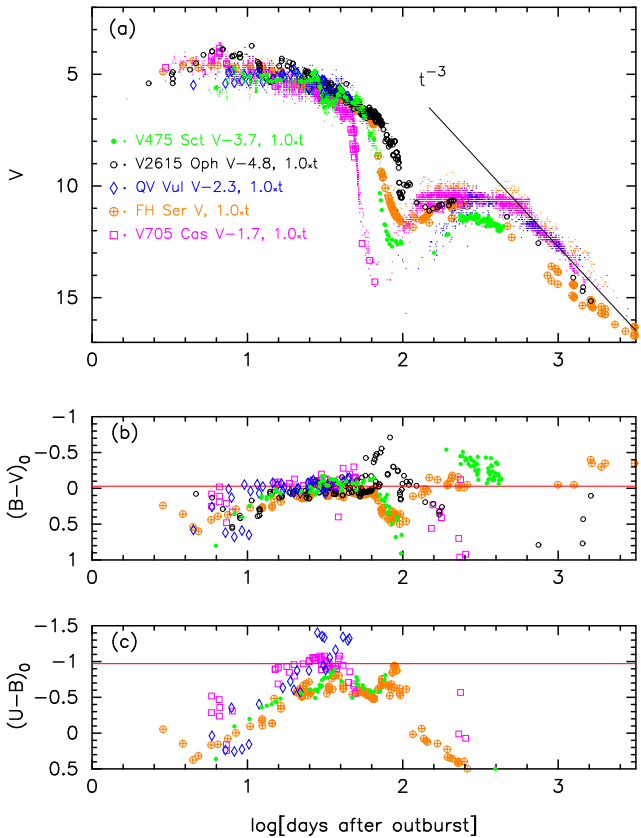}
%\plotone{v2615_oph_v475_sct_v705_cas_qv_vul_fh_ser_v_bv_ub_color_logscale.epsi}
%\plotfiddle{evolution1.ps}{5.0cm}{270}{0.4}{0.4}{-170}{220}
\caption{
Same as Figure \ref{t_pyx_pw_vul_nq_vul_dq_her_v_bv_ub_color_logscale_no6},
but for V2615~Oph (open black circles) and V475~Sct (filled green circles).
We also added V705~Cas, FH~Ser, and QV~Vul.
%The $BV$ data of V2615~Oph are from \citet{mun08a}.
%The $V$ light curves of V2615~Oph, V475~Sco, V705~Cas, and QV~Vul are
%shifted up by 4.8, 3.9, 1.7, and 2.3 mag, respectively, against that of
%FH~Ser.  See the main text for more detail.
\label{v2615_oph_v475_sct_v705_cas_qv_vul_fh_ser_v_bv_ub_color_logscale}}
\end{figure}

We plot the $V$ light curve and $B-V$ color evolution of V2615~Oph 
on a logarithmic timescale  in Figure 
\ref{v2615_oph_v475_sct_v705_cas_qv_vul_fh_ser_v_bv_ub_color_logscale}(a),
as well as those of FH~Ser, QV~Vul, V705~Cas, and V475~Sct,
because these light curves are very similar to each other. 
Since the timescales of these novae are almost the same, we regard that
their brightnesses are the same as that of V2615~Oph.
Then we obtain the apparent distance modulus of V2615~Oph as
\begin{eqnarray}
(m-M)_{V,\rm V2615~Oph} &=& 16.5 \cr
&=& (m-M+ \Delta V)_{V,\rm FH~Ser}  \cr 
&=& 11.7 + (-0.0 + 4.8) = 16.5 \cr
&=& (m-M+ \Delta V)_{V,\rm QV~Vul} \cr 
&=& 14.0 + (-2.3 + 4.8) = 16.5 \cr
&=& (m-M+ \Delta V)_{V,\rm V705~Cas} \cr 
&=& 13.4 + (-1.7 + 4.8) = 16.5 \cr
&=& (m-M+ \Delta V)_{V,\rm V475~Sct} \cr
&=& 15.4 + (-3.7 + 4.8) = 16.5,
\end{eqnarray}
where we use $(m-M)_{V,\rm FH~Ser}=11.7$ from Section \ref{fh_ser_cmd},
$(m-M)_{V,\rm QV~Vul}=14.0$ from Section \ref{qv_vul}, 
$(m-M)_{V,\rm V705~Cas}=13.4$ from Section \ref{v705_cas}, and 
$(m-M)_{V,\rm V475~Sct}=15.4$ from Section \ref{v475_sct}.
We also obtained $E(B-V)=0.95\pm0.05$ by assuming that
the intrinsic $(B-V)_0$ color of V2615~Oph are similar to those of
FH~Ser, QV~Vul, and V705~Cas until the dust blackout started,
as shown in Figure
\ref{v2615_oph_v475_sct_v705_cas_qv_vul_fh_ser_v_bv_ub_color_logscale}(b).
The NASA/IPAC galactic dust absorption map gives $E(B-V)=0.87 \pm 0.02$
in the direction toward V2615~Oph, being roughly consistent with 
our obtained value of $E(B-V)=0.95\pm0.05$.  
Thus, we adopt $E(B-V)=0.95\pm0.05$.

Figure \ref{distance_reddening_v1280_sco_v2467_cyg_v2615_oph_v458_vul}(c)
shows plot of various distance-reddening relations for V2615~Oph,
$(l, b)= (4\fdg1475,+3\fdg3015)$.
Here we plot the distance-reddening relations given by \citet{mar06},
i.e., $(l, b)= (4\fdg00,3\fdg25)$ denoted by open red squares,
$(4\fdg25,3\fdg25)$ by filled green squares, 
$(4\fdg00,3\fdg50)$ by blue asterisks, and
$(4\fdg25,3\fdg50)$ by open magenta circles.
The closest one is the one denoted by filled green squares.
We also added the distance-reddening relation given by \citet{gre15}.
These four trends, Marshall et al.'s relations,
Green et al.'s relation (solid black line),
$(m-M)_V=16.5$ (thick solid blue line),
and $E(B-V)=0.95$ (vertical solid red line), consistently cross at 
$d=5.1$~kpc.

Using $E(B-V)=0.95$ and $(m-M)_V =16.5$, we plot the color-magnitude
diagram of V2615~Oph in Figure
\ref{hr_diagram_v2467_cyg_v2615_oph_v458_vul_v2468_cyg_outburst}(b).
The track of V2615~Oph is similar to that of FH~Ser 
(solid orange lines) in the very early phase and then makes a loop
clockwise near the track of LV~Vul (solid magenta lines),
corresponding to the oscillatory feature of the $V$ light curve,
and then goes down along the track of FH~Ser.
However, note that there are two different $B-V$ color data sets in 
\citet{mun08a} from JD~2454220 to JD~2454250 
as shown in Figure \ref{v2615_oph_v_bv_ub_color_curve}(b).  
Between $M_V=-6$ and $-4$ in Figure 
\ref{hr_diagram_v2467_cyg_v2615_oph_v458_vul_v2468_cyg_outburst}(b),
one goes down along the track of FH~Ser 
i.e., $(B-V)_0\sim+0.06$, and the other goes down along
that of LV~Vul, i.e., $(B-V)_0\sim-0.10$, until the dust-blackout started.
These two different $B-V$ color data set were obtained at two different
observatories, so we suppose that the response functions of the
$V$ filters differ between them. 
After the recovery of the dust blackout, the V2615~Oph data follows those 
of LV~Vul (magenta thick solid).
If we adopt the redder side data,
we regard V2615~Oph as a FH~Ser type in the color-magnitude diagram.
A dust shell was formed when the brightness declined to $m_V=12.0$
\citep{rud07a} as indicated by a large open red square in Figure
\ref{hr_diagram_v2467_cyg_v2615_oph_v458_vul_v2468_cyg_outburst}(b).
We specify this starting point by $(B-V)_0=-0.06$ and $M_V=-4.58$.

\clearpage

%Fig.63
%\placefigure{v2615_oph_v_bv_ub_color_curve}

%Fig.64 
%\placefigure{v2615_oph_v475_sct_v705_cas_qv_vul_fh_ser_v_bv_ub_color_logscale}

%Fig.65
%\placefigure{v458_vul_v_bv_ub_color_curve}

\begin{figure}
%%\epsscale{0.60}
%\epsscale{0.75}
%\epsscale{1.0}
\epsscale{1.15}
\plotone{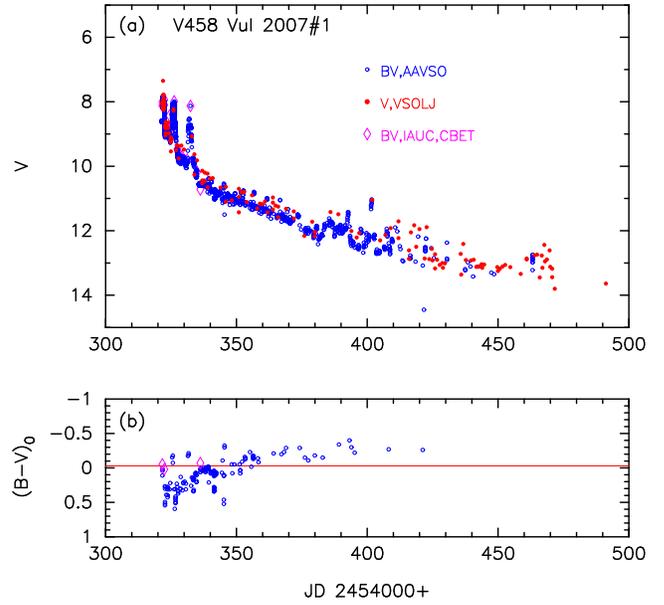}
%\plotone{v458_vul_v_bv_ub_color_curve.epsi}
%\plotfiddle{evolution1.ps}{5.0cm}{270}{0.4}{0.4}{-170}{220}
\caption{
Same as Figure \ref{v1065_cen_v_bv_ub_color_curve}, but
%%(a) $V$ band and visual light curves and (b) $(B-V)_0$ color evolution 
for V458~Vul.  
%The data are taken from the AAVSO and
%VSOLJ archives, and IAU Circular No.\ 8863, CBET Nos.\ 1029, 1035, and 1038.
We de-reddened $(B-V)_0$ color with $E(B-V)=0.50$.
\label{v458_vul_v_bv_ub_color_curve}}
\end{figure}

%Fig.66
%\placefigure{v458_vul_v443_sct_pw_vul_v_bv_ub_color_logscale}

\begin{figure}
%\epsscale{0.75}
%%\epsscale{0.8}
%\epsscale{1.0}
\epsscale{1.15}
\plotone{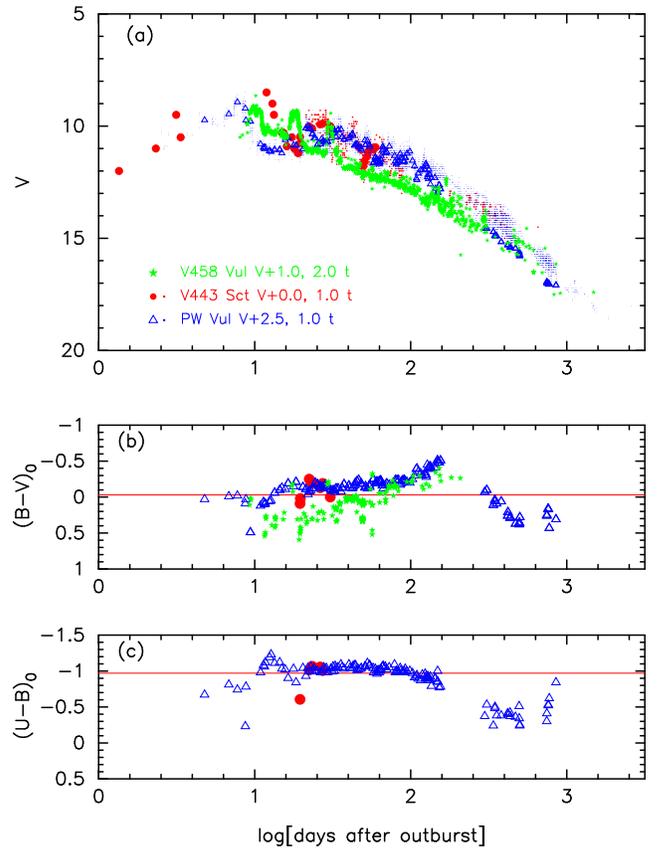}
%\plotone{v458_vul_v443_sct_pw_vul_v_bv_ub_color_logscale.epsi}
%\plotfiddle{evolution1.ps}{5.0cm}{270}{0.4}{0.4}{-170}{220}
\caption{
Same as Figure \ref{t_pyx_pw_vul_nq_vul_dq_her_v_bv_ub_color_logscale_no6},
but for V458~Vul (filled green stars).
We also added V443~Sct (filled red circles) and PW~Vul (open blue triangles).
The $BV$ data of V458~Vul are the same as those in Figure
\ref{v458_vul_v_bv_ub_color_curve}.
%The $V$ light curves of V458~Vul and PW~Vul are
%shifted down by 1.0 and 2.5 mag, respectively, against that of
%V443~Sct.  The timescale of V458~Vul is stretched by a factor of two
%against that of V443~Sct.
%In panel (a) we plot the absolute magnitude of our $0.95~M_\sun$ WD 
%model (thick solid red line) with the envelope chemical composition of
%$X=0.55$, $X_{\rm CNO}=0.20$, $X_{\rm Ne}=0.0$, and $Z=0.02$
%(CO nova 4).
%See the main text for more detail.
\label{v458_vul_v443_sct_pw_vul_v_bv_ub_color_logscale}}
\end{figure}

%Fig.67 
%\placefigure{all_mass_v458_vul_x55z02o10ne03_x55z02c10o10_new}

\begin{figure}
%\epsscale{0.70}
%\epsscale{1.0}
\epsscale{1.15}
\plotone{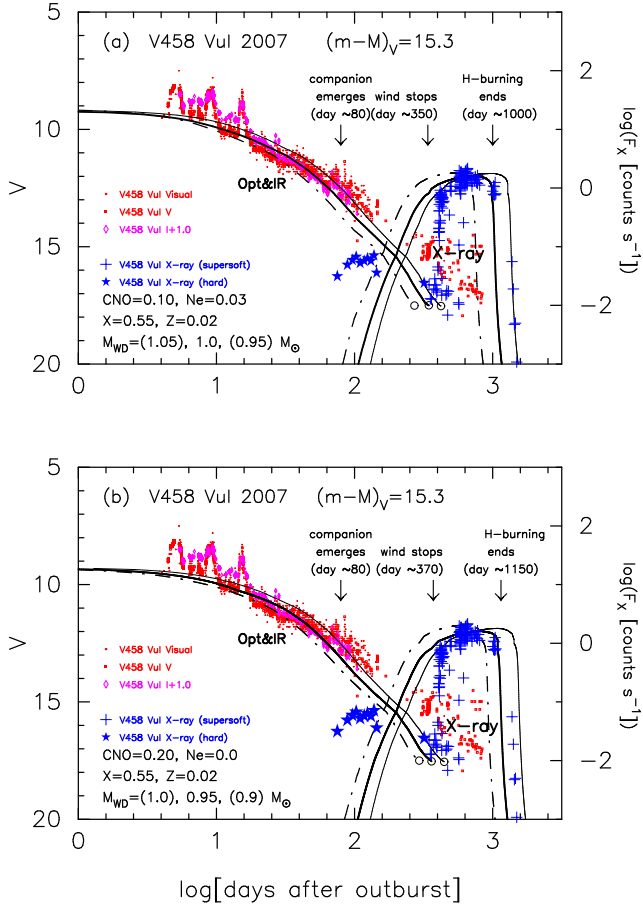}
%\plotone{all_mass_v458_vul_x55z02o10ne03_x55z02c10o10_new.epsi}
%\plotfiddle{evolution1.ps}{5.0cm}{270}{0.4}{0.4}{-170}{220}
\caption{
Optical and supersoft X-ray light curves of V458~Vul on a logarithmic
timescale.  Each symbols represent various bands as shown in the figure.
The optical and NIR data are taken from the AAVSO and VSOLJ archives.
The X-ray count rate data are taken from \citet{nes09} and from
an automatic analyzer of the {\it Swift} web page \citep{eva09}.
Assuming that the distance modulus in the $V$ band is $(m-M)_V=15.3$, 
we plot each of the three model light curves of (a)
$M_{\rm WD}= 1.05$ (thick dash-dotted lines), 
$1.0$ (thick solid lines), and
$0.95~M_\sun$ (thin solid lines) WDs
for the envelope chemical composition of Ne nova 2
\citep[see Table 1 of][for more detail]{hac10k} and (b)
$M_{\rm WD}= 1.0$ (thick dash-dotted lines), 
$0.95$ (thick solid lines), and
$0.9 ~M_\sun$ (thin solid lines) WDs
for the envelope chemical composition of CO nova 4.
\label{all_mass_v458_vul_x55z02o10ne03_x55z02c10o10_new}}
\end{figure}

\subsection{V458~Vul 2007\#1}
\label{v458_vul}
This nova is not studied in Paper I.
Figure \ref{v458_vul_v_bv_ub_color_curve} shows the $V$ and $(B-V)_0$ 
evolutions of V458~Vul on a linear timescale.
The $BV$ data are taken from IAU Circular No.\ 8863, CBET No.\ 1029,
and the AAVSO archive.  The $V$ data are from VSOLJ and IAU Circular
Nos.\ 8861 and 8899, CBET Nos.\ 1029 and 1035.
V458~Vul reached $m_{V,\rm max}=8.1$ at optical maximum on August 9.43 UT 
\citep{tar07}.   
Subsequently, it underwent two major rebrightenings with an amplitude
of one magnitude on August 13.5 UT and August 19.5 UT.
Then it declined with $t_2=8$ and $t_3=21$~days \citep[e.g.,][]{wes08},
showing small amplitude fluctuations.

We also plot the light curve and color evolutions of V458~Vul 
on a logarithmic timescale in Figure
\ref{v458_vul_v443_sct_pw_vul_v_bv_ub_color_logscale} together with
V443~Sct and PW~Vul.  The shape of the light curve and color evolution
of V458~Vul are similar to those of V443~Sct and PW~Vul.
In the figure, we stretch the time of V458~Vul
by a factor of 2.0 and shift the $V$ light curve down by 1.0 mag in order
to make them overlap.  In Paper I, we determined
the apparent distance modulus of V458~Vul as $(m-M)_V=15.5$ 
by the time-stretching method.  We reanalyzed the data and obtained
$(m-M)_V=15.3$ from Figure 
\ref{v458_vul_v443_sct_pw_vul_v_bv_ub_color_logscale}(a), i.e.,
\begin{eqnarray}
(m-M)_{V,\rm V458~Vul} &=& 15.3 \cr
&=& (m-M+ \Delta V)_{V,\rm PW~Vul}  - 2.5 \log 1.0/2.0\cr 
&\approx& 13.0 + (+2.5 - 1.0) + 0.75 = 15.25 \cr
&=& (m-M+ \Delta V)_{V,\rm V443~Sct} - 2.5 \log 1.0/2.0\cr 
&\approx& 15.5 + (+0.0 - 1.0) + 0.75 = 15.25, 
\end{eqnarray}
where we use $(m-M)_{V,\rm PW~Vul}=13.0$ from Section \ref{pw_vul_cmd} and
$(m-M)_{V,\rm V443~Sct}=15.5$ from Section \ref{v443_sct}.
Thus, we adopt $(m-M)_{V,\rm V458~Vul}=15.3$ in the present paper.

The reddening for V458~Vul was obtained as $E(B-V)=0.6$ 
\citep{lyn07} from \ion{O}{1} lines, 
as $E(B-V)=0.55\pm0.12$ \citep{pog08} from $E(B-V)= (B-V)_{t2}
- (B-V)_{0, t2} = 0.53 - (-0.02\pm 0.12) = 0.55\pm 0.12$, 
as $E(B-V)=0.63$ \citep{wes08} from the Balmer line ratio 
H$\alpha$/H$\beta$ of the south-west knot of the nebula associated
with the nova.  The NASA/IPAC galactic dust absorption map
gives $E(B-V)=0.54 \pm 0.03$ in the direction toward V458~Vul.
We obtain the reddening of $E(B-V)=0.50\pm0.05$ by assuming that
the intrinsic $(B-V)_0$ color evolution of V458~Vul is similar to
that of PW~Vul as shown in Figure 
\ref{v458_vul_v443_sct_pw_vul_v_bv_ub_color_logscale}(b).
Then the distance is calculated as $d=5.6$~kpc for $E(B-V)=0.50$
and $(m-M)_V=15.3$. 

The distance to V458~Vul was obtained as $d=6.7$--10.3~kpc
\citep{pog08} from various MMRD relations for $t_2=7\pm2$
and $t_3=15\pm2$ days and $M_{V,\rm 15}$ magnitude,
and $d=13$~kpc \citep{wes08} from the MMRD relation, 
galactic rotation, and light echo time of nearby nebula
associated with the nova.
It is well known that the MMRD and $M_{V,15}$ relations represent
statistical tendencies and do not give correct
brightness for individual novae.  \citet{roy12} claimed that the nova 
could be as close as $d\sim6.5$~kpc or more from an analysis
of the \ion{H}{1} cloud associated with the nova.

Figure \ref{distance_reddening_v1280_sco_v2467_cyg_v2615_oph_v458_vul}(d)
shows various distance-reddening relations for V458~Vul,
$(l, b)= (58\fdg6331,-3\fdg6171)$.
Here we plot the distance-reddening relations given by \citet{mar06},
i.e., $(l, b)= (58\fdg50,-3\fdg50)$ denoted by open red squares,
$(58\fdg75,-3\fdg50)$ by filled green squares, 
$(58\fdg50,-3\fdg75)$ by blue asterisks, and
$(58\fdg75,-3\fdg75)$ by open magenta circles.
The closer ones are those denoted by filled green squares and
open magenta circles.  We also plot the distance-reddening relation
given by \citet{gre15}.  These trends cross consistently at the point
$E(B-V)=0.50$ and $d=5.6$~kpc.

Assuming $E(B-V)=0.50$ and $(m-M)_V=15.3$,
we plot the color-magnitude diagram of V458~Vul in Figure
\ref{hr_diagram_v2467_cyg_v2615_oph_v458_vul_v2468_cyg_outburst}(c).
The basic part of the track is close to that of PU~Vul (thick solid
blue lines).  Therefore, we regard V458~Vul as a PU~Vul type.
This is consistent with the fact that this type of novae, including
RR~Pic, V723~Cas, HR~Del, and V5558~Sgr, have multiple peaks (or flares)
in the very early phase \citep[see][for the reason behind
multiple flaring pulses]{kat09h,kat11h}.   
We connect the first maximum by red lines, second by blue lines,
and third by black lines.  V458~Vul moves clockwise in the color-magnitude
diagram during these three pulses (Figure
\ref{v458_vul_v_bv_ub_color_curve}) as indicated by blue arrows.
This is very similar to that of V5558~Sgr in Figure
\ref{hr_diagram_pu_vul_v723_cas_hr_del_v5558_sgr_outburst}(d).

Finally, we revisit the model light curve analysis of V458~Vul,
which was analyzed in our previous paper \citep{hac10k}.
After that study, the end of the supersoft X-ray phase of V458~Vul
was reported \citep[e.g.,][]{sch11}, which enables us to determine
the WD mass more accurately.  Our new model light curves are presented
in Figure \ref{all_mass_v458_vul_x55z02o10ne03_x55z02c10o10_new}.
The detail of our numerical methods are described in \citet{hac16k}.  
We add calculated supersoft X-ray fluxes together with the observational
X-ray data (filled blue stars for the hard X-ray phase and 
blue plus symbols for the soft X-ray phase).
The X-ray count rates are taken from the {\it Swift} 
web page.  %%%\footnote{http://www.swift.ac.uk/} \citep{eva09}.
We think that the steep rise in the X-ray flux near 380 days after
the outburst corresponds to the epoch when the optically thick winds
stopped \citep[see, e.g.,][]{hac10k}. 
The optical and near infrared (NIR) light curves of free-free emission
depend mainly on the WD mass and weakly on the chemical composition of
the envelope.  For the chemical composition $X=0.55$, $Y=0.30$, 
$X_{\rm CNO}=0.10$, $X_{\rm Ne}=0.03$, and $Z=0.02$ (Ne nova 2)
for a typical neon nova,
we have a best fit model with $M_{\rm WD}=1.0~M_\sun$ (black thick solid
line) among $0.95$, $1.0$, and $1.05~M_\sun$ WDs as shown in Figure
\ref{all_mass_v458_vul_x55z02o10ne03_x55z02c10o10_new}(a).
If we assume $X=0.55$, $Y=0.23$, $X_{\rm CNO}=0.20$, and $Z=0.02$
(CO nova 4) for a typical carbon-oxygen nova, on the other hand,
we have a best fit model with
$M_{\rm WD}=0.95~M_\sun$ among 0.9, 0.95, and $1.0~M_\sun$ WDs
as shown in Figure
\ref{all_mass_v458_vul_x55z02o10ne03_x55z02c10o10_new}(b).

Our previous model was $M_{\rm WD}=0.95~M_\sun$ 
for the chemical composition of Ne nova 2 \citep[see Figure 27
of][]{hac10k} or $M_{\rm WD}=0.93~M_\sun$ for the chemical composition
of CO nova 4 \citep[see Figure 26 of][]{hac10k}.  The $V$ light curve
was fitted with the free-free emission model light curve and
it gave a distance modulus of $(m-M)_V=17.0$ 
\citep[see Equation (86) of][]{hac10k}.
Here, we revised this fitting using new estimates of the WD mass,
$1.0~M_\sun$ and $0.95~M_\sun$ for Ne nova 2 and CO nova 4,
respectively, and new distance modulus,
$(m-M)_V=15.3$, as shown in Figure
\ref{all_mass_v458_vul_x55z02o10ne03_x55z02c10o10_new}(a) and
\ref{all_mass_v458_vul_x55z02o10ne03_x55z02c10o10_new}(b).
This is consistent with the end of hydrogen shell burning.
We included the contribution from the photospheric emission 
(blackbody approximation) in the $V$ light curve model 
in addition to the free-free emission \citep[see, e.g.,][]{hac15k}.

Hard X-rays from novae are considered to originate from internal
shocks in the ejecta \citep[e.g.,][]{muk01} or shocks between the ejecta and
circumstellar matter that is fed by the cool wind from a red-giant companion
\citep[e.g.,][]{sok06}.  
\citet{hac10k} claimed that the hard X-ray emission from V458~Vul
comes from the shock between the ejecta and the companion.
If that is the case, we may observe hard X-rays when the companion
appears out of the optically thick nova envelope.
We again examine this possibility.
The orbital period of V458~Vul was first reported as
$P_{\rm orb}=0.589$~days by \citet{gor08}, but was later revised
to $P_{\rm orb}= 0.0681$~days by \citet{rod11}.
Assuming that the companion is a $0.6~M_\sun$ CO core
\citep[e.g.,][]{wes08}, we obtain
a separation of $0.82~R_\sun$ for the primary of
$M_{\rm WD}=1.0~M_\sun$.  Then the emergence time of the secondary
from the optically thick nova envelope is recalculated as
$\sim 80$~days after the outburst based on our new nova models.
This emergence time of the companion is coincident with the start
of the hard X-ray (filled blue stars) increase.

%Fig.68 
%\placefigure{v2468_cyg_v_bv_ub_color_curve}

\begin{figure}
%%\epsscale{0.60}
%\epsscale{0.75}
%\epsscale{1.0}
\epsscale{1.15}
\plotone{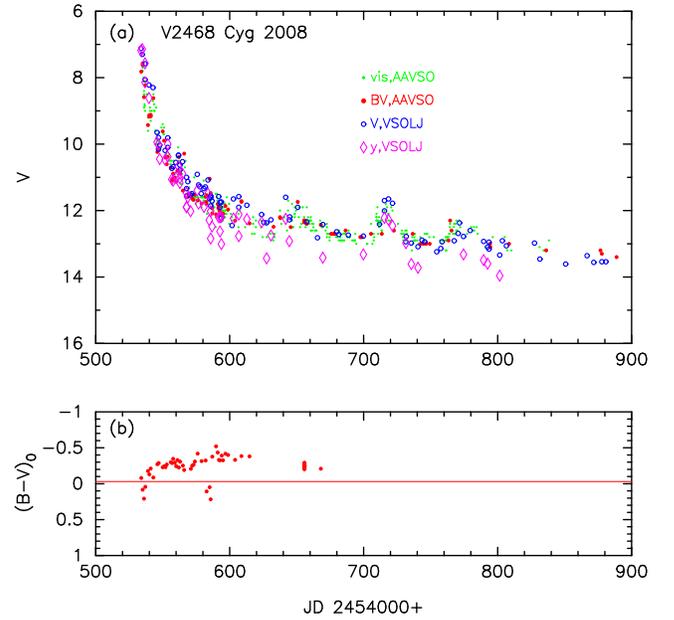}
%\plotone{v2468_cyg_v_bv_ub_color_curve.epsi}
%\plotfiddle{evolution1.ps}{5.0cm}{270}{0.4}{0.4}{-170}{220}
\caption{
(a) $V$ and $y$ bands and visual light curves 
and (b) $(B-V)_0$ color evolution for V2468~Cyg.
%%The data are taken from the AAVSO and VSOLJ archives.
We de-reddened $(B-V)_0$ color with $E(B-V)=0.75$.
\label{v2468_cyg_v_bv_ub_color_curve}}
\end{figure}

%Fig.69 
%\placefigure{distance_reddening_v2468_cyg_v2491_cyg_v496_sct_ci_aql}

\begin{figure*}
%\begin{figure}
\epsscale{0.75}
%%\epsscale{0.8}
%%\epsscale{1.0}
%%\epsscale{1.15}
\plotone{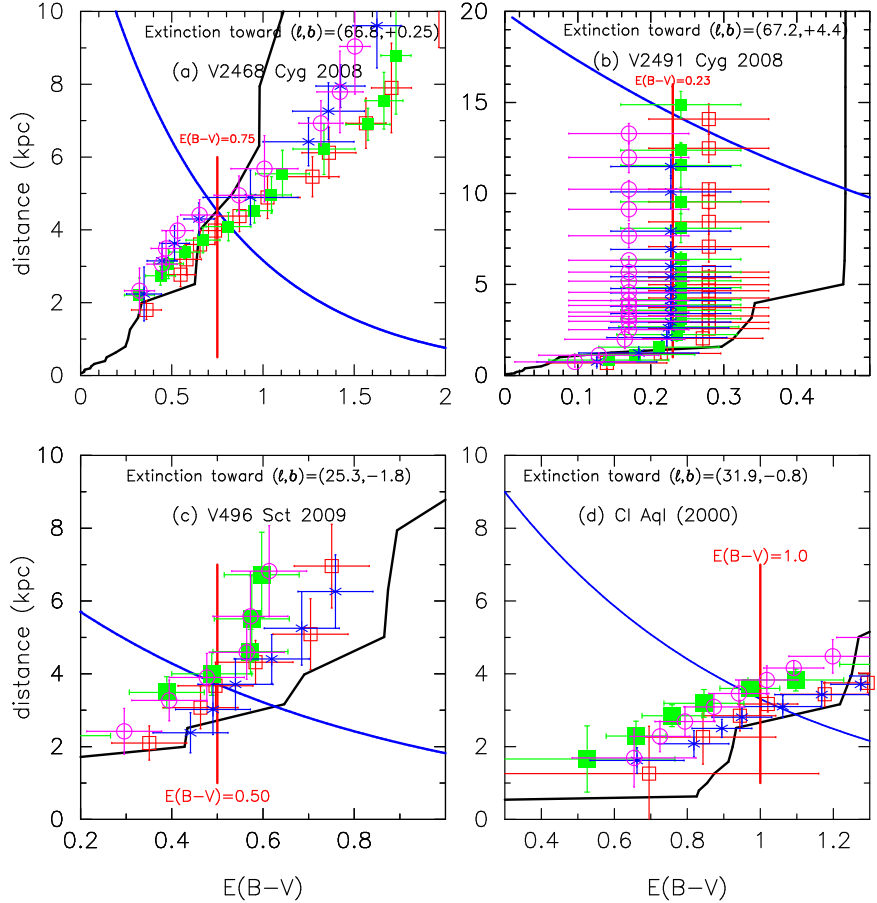}
%\plotone{distance_reddening_v2468_cyg_v2491_cyg_v496_sct_ci_aql.epsi}
%\plotfiddle{evolution1.ps}{5.0cm}{270}{0.4}{0.4}{-170}{220}
\caption{
Same as Figure \ref{distance_reddening_fh_ser_pw_vul_v1500_cyg_v1974_cyg},
but for (a) V2468~Cyg, (b) V2491~Cyg, (c) V496~Sct, and (d) CI~Aql.
The thick solid blue lines denote 
%%the distance-reddening relation calculated from the distance modulus 
%%in the $V$ band, i.e., 
(a) $(m-M)_V=15.6$, 
(b) $(m-M)_V=16.5$, (c) $(m-M)_V=14.4$,  and (d) $(m-M)_V=15.7$.
%The vertical solid red lines represent the color excesses of 
%(a) $E(B-V)=0.75$, (b) $E(B-V)=0.23$,
%(c) $E(B-V)=0.50$, and (d) $E(B-V)=1.0$.
%The black solid lines denote the distance-reddening relation given
%by \citet{gre15}.
%In panel (a), the magenta thick solid line represents
%the distance-reddening relation calculated from the UV~1455 \AA\  flux
%fitting with the $0.51~M_\sun$ WD model \citep{hac15k}.
%In panels (a), (b), and (d), two or 
%four sets of data with error bars show distance-reddening relations
%in two or four directions close to each nova, the data of which are taken
%from \citet{mar06}.  
%See the main text for more detail.
\label{distance_reddening_v2468_cyg_v2491_cyg_v496_sct_ci_aql}}
%\end{figure}
\end{figure*}

\subsection{V2468~Cyg 2008\#1}
\label{v2468_cyg}
This nova is not studied in Paper I.
Figure \ref{v2468_cyg_v_bv_ub_color_curve} shows the visual, $V$,
and $y$ light curves and $(B-V)_0$ color evolution of V2468~Cyg.
The $BV$ data are taken from the AAVSO archive, and the $V$ and $y$ data are
from the VSOLJ archive.  V2468~Cyg reached $m_{V,\rm max}=7.6$
at optical maximum on UT 2008 March 9.5 \citep[e.g.,][]{cho12}.   
Then it declined with $t_2=9$ and $t_3=20$~days, showing
fluctuations with an amplitude of one magnitude and a period of a few days
\citep{cho12}.  An orbital period of 3.49~hr was detected
by \citet{cho13}.

The reddening for V2468~Cyg was obtained as $E(B-V)=0.77$ 
\citep{rud08a} from \ion{O}{1} lines, $E(B-V)=0.80$ 
\citep{iij11} from hydrogen column density, $E(B-V)=0.8$ 
\citep{sch09} from Balmer decrements, $E(B-V)=0.79\pm 0.01$ 
\citep{cho12} from a simple average of
$E(B-V)=0.78$ from $E(B-V)= (B-V)_{t2} - (B-V)_{0, t2} =
0.76 - (-0.02\pm 0.04) = 0.78$,
$E(B-V)=0.77$ from \ion{O}{1} lines \citep{rud08a},
$E(B-V)=0.8$ from the Balmer decrements \citep{sch09},
and $E(B-V)=0.80$ from the hydrogen column density \citep{iij11}.
These values are all consistent with each other, so we adopt
$E(B-V)=0.75\pm0.05$ in this paper.

The distance to V2468~Cyg was estimated mainly with the MMRD relations.
\citet{iij11} estimated
$(m-M)_V= m_{V,\rm max} - M_{V,\rm max}= 7.3 - (-8.8\pm0.3)=16.1\pm0.3$
from $t_2=7.8\pm0.5$~days \citep{del95} and calculated 
$d=5.5\pm0.8$~kpc for $E(B-V)=0.8\pm0.1$.  \citet{cho12} also obtained
$(m-M)_V= m_{V,\rm max} - M_{V,\rm max}= 7.57 - (-8.7\pm0.07)=16.27\pm0.07$
for $t_2=9$~days and $t_3=22$~days \citep{del95,dow00}
and $d=5.4\pm0.4$~kpc for $E(B-V)=0.79\pm0.01$.

In Paper I, we obtained the distance modulus of $(m-M)_V=15.6$ for V2468~Cyg
by the time-stretching method.  We confirmed this result of Paper I in Figure
\ref{v2362_cyg_v2468_cyg_v1500_cyg_v1668_cyg_v_bv_ub_color_logscale}(a)
and Equation (\ref{v2362_cyg_v2468_cyg_v1500_cyg_v1668_cyg}), and also in 
Figure \ref{v2467_cyg_v2468_cyg_v1668_cyg_iv_cep_v_color_logscale}(a)
and Equation (\ref{v2467_cyg_v1668_cyg_iv_cep_v2468_cyg}).
Assuming that the $(B-V)_0$ colors of V2468~Cyg, V1500~Cyg, V1668~Cyg, 
IV~Cep, V2362~Cyg, and V2467~Cyg are very similar to each other,
we obtain the color excess of V2468~Cyg $E(B-V)=0.75\pm0.05$
as shown in Figures 
\ref{v2362_cyg_v2468_cyg_v1500_cyg_v1668_cyg_v_bv_ub_color_logscale}(b)
and
\ref{v2467_cyg_v2468_cyg_v1668_cyg_iv_cep_v_color_logscale}(b).
This estimate is consistent with the previous results mentioned above.

Figure \ref{distance_reddening_v2468_cyg_v2491_cyg_v496_sct_ci_aql}(a)
shows various distance-reddening relations for V2468~Cyg,
$(l,b)=(66\fdg8084, +0\fdg2455)$.
Here we plot the relations given by \citet{mar06}, i.e.,
$(l, b)= (66\fdg75,0\fdg00)$ denoted by open red squares,
$(67\fdg00,0\fdg00)$ by filled green squares, 
$(66\fdg75,0\fdg25)$ by blue asterisks, and
$(67\fdg00,0\fdg25)$ by open magenta circles.
The closest one is denoted by blue asterisks.
We also plot the distance-reddening relation given by \citet{gre15}.
These distance-reddening relations cross at the same point, i.e.,
at the cross point of the two lines, $(m-M)_V=15.6$ (thick solid blue line) 
and $E(B-V)=0.75$ (vertical solid red line).
The distance is calculated as $d=4.5$~kpc for $E(B-V)=0.75$
and $(m-M)_V=15.6$.

Using $E(B-V)=0.75$ and $(m-M)_V= 15.6$,
we plot the color-magnitude diagram of V2468~Cyg in Figure
\ref{hr_diagram_v2467_cyg_v2615_oph_v458_vul_v2468_cyg_outburst}(d).
The track of V2468~Cyg is similar to that of V1500~Cyg 
(thick solid green line).
Therefore, we regard V2468~Cyg as a V1500~Cyg type in the color-magnitude
diagram.  The nova entered the nebular phase on UT 2008 July 8,
about 125~days after the outburst, i.e., at $m_V=12.2$ \citep{iij11}
as indicated by the large open red square at $(B-V)_0=-0.38$ and 
$M_V=-3.80$ in Figure 
\ref{hr_diagram_v2467_cyg_v2615_oph_v458_vul_v2468_cyg_outburst}(d).

%Fig.70 
%\placefigure{v2491_cyg_v_bv_color_curve}

\begin{figure}
%%\epsscale{0.60}
%\epsscale{0.75}
%\epsscale{1.0}
\epsscale{1.15}
\plotone{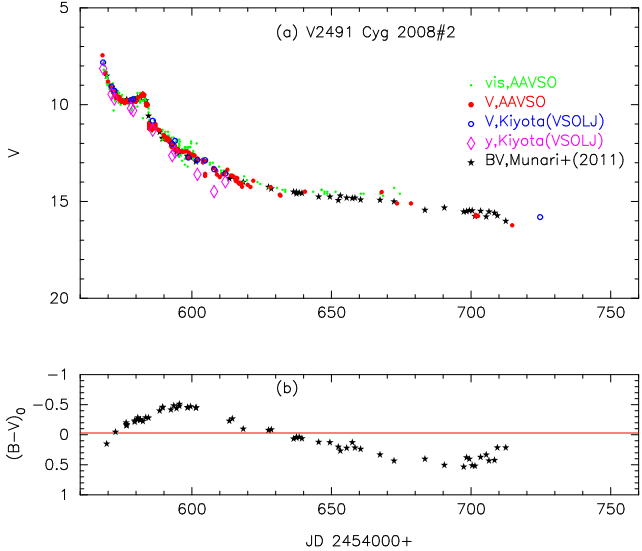}
%\plotone{v2491_cyg_v_bv_color_curve.epsi}
%\plotfiddle{evolution1.ps}{5.0cm}{270}{0.4}{0.4}{-170}{220}
\caption{
(a) $V$, $y$, and visual light curves 
and (b) $(B-V)_0$ color evolution for V2491~Cyg.
%The data are taken from \citet{mun11} and
%the AAVSO and VSOLJ archives.
We de-reddened $(B-V)_0$ color with $E(B-V)=0.23$.
\label{v2491_cyg_v_bv_color_curve}}
\end{figure}

%Fig.71 
%\placefigure{v2491_cyg_v1500_cyg_v1668_cyg_v_color_logscale}

\begin{figure}
%%\epsscale{0.60}
%\epsscale{0.75}
%\epsscale{1.0}
\epsscale{1.15}
\plotone{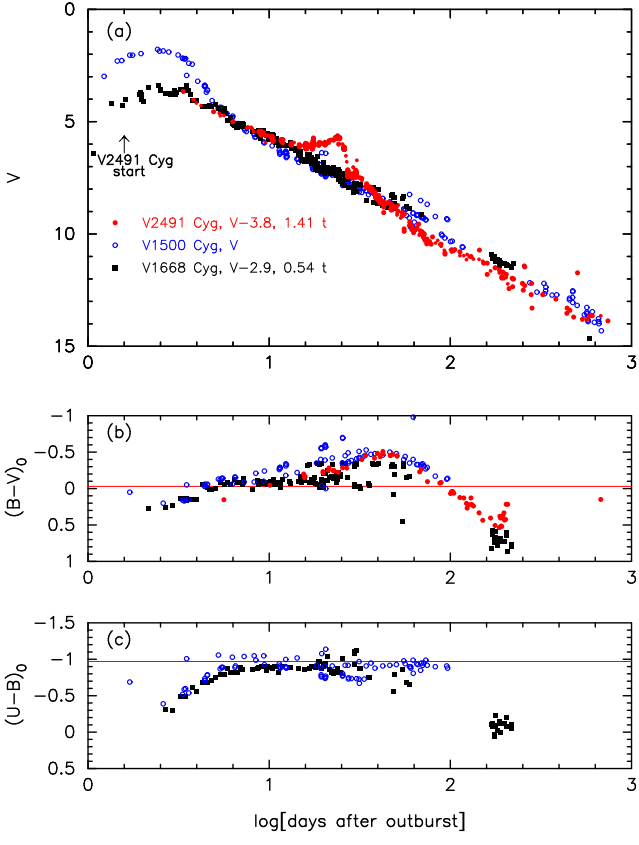}
%\plotone{v2491_cyg_v1500_cyg_v1668_cyg_v_color_logscale.epsi}
%\plotfiddle{evolution1.ps}{5.0cm}{270}{0.4}{0.4}{-170}{220}
\caption{
Same as Figure \ref{t_pyx_pw_vul_nq_vul_dq_her_v_bv_ub_color_logscale_no6},
but for V2491~Cyg (filled red circles).
We also added V1500~Cyg (open blue circles) and V1668~Cyg (filled black
squares).  The $BV$ data of V2491~Cyg are the same as those in Figure
\ref{v2491_cyg_v_bv_color_curve}.
%The $V$ light curves of V2491~Cyg and V1668~Cyg are
%shifted up by 3.8 and 2.9 mag, respectively, against that of
%V1500~Cyg.  The timescales of V2491~Cyg and V1668~Cyg are
%stretched by a factor of 1.41 and 0.54, respectively, against
%that of V1500~Cyg.  See the main text for more detail.
\label{v2491_cyg_v1500_cyg_v1668_cyg_v_color_logscale}}
\end{figure}

%Fig.72
%\placefigure{hr_diagram_v2491_cyg_v496_sct_ci_aql_u_sco_outburst}

\begin{figure*}
%\begin{figure}
%\epsscale{0.75}
\epsscale{0.8}
%%\epsscale{1.0}
\plotone{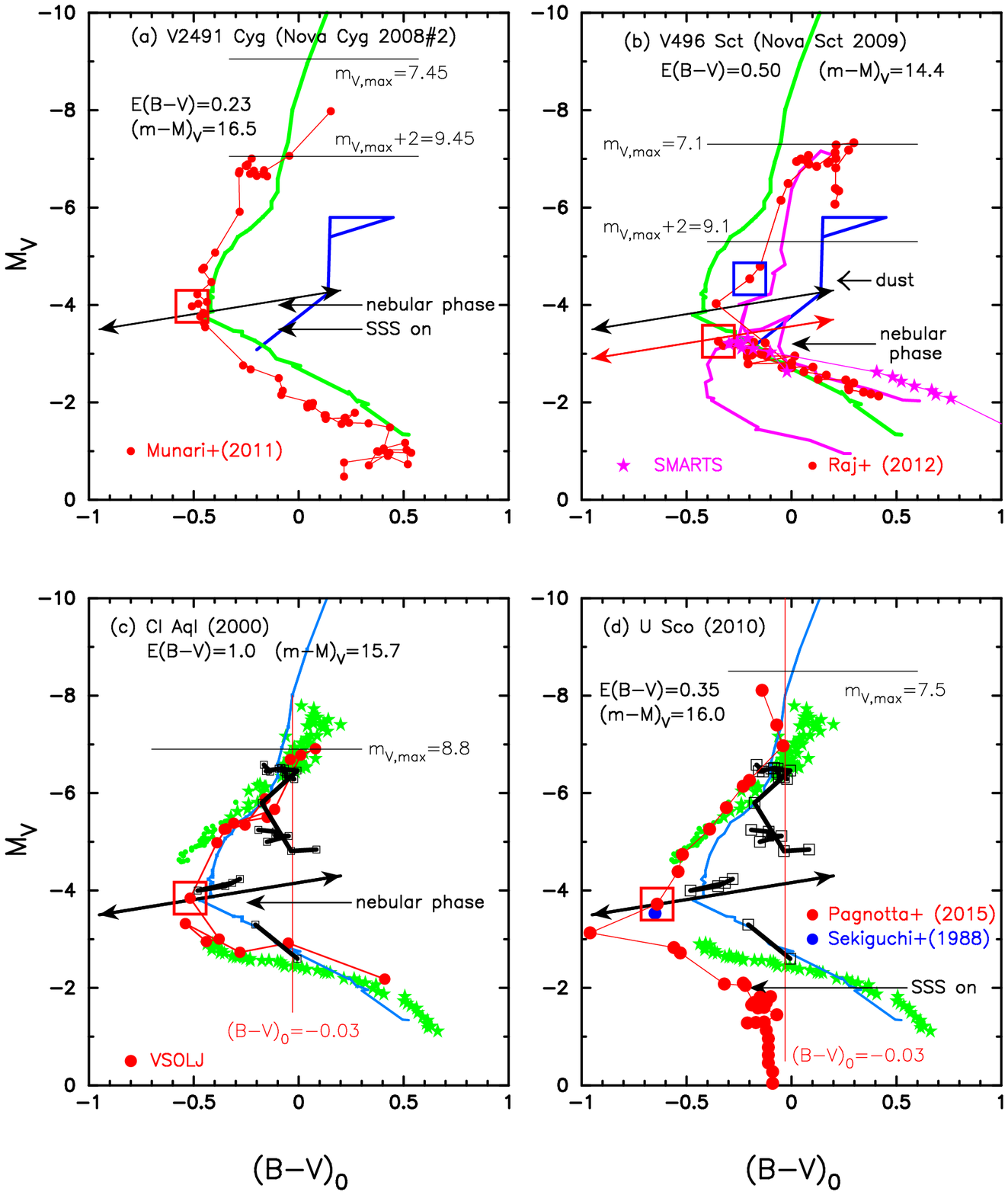}
%\plotone{hr_diagram_v2491_cyg_v496_sct_ci_aql_u_sco_outburst.epsi}
%\plotfiddle{evolution1.ps}{5.0cm}{270}{0.4}{0.4}{-170}{220}
\caption{
Same as Figure 
\ref{hr_diagram_rs_oph_v446_her_v533_her_t_pyx_outburst}, but
%%Color-magnitude diagrams 
for (a) V2491~Cyg, (b) V496~Sct, (c) CI~Aql, and (d) U~Sco.  
Filled red circles connected by
thin solid red lines are the color-magnitude data of each nova.
In panels (a) and (b), thick solid green and blue lines denote
the tracks of V1500~Cyg and PU~Vul, respectively.
In panel (b), thick solid magenta lines that of LV~Vul.  
The SMARTS data (magenta stars connected by thin solid magenta line)
are added.  In panels (c) and (d), we plot the track
of M31N 2008-12a, adopting $(m-M)_V=25.1$ and $E(B-V)=0.21$, 
by open black squares with connected solid black lines.
The green symbols denote the track of T~Pyx (1966, 2011),
thick solid sky-blue lines that of V1500~Cyg.
\label{hr_diagram_v2491_cyg_v496_sct_ci_aql_u_sco_outburst}}
%\end{figure}
\end{figure*}

\subsection{V2491~Cyg 2008\#2}
\label{v2491_cyg}
This nova is not studied in Paper I.
Figure \ref{v2491_cyg_v_bv_color_curve} shows the $y$, $V$, and visual light
curves and $(B-V)_0$ color evolution of V2491~Cyg.
This nova reached $m_{V,\rm max}=7.45\pm0.05$ at optical maximum
on UT 2008 April 11.37 \citep{mun11}.   
Then it declined with $t_2=4.8$~days, but rose again
to $m_V=9.5$ about 15 days after maximum, which is the secondary maximum
similar to those of V1493~Aql (Figure \ref{v1493_aql_v_bv_ub_color_curve})
and V2362~Cyg (Figure \ref{v2362_cyg_v_bv_ub_color_curve}).
\citet{hac09ka} proposed a mechanism for the secondary maximum on the basis
of a strong magnetic field on the WD.  Although \citet{pag10} 
discussed various X-ray properties of V2491~Cyg against a strong
magnetic field, recently \citet{zem15} found a 38~min periodicity
and a possibility of a soft intermediate polar.

The reddening for V2491~Cyg was obtained as $E(B-V)=0.3$ 
\citep{lyn08b} from \ion{O}{1} lines, which was revised by \citet{rud08b}
to be $E(B-V)=0.43$ from the \ion{O}{1} lines at 0.84 and $1.13~\mu$m, and
$E(B-V)=0.23\pm0.01$ \citep{mun11} from an average of
$E(B-V)=0.24$ from \ion{Na}{1} 5889.953 line profiles, 
$E(B-V)= (B-V)_{\rm max} - (B-V)_{0, \rm max} = 
0.46 - (0.23 \pm 0.06) = 0.23\pm 0.06$,
and $E(B-V)= (B-V)_{t2} - (B-V)_{0, t2} = 0.20 - (-0.02\pm 0.04) 
= 0.22\pm 0.04$.
The distance modulus and distance to V2491~Cyg were estimated
by \citet{mun11} as 
$(m-M)_V= m_{V,\rm max} - M_{V,\rm max}= 7.45 - (-9.06)=16.51$
from the MMRD relation \citep{coh88} together with $t_2=4.8$, and
then derived as $d=14$~kpc.   In this paper, we adopt $E(B-V)=0.23$
after Munari et al. 
because the $(B-V)_0$ color de-reddened with $E(B-V)=0.23$
almost overlaps the other color evolution curves of
V1500~Cyg and V1668~Cyg as seen in Figure
\ref{v2491_cyg_v1500_cyg_v1668_cyg_v_color_logscale}(b).

Using the time-stretching method (Figure
\ref{v2491_cyg_v1500_cyg_v1668_cyg_v_color_logscale}(a)),
we obtain the distance modulus of V2491~Cyg in the $V$ band  
\begin{eqnarray}
(m-M)_{V, \rm V2491~Cyg}&=&16.5\cr 
&=& (m-M+ \Delta V)_{V,\rm V1500~Cyg}  - 2.5 \log 1.0/1.41 \cr
&\approx& 12.3 + (+0.0 + 3.8) + 0.37 = 16.47\cr
&=& (m-M+ \Delta V)_{V,\rm V1668~Cyg} - 2.5 \log 0.54/1.41 \cr
&\approx& 14.6 + (-2.9 + 3.8) + 1.04 = 16.54
\end{eqnarray}
where we use the apparent distance moduli of 
$(m-M)_{V,\rm V1500~Cyg}=12.3$ from Section \ref{v1500_cyg_cmd}
and $(m-M)_{V,\rm V1668~Cyg}=14.6$ 
from Sections \ref{v1668_cyg_cmd}.

Figure \ref{distance_reddening_v2468_cyg_v2491_cyg_v496_sct_ci_aql}(b) 
shows various distance-reddening relations for V2491~Cyg,
$(l, b)= (67\fdg2287,+4\fdg3531)$.
We plot the distance-reddening relations given by \citet{mar06}, i.e.,
$(l, b)= (67\fdg00,4\fdg25)$ denoted by open red squares,
$(67\fdg25,4\fdg25)$ by filled green squares, 
$(67\fdg00,4\fdg50)$ by blue asterisks, and
$(67\fdg25,4\fdg50)$ by open magenta circles.
The closest one is that denoted by filled green squares.
We also plot the distance-reddening relation given by \citet{gre15}.
The extinction denoted by filled green squares nicely match 
the value of $E(B-V)=0.23$.  The NASA/IPAC galactic
dust absorption map gives $E(B-V)=0.48 \pm 0.03$ in the direction
toward V2491~Cyg, being consistent with Green et al.'s relation.
However, such large values of the reddening are inconsistent with
those mentioned above.
We obtain the cross point of the two trends,
i.e., $(m-M)_V=16.5$ (thick solid blue line)
and $E(B-V)=0.23$ (vertical solid red line), i.e., 
$d=14$~kpc at $E(B-V)=0.23$.

Using $E(B-V)=0.23$ and $(m-M)_V=16.5$,
we plot the color-magnitude diagram of V2491~Cyg in Figure 
\ref{hr_diagram_v2491_cyg_v496_sct_ci_aql_u_sco_outburst}(a).
V2491~Cyg is located very close to the track of V1500~Cyg.
This similarity supports that our adopted values of
$E(B-V)=0.23$ and $(m-M)_V=16.5$ ($d=14$~kpc) are reasonable.
Therefore, we regard V2491~Cyg as a V1500~Cyg type
in the color-magnitude diagram.
It is interesting that V1500~Cyg is also a strong magnetic system,
a polar \citep{sto88}. 
The nova possibly entered the nebular phase around $\sim30$ days after
maximum at $m_V=12.5$ \citep[e.g.,][]{mun11, tar14}.
We specify the turning point $m_V\approx12.5$ and $B-V=-0.28$,
i.e.,  $M_V\approx-4.0$ and $(B-V)_0=-0.51$,
denoted by a large open red square in Figure 
\ref{hr_diagram_v2491_cyg_v496_sct_ci_aql_u_sco_outburst}(a) and
tabulated in Table \ref{color_magnitude_turning_point}.
We add the epoch when the SSS phase started
at $m_V=13.0$, about 40 days after the outburst \citep{pag10}.

%Fig.73
%\placefigure{v496_sct_v_bv_ub_color_curve}

\begin{figure}
%%\epsscale{0.60}
%\epsscale{0.75}
%\epsscale{1.0}
\epsscale{1.15}
\plotone{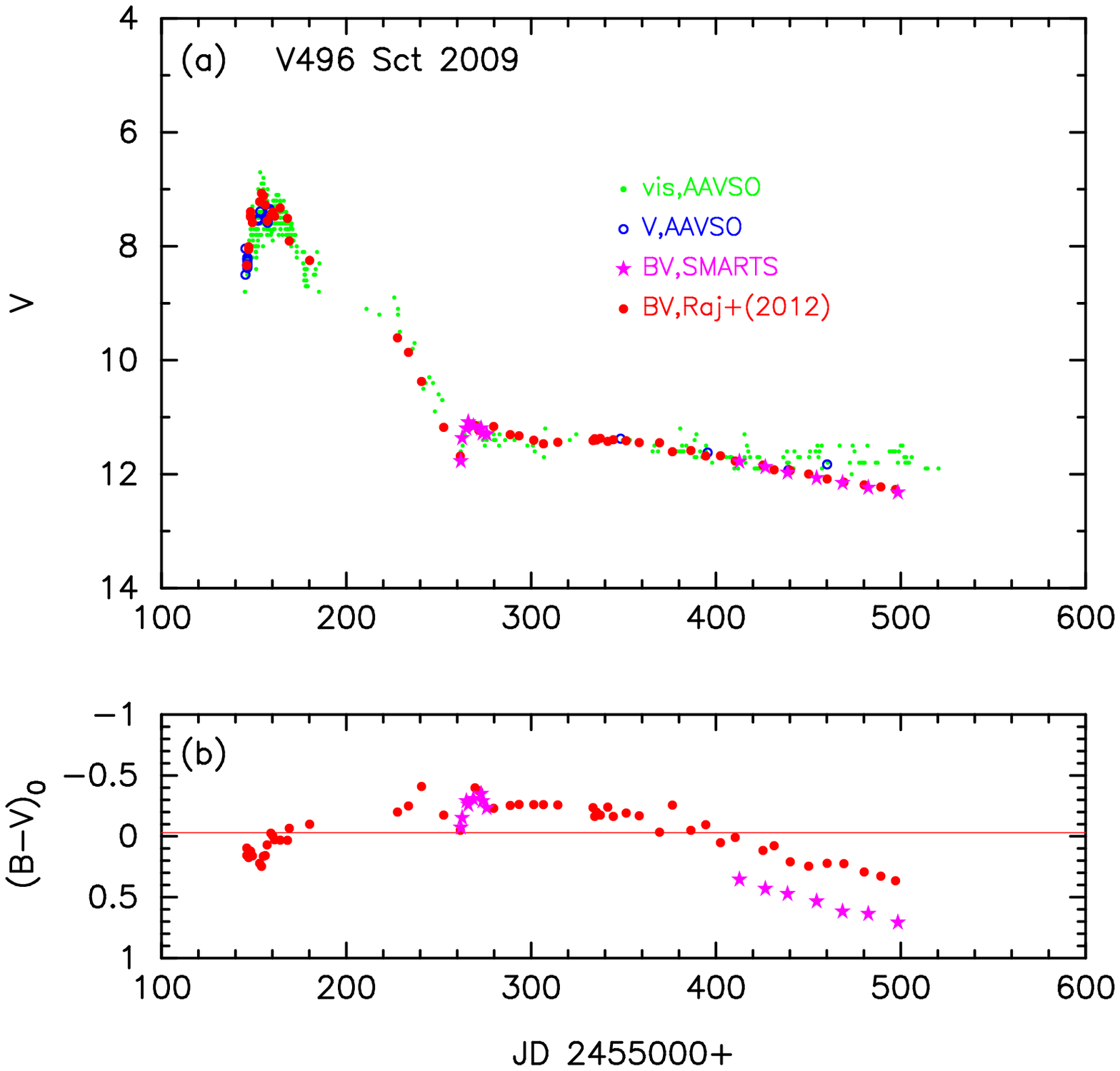}
%\plotone{v496_sct_v_bv_ub_color_curve.epsi}
%\plotfiddle{evolution1.ps}{5.0cm}{270}{0.4}{0.4}{-170}{220}
\caption{
Same as Figure \ref{v1065_cen_v_bv_ub_color_curve}, but
%(a) $V$ and visual light curves and (b) $(B-V)_0$ color evolution 
for V496~Sct.
%The data are taken from \citet{raj12} and
%the AAVSO (open blue circles) and SMARTS (magenta stars) archives.
We de-reddened $(B-V)_0$ color with $E(B-V)=0.50$.
\label{v496_sct_v_bv_ub_color_curve}}
\end{figure}

\subsection{V496~Sct 2009}
\label{v496_sct}
This nova is not studied in Paper I.
Figure \ref{v496_sct_v_bv_ub_color_curve} shows the visual and $V$,
and $(B-V)_0$ evolutions of V496~Sct on a linear timescale.
This nova reached $m_{V,\rm max}=7.07$ at optical maximum
on UT 2009 November 18.716 \citep{raj12}.
Then it declined with $t_2=59\pm5$~days.  There is
lack of data between mid-December 2009 (JD 2455180)
and early February 2010 (JD 2455220) due to 
a solar conjunction of V496~Sct.  
The $BV$ data are taken from \citet{raj12} (filled red circles)
and the SMARTS archive (magenta stars) 
while the visual (green dots) and $V$ (open blue circles)
data are from the AAVSO archive.
It showed a shallow dip of dust
shell formation about 90 days after optical maximum \citep[e.g.,][]{raj12}.
Figure \ref{v496_sct_v_bv_ub_color_curve} has no $U-B$ color data
because we could not find any $U$ band observation in the literature.

The reddening for V496~Sct was obtained by \citet{raj12}
as $E(B-V)= (B-V)_{\rm max} - (B-V)_{0, \rm max} 
= 0.797\pm0.014 - (0.23 \pm 0.06) = 0.57\pm 0.06$ 
from the intrinsic $B-V$ color at maximum
and $E(B-V)=0.65$ from interstellar \ion{Na}{1} line profile. 
They also estimated the distance modulus as
$(m-M)_V= m_{V,\rm max} - M_{V,\rm max}= 7.07 - (-7.0\pm0.2)=14.1$
from the MMRD relation \citep{del95} together with $t_2=59$
and then derived $d=2.9\pm0.3$~kpc.

Using the time-stretching method, 
we have already obtained $(m-M)_V=14.4$ as shown in Figure
\ref{v496_sct_v1065_cen_v1419_aql_v1668_cyg_lv_vul_v_bv_ub_color_logscale}(a)
together with Equation 
(\ref{v1065_cen_v1668_cyg_lv_vul_v1419_aql_v496_sct}).
The overall behavior of the V496~Sct $V$ light curve 
is similar to those of FH~Ser and NQ~Vul as shown in Figure 
\ref{v2274_cyg_v496_sct_fh_ser_nq_vul_v_bv_ub_color_logscale}(a).
Assuming that these novae have the same brightness, we estimate
the distance modulus as
\begin{eqnarray}
(m-M)_{V, \rm V496~Sct}&=&14.4 \cr 
&=& (m-M+ \Delta V)_{V,\rm FH~Ser} - 2.5\log 0.7 \cr
&=& 11.7 + (+1.9 + 0.4) + 0.39 = 14.39 \cr
&=& (m-M+ \Delta V)_{V,\rm NQ~Vul} - 2.5\log 0.7 \cr
&=& 13.6 + (+0.0 + 0.4) + 0.39 = 14.39,
\end{eqnarray}
where we use $(m-M)_{V,\rm FH~Ser}=11.7$ from Sections \ref{fh_ser_cmd} and
$(m-M)_{V,\rm NQ~Vul}=13.6$ from Sections \ref{nq_vul}.
The reddening is also estimated by assuming that the intrinsic $B-V$ color
of V496~Sct is the same as that of FH~Ser and NQ~Vul.  
We obtain $E(B-V)=0.50\pm0.05$, as shown in Figure 
\ref{v2274_cyg_v496_sct_fh_ser_nq_vul_v_bv_ub_color_logscale}(b) and 
\ref{v496_sct_v1065_cen_v1419_aql_v1668_cyg_lv_vul_v_bv_ub_color_logscale}(b).
We adopt $E(B-V)=0.50$ and $(m-M)_V=14.4$.
Then the distance is calculated to be $d=3.7$~kpc.

Figure \ref{distance_reddening_v2468_cyg_v2491_cyg_v496_sct_ci_aql}(c)
shows various distance-reddening relations for V496~Sct,
$(l, b)= (25\fdg2838,-1\fdg7678)$.
We plot the distance-reddening relation given by \citet{mar06}, i.e.,
$(l, b)= (25\fdg25,-1\fdg75)$ denoted by open red squares,
$(25\fdg50,-1\fdg75)$ by filled green squares, 
$(25\fdg25,-2\fdg00)$ by blue asterisks, and
$(25\fdg50,-2\fdg00)$ by open magenta circles.
The closest one is the one denoted by open red squares.
We also plot the distance-reddening relation given by \citet{gre15}.
Marshall et al.'s distance-reddening relation, denoted by open red squares, 
cross consistently our extinction of 
$E(B-V)\approx 0.50$ at $d\approx3.7$~kpc.

Using $E(B-V)=0.50$ and $(m-M)_V=14.4$,
we plot the color-magnitude diagram of V496~Sct in Figure
\ref{hr_diagram_v2491_cyg_v496_sct_ci_aql_u_sco_outburst}(b).
The track of V496~Sct is very similar to that of LV~Vul.
This confirms that our values of $E(B-V)=0.50$ and $(m-M)_V=14.4$
are reasonable.
Therefore, we regard V496~Sct as an LV~Vul type in the color-magnitude
diagram.
The dust blackout started $\sim90$~days after the outburst
at $m_V\approx9.9$ \citep[e.g.,][]{raj12}.
We denote the start of the dust blackout by a large open blue square.
The nebular phase possibly started $\sim130$~days after the
outburst at $m_V\approx11.2$ \citep[e.g.,][]{raj12}.
We specify it by a large open red
square on the track, i.e., $(B-V)_0=-0.35$ and $M_V=-3.26$.

%Fig.74
%\placefigure{hr_diagram_turning_3type_points_no2}

\begin{figure}
%\epsscale{0.25}
\epsscale{0.75}
%\epsscale{0.8}
%%\epsscale{1.0}
%%\epsscale{1.15}
\plotone{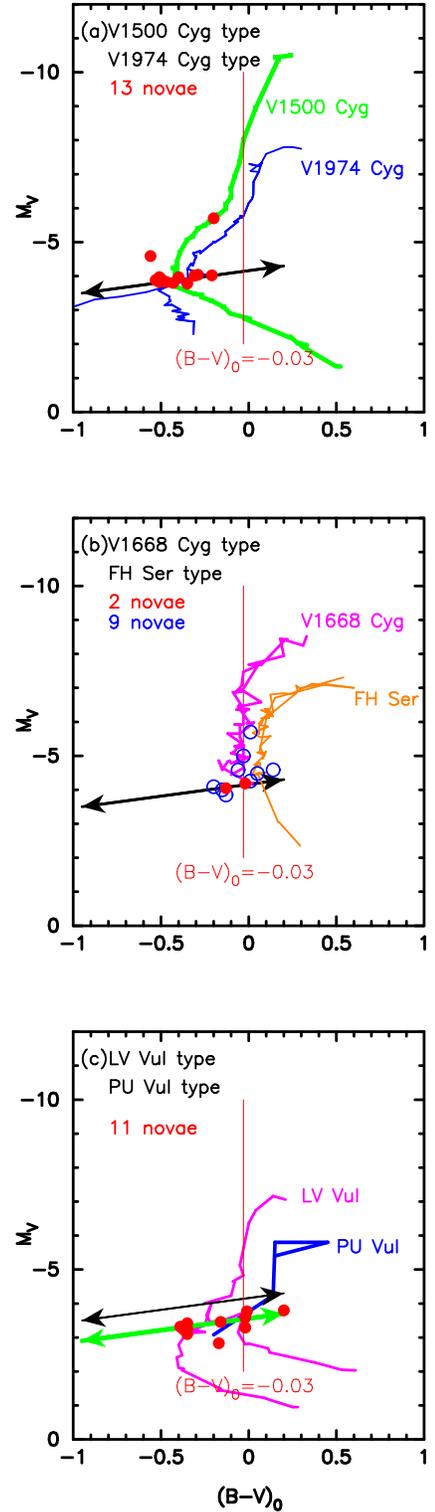}
%\plotone{hr_diagram_turning_3type_points_no2.epsi}
%\plotfiddle{evolution1.ps}{5.0cm}{270}{0.4}{0.4}{-170}{220}
\caption{
Starting points of nebular phase or dust blackout on nova tracks
in the color-magnitude diagram.  
(a) V1500~Cyg (thick solid green line) and V1974~Cyg (thin solid blue line)
types, where the starting points of the nebular phase (filled red circles)
of 13 novae are plotted.
(b) V1668~Cyg (thin solid magenta lines) and FH~Ser (thin solid orange
lines) types, where the starting points of the nebular phase (filled red 
circles) of 2 novae and the starting points of the dust blackout
(open blue circles) of 9 novae are plotted.
(c) LV~Vul (solid magenta line) and PU~Vul (thick solid blue line) types,
where the starting points of the nebular phase (filled red circles)
of 11 novae are plotted.  The data of these points are
tabulated in Table \ref{color_magnitude_turning_point}.
%%We also added the track of PW~Vul (orange thin solid line).  
%Other various tracks and symbols are the same as those in Figures
%\ref{hr_diagram_v1668_cyg_only_outburst},
%\ref{hr_diagram_lv_vul_lv_vul_2fig_outburst},
%\ref{hr_diagram_fh_ser_pw_vul_v1500_cyg_v1974_cyg_outburst}, and
%\ref{hr_diagram_pu_vul_v723_cas_hr_del_v5558_sgr_outburst}.
The two-headed black arrows represent Equations 
(\ref{absolute_magnitude_cusp}), and the lower two-headed green 
arrows in panel (c) denote Equation (\ref{absolute_magnitude_cusp_low_red}).
\label{hr_diagram_turning_3type_points_no2}}
\end{figure}

\section{Discussion}
\label{discussion}
\subsection{Categorization of nova tracks}
\label{six_typical_tracks}
Figure \ref{hr_diagram_6types_novae_one} depicts the color-magnitude
tracks of six templates, from left to right: 
V1500~Cyg (thick solid green), V1668~Cyg (magenta), 
V1974~Cyg (sky blue), LV~Vul (black), FH~Ser (orange), and PU~Vul (blue).
We have collected 40 nova tracks in the color-magnitude diagram
(Figures 
\ref{hr_diagram_v1668_cyg_only_outburst},
\ref{hr_diagram_lv_vul_lv_vul_2fig_outburst},
\ref{hr_diagram_fh_ser_pw_vul_v1500_cyg_v1974_cyg_outburst},
\ref{hr_diagram_pu_vul_v723_cas_hr_del_v5558_sgr_outburst},
\ref{hr_diagram_rs_oph_v446_her_v533_her_t_pyx_outburst},
\ref{hr_diagram_iv_cep_nq_vul_v1370_aql_gq_mus_outburst},
\ref{hr_diagram_qu_vul_os_and_qv_vul_v443_sct_outburst},
\ref{hr_diagram_v1419_aql_v705_cas_v382_vel_v1493_aql_outburst},
\ref{hr_diagram_v1494_aql_v2274_cyg_v2275_cyg_v475_sct},
\ref{hr_diagram_v5114_sgr_v2362_cyg_v1065_cen_v1280_sco_outburst},
\ref{hr_diagram_v2467_cyg_v2615_oph_v458_vul_v2468_cyg_outburst},
and \ref{hr_diagram_v2491_cyg_v496_sct_ci_aql_u_sco_outburst}),
compared them with the template tracks in Figure
\ref{hr_diagram_6types_novae_one}, and categorized
them into one of the six
as listed in Table \ref{color_magnitude_turning_point}.
The V1500~Cyg type includes 8 novae, i.e.,
V1500~Cyg, T~Pyx, GQ~Mus, 
V1494~Aql, V2275~Cyg, V2362~Cyg, V2468~Cyg, and V2491~Cyg,
mainly fast novae. 
The V1974~Cyg type includes 7 novae, i.e., V1974~Cyg, V2467~Cyg, 
V533~Her, QU~Vul, V1493~Aql, V382~Vel, and V5114~Sgr. 
The V1668~Cyg type includes 8 novae, i.e.,
V1668~Cyg, V446~Her, OS~And, V705~Cas, V1419~Aql, V2274~Cyg, V475~Sct,
and V1280~Sco.   
The FH~Ser type includes 4 dust formation novae, i.e.,
FH~Ser, NQ~Vul, QV~Vul, and V2615~Oph.
The LV~Vul type includes 8 novae, i.e., LV~Vul, RS~Oph, PW~Vul, 
IV~Cep, V443~Sct, V1370~Aql, V1065~Cen, and V496~Sct.
The PU~Vul type includes 5 novae, i.e.,
PU~Vul, HR~Del, V723~Cas, V5558~Sgr, and V458~Vul.
Some of these novae show a hybrid signature, e.g., V1065~Cen 
evolved along the path of V1668~Cyg in the early phase
until $M_V \sim -4$ and then followed LV~Vul as shown in Figure
\ref{hr_diagram_v5114_sgr_v2362_cyg_v1065_cen_v1280_sco_outburst}(c).
In this paper, therefore, we put V1065~Cen into the LV~Vul type,
although V1065~Cen underwent a shallow dust blackout like V1668~Cyg.

The different $B-V$ color position of each track in the color-magnitude
diagram can be understood as a difference in the envelope mass.
In general, novae on the bluer side have smaller envelope masses.
For example, the V1500~Cyg and V1974~Cyg types differ in the $B-V$ color
as shown in Figure \ref{hr_diagram_turning_3type_points_no2}(a).
This color difference originates from the difference in the envelope mass,
i.e., V1500~Cyg has a less massive envelope and evolves more rapidly
\citep[e.g., $M_{\rm env}=0.53\times10^{-5}M_\sun$ for a $1.2~M_\sun$ WD model
of V1500~Cyg versus $M_{\rm env}=1.2\times10^{-5}M_\sun$
for a $0.98~M_\sun$ WD model of V1974~Cyg in][]{hac16k}.
A good example of this difference can be seen in the track of V2362~Cyg.
In the early decline phase, V2362~Cyg evolves downward
along the track of FH~Ser until $M_V\sim-4$, as
depicted by filled red circles in Figure
\ref{hr_diagram_v5114_sgr_v2362_cyg_v1065_cen_v1280_sco_outburst}(b).
After that, the $V$ magnitude rises up to the secondary maximum 
($M_V\sim-6$) and sharply drops down to $M_V\sim-2$, as depicted
by open magenta circles.  A large amount of envelope mass was lost
during this secondary maximum \citep{mun08b}.  The position of the secondary 
maximum in the color-magnitude diagram is about 0.2 mag bluer 
than that of the first maximum.  This track shift toward blue 
is caused by the loss of envelope mass. 

The difference between the V1668~Cyg and FH~Ser types shown in Figure
\ref{hr_diagram_turning_3type_points_no2}(b) can also be understood
as a difference in envelope mass.
Both types have a dust formation episode, but the dust blackout
is much shallower in the V1668~Cyg type than in the FH~Ser type.
This indicates that the envelope mass is more massive in the FH~Ser
type novae.

In PW~Vul, a massive envelope could be ejected during the very
early phase as represented by a large clockwise circle in the
color-magnitude diagram \citep[e.g.,][]{kol86}.
After that, it follows a track similar to 
that of LV~Vul in the mid and late decline phase.  
Such a shift toward blue is also observed in the multiple peaks of
V5558~Sgr (Figure 
\ref{hr_diagram_pu_vul_v723_cas_hr_del_v5558_sgr_outburst}(d)).
This also suggests a large decrease in the envelope mass.

\clearpage

\subsection{Absolute magnitude at nebular phase}
\label{absolute_mag_novae}
The color-magnitude diagram (HR diagram) is an excellent tool
for understanding the nature of stars (e.g., their evolution
and absolute magnitudes), but it has not yet been widely applied to
nova outbursts.  In Paper I,
we discussed a new way to estimate the distance to a nova.
The method in Paper I was as follows:
(1) we first determine $E(B-V)$ by fitting the color-color evolution 
of a target nova with the general course of novae,
(2) choose a reference nova with known distance and extinction and
compare it with our target nova,
(3) obtain $(m-M)_V$ by the time-stretching method of light curves,
i.e., using the relation $m'_V = m_V - 2.5 \log f_s$ \citep{hac10k},
and (4) calculate the distance with Equation (\ref{v_distance_modulus}).

Here, we propose another new method based on the type of nova tracks
and the starting points of the nebular phase in the color-magnitude diagram.
Figure \ref{hr_diagram_turning_3type_points_no2}
summarizes the positions of starting points of the nebular phase
or dust blackout of each nova in the color-magnitude diagram.
In the V1500~Cyg and V1974~Cyg type novae, i.e., the V1500~Cyg family,
the gradual change toward blue in the very early phase ($M_V < -6$)
corresponds to continuum free-free emission phase of the nova spectrum.
Then some emission lines become stronger in the $B$ band.  The
contribution of these emission lines causes an excursion toward blue 
between $-6 < M_V < -4$.  After the turning point near $M_V\sim-4$,
it went toward red due to a large contribution of emission lines [\ion{O}{3}]
to the $V$-band.  This turning point (or cusp) can be clearly identified
in many of the V1500~Cyg family novae.
We found that the start of the nebular phase almost coincides
with the turning point (or cusp) and are already showing its position 
as a large red square in each color-magnitude diagram.  
We plot this turning point (or cusp) as an indication of the development
of the nebular stage by filled red circles in 
Figure \ref{hr_diagram_turning_3type_points_no2}(a).
These filled red circles are located on the two-headed black arrow of
Equation (\ref{absolute_magnitude_cusp}) except for two novae,
T~Pyx and GQ~Mus. 

For the V1668~Cyg and FH~Ser types of novae, i.e., the V1668~Cyg family,
they often show a dust blackout at $M_V\sim-4.5$
before the above excursion toward red or blue.  
We plot the starting points of the dust blackout by open blue circles
in Figure \ref{hr_diagram_turning_3type_points_no2}(b).
They are rather scattered.  On the other hand, we plot the starting
points of nebular phase by filled red circles for the two novae
V1668~Cyg and V446~Her.  These points are located on the two-headed
arrow of Equation (\ref{absolute_magnitude_cusp}).
The filled red circles in Figure \ref{hr_diagram_turning_3type_points_no2}(a) and 
\ref{hr_diagram_turning_3type_points_no2}(b) are located on 
the two-headed arrows represented by Equation (\ref{absolute_magnitude_cusp})
except for T~Pyx and GQ~Mus.   Inversely, we obtain 
Equation (\ref{absolute_magnitude_cusp}) from these 13 
red points, excluding T~Pyx and GQ~Mus.

The LV~Vul and PU~Vul type novae, i.e., the LV~Vul family,
have slightly different starting positions
for the nebular phase.  We plot also the starting points 
for the LV~Vul family novae in Figure
\ref{hr_diagram_turning_3type_points_no2}(c) by filled red circles.
They are located on the two-headed green arrow
represented by Equation (\ref{absolute_magnitude_cusp_low_red}), which is
0.6 mag below the two-headed black arrow represented by Equation
(\ref{absolute_magnitude_cusp}).
We obtained this lower line from these 11 red points in Figure
\ref{hr_diagram_turning_3type_points_no2}(c).

In this way, we found the characteristic epochs corresponding
to a certain magnitude given by Equation (\ref{absolute_magnitude_cusp})
or (\ref{absolute_magnitude_cusp_low_red}).
Using this property, 
we are able to obtain the absolute magnitude of the target nova
by placing the starting point of the nebular phase on the line of
Equation (\ref{absolute_magnitude_cusp}) or
Equation (\ref{absolute_magnitude_cusp_low_red}) depending
on the nova type.  This is a new method for determining the absolute
magnitude of a nova.

%Fig.75
%\placefigure{v745_sco_v_bv_ub_color_curve1989_2013}

\begin{figure}
%%\epsscale{0.60}
%\epsscale{0.75}
%\epsscale{1.0}
\epsscale{1.15}
\plotone{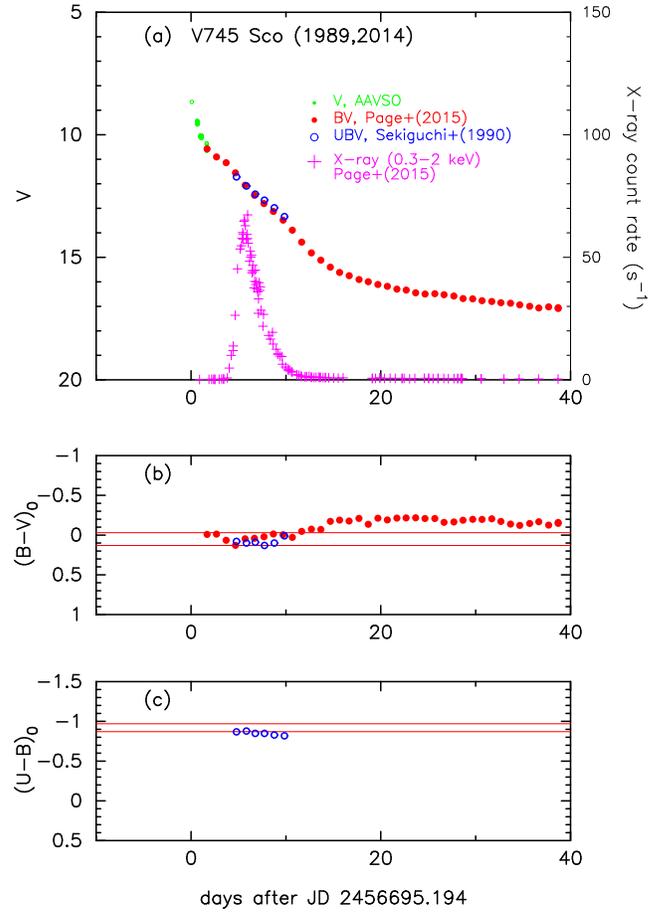}
%\plotone{v745_sco_v_bv_ub_color_curve1989_2013.epsi}
%\plotfiddle{evolution1.ps}{5.0cm}{270}{0.4}{0.4}{-170}{220}
\caption{
Same as Figure \ref{v446_her_v_bv_ub_color_curve}, but for V745~Sco.
We de-reddened $(B-V)_0$ and $(U-B)_0$ colors with $E(B-V)=0.70$.
In panel (b), the two horizontal thin solid red lines denote 
$(B-V)_0=-0.03$ and $+0.13$.  In panel (c), they represent
$(B-V)_0=-0.97$ and $-0.87$.  These correspond to the intrinsic
colors of optically thick free-free ($F_\nu \propto \nu^{2/3}$)
and optically thin free-free ($F_\nu \propto \nu^0$)
emissions \citep{hac14k}, respectively. 
%%See the main text for more detail.
\label{v745_sco_v_bv_ub_color_curve1989_2013}}
\end{figure}

%Fig.76
%\placefigure{hr_diagram_v745_sco_distance_reddening_outburst}

\begin{figure}
%\epsscale{0.50}
\epsscale{1.0}
%\epsscale{1.15}
\plotone{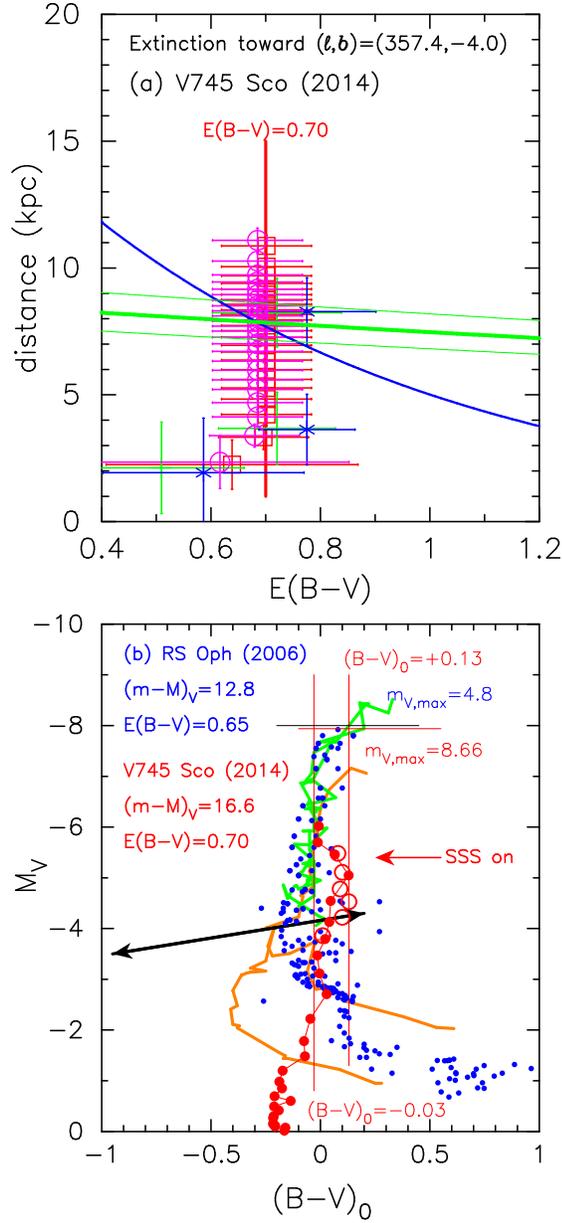}
%\plotone{hr_diagram_v745_sco_distance_reddening_outburst.epsi}
%\plotfiddle{evolution1.ps}{5.0cm}{270}{0.4}{0.4}{-170}{220}
\caption{
(a) Distance-reddening relation for V745~Sco.
A thick solid green line flanked by two thin solid green lines
denotes the distance-reddening relation calculated from
the period-luminosity relation of semi-regular variables,
i.e., Equation (\ref{equation_distmod_IR2}).
A thick solid blue line denotes $(m-M)_V=16.6$.
(b) Color-magnitude diagram of the recurrent nova V745~Sco (2014)
(filled and open red circles) together with the recurrent novae
RS~Oph (2006) (filled blue circles).
The data of V745~Sco are taken from \citet{pag15} (filled red circles)
and \citet{sek90} (open red circles).
%%the archives of the American Association of Variable Star Observers (AAVSO)
%%and Variable Star Observers League in Japan (VSOLJ).
%%%For T~Pyx, we used the 1967 outburst data from \citet{lan70}.
The thick two-headed black arrow indicates the relationship of
Equation (\ref{absolute_magnitude_cusp}).
Thick solid green lines denote the tracks of V1668~Cyg. 
Thick solid orange lines denote that of LV~Vul.
%%Green solid line represent part of the V1500~Cyg track.
%%Blue solid line the V1974~Cyg track.
The start of the supersoft X-ray source phase is denoted by a red arrow
labeled ``SSS on.''   
%%See the main text for more detail.
\label{hr_diagram_v745_sco_distance_reddening_outburst}}
\end{figure}

\subsection{Application to recurrent novae}
\label{hrd_recurrent_novae}
Using the color-magnitude diagram method summarized
above in Section \ref{absolute_mag_novae}, we can estimate the
absolute magnitudes of novae.  This is essentially the same method
as the main-sequence fitting or horizontal-branch fitting in the HR diagram
to estimate the distances to star clusters.  In this subsection, we first
confirm our method by applying it to the M31 1~yr recurrent nova,
M31N 2008-12a, and then estimate the absolute magnitudes of 
three well-observed recurrent novae, CI~Aql, U~Sco, and V745~Sco. 

\subsubsection{M31N 2008-12a}
\label{m31n2008-12a}
The 1~yr recurrence period nova M31N 2008-12a is an excellent
example of a recurrent nova 
because the distance and extinction are well determined
\citep{dar15, hen15, kat15sh}.  The nova has $t_2=1.77$ days and
$t_3=3.84$ days in the $V$ band \citep{dar15}, being a very fast nova.
\citet{kat15sh} concluded, on the basis of their multi-wavelength
light curve model, that the WD mass is close to $1.38~M_\sun$. 
The distance to M31 is $d\approx780$~kpc and the extinction is 
$E(B-V)\approx0.21$ toward the nova \citep[e.g.,][]{hen15}.
Then the distance modulus is calculated to be $(m-M)_V=25.1$. 

Adopting $E(B-V)=0.21$ and $(m-M)_V=25.1$, we plot the color-magnitude
diagram of M31N 2008-12a in Figure 
\ref{hr_diagram_v2491_cyg_v496_sct_ci_aql_u_sco_outburst}(c) and 
\ref{hr_diagram_v2491_cyg_v496_sct_ci_aql_u_sco_outburst}(d)
by open black squares with connected solid black lines.
The data for the 2014 outburst are taken from \citet{dar15} and
for the 2015 outburst from ATel Nos.\
7974, 7965, 7967, 7969, 8033, 8029, and 8038.  
The peak brightness is about $M_V\approx-6.6$,
much fainter than typical classical novae.  \citet{dar15} 
estimated that the peak brightness is $-6.8 < M_V < -6.3$.
The nova almost goes down along the line of $(B-V)_0=-0.03$
in the early phase, then turns to the left (toward blue) in the middle
phase, and comes back to the right in the later phase.
%%% (see Figure \ref{hr_diagram_rs_oph_v446_her_v533_her_t_pyx_outburst}(a)).
The turning point is not clearly identified
but possibly located near the two-headed arrow.
This supports our method of absolute magnitude determination
discussed in the previous subsection 
(Section \ref{absolute_mag_novae}).

\subsubsection{CI~Aql}
CI~Aql is also a recurrent nova with recorded outbursts in
1917, 1941, and 2000.  \citet{sch01c} proposed that CI~Aql could
have an outburst every $\sim20$~yr.
\citet{kis01} estimated the reddening of CI~Aql 
$E(B-V)=0.85\pm0.3$ from an average of various methods.
\citet{hac01ka}, \citet{hac03a},
and \citet{led03} derived $E(B-V)\approx1.0$ based on their 
binary model light curve fittings.  \citet{lyn04} obtained 
$E(B-V)=1.5\pm0.1$ from the \ion{O}{1} line ratios.
\citet{iij12} obtained $E(B-V)=0.92\pm0.15$ from the equivalent
widths of the diffuse interstellar absorption bands.

Figure \ref{hr_diagram_v2491_cyg_v496_sct_ci_aql_u_sco_outburst}(c) shows
the data of CI~Aql (filled red circles connected with a solid red line),
taken from the VSOLJ archive (mainly from Kiyota's data).
\citet{iij12} found that CI~Aql entered the nebular phase
between UT 2000 May 31 and June 8 at $m_V\approx11.8$.
We identify a turning point
(or cusp) at the starting point of the nebular phase,
i.e., at the point $(B-V)_0=-0.52$ and $M_V=-3.84$
(large open red square).
Adopting $E(B-V)=1.0$, we obtain the distance modulus in the $V$ band
from a fit with Equation (\ref{absolute_magnitude_cusp}).
The derived distance modulus is $(m-M)_V=11.9-(-3.8)=15.7$
at the turning point.  
We can regard CI~Aql as a V1500~Cyg type in the color-magnitude diagram,
so the fit with Equation (\ref{absolute_magnitude_cusp}) is justified.
We plot the track of V1500~Cyg by thick solid sky-blue lines
and that of the recurrent nova T~Pyx by green stars in
Figure \ref{hr_diagram_v2491_cyg_v496_sct_ci_aql_u_sco_outburst}(c).
The track of CI~Aql almost follows these trends.
This also confirms that our adopted values of $E(B-V)=1.0$ and 
$(m-M)_V=15.7$ are reasonable.

The distance is then calculated to be $d\approx3.3$~kpc.  This value
is shorter than the $\sim 5.0$~kpc obtained by \citet{sch10a}
from the MMRD relation \citep{dow00}, but longer than the value of 
$d\sim1.5$~kpc estimated from the blackbody binary model light curves
\citep{hac01ka,hac03a,led03} together with $E(B-V)=1.0$.
In general, one should not use the MMRD relations to obtain 
the maximum magnitudes of recurrent novae because they are 
faint objects as already discussed
in our previous papers \citep{hac10k, hac14k, hac15k, hac16k}.
\citet{sah13} obtained $d=1.3\pm0.2$~kpc from the spectral type
(F8IV) of the companion star together with Lynch et al.'s 
estimate of $A_V=4.6\pm0.5$.  If we adopt $E(B-V)=1.0$ ($A_V=3.1$),
Sahman et al.'s distance could be larger than $d\gtrsim2.6$~kpc.

The set $d=3.3$~kpc and $E(B-V)=1.0$ is consistent with
the distance-reddening relations for CI~Aql, 
$(l,b)=(31\fdg6876,-0\fdg8120)$, in Figure
\ref{distance_reddening_v2468_cyg_v2491_cyg_v496_sct_ci_aql}(d).
Here, we plot $(m-M)_V=15.7$, $E(B-V)=1.0$, and
four nearby distance-reddening relations given by \citet{mar06},
i.e., $(l, b)= (31\fdg50,-0\fdg75)$ denoted by open red squares,
$(31\fdg75,-0\fdg75)$ by filled green squares, 
$(31\fdg50,-1\fdg00)$ by blue asterisks, and
$(31\fdg75,-1\fdg00)$ by open magenta circles.
The closest one is that of the filled green squares.
We also add the relation (solid black line) given by \citet{gre15}.
These trends cross consistently at $d\approx3.3$~kpc and $E(B-V)\approx1.0$.

To summarize, the color-magnitude track of CI~Aql
almost overlaps with the tracks of T~Pyx and V1500~Cyg and
its distance-reddening relation is consistent
with the Marshall et al. relation.  Thus, we may conclude that
the set of $E(B-V)=1.0$ and $(m-M)_V=15.7$ ($d\approx3.3$~kpc)
is reasonable.  We summarize the results in Table
\ref{physical_properties_recurrent_novae}.

\subsubsection{U~Sco}
     U~Sco is also a recurrent nova with ten recorded outbursts
in 1863, 1906, 1917, 1936, 1945, 1969, 1979, 1987, 1999, and 2010,
almost every ten years \citep{sch10a}.
It is located at a high galactic latitude, 
$(l,b)=(357\fdg6686,+21\fdg8686)$.  %%% 357.6686 +21.8686
Thus, it is far from the galactic plane ($z > 1$~kpc)
if its distance is large enough ($d\gtrsim3$~kpc),
as suggested in the literature.
Therefore, the extinction for U~Sco should be close to
the galactic dust extinction.  The NASA/IPAC galactic dust absorption
map gives $E(B-V)=0.32\pm0.04$ for U~Sco.  Unfortunately,
direct measurements of reddening show a scatter between 0.1 and 0.35
\citep[e.g.,][]{sch10a}.
For example, \citet{bar81} obtained two different absorptions
toward U Sco in the 1979 outburst,
$E(B-V)\sim0.2$ from the line ratio of \ion{He}{2}
at an early phase of the outburst ($\sim12$ days after maximum)
and $E(B-V)\sim0.35$ from the Balmer line ratio at a late phase
of the outburst ($\sim 33$---34 days after maximum).
The latter value is consistent with that of the galactic dust absorption map.  

Adopting $E(B-V)=0.35$, we plot the color-color diagram of the U~Sco
2010 outburst in Figure 
\ref{color_color_diagram_v1419_aql_v705_cas_u_sco_v745_sco_no2}(c).
The $B-V$ and $U-B$ color data are taken from \citet{pagnotta15}.
The track of U~Sco almost overlaps with the general track of novae
(solid green lines) in the color-color diagram.
Thus, we confidently determine the reddening of U~Sco as
$E(B-V)=0.35\pm0.05$.  

Figure \ref{hr_diagram_v2491_cyg_v496_sct_ci_aql_u_sco_outburst}(d)
shows the color-magnitude diagram of U~Sco.  Here we adopt
$E(B-V)=0.35$ and then determine the distance modulus in the $V$ band
from the fit with Equation (\ref{absolute_magnitude_cusp}).
We identify a turning point (or cusp) at the point $(B-V)_0=-0.64$
and $M_V=-3.72$ (large open red square) as shown in 
Figure \ref{hr_diagram_v2491_cyg_v496_sct_ci_aql_u_sco_outburst}(d).
The distance modulus is then calculated to be $(m-M)_V=12.3-(-3.7)=16.0$
at the turning point (large open red square).
Then the peak is as bright as $M_{V,\rm max}=7.5-16.0=-8.5$,
about 0.5 mag brighter than the peaks of RS~Oph and T~Pyx (see Figure
\ref{hr_diagram_rs_oph_v446_her_v533_her_t_pyx_outburst}(a) and 
\ref{hr_diagram_rs_oph_v446_her_v533_her_t_pyx_outburst}(d)),
and 1.5 mag brighter than the peak of CI~Aql (Figure
\ref{hr_diagram_v2491_cyg_v496_sct_ci_aql_u_sco_outburst}(c)). 
After the turning point (large open red square), U~Sco goes down further,
crossing the two-headed arrow, and jumps 
to the left (toward blue) up to $(B-V)_0\sim-1.0$.
We neglect this bluest point in our determination of the 
turning point, because it could be affected by the first flare
in the 2010 outburst \citep[e.g.,][]{scheafer11, max12, anu13, pagnotta15}.
Below $M_V > -3.0$, we suppose that the $B$ and $V$ magnitudes are affected
by a large irradiated disk around the WD \citep[e.g.,][]{hkkm00}
and, as a result, the $B-V$ color is 
contaminated by the disk radiation.  This is the reason
why the color-magnitude track of U~Sco stays at $(B-V)_0\approx-0.1$
between $-2.0 \lesssim M_V \lesssim -0.0$.
We added the epoch when the SSS phase started
at $m_V=14.0$, about 14 days after the outburst \citep{sch11b}.

The distance to U~Sco is calculated to be $d=9.6\pm0.6$~kpc.
This is larger than the $d=6$--7~kpc calculated
from the nova explosion/quiescent models of \citet{hkkm00, hkkmn00}.
Assuming totality at mid-eclipse,
\citet{sch10a} proposed a larger value of $d=12\pm2$~kpc
estimated from the spectral type (G5IV) of the companion star.
We should note that a different spectral type (K2IV)
of the companion was proposed by \citet{anu00}. 
\citet{anu13} revised their spectral type to K0$\sim$K1.
\citet{mas12} proposed that the spectral type of the secondary star is
not earlier than F3 and not later than G, but concluded that they cannot  
confidently exclude an early K spectral type for U~Sco.  Thus,
we must be careful with the distance determination from only
the spectral type of the secondary.  
We summarize our results in Table
\ref{physical_properties_recurrent_novae}.

We added the $B-V$ and $U-B$ color evolution of T~Pyx (1966) in Figure
\ref{color_color_diagram_v1419_aql_v705_cas_u_sco_v745_sco_no2}(c)
by filled blue circles. The data are the same as
those in Figure 30(a) of Paper I.  
The color-color evolutions of U~Sco (2010) and T~Pyx (1966)
overlap each other.  
%%This also supports that $E(B-V)=0.35$ is reasonable for U~Sco.  
The color-magnitude track of U~Sco (filled red circles connected
by solid red line) in Figure 
\ref{hr_diagram_v2491_cyg_v496_sct_ci_aql_u_sco_outburst}(d)
follows that of T~Pyx (green stars) in the early phase.
The color-color and color-magnitude evolutions of U~Sco are
very similar to those of T~Pyx.

%Table 3
%\placetable{physical_properties_recurrent_novae}

\begin{deluxetable*}{llcccll}
%\begin{deluxetable}{llcccll}
\tabletypesize{\scriptsize}
\tablecaption{Physical properties of selected recurrent novae
\label{physical_properties_recurrent_novae}}
\tablewidth{0pt}
\tablehead{
\colhead{Object} & \colhead{$E(B-V)$} & \colhead{$(m-M)_V$} 
& \colhead{Distance} & \colhead{$M_{V,\rm max}$} & \colhead{$P_{\rm orb}$} 
& \colhead{RN type\tablenotemark{a}} \\
\colhead{} & \colhead{} & \colhead{} 
& \colhead{(kpc)} & \colhead{} & \colhead{(day)} 
& \colhead{} 
} 
\startdata
RS~Oph   & $0.65$ & $12.8$ & $1.4$ & $-8.0$ & $457$\tablenotemark{b} & RS~Oph \\
T~Pyx   & $0.25$  & $14.2$ & $4.8$\tablenotemark{c} & $-7.9$ & $0.076$\tablenotemark{d} & T~Pyx \\
CI~Aql   & $1.0$  & $15.7$ & $3.3$ & $-6.9$ & $0.618$\tablenotemark{e} & U~Sco  \\
U~Sco    & $0.35$ & $16.0$ & $9.6$ & $-8.5$ & $1.23$\tablenotemark{f} & U~Sco \\
V745~Sco & $0.70$ & $16.6$ & $7.8$ & $-7.9$ & --- & RS~Oph 
\enddata
\tablenotetext{a}{Recurrent novae (RNe) are divided into three types, 
depending on the nature of the companion (or the orbital period), i.e., 
T~Pyx ($P_{\rm orb}\sim$ a few hours), 
U~Sco ($P_{\rm orb}\sim$ a day), and RS~Oph ($P_{\rm orb}\sim$ a few to
several hundred days) types \citep[e.g.,][]{anu08}.} 
\tablenotetext{b}{taken from \citet{fek00}.} 
\tablenotetext{c}{taken from \citet{sok13}.} 
\tablenotetext{d}{taken from \citet{uth10}.} 
\tablenotetext{e}{taken from \citet{men95}.} 
\tablenotetext{f}{taken from \citet{sch95r}.} 
%\end{deluxetable}
\end{deluxetable*}

\subsubsection{V745~Sco}
V745~Sco is also a recurrent nova with three recorded outbursts,
in 1937, 1989, and 2014.
Figure \ref{v745_sco_v_bv_ub_color_curve1989_2013} shows the $V$ light
curve, supersoft X-ray count rate,
$(B-V)_0$, and $(U-B)_0$ color evolutions of V745~Sco
for the 1989 and 2014 outbursts.  The $UBV$ data of the 1989 outburst
are taken from \citet{sek90} and the $BV$ and X-ray data of the 2014
outburst are from \citet{pag15}.  If the 2014 outburst is the same as
the 1989 outburst, the $V$ magnitude (open blue circles)
of \citet{sek90} are systematically 0.3 mag brighter than those 
(filled red circles) of \citet{pag15}.  Therefore, we shift the 
$V$ magnitudes down by 0.3 mag and confirm that these two
$V$ light curves overlap each other as shown in 
Figure \ref{v745_sco_v_bv_ub_color_curve1989_2013}(a).
Such a difference among observers was already discussed in our
previous sections and attributed to slightly different responses
of the $V$ filters.
When Sekiguchi et al.\ started $UBV$ photometry about four days
after the outburst; the supersoft X-ray source (SSS) phase just began,
indicating that the ejecta had already become optically thin.
The spectra of the 1989 outburst show that the nebular [\ion{O}{3}]
lines had already developed at this time \citep[e.g.,,][]{due89}.
If we shift Sekiguchi et al.'s $V$ down by 0.3 mag, then Sekiguchi et al.'s 
$(B-V)_0$ data are consistently overlapping those of Page et al.\
as shown in Figure \ref{v745_sco_v_bv_ub_color_curve1989_2013}(b).
The 2014 outburst shows $t_2=2$ and $t_3=4$~days \citep[e.g.,][]{ori14}.

The galactic coordinates of V745~Sco is $(l,b)=(357\fdg3584, -3\fdg9991)$.
The NASA/IPAC galactic dust absorption
map gives $E(B-V)=0.71\pm0.02$ for V745~Sco.  
\citet{ban14} suggested that the extinction for V745~Sco is
$E(B-V)=0.70$ on the basis of galactic dust extinction of \citet{schl11}
and \citet{mar06}. 
\citet{ori14} fitted the X-ray spectrum 10 days after the discovery of
the 2014 outburst with a model spectrum and obtained the hydrogen
column density of $N_{\rm H}=(6.9\pm0.9) \times 10^{21}$ cm$^{-2}$.
They suggested an extinction of $E(B-V)\approx1.0$ from
the relationship $E(B-V)=N_{\rm H}/6.8 \times 10^{21}$ cm$^{-2}$
\citep{gue09}.  However, if we use the relation $E(B-V)= N_{\rm H}/ 
8.3 \times 10^{21}$ proposed by \citet{lis14}, we obtain $E(B-V)=0.83\pm0.1$.
Thus, we adopt $E(B-V)=0.70\pm0.1$, mainly from the results of
\citet{ban14} and the NASA/IPAC galactic dust absorption map.

Adopting $E(B-V)=0.70$, we plot the color-color diagram of V745~Sco in 
Figure \ref{color_color_diagram_v1419_aql_v705_cas_u_sco_v745_sco_no2}(d).
There are only six data (open magenta circles)
but five of the six data are located at or near
the point $(B-V)_0=+0.13$ and $(U-B)_0=-0.87$ denoted by
the open black square labeled ``free-free ($F_\nu\propto \nu^0$).''
This point corresponds to the position of optically thin free-free
emission (Paper I).  These positions of the data in the color-color
diagram are consistent with the fact that the ejecta had already been
optically thin when these data were obtained.  This supports our value of
$E(B-V)=0.70$.

Next we estimate the distance to V745~Sco. 
The orbital period of $P_{\rm orb}=510\pm20$~days proposed by
\citet{scha09} is not confirmed by \citet{mro14}, so we do not use
the method of distance estimate proposed by \citet{scha09} based on 
the Roche lobe size.  \citet{mro14} detected semi-regular pulsations of
the red giant companion (with periods of 136.5~days and 77.4~days).  
Thus, we use a distance estimating method for the pulsating red giant
companion. There is a well-known relation between the pulsation period
and its luminosity for Mira variables, which is applicable also to
semi-regular variables pulsating in the fundamental mode,
\begin{equation}
M_K= -3.51 \times (\log P ({\rm day})- 2.38)-7.25,
\label{equation_MkP}
\end{equation}
with an error of $\sim$ 0.2 mag \citep{whitelock08}.
For the fundamental 136.5 day pulsation, we get the absolute $K$ magnitude
of $M_K=-6.39 \pm 0.2$.
The average $K$ mag is $m_K=8.33$~mag \citep{hoa02}, so we have
\begin{equation}
(m - M)_K= 0.353 \times E(B-V) + 5 \log ~(d/1~{\rm kpc}) +10,
\label{equation_distmod_IR}
\end{equation}
where we adopt the reddening law of $A_{\rm K}=0.353 \times E(B-V)$ 
\citep{cmm89}.
We plot the distance-reddening relation (thick solid green line flanked
with thin solid green lines), 
\begin{equation}
14.72\pm0.2= 0.353 \times E(B-V) + 5 \log ~(d/1~{\rm kpc}) +10,
\label{equation_distmod_IR2}
\end{equation}
in Figure  
\ref{hr_diagram_v745_sco_distance_reddening_outburst}(a).
For a particular value of $E(B-V)=0.70$, we get $d=7.8\pm0.8$ kpc.
Then we have $(m-M)_V=16.6$ for the distance modulus in the $V$ band.

Using $E(B-V)=0.70$ and $(m-M)_V=16.6$, 
we plot the color-magnitude diagram of V745~Sco in Figure
\ref{hr_diagram_v745_sco_distance_reddening_outburst}(b).
The $BV$ data of V745~Sco (2014) are taken from \citet{pag15} 
(filled red circles connected with a red line),
while the $BV$ data of the 1989 outburst
are from \citet{sek90} (open red circles).
The track of V745~Sco follows the line of $(B-V)_0=-0.03$ in the
early phase (top two data points), being consistent with optically thick
free-free emission ($F_\nu\propto \nu^{2/3}$).  Then, it jumps to
$(B-V)_0=+0.13$ about four days after the outburst, being consistent
with the optically thin free-free emission ($F_\nu\propto \nu^0$).
The SSS phase started four days after the outburst and 
the ejecta had already become optically thin at this time.
After that, the track almost follows that of RS~Oph 
until $M_V \sim -2.5$.  This agreement supports our adopted values of
$E(B-V)=0.70$ and $(m-M)_V=16.6$.
We regard V745~Sco as an LV~Vul type in the color-magnitude diagram.
We summarize the results in Table \ref{physical_properties_recurrent_novae}.


%% Included in this acknowledgments section are examples of the
%% AASTeX hypertext markup commands. Use \url without the optional [HREF]
%% argument when you want to print the url directly in the text. Otherwise,
%% use either \url or \anchor, with the HREF as the first argument and the
%% text to be printed in the second.

\section{Conclusions}
\label{conclusions}
In this series of papers, 
we have extensively examined $UBV$ color-color and color-magnitude
evolutions of classical novae and found several important properties.
In Paper I, we discussed the color-color evolution of novae and 
found the general track of novae in the color-color evolutions.
Thus, we developed a way to determine the color excess $E(B-V)$ of
a target nova by fitting its color evolution track with the general
course in the color-color diagram.  Using this new and convenient
method, we redetermined the color excesses of 27 novae.
In the present paper, we have focused on the color-magnitude diagram
and identified some general trends in the color-magnitude evolutions of novae.
Our results are summarized as follows:\\

\noindent
{\bf 1.} We redetermined the color excesses of novae, mainly by the method of
the general track in the color-color diagram and partly by other methods
when sufficient $UBV$ data were not available.  \\

\noindent
{\bf 2.} Using the time-stretching method of nova light curves \citep{hac10k},
we estimated the distance modulus $(m-M)_V$ of a target nova.
Thus, we determined the distance moduli and color excesses of 
40 novae and plotted their tracks in the color-magnitude diagram.\\

\noindent
{\bf 3.} We reduced 10 well-observed nova tracks into the six template tracks
of V1500~Cyg, V1668~Cyg, V1974~Cyg, LV~Vul, FH~Ser, and PU~Vul
in the color-magnitude diagram (Figure \ref{hr_diagram_6types_novae_one}).
These six template tracks are almost parallel
and spread out from bluer to redder
in the order of nova speed class (from faster to slower).
In other words, the order indicates the envelope mass (from less to more 
massive).  A redder nova has a more massive envelope and 
belongs to a slower speed class.
Based on these six template tracks, we categorized our 40 novae into
six types as listed in Table \ref{color_magnitude_turning_point}.\\

\noindent
{\bf 4.} We found a turning/cusp point, which corresponds to the onset
of the nebular phase, on the track in the color-magnitude diagram.  
Such a turning point appears when the absolute $V$ magnitude fades to 
$M_V \sim -4$ (V1500~Cyg and V1974~Cyg types) or slightly below
(LV~Vul and PU~Vul types).  
The positions of these points are expressed 
by Equation (\ref{absolute_magnitude_cusp}) with a standard deviation
of $\sim0.1$ mag for the V1500~Cyg and V1974~Cyg types 
and the V1668~Cyg and FH~Ser types, or 
by Equation (\ref{absolute_magnitude_cusp_low_red}) with a standard deviation
of $\sim0.2$ mag for the LV~Vul and PU~Vul types.

\noindent
{\bf 5.}  Using this property, 
we can estimate the distance modulus of a nova by placing
its turning/cusp point on the line of Equation 
(\ref{absolute_magnitude_cusp}) or (\ref{absolute_magnitude_cusp_low_red})
depending on the type. 
This is a new method for obtaining the absolute magnitudes of novae.\\

\noindent
{\bf 6.}  We applied this method to three recurrent novae and
redetermined their color excesses and absolute magnitudes.

\acknowledgments
     We are grateful to T. Kato for sending us the VSNET data for 
V1494~Aql 1999\#2, to K. Page for sending us the data for V745~Sco (2014),
to S. Shugarov for providing us with the data of PU~Vul,
to U. Munari for sending us the data of V2362~Cyg,
and to the late A. Cassatella for providing us with  
UV 1455 \AA~data for {\it IUE} novae.
     We also thank
the American Association of Variable Star Observers
(AAVSO) and the Variable Star Observers League of Japan (VSOLJ)
for the archival data of various novae.
We are also grateful to the anonymous referee for useful comments
regarding how to improve the manuscript.
This research has been supported in part by Grants-in-Aid for
Scientific Research (24540227, 15K05026) 
from the Japan Society for the Promotion of Science.

\end{document}